\documentclass[11pt,twoside, openright]{report}

\usepackage{epsfig}
\usepackage{graphicx}
\usepackage{rotating}
\usepackage{./titlepage}
\usepackage{caption}
\usepackage{morefloats}
\usepackage[inner=3.5cm, outer=2.5cm, bottom=2.5cm, top=2.5cm, twoside]{geometry}

\usepackage{fancyhdr}

\begin{document}
\begin{normalsize}
\title{Design studies for a multi-TeV $\gamma$-ray telescope array : PeX (PeV eXplorer)}
\author{Jarrad Denman}
\date{July 16, 2012} 
\maketitle


\addcontentsline{toc}{chapter}{\textbf{Contents}}
\tableofcontents

\begin{center}
\chapter*{\textbf{Abstract}}
\end{center}
\addcontentsline{toc}{chapter}{\textbf{Abstract}}
\linespread{1.2}
This thesis presents work towards the design of a new array of Image Atmospheric Cherenkov Telescopes (IACTs) to detect multi-TeV (E $>$ 10$^{12}$ eV) $\gamma$-ray sources. The array consists of 5 telescopes in a square layout with one central telescope, known as the \textit{Pevatron eXplorer} or \textit{PeX}. PeX is a PeV (10$^{15}$ eV) cosmic ray explorer that aims to study and discover $\gamma$-ray sources in the 1 to 500 TeV range. The initial PeX design has been influenced by the HEGRA CT-System and H.E.S.S. configurations. One important feature of multi-TeV air showers is their ability to trigger telescopes at large core distance ($>$ 400 \rm{m}). PeX will utilise large core distance events to improve the performance and illustrate the viability of a sparse array for multi-TeV $\gamma$-ray astronomy.

In Chapter 1, I will discuss the astrophysical motivation behind multi-TeV observations. A number of $\gamma$-ray sources have shown emission that extends above 10 TeV, for example unidentified source HESS J1908-063. A new multi-TeV detector can provide a new look at the Galactic plane and work towards uncovering the origin of Galactic cosmic ray acceleration.

In Chapter 2, I will look at the physics of air showers, which involves the interaction of protons and $\gamma$-rays with the atmosphere to form a cascade of particles. I will discuss the lateral distribution for $\gamma$-rays and show the importance of large core distance shower for multi-TeV events. Gamma-ray showers with an image \textit{size} $>$ 60\textit{pe} can be detected up to 700 \rm{m} away from PeX for 500 TeV showers.

In Chapter 3, I introduce PeX in detail along with the simulation programs used to model it. I discuss the standard shower reconstruction algorithm (Algorithm 1) and an advanced shower reconstruction algorithm (Algorithm 3). I also introduce the image parameters that I will investigate while optimising PeX, which include; site altitude, image triggering conditions, image cleaning conditions, telescope separation and image \textit{size} cut.

In Chapter 4, I have optimised the PeX cell for a low altitude (0.22 \rm{km}) observational site using Algorithm 1. Parameters such as telescope separation, triggering combination, cleaning combination and image \textit{size} cut have been varied over a range of values to provide the optimum results for PeX. 

In Chapter 5, I have optimised the PeX cell for a higher altitude (1.8 \rm{km}) observational site using Algorithm 1. The same parameter variations considered in Chapter 4 have been used in Chapter 5. It appears that scaling the H.E.S.S. values to appropriate values for PeX provides the near optimum results. A comparison between the site altitudes suggests that a 0.22 \rm{km} altitude provides the slightly better performance for energy $>$ 10 TeV.

In Chapter 6, a new time cleaning cut has been investigated. The arrival time between photons in two adjacent pixels in the camera is used to apply an extra cut which helps mitigate night sky background. To illustrate the robustness of the time cleaning cut, various level of night sky background have been considered. These levels include: off-Galactic plane, on-Galactic plane and towards the Galactic centre. The most important result is that PeX performance with a time cleaning cut improves results when a high level of night sky background is present. For a Galactic centre level of night sky background there is a factor of 1.5 improvement in angular resolution, effective area and quality factor when a time cleaning cut is applied compared to using no time cleaning cut. 

In Chapter 7, Algorithm 3 has been considered. A smaller sample of parameter variations has been simulated to confirm that the same trends found in Chapters 4 and 5 appear for Algorithm 3. The site altitude and time cleaning cut have also been considered. Algorithm 3 provides a direction reconstruction improvement over Algorithm 1 especially for large core distance events which are important for PeX.

In Chapter 8, I consider some possible enhancements to PeX. These enhancements include: varying pixel size and pixel arrangement in the camera, further cuts to rejection proton events and possible separation between proton and $\gamma$-ray pulses. Chapter 8 also provides the flux sensitivity results for multiple PeX configurations. The final configuration and flux sensitivity for PeX is presented in this Chapter. This work shows the value of a sparse array of Cherenkov telescopes to open up the $>$ 10 TeV energy regime.

\begin{center}
\chapter*{\textbf{Declaration of Originality}}
\end{center}
\addcontentsline{toc}{chapter}{\textbf{Declaration of Originality}}
\linespread{1.2}

I, Jarrad Denman certify that this work contains no material which has been accepted for the award of any other degree or diploma in any university or other tertiary institution and, to the best of my knowledge and belief, contains no material previously published or written by another person, except where due reference has been made in the text. I give consent to this copy of my thesis, when deposited in the University Library, being made available for loan and photocopying, subject to the provisions of the Copyright Act 1968. The author acknowledges that copyright of published works contained within this thesis (as listed below*) resides with the copyright holder(s) of those words. I also give permission for the digital version of my thesis to be made available on the web, via the University's digital research repository, the Library catalogue and also through web search engines, unless permission has been granted by the University to restrict access for a period of time.\\

\noindent{*Published works contained within this thesis:}\\

\noindent{Denman, J., Rowell, G., Stamatescu, V., Thornton, G., Dunbar, R., Clay, R., Dawson, B., Smith, A., Wild, N., Protheroe, R., 2008, American Institute of Physics (AIP) Conference Proceedings, Vol. 1085, p838}\\

\noindent{Rowell, G., Denman, J., Stamatescu, V., Thornton, G., Dunbar, R., Clay, R., Dawson, B., Smith, A., Wild, N., Protheroe, R., \textit{In preparation}}\\

\noindent{\bfseries\large Jarrad Denman\hfill 13th January, 2013}

\begin{center}
\chapter*{\textbf{Acknowledgements}}
\end{center}
\addcontentsline{toc}{chapter}{\textbf{Acknowledgements}}

\linespread{1.2}
Firstly I would like to thank my supervisors Dr. Gavin Rowell and Prof. Bruce Dawson for all their support, guidance and feedback over the last few years. I appreciate all the time my supervisors and Prof. Roger Clay have spent reading and correcting my thesis. I would also like the thank the rest of the \textit{TenTen}/\textit{PeX} collaboration: Dr. Greg Thornton, Dr. Andrew Smith, Tristan Sudholz, Neville Wild and Assoc. Prof. Ray Protheroe for all their helpful suggestions and comments. A special thanks to Dr. Victor Stamatescu for all the discussions we have had about different ideas and reconstruction algorithms.\\

I would like the thank my office-mate Dr. Vanessa Holmes for all her support and for putting up with me for all those years. I would like to thank the University of Adelaide and the School of Chemistry and Physics for the resources provided and Ramona Adorjan for the computer assistance. I would like to thank eResearch SA for the use of the super computer, which allowed me to complete my research. A special thank you to Dr. Ben Whelan, Dr. Brent Nicholas, Nigel Maxted and the rest of my friends who have provided some much needed social entertainment during my PhD.\\

Finally I would really like to thank my parents Helen and Peter, and my brother Christien for all the support they have given me during my years of studying. I could not have done this without them.\\

\begin{center}
\chapter*{\textbf{Public Display of Results}}
\end{center}
\addcontentsline{toc}{chapter}{\textbf{Public Display of Results}}

Here I summarise the contributions of conference posters, conference proceedings and published works of which I have been a part.\\

\begin{itemize}
\item \textbf{Conference Poster}:\\
 \textit{Optimising parameters for a multi-TeV IACT cell}

 4$^{th}$ Heidelberg International Symposium on High Energy Gamma-Ray Astronomy, Heidelberg, Germany (2008).

\textbf{$http://www.mpi-hd.mpg.de/hd2008/pages/news.php$}\\

\item \textbf{Conference Proceedings}:\\
 \textit{Optimising parameters for a multi-TeV IACT cell}

 Denman, J., Rowell, G., Stamatescu, V., Thornton, G., Dunbar, R., Clay, R., Dawson, B., Smith, A., Wild, N., Protheroe, R., American Institute of Physics (AIP) Conference Proceedings, Vol. 1085, p838 (2008).\\
 
 \item \textbf{Conference Poster}:\\
 \textit{Design studies for a new multi-TeV gamma-ray telescope array}

 Astronomical Society of Australia - Annual Scientifc Meeting, University of Melbourne, Melbourne, Australia (2009).

 \textbf{$http://asa2009.science.unimelb.edu.au/Site/Home_Page.html$}\\

\item \textbf{Conference Poster}:\\
 \textit{PeV explorer (PeX): A new multi-TeV gamma-ray telescope}

 Astronomical Society of Australia - Annual Scientifc Meeting, University of Adelaide, Adelaide, Australia (2010).

\textbf{$http://www.physics.adelaide.edu.au/astrophysics/asa2011/$}\\

\item \textbf{In Preparation}:\\
 Rowell, G.,  Denman, J., Stamatescu, V., Thornton, G., Dunbar, R., Clay, R., Dawson, B., Smith, A., Wild, N., Protheroe, R.\\
\end{itemize}

\newpage

\pagenumbering{arabic}

\fancyhead[L]{}
\pagestyle{fancy}
\renewcommand{\chaptermark}[1]{
\markboth{\chaptername\
\thechapter.\ #1}{}}

\chapter{TeV Gamma-Ray Astronomy and its Motivation}

	It was discovered over 100 years ago that a mysterious phenomenon caused the air in an electroscope to become electrically charged or ionized \cite{Pacini}. The initial idea was that the radiation from materials in the ground must cause this ionization. Therefore, it was predicted that at high altitudes the ionization level should drop. Multiple ionization experiments were conducted at different altitudes and the results indicated that the ionization increased as the altitude increased \cite{Pacini}. It was concluded that the radiation must be coming from above based on these results. To prove the concept, in 1912 Victor Hess conducted a number of balloon ascents which proved that the rate of ionization in the atmosphere increased with altitude. At an altitude of 5000 \rm{m}, the ionization rate was twice as fast as that seen at sea level \cite{HESS}. His conclusion was \cite{HESS}; \begin{quotation} \textit{The results of my observations are best explained by the assumption that a radiation of very great penetrating power enters our atmosphere from above.} \end{quotation} It was not until 10 years later that R. Millikan proposed the name \textit{Cosmic Rays}, due to the fact that the penetrating radiation was coming from extraterrestrial origins. Victor Hess solved the mystery of where the penetrating radiation was coming from but at the same time he created a number of new questions; what are cosmic rays? what extraterrestrial origins do cosmic rays come from? and how do they gain so much energy?

	Today, cosmic rays are known to be relativistic charged subatomic particles; protons, electrons and heavy nuclei, believed to originate from Galactic and extragalactic sources. The composition of cosmic rays is $98\%$ hadronic ($85\%$ protons, $12\%$ alpha particles, and $3\%$ heavier nuclei (all remaining elements)), and $2\%$ electrons \cite{CRtalk}. However, the exact origin of cosmic rays is unknown even after 100 years of ongoing research in the field. As well as the unknown origins, the mechanisms which accelerate cosmic rays up to the highest energies are uncertain. Theoretical studies of diffusive shock acceleration, suggest that Galactic particles could be accelerated up to 10$^{15}$ eV but for higher energies, neither the acceleration mechanisms nor the acceleration sites are clearly understood \cite{DSAenergy}. Since the energy range of the incoming flux of particles extends from $10^{9} $ eV to $10^{20}$ eV (Figure~\ref{fig:Screenshot}) and diffusive shock theory can only account for some of the acceleration of the particles in this range, it suggests that some cosmic ray acceleration mechanisms and origins are still a mystery. These unknowns form some of the main motivations for studying cosmic rays. Understanding the acceleration mechanisms and the origin of cosmic rays would provide valuable information about Galactic and extragalactic sources, the intergalactic and extragalactic magnetic field strengths and the composition of particles in the Universe.


	The motivation comes from the origin of cosmic rays and the shape of the spectrum \cite{spectrum}, which follows a general power law (Figure~\ref{fig:Screenshot}). The basic power-law is expressed as $I(E)= C E^{-\Gamma}$ where $\Gamma$ is the spectral index, E is energy, C is a normalisation constant and I(E) is the flux in ph (cm$^{2}$ s TeV)$^{-1}$ of the incoming particles with energy. Within the cosmic ray spectrum there are three distinguishing features where the spectral index changes from a pure power law. These features fuel the investigation into cosmic rays and their origins. 
For energies $10^{9}$ eV to $10^{11}$ eV the energy spectrum turns over due to the shielding effect of the Sun's magnetosphere. The turn over changes due to the current activity of the Sun. Above $10^{11}$ eV, the spectrum follows a power law with index $\Gamma \approx 2.7$. At $4 \times 10^{15}$ eV there is the first \textit{knee}, at which the spectral index changes from $\Gamma = 2.7$ to $3.0$. The second \textit{knee} appears at $4 \times 10^{17}$ eV and the spectrum steepens to $\Gamma = 3.3$ \cite{spectrum}. The first \textit{knee} is believed to be the change over in cosmic ray composition, from protons to heavier nuclei such as iron. The origins of cosmic rays which produce the second \textit{knee} are still under investigation. The last feature in the cosmic ray spectrum is the ankle which occurs at  $10^{18.6}$ eV where the spectral index changes back to $\Gamma = 2.6$ \cite{Auger, spectrum2}. The ankle is thought to be the change over from Galactic cosmic rays to extragalactic cosmic rays. At $10^{19.5}$ eV, the cosmic ray spectrum appears to suffer a steepening possibly due to interactions with the cosmic microwave background, known as the Greisen-Zatsepin-Kuz'min (GZK) effect \cite{GZK, GZK2}, or due to a natural acceleration limit in sources.\\

\begin{figure}
\begin{centering}
\includegraphics[scale=0.7]{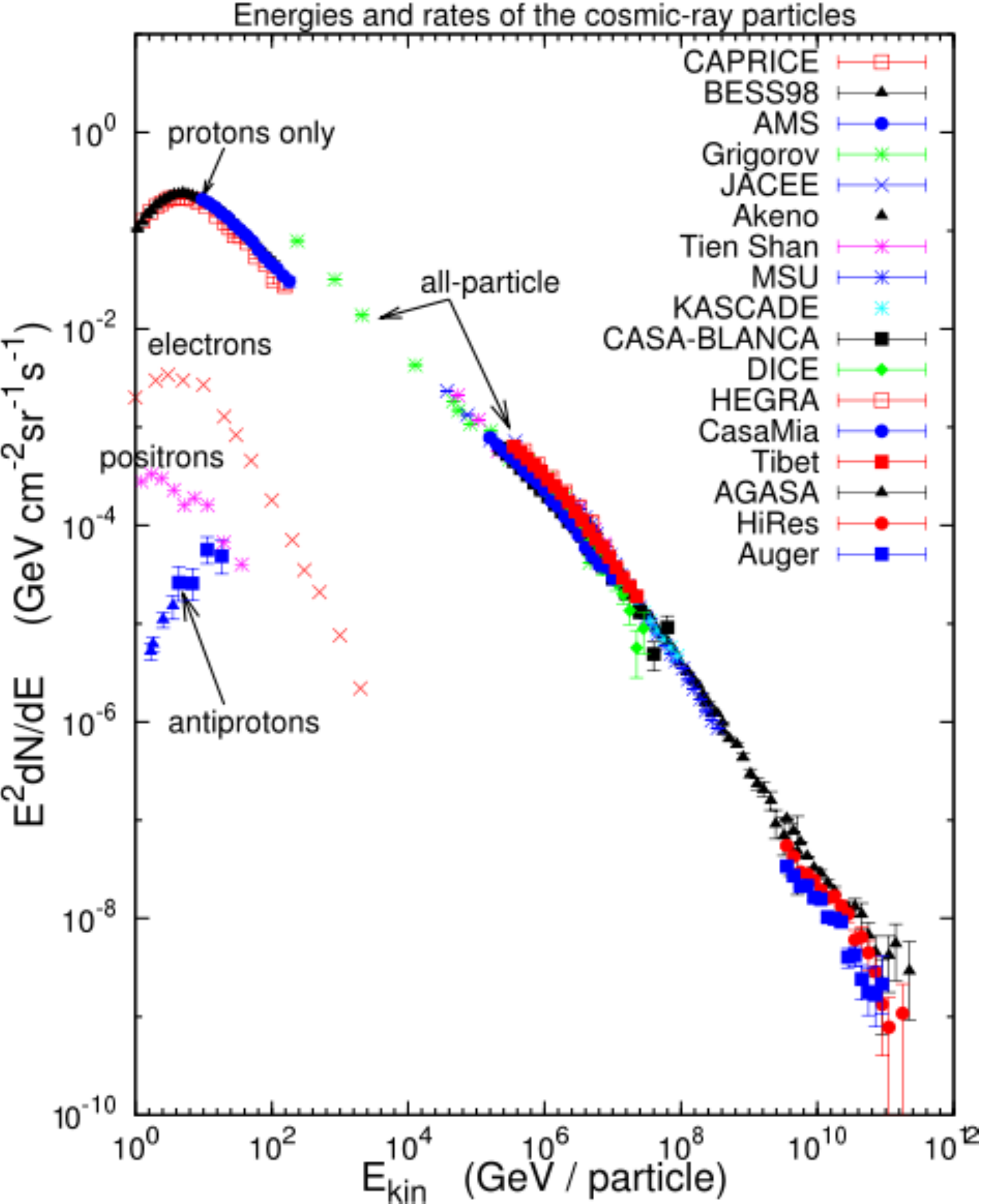}
\captionsetup{width=13cm}  
\caption{The cosmic ray spectrum from \cite{CRspectrum}. The results from previous and current experiments are shown. The distinguishing features known as the \textit{knee}, the second \textit{knee} and ankle are at $4 \times 10^{15}$ eV, $4 \times 10^{17}$ eV and $10^{18.6}$ eV respectively \cite{Stamatescu}.}
 \label{fig:Screenshot}
\end{centering}
\end{figure}

	The Pierre Auger Observatory \cite{PAO} is currently investigating ultra high energy cosmic rays, E $>$ 10$^{18}$ eV, to determine the origin and composition of the particles at the highest energies. The collaboration has produced some exciting new results involving the origin of cosmic rays. The most recent results from Auger have shown a possible correlation between the origin of the highest energy cosmic rays and the direction of Active Galactic Nuclei (AGN) \cite{Augeragn}.  Previous Auger results have shown that the highest energy cosmic rays can be traced back to AGNs within a 3.1$^{\circ}$ circle around the source \cite{Augeragn}. The highest energy cosmic ray origins can be traced directly but the error circles on the origin of these high energy cosmic rays is around 5$^{\circ}$. Therefore, the deflection is small. The origins of cosmic rays below 10$^{17}$ eV are more difficult to study. Since cosmic rays are charged particles, their trajectories are affected by galactic magnetic fields which makes the detected direction useless in determining the source of origin.

	The effect of the galactic magnetic field is characterised by the gyroradius, \textit{$r_{g}$} of the helical motion caused by the presence of a magnetic field:

\begin{eqnarray}
r_{g} = \frac{p}{Bq} = \frac{R}{Bc}  \;[\rm{m}]
\end{eqnarray}
where \textit{$R = pc/Ze$} is the magnetic rigidity, \textit{Z} is the atomic number of the nuclei, \textit{e} is the charge of an electron, \textit{c} is the speed of light, \textit{B} is the magnetic field, \textit{q} is the charge of the particle and \textit{p} is the momentum of the particle. To consider the effect of the magnetic field on particles with different energies, a few examples can be considered. A low energy cosmic ray proton with $E = 10^{12}$ \rm{eV} in the presence of a magnetic field of \textit{$\sim10$} \rm{$\mu$G} will have a gyroradius of $10^{-4}$ \rm{pc}. Therefore, any previous directional information is lost to an observer on Earth. An ultra high energy cosmic ray proton with $E = 10^{20}$ \rm{eV} in the presence of the same magnetic field would have a gyroradius of 10 \rm{kpc}. Therefore the high energy particle experiences only a small deviation from its original trajectory and will retain information concerning its general direction from the cosmic ray accelerator or source when detected on Earth, assuming the field is only experienced within our galaxy.


Detecting cosmic rays with energies $<$ 10$^{17}$ eV will not provide information on the origins of cosmic ray acceleration. However, secondary particles from the interaction of cosmic rays with the interstellar medium could provide a solution for identifying the cosmic ray origins. The interaction with the interstellar medium produces $\gamma$-rays via secondary neutral pion decay at or near a site of cosmic ray acceleration. It was S. Hayakawa \cite{Hayakawa} and P. Morrison \cite{Morrison} in the 1960's who first predicted that the sites of cosmic ray acceleration could be detected via secondary particles emitted from cosmic ray interactions. 
The $\gamma$-rays have no charge, hence they propagate through the Galactic magnetic field without deviating, thereby preserving the trajectory from their origin. From this fact alone, $\gamma$-ray detection provides a means for tracing the origin of Galactic cosmic rays and one may then discover the origin of cosmic rays up to the \textit{knee} in the spectrum.

\section{TeV $\gamma$-ray Production Mechanisms}

	For this section we consider $\gamma$-rays with energies above $10^{10}$ eV or 10 GeV \footnote[1]{ Prefixes for energies:
1 GeV = $10^{9}$ eV, 1 TeV = $10^{12}$ eV, 1 PeV = $10^{15}$ eV, 1 EeV = $10^{18}$ eV, 1 ZeV = $10^{21}$ eV}, which are known as Very High Energy (VHE) $\gamma$-rays. The main mechanisms for GeV to TeV $\gamma$-ray production come from both hadronic cosmic ray and electron origins. Two processes dominate the production of $\gamma$-rays from electrons, inverse Compton scattering and non-thermal Bremsstrahlung.
	Inverse Compton (IC) scattering is often the dominant process for $\gamma$-ray production in many astrophysical settings. IC scattering occurs in the presence of soft photons, for example molecular clouds or the Cosmic Microwave Background (CMB), which is isotropic and has an approximately constant energy density \cite{longair}. Bremsstrahlung occurs in the presence of dense material such as molecular gas clouds. For both these processes, a relativistic electron may have an origin in a number of different possible sources e.g: supernova remnants, pulsar wind nebulae and active galactic nuclei. The electron, $e^{-}$, interacts with the soft photon, ${\gamma}_{soft}$, via a collision, which transfers part of the electron's initial energy to the soft photon. The interaction leads to a lower energy electron, $e^{-*}$, and a boosted soft photon to $\gamma$-ray energies, ${\gamma}_{TeV}$, possibly with GeV to TeV energies depending on the initial electron energy

\begin{eqnarray}
e^{-} + {\gamma}_{soft} \rightarrow e^{-*} + {\gamma}_{TeV}
\end{eqnarray}

The hadronic production of TeV $\gamma$-rays comes from hadron-hadron or pp collisions.  An accelerated cosmic ray proton, $p_{cr}$, may collide with an ambient particle, $p_{ap}$, through an inelastic collision. The ambient particle could be in a molecular gas cloud. The collision produces secondary particles with a neutral pion, ${\pi}^{o}$, being the key particle for $\gamma$-ray production. The lifetime of a neutral pion is extremely short,  $t = (8.4 \pm 0.6) \times 10^{-17}$ \rm{s} \cite{pion}, and it decays into two $\gamma$-rays with equal energies

\begin{eqnarray}
{p}_{cr} + {p}_{ap} \rightarrow {\pi}^{o} + X
where
{\pi}^{o} \rightarrow {\gamma} + {\gamma}
\end{eqnarray}

In some sources a number of production mechanisms could occur for protons and/or electrons. However, the process which dominates the production of $\gamma$-rays can be characterised by the cooling time for each process. We define the cooling time $\tau$ = E/(dE/dt), for a particle of energy E and energy loss rate dE/dt, as the characteristic time taken for the particle to lose its energy. Usually there are four main energy loss mechanisms which need to be considered: synchrotron emission, IC scattering, bremsstrahlung and pp collisions. The cooling times for these are given by:
\begin{eqnarray}
\tau_{sync} \;=\; \frac{E_{e}}{4/3\; {\sigma}_{T}\; c\; U_{mag}\; {\gamma}^{2}}\; \rm{[s]}
 \label{equation:cooling1}
\end{eqnarray}
\begin{eqnarray}
\tau_{IC} \; =\; \frac{E_{e}}{4/3\; {\sigma}_{T}\; c\; U_{rad}\;  {\gamma}^{2}}\; \rm{[s]}
 \label{equation:cooling2}
\end{eqnarray}
\begin{eqnarray}
\tau_{brem} \; =\; \frac{X_{0}}{ \rho\; \;c}\; \rm{[s]}
 \label{equation:cooling3}
\end{eqnarray}
\begin{eqnarray}
\tau_{pp} \;=\; \frac{1}{n_{H}\; \sigma_{pp}\; c \;f}\; \rm{[s]}
 \label{equation:cooling4}
\end{eqnarray}
respectively, where ${\sigma}_{T}$ is the Thomson cross-section, $c$ is the speed of light, $U_{mag} = B^{2}/2{\mu}_{o}$ is the energy density of the magnetic field, B is the magnetic field in \rm{Tesla}, $U_{rad}$ is the energy density of the background electromagnetic field, ${\beta}^{2} = v^{2}/c^{2}$, ${\gamma^{2}} = 1/\sqrt{1-{\beta^{2}}}$, $X_{0}$ is the radiation length in a particular medium in units of column density, $n_{H}$ is the gas number density of the medium, $\sigma_{pp}$ is the inelastic cross-section of pp interactions and f = $\Delta$E/E is the fraction of energy lost per interaction.

Comparing the versions of the cooling time equations in convenient units for all production mechanisms, the dominant factors which affect each mechanism become apparent:


\begin{eqnarray}
\tau_{sync}\; =\; 12 \times 10^{6}\; (\rm{B/ \mu G})^{-2}(E_{e}/\rm{TeV})^{-1}\; \rm{yr}\;
\end{eqnarray}
\begin{eqnarray}
\tau_{IC}\; =\; 3 \times 10^{8}\; (U_{rad}/\rm{eV/cm^{3}})^{-1}(E_{e}/\rm{GeV})^{-1}\; \rm{yr}\;
\end{eqnarray}
\begin{eqnarray}
\tau_{brem}\; =\; 4 \times 10^{7}\; (\rm{n/cm^{3}})^{-1} \;\rm{yr}\;
\end{eqnarray}
\begin{eqnarray}
\tau_{pp}\; =\; 5.3 \times 10^{7}\; (\rm{n/cm^{3}})^{-1} \;\rm{yr}\;
\end{eqnarray}

Both synchrotron and IC scattering are energy dependent and as the energy increases the cooling time for both mechanisms decreases. If the environment has a high magnetic field, then synchrotron emission will become the dominant energy loss process for electrons over IC scattering. Therefore, synchrotron emission will provide the X-ray production mechanism and produce X-ray emission in regions of high magnetic fields. The magnetic field is usually strong at the centre of a source, while, moving further away from the source, the magnetic field strength will decrease. As the magnetic field strength decreases, synchrotron emission loses its dominance over IC scattering. IC scattering produces an extended $\gamma$-ray morphology.  Therefore, IC scattering and synchrotron emission provide good indications of the magnetic field strength in the source e.g. $\tau_{IC}/\tau_{sync} \propto B^{2}$.

In environments with high gas density, pp collisions and bremsstrahlung become the dominant $\gamma$-ray production mechanisms. When the magnetic field strength is low and the gas density is high, bremsstrahlung is the dominant energy loss process for electrons, although this is generally uncommon.


\section{Established sources of relativistic particles}

	The main way to detect high energy, $>$ 100 GeV, $\gamma$-rays is through ground-based detectors. Ground-based $\gamma$-ray detection took off with the development the Whipple telescope. It was a 10 \rm{m} single telescope detector in Arizona with an energy range from 300 GeV to 10 TeV. It was the first large mirror ground-based detector. The Whipple telescope in 1989 provided the first TeV $\gamma$-ray detection from the Crab Nebula \cite{whipple}. The strong 9$\sigma$ detection demonstrated the power of ground-based detectors. Since Whipple, several Imaging Atmospheric Cherenkov Telescopes (IACT) have employed multiple telescopes in a stereoscopic technique, such as the High Energy Gamma Ray Array (HEGRA). The design and results from HEGRA has encouraged similar designs for the current IACTs, which are continually providing new results in the field of $\gamma$-ray astronomy.

Gamma-ray detectors such as the High Energy Stereoscopic System (H.E.S.S.), the Collaboration of Australia and Nippon (Japan) Gamma Ray Observatory in the Outback (CANGAROO), the Very Energetic Radiation Imaging Telescope Array System (VERITAS) and the Major Atmospheric Gamma-ray Imaging Cherenkov Telescopes (MAGIC-II) have confirmed a number of TeV (1 TeV = 10$^{12}$ eV) $\gamma$-ray emitters; supernova remnants (SNR), pulsars and pulsar wind nebulae (PWN), stellar clusters, X-ray binaries (XRB) and active galactic nuclei (AGN), which suggests that cosmic rays could be accelerated at these sources. 
	A list of well known TeV sources and links to specific papers can be found at \cite{Sourcelist}. A summary of the types of TeV sources and the numbers of known sources are provided in Table~\ref{table:sources}.

\begin{table}[h]
\centering
\begin{tabular}{lrr}
\hline
TeV source Type & Number Detected & G or EG\\
\hline
SNR & 13 & G\\
Pulsar & 4 & G\\
PWN & 28 & G\\
XRB & 6 & G\\
Stellar Clusters & 3 & G\\
AGN & 35 & EG\\
Starburst Galaxies & 3 & EG \\
Unidentified & 32 & G\\
\hline
\end{tabular}
\caption{Table of current TeV $\gamma$-ray sources. The table is constructed from the TeV emission catalogue version 3.1 \cite{Sourcelist}. It shows the source types, the number of detected sources and whether the source is Galactic (G) or extra-galactic (EG). Almost one third of the total TeV sources remain unidentified.}
 \label{table:sources}
\end{table}

	This table shows a list of detected TeV sources containing a variety of Galactic and extra-galactic sources. The list includes SNRs, pulsars and PWNs, XRBs, starburst galaxies and stellar clusters for Galactic TeV sources. The extra-galactic sources comprise a number of different AGN types, such as low, intermediate and high frequency BL Lac objects (LBL, IBL and HBL), radio loud galaxies and quasars.

The TeV source catalogue also contains 32 unidentified Galactic sources, which show no known counterpart at lower energies thus proving difficult to identify. These unidentified sources provide enough interest to warrant further observations in the yet unobserved multi-TeV energy regime since current and past detectors can only observe in low TeV energies (Table~\ref{table:IACT2}). The multi-TeV energy regime is from 10 to 500 TeV. There are many different sources that produce TeV $\gamma$-rays and they will be discussed below.

\begin{table}[h]
\centering
\begin{tabular}{lrrrrrrr}
\hline
IACT &  Site/Hemisphere & Energy Range \\
\hline
MAGIC-II & La Palma, Canary Islands (NH) & $\sim$60 GeV to $\sim$10 TeV\\
H.E.S.S. & Namibia, Africa (SH) & 100 GeV to $\sim$10 TeV\\
CANGAROO-III & Woomera, Australia (SH) & $\sim$400 GeV to 10 TeV\\
VERITAS & Mount Hopkins, Arizona (NH) & 50 GeV to $\sim$10 TeV\\
HEGRA & La Palma, Canary Islands (NH) & 500 GeV to $\sim$100 TeV\\
PeX/TenTen & possibly Australia (SH) & $\sim$1 TeV to $\sim$500 TeV\\
CTA & undecided (SH and NH) & $\sim$10 GeV to $\sim$100 TeV\\
\hline
\end{tabular}
\caption{A combination of IACTs including past, present and future arrays. The table displays the site where the IACT is located and the energy range of the respective IACT.}
 \label{table:IACT2}
\end{table}

A TeV sky map including the Galactic plane is presented in Figure~\ref{fig:TeVskymap}. The sky map shows the position of various sources along the Galactic plane. Many $\gamma$-ray detectors have conducted Galactic Plane surveys including H.E.S.S., CANGAROO-III, VERITAS and Milagro (the large field of view Cherenkov extensive air shower array) \cite{HESSplane, veritasplane, CANGAROOplane, Milagroplane}.

\begin{figure}[h]
\begin{centering}
\includegraphics[scale=0.5]{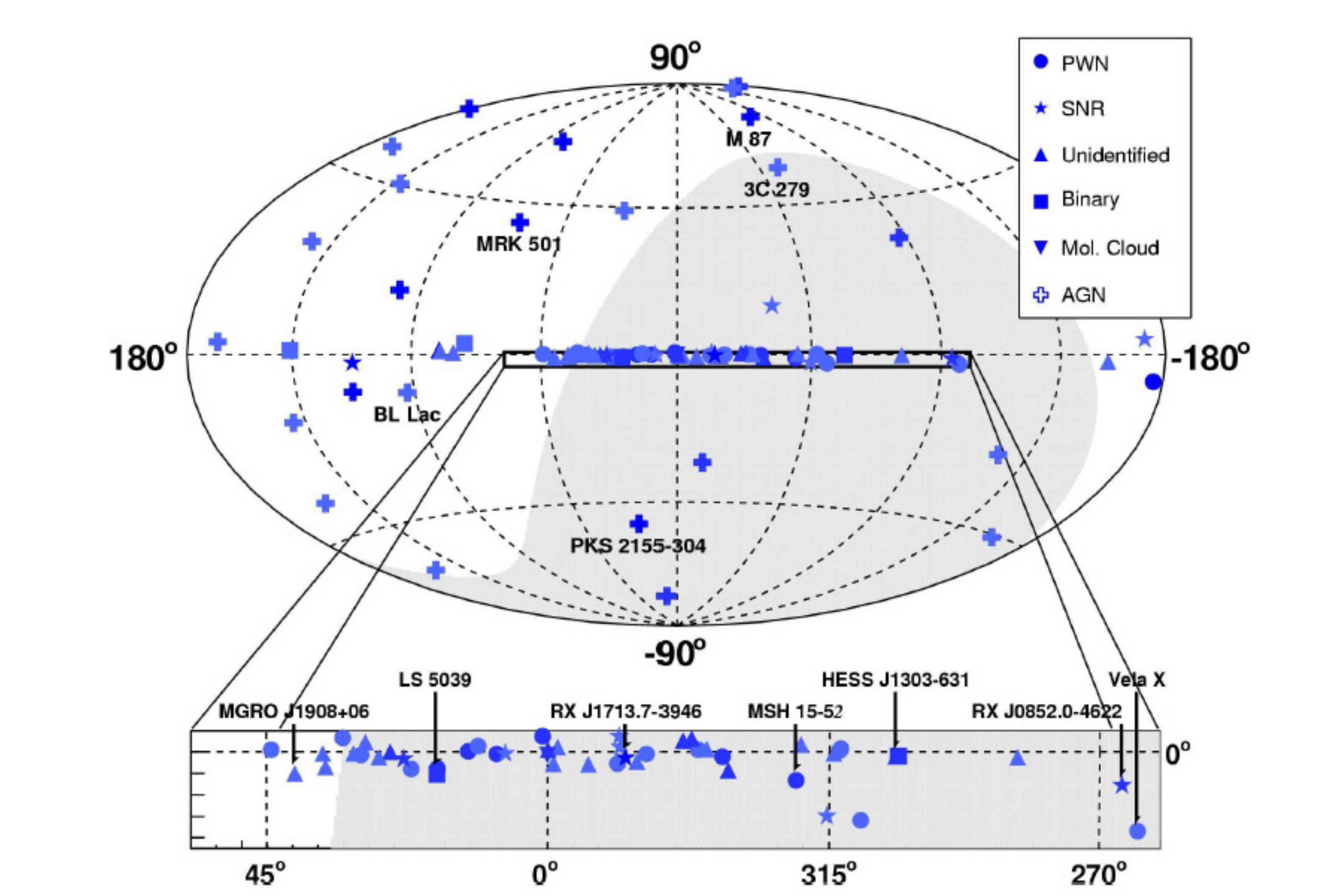}
\captionsetup{width=13cm}  
\caption{A plot in Galactic coordinates of the TeV $\gamma$-ray sky. The different symbols represent the different type of sources \cite{teraelectron}.}
 \label{fig:TeVskymap}
\end{centering}
\end{figure}

\subsubsection{Supernova Remnants (SNR)}

A star with a mass $>$ 3 $M_{\odot}$ ends its lifetime with a supernova explosion. The outer layers of the star are ejected leaving a compact core, which is usually a neutron star or a black hole. If the massive star is $> 10 M_{\odot}$, the resultant supernova explosion will likely produce a black hole.

An expanding shockwave forms as the ejected mass moves outward from the site of the explosion. The expanding shockwave forms a boundary for the SNR. A possible acceleration process in such a shockwave, is diffusive shock acceleration (DSA) \cite{longair}. However, the upper limit on the energy of accelerated particles that leave the shock region is approximately $0.1$ PeV \cite{longair} but more recent theory can suggest $>$ 10 to 100 PeV \cite{Lagage}. These upper limits are just below the energy of the cosmic ray spectral knee. Particles gain energy when crossing the shock front and they can make another crossing after interacting with turbulent magnetic fields on each side of the shock. As a particle repeatedly crosses the shock front, energy is gained. The particle escapes the shock roughly when its gyroradius exceeds the shock radius \cite{Lagage}. 

	TeV $\gamma$-ray observations from CANAGROO-III and H.E.S.S. have revealed that SNR RX J1713.7-3946 produces multi-TeV $\gamma$ ray emission \cite{Aharonian1, CANG1713}. The H.E.S.S. collaboration analysed RX J1713.7-3946 through its spectral energy distribution at multi-TeV energies and tested two $\gamma$-ray production scenarios. The leptonic scenario requires a  magnetic field of 10\rm{$\mu$G} to reproduce the fluxes seen in X-rays and $\gamma$-rays. However, the proposed magnetic field strength of RX J1713.7-3946 exceeds that expected in a SNR \cite{Koyama}. This suggests that other $\gamma$-ray production mechanisms may provide the more appropriate model for the $\gamma$-ray emission \cite{FUKUI}. Fermi-LAT (Fermi Gamma-ray Space Telescope - Large Area Telescope) has also conducted observations of RX J1713.7-3946 which suggest that leptonic origins dominate $\gamma$-ray emission \cite{fermi-latSNR}. Further investigations are required to determine the true origin of TeV $\gamma$-rays.

	Testing the proton scenario required an accurate measurement of the mean target density or mean hydrogen density. The H.E.S.S. observations suggest an approximate density of 1 \rm{cm$^{-3}$}, which provides the correct proton energetics to satisfy the SNR origin for Galactic cosmic rays with the canonical $\approx 10\%$ conversion efficiency of the total supernova explosion energy \cite{Aharonian1}. The large uncertainties in distance to the source and in the target density affect the estimation. 
	
\subsubsection{Pulsar and Pulsar Wind Nebulae (PWN)}

A pulsar wind nebula is thought to be powered by a relativistic outflow of particles primarily consisting of electrons and positrons. The outflow of particles comes from the rapidly rotating neutron star or pulsar at the centre of the source. The particle outflow from the pulsar interacts with the material from the surrounding medium producing a standing shock wave, which is the location of particle acceleration \cite{Aharonian2}. The lifetime of the PWN exceeds that of a SNR due to the constant replenishment of particles from the pulsar. The outflow of particles produces $\gamma$-rays via scattering on a target photon field. 

	HESS J1825-137 is a confirmed TeV $\gamma$-ray emitter as well as an X-ray emitter. The cooling time scenarios outlined in Eq~\ref{equation:cooling1} to Eq~\ref{equation:cooling4} are consistent with the observed TeV and X-ray morphologies of HESS J1825-137 for the leptonic model \cite{Aharonian2}.
	
\subsubsection{Active Galactic Nuclei (AGN)}

	An AGN is powered by a central super-massive black hole with mass $> 10^{7}$ $M_{\odot}$ surrounded by an accretion disk \cite{Aharonian3}. A larger dust torus also forms around both super-massive black hole and accretion disk. The matter within the accretion disk is believed to be responsible for the two jets of relativistic particles that are perpendicular to the plane of the accretion disk. The photon emission extends from radio to TeV $\gamma$-rays. The origin of the TeV $\gamma$-ray emission from AGNs remains uncertain.

	Electrons could be accelerated in the jets and then up-scatter photons to high energies, which provides the leptonic origin. Neutral pion decay, from proton proton collisions, which provides the hadronic origin.  

	BL lac objects belong to the class of AGNs called Blazars, which are distinguished by the lack of emission or absorption lines in their spectra. There are many types of Blazers: high frequency BL lac objects (HBL), intermediate frequency BL lac object (IBL), low frequency BL lac object (LBL) and X-ray BL Lac objects. These sub-classes are separated based on the frequencies at which the two spectral energy distribution peaks occur. Blazars have relativistic jets which point a few degrees off the observer's line of sight. One confirmed BL Lac TeV emitter is H 2356-309 \cite{Aharonian4}. TeV emission has also been observed from Markarian 421. CANGAROO conducted observations of Markarian 421, which is another BL Lac object. Their results indicated that the spectrum is steeper in the 10 to 30 TeV range compared to the 1 TeV range, which agrees with cutoff spectrum of Markarian 421 in the 0.2 to 10 TeV range measured by other groups \cite{CANGAROOmkn421}. CANGAROO has also conducted observations on four other Blazars between 2005 and 2009, where they narrow down the possible TeV $\gamma$-ray emission mechanisms \cite{BlazarsCANG}. 
	
	VERITAS has detected some of the first TeV $\gamma$-ray emissions from IBL objects such as W-comae and 3C 66A \cite{veritasIBL}. The observations from both sources provide statistical significance levels of 4.9$\sigma$ and 4.8$\sigma$ respectively for the entire data sets. Blazar W-Comae has shown flaring activity which has produced 70$\%$ of the signal and has a soft spectrum with photon index of 3.8 \cite{wcomae}. For W-Comae, the emission from the flares is best represented by synchrotron self Compton or external Compton leptonic jet models. Blazar 3C 66A has also shown strong flaring although the spectrum is slightly softer than W-Comae with a photon index of only 4.1 $\pm$ 0.4$_{stat}$ $\pm$ 0.6$_{sys}$ \cite{3C66A}. The detection of TeV $\gamma$-rays is strong and with more sources providing TeV emission, further modelling of $\gamma$-ray production from Blazars can be performed. With a larger AGN catalogue for TeV $\gamma$-rays, certain ideas and models can be dismissed which will help identify the origin of TeV $\gamma$-rays from Blazars.

\subsubsection{X-ray Binaries (XRBs)}

	XRBs consist of two objects, namely a compact object and a companion star, which orbit around a common centre of mass. The compact object, usually a neutron star or black hole, accretes matter from the companion star forming an accretion disc. XRBs are split into two different categories depending on the mass of the companion star. If the companion star has a mass $\leq$ 1 $M_{\odot}$ then the binary is a low mass X-ray binary (LMXB) and if the companion star has a mass $>$ 10 $M_{\odot}$ then the binary is a high mass X-ray binary (HMXB). High resolution radio maps for XRBs reveal similar characteristics to AGNs but on a much smaller scale. XRB have a relativistic outflow of particles similar to the relativistic jets from AGNs, which may be formed from the accreting material and the compact object. XRBs or microquasars (radio emitting X-ray Binary) can emit TeV $\gamma$-rays with the relativistic jets from the compact object being the origin. H.E.S.S. detected TeV emission from compact binary LS 5039 \cite{microquasar}. A compact binary suggests that one of the stars in the system is a compact object but also the distance between the stars is small such that the compact object is accreting matter from the companion star. An orbital modulation in the $\gamma$-ray flux of 3.9 days was detected for LS 5039, which suggests that $\gamma$-ray absorption can occur within the source. As well as absorption, the dominant radiative mechanism and the inverse Compton scattering angles could be affected by the modulation in maximum electron energy \cite{Aharonian6}. This will affect the main $\gamma$-ray production mechanisms for XRBs. 

Another interesting source is the HMXB LSI +61$^{\circ}$303, which has been observed in multiple wavelengths from radio to $\gamma$-rays \cite{LSIbinary}. The source emits a bright radio outburst every 26.5 days. However, the outburst and the high energy emission still remain a mystery that requires further observations with higher sensitivity detectors.

The compact binary system PSR B1259 has provided strong evidence for particle acceleration up to multi-TeV energies \cite{1259pulsar}. The system consists of a radio pulsar orbiting a massive luminous Be star. This binary system was the first detected Galactic variable source of VHE $\gamma$-rays, where the variability of the system is on a time scale of a few days.

\subsubsection{Massive Stellar Clusters}

Stellar clusters are characterised by ongoing star formation or the presence of massive stars near the end of their life cycle but before their explosion as a supernovae. It has been speculated that cosmic rays could be accelerated in the stellar winds either around massive stars before a supernova explosion or around young star clusters \cite{teraelectron}. High energy particle acceleration connected with stellar winds is a phenomenon becoming more widely considered with current $\gamma$-ray detectors. H.E.S.S. has discovered high energy $\gamma$-rays coinciding with the young stellar cluster Westerlund 2, HESS J1023-575, which is embedded in the giant HII hydrogen cloud region RCW49 \cite{Aharonian7}. Westerlund 1 and Cygnus OB2 have proven to be TeV $\gamma$-ray emitters \cite{Aharonianwester1, cygnusob2}. However, the emission from these regions many be due to SNRs within the clusters.

\subsubsection{Galactic Centre}

The Galactic Centre has been observed in multiple wavelengths and is still a point of extreme interest due to the fact that so much about it is still unknown. One main motivation for studying the Galactic Centre is to measure the emission from the super massive black hole which is associated with the compact radio source Sagittarius A$^{*}$. Since the Galactic Centre is obscured by dust in the plane, the optical and UV observations are not so effective for for observing Sgr A$^{*}$. To provide the best observations, the Galactic Centre is observed in radio to infra-red wavelengths. 

There are so far two localised TeV emissions near the Galactic Centre, one is a PWN inside of SNR G0.9+0.1 and the other includes both Sgr A$^{*}$ and PWN G359.95-0.04 \cite{Galactic_Centre}. Further observations of these regions will unlock the mystery behind the Galactic Centre. With improved sensitivity and resolution, multi-TeV observations will provide exciting new results on particle acceleration, $\gamma$-ray production near black holes and particle propagation in the centre of dense molecular clouds. Combining new multi-TeV results with other multi-wavelength observations, should unlock the key physics behind Sgr A$^{*}$ and provide exciting new discoveries.

\subsubsection{Molecular Clouds}

An additional source of $\gamma$-ray emission comes from the interaction of cosmic rays with molecular clouds. The neutral pions from the interactions between cosmic rays and the interstellar gas are believed to be the cause of the diffuse MeV to GeV $\gamma$-ray emission seen over the Galactic plane by Fermi-LAT and the Energetic Gamma Ray Experiment Telescope (EGRET) \cite{Gabici_2008}. Molecular gas can be useful for the detection of cosmic ray acceleration sites since the molecular gas can enhance the emission of $\gamma$-rays. As a simple approximation, if the cosmic ray spectrum is the same throughout the Galaxy then it can be assumed that the column density of gas in a particular direction should be proportional to the $\gamma$-ray emission \cite{molecularcloud}.

One region of extreme interest is the W 28 SNR TeV source. These new sources coincide with molecular clouds over the W 28 SNR region. The most exciting point to note is that the W 28 region is an old SNR so electrons should have lost most of their energy to radiative energy loses. Therefore, if the $\gamma$-ray emission does come from interactions with molecular clouds, then there could be a higher chance that the origin is hadronic \cite{FUKUI}. This could provide evidence for sites of Galactic cosmic ray acceleration and will motivate further detections at multi-TeV energies.

\subsubsection{Starburst Galaxies}

Starburst galaxies exhibit strong high-mass star formation and have an increased rate of supernovae. Multiple regions in a starburst galaxy experience these traits but most of the starburst activity is seen in the centre of the galaxy. These objects provide a prime location for the acceleration of cosmic rays due to the high supernovae rate. Another key feature of starburst galaxies is the high density of gas in the galaxy. This gas can provide the necessary target material for the production of $\gamma$-rays. Hence, starburst galaxies are the perfect candidates for $\gamma$-ray detection.

VERITAS detected the first $\gamma$-ray emission from the Messier 82 (M82) with 5$\sigma$ significance \cite{M82}. This exciting discovery helped provide evidence for cosmic ray acceleration from exploding stars and stellar winds. Therefore, $\gamma$-ray emission from extra-galactic sources is not limited to AGNs. The signal detection was 5$\sigma$, which established starburst galaxies as a new class of $\gamma$-ray emitters \cite{Karlsson_Collaboration_2009}. H.E.S.S. detected high energy $\gamma$-rays from the first known starburst galaxy M82 \cite{NGC253}. 

\subsubsection{Unidentified Sources}

Approximately 1/3 of the current detected TeV $\gamma$-ray sources are unknown. These unknown sources usually have no clear counterpart in any other wavelength. Since they lack counterparts at low frequencies, the TeV sources are sometimes referred to as \textit{Dark Accelerators} \cite{teraelectron}. A number of questions are associated with these sources; do these objects belong to a new class or an existing class, why do they have no counterpart and is the origin of the $\gamma$-rays hadronic or leptonic?

For example, observations from H.E.S.S. \cite{unidentified} have confirmed HESS J1908+063 as another unidentified $\gamma$-ray source above 30 TeV. This observation agrees with the earlier detection by Milagro of MGRO 819808+06 at 20 TeV \cite{MilagroUn}. More multi-wavelength observations are required to confirm or rule out a link between J1908+06 and any of the nearby sources such as SNR G40.5-0.5, open cluster DSH J1907.4+0549 or the unidentified source GRO J1908+0556. Fermi-LAT have detected a pulsar 0FGL J1907.5+0617 that could be associated with the unidentified TeV source \cite{unidentified}.

There is also a second class known as \textit{no-longer unidentified}. These objects were identified with longer follow up observations using other wavelength instruments. One example of this includes HESS J1813-178, which is being established as a new composite SNR \cite{teraelectron}. Further observations in E $>$ 10 TeV may provide useful information to shed light on unidentified sources. Many of the extended sources are larger than the field of view of current detectors like H.E.S.S. so an increased field of view with improved resolution is required.

\section{Motivation for a Multi-TeV ( $>$ 10 TeV) $\gamma$-ray detector}

\subsection{Astrophysical Motivations}

	One thing that needs to be considered is the attenuation of $\gamma$-rays at high energies. At high energies, E $>$ 100 TeV, the incoming $\gamma$-ray flux is attenuated by pair production on the CMB and the Galactic interstellar radiation field (ISRF). The energy threshold for the attenuation on the CMB is roughly 100 TeV and reaches a maximum at 2000 TeV \cite{attenuation}. For the ISRF the attenuation starts at roughly 100 GeV and reaches a maximum attenuation at 100 TeV. Moskalenko et. al. \cite{attenuation} have made a calculation of the attenuation of $\gamma$-rays from the interaction with the ISRF. They found that the attenuation at 30 TeV is approximately 12$\%$ and the attenuation at 100 TeV is approximately 25$\%$ from sources at the Galactic centre. Therefore, as the energy increases the attenuation increases which limits the detectable sources to Galactic or close extra-galactic sources. So a new multi-TeV detector could provide useful observations of Galactic sources.

	As described earlier, for sources such as SNR, the current theory for DSA in SNRs limits particle acceleration to $\approx0.1$ PeV. A number of known TeV sources show $\gamma$-ray emission exceeding E $\approx$ 10 TeV.  
Most Galactic TeV sources have a hard or flat power law photon spectra, $dN/dE \sim kE^{-\Gamma}$ with index $\Gamma < 2.5$. Many sources show an emission cut-off or turn-over suggesting that the emission dies off as energy increases. For HESS RXJ1713.7-3946 \cite{Aharonian5} a 4$\sigma$ $\gamma$-ray detection is seen above 30 TeV in Figure~\ref{fig:hesssourcespectrum}. To investigate this source further a higher energy detector is required.

Another source that has been classed as a pevatron emitter is HESS J1908+063 / MGRO J1908+06. It has been observed with Milagro, H.E.S.S. and VERITAS. H.E.S.S. results have shown a hard spectrum with photon index of 2.10 which provides strong emission at the energy limits of current detectors \cite{unidentified}. VERITAS have confirmed the size and position of the very high energy $\gamma$-ray emitter with a 4.9$\sigma$ level \cite{veritas1908, veritasstatus} which matches the H.E.S.S. detection. All three detectors have seen emission at 20 TeV with evidence that the emission extends above this energy. The hard spectrum with possible emission at multi-TeV energies warrants future observations with a new multi-TeV $\gamma$-ray detector.

\begin{figure}[h]
\begin{centering}
\includegraphics[scale=0.65]{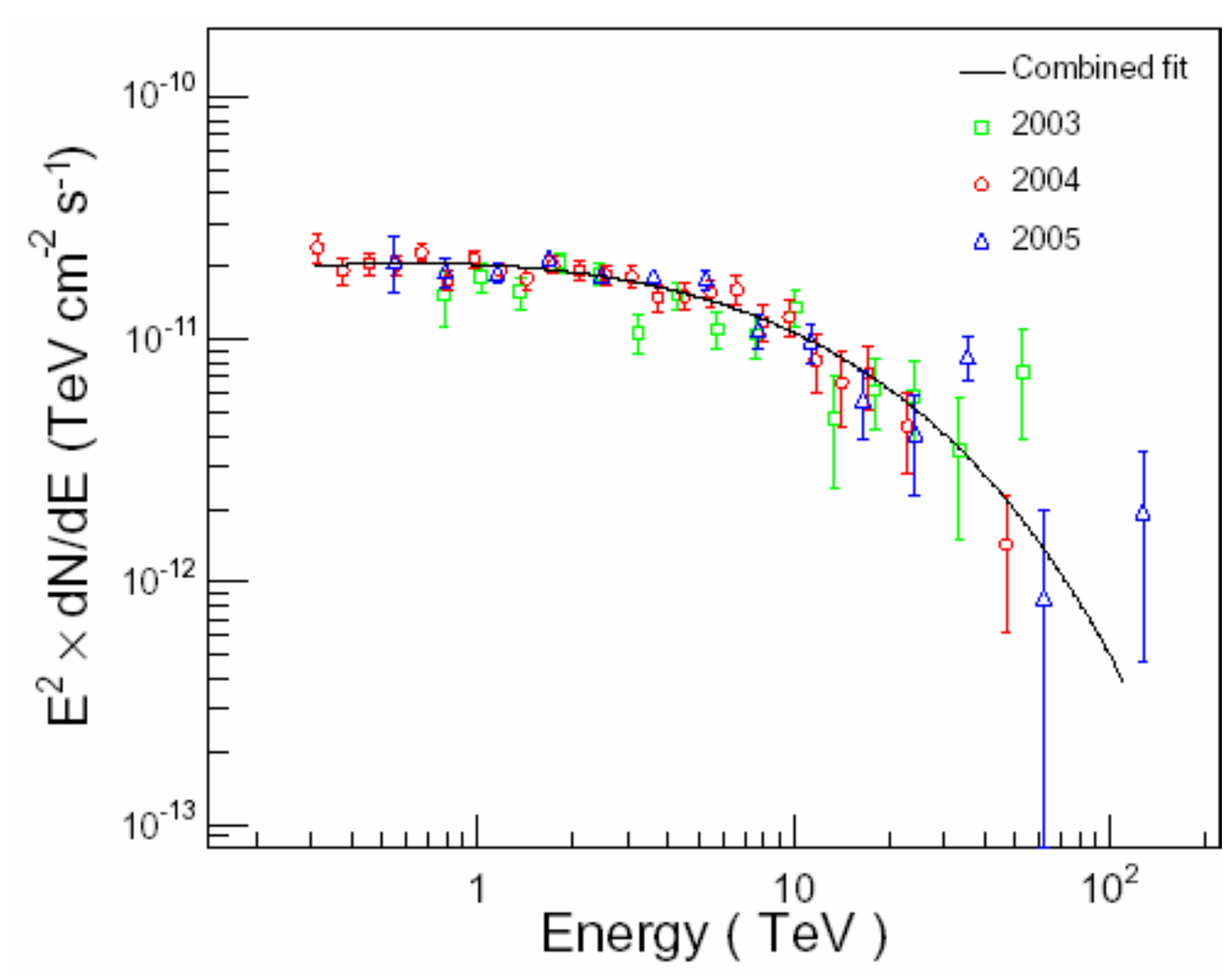}
\captionsetup{width=13cm}  
\caption{TeV Spectral data for RXJ1713.7-3946 \cite{Aharonian5}. The observations indicate that the spectrum extends above E $>$ 10 TeV. At $\approx$ 30 TeV, the $\gamma$-ray sigma significance is 4.3$\sigma$. However, further detection and confirmation above 30 TeV is limited by the energy range of the H.E.S.S. telescopes. A multi-TeV detector would provide vital information above 30 TeV.}
 \label{fig:hesssourcespectrum}
\end{centering}
\end{figure}

	For E $>$ 10 TeV, the hadronic or leptonic origin of $\gamma$-ray emission can be easier to distinguish. This is due to leptonic origins suffering from a number of competing effects. The cooling time for electrons is strongly dependent on the magnetic field strength in the surrounding environment. The high magnetic field environments cause electrons to lose energy to synchrotron emission which suppresses the production of TeV $\gamma$-rays since the electrons do not have enough energy to produce TeV $\gamma$-rays. IC scattering is dampened by the Klein-Nishina effect \cite{nuclear, KNenergy_spectrum}, where the efficiency of IC scattering reduces as particle energy increases. Therefore, SNR sources which emit $\gamma$-rays at multi-TeV energies may have a higher chance that they have hadronic origins over leptonic origins since there is no continuous injection of electrons as in PWNe. In short, SNRs are impulsive cosmic ray accelerators, while PWNe are continuous accelerators. This is true for old or relic SNRs where the age of the remnant is more than $\approx$ 10000 \rm{yrs} \cite{molecularcloud}. The electrons in relic SNRs have undergone synchrotron emission in the presence of a high magnetic field at an earlier epoch. As the SNR ages, the hadronic component is likely to remain while the electron component dies out. 
	
Above 10 TeV, $\gamma$-rays can also probe particle acceleration to E $>$ 500 TeV energies which approaches the cosmic ray \textit{knee} energy. The highest energy $\gamma$-rays observed from RXJ1713 at $\sim$30 TeV, with 3 $\sigma$ significance, represents either $\sim$200 TeV protons or $\sim$110 TeV electrons respectively 
\cite{Aharonian5}.



A source that provides interesting results to motivate a new multi-TeV detector is HESS J1908+063. The observation of this unidentified TeV emitter provides the first confirmation of the TeV source detected by Milagro \cite{MilagroUn}. Milagro has highlighted the potential of a multi-TeV detector and an IACT array that can work in the E $>$ 10 TeV energy range will be able to complement the observations from the Milagro all-sky survey. The spectra of J1908+063 from H.E.S.S. and Milagro is shown in Figure~\ref{fig:hesssourcespectrum2}. The spectrum has a spectral index of 2.1 and there appears to be no strong turn-off at high energies. At present this source requires further investigation using multi-wavelength observations. An improved sensitivity and resolution from an IACT array above 10 TeV could help associate the unidentified source J1908+063 with a number of nearby sources \cite{unidentified}. 

\begin{figure}[h]
\begin{centering}
\includegraphics[scale=0.65]{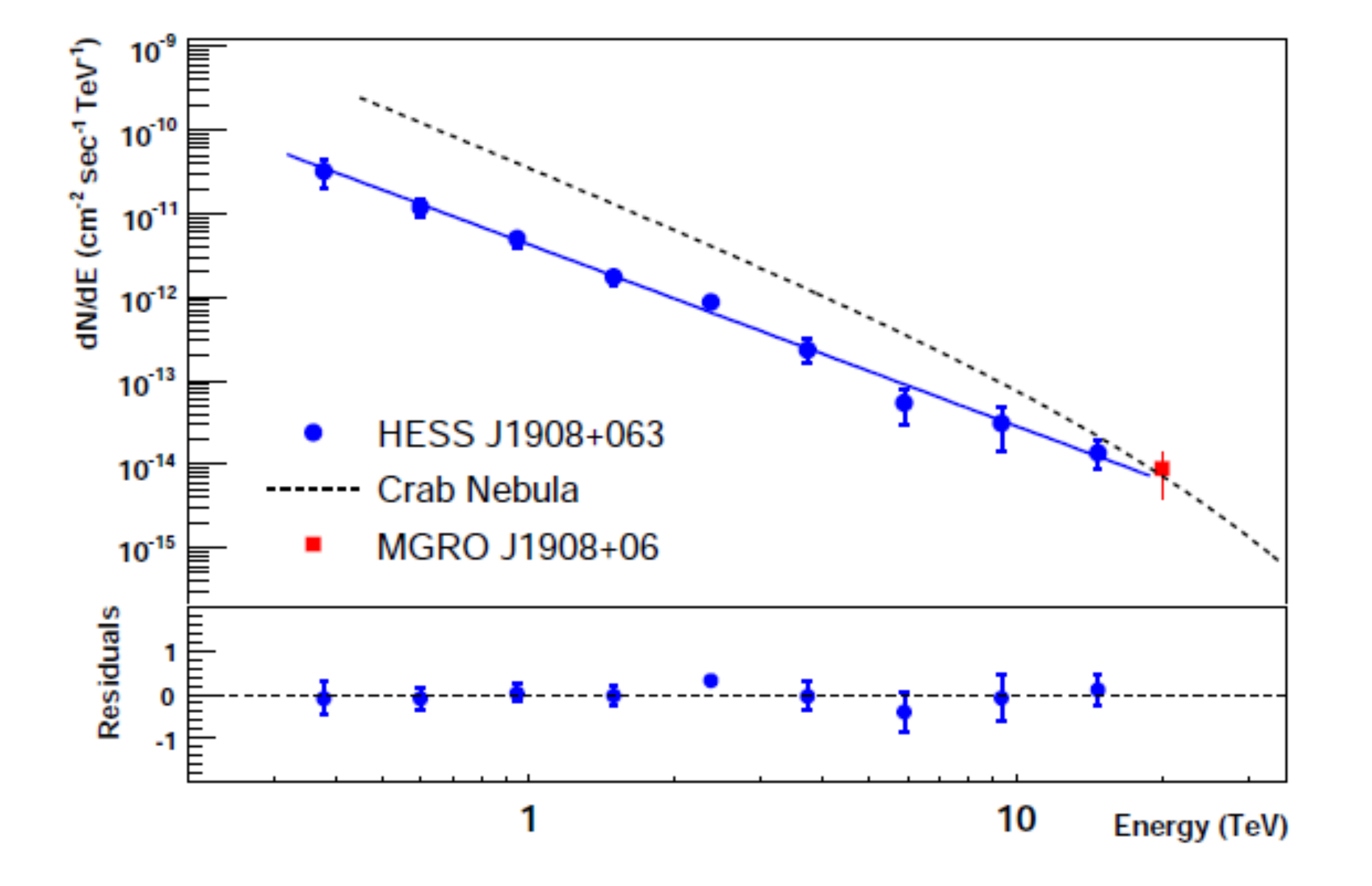}
\captionsetup{width=13cm}  
\caption{Spectral data for J1908-063 from H.E.S.S. and Milagro \cite{unidentified}. The observations indicate that the spectrum extends above E $>$ 10 TeV. The blue points represent the H.E.S.S. spectrum, the red point represents the Milagro flux and the black dotted line represents the Crab Nebula energy spectrum from H.E.S.S.. The flux from J1908-063 shows no sign of turn-off at high energies.}
 \label{fig:hesssourcespectrum2}
\end{centering}
\end{figure}

HESS J1825-137 is a PWN that is part of a growing list of $\gamma$-ray sources that exhibit emission above 10 TeV. Recent observations of PWN J1825-137 have established that this source has the first energy dependent morphology within the $\gamma$-ray regime to be detected by the H.E.S.S. collaboration \cite{Aharonian2}. Current IACTs such as H.E.S.S., VERITAS and MAGIC-II all have an upper energy limit of 20 - 30 TeV where the sensitivity at the upper limit is not as strong as in the middle of the energy range. Observations above these energy limits require very deep observations, usually of the order of hundreds of hours. A new detector with the same sensitivity as current detectors but in the 10 to 500 TeV range will reduce the number of observation hours required to observe sources that exhibit emission above 10 TeV. This multi-TeV component may provide useful information regarding the acceleration mechanisms within sources that have emission extending above 10 TeV. It could also test for the energy dependent morphology for energies $>$ 100 TeV. 

Recently, GeV $\gamma$-ray flares have been observed from the Crab Nebula. The spectral fit has been remodelled to include the new observed flaring from the Crab. They consider the relativistic model of a small blob that is Lorentz-boosted towards us and thus emits synchrotron radiation beyond the maximum energy \cite{2012arXiv1202.6439K}. The model agrees well with the GeV emission but is unconfirmed at TeV energies. Future TeV detectors like CTA, Tibet AS + MD and Large High Altitude Air Shower Observatory (LHAASO) will be able to narrow down the range of Lorentz factors that could model the emission at TeV energies. This new emission from the Crab has provided another strong motivation for the study of multi-TeV $\gamma$-rays.



\subsection{Technical Motivations}
	The motivations for multi-TeV $\gamma$-ray detectors extend beyond the astrophysical reasons. Additional technical reasons include improvements to analysis techniques, new methods to separate $\gamma$-ray and cosmic ray events and even improvements to camera technology e.g: photomultiplier tubes. New technical improvements for a multi-TeV detectors will be discussed. 

A wide field of view is essential for multi-TeV observations. H.E.S.S. and Milagro have proven the worth of a large field of view by showing that most Galactic sources are large and can extend up to 2$^{\circ}$ (Figure~\ref{fig:background_2} and Figure~\ref{fig:hesssource2}). It will enable a larger portion of any extended morphology to be detected by the camera. As the energy of particles from a source increases, the particles penetrate further before interacting and producing $\gamma$-rays. At higher energies the $\gamma$-ray emission is usually expanding a few degrees (Figure~\ref{fig:Figure8}). Gabici \cite{Casanova} modelled a supernova which has exploded into the interstellar medium. With this model Gabici has shown that the high energy sources produce larger $\gamma$-ray morphologies, which require larger field of views to detect. This model confirms the results presented in \cite{GABICI} which state that TeV cosmic rays can diffuse further into the interstellar medium compared to GeV cosmic rays.

\begin{figure}[h]
\begin{centering}
\includegraphics[scale=0.62, angle=270]{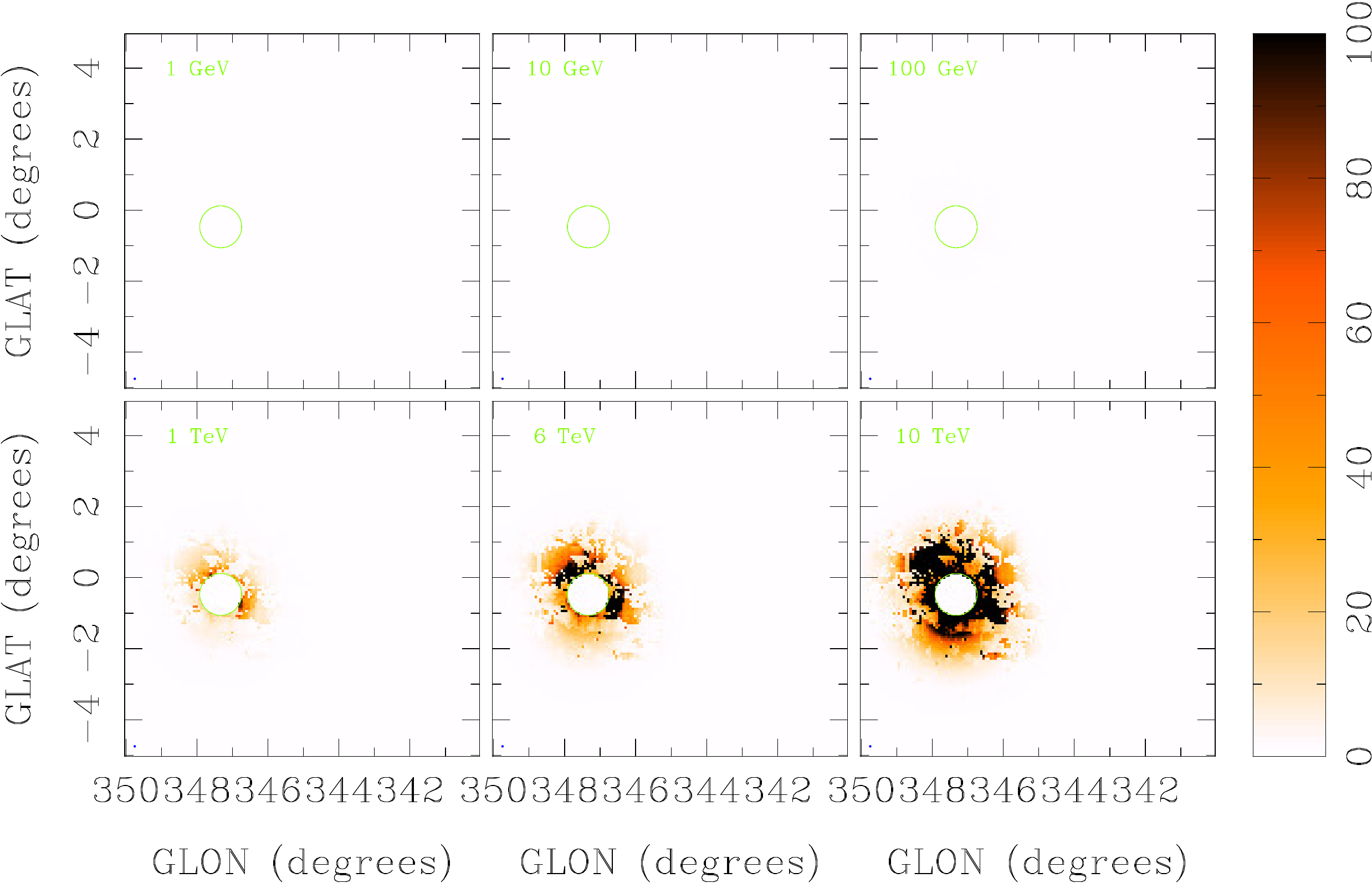}
\captionsetup{width=13cm}  
\caption{The ratio of hadronic $\gamma$-ray emission due to total cosmic ray spectrum to that of the background cosmic rays for the entire region (taken from \cite{Casanova}). The model shows a SNR (green circle) with a 10$^{26}$ \rm{cm$^{2}$s$^{-1}$} diffusion coefficient which is quite strong. The cosmic rays penetrate further into the surrounding dense gas as the energy increases. The 10 TeV panel shows that the emission would extend a few degrees, which would be larger for 100 TeV sources.}
 \label{fig:Figure8}
\end{centering}
\end{figure}


For a point source observation, the large field of view will provide enough background sky to give an accurate value for the level of noise. This helps calculate an improved signal significance since the significance in Eq~\ref{eqn:sig_li_ma} is dependant on the number of on- and off-source regions or time used to observe the on- and off-source regions, $\alpha = N_{on}/N_{off}$. The more off-source regions used in the calculation, the better the estimation of the hadronic background. Figure~\ref{fig:background_2} provides a good example of the on-source region (white circle) and the multiple off-source regions (red circles) for a point source observation. A new detector with a large field of view will be able to shed new light on sources that are too large for observations with current detectors. For example, Figure~\ref{fig:hesssource2} shows a large extended morphology of the Cygnus region viewed by the all sky detector Milagro. H.E.S.S. would require multiple observations of the Cygnus region to produce the same image as Milagro (Figure~\ref{fig:hesssource2}). The H.E.S.S. field of view would only encompass the bright section centered at (l, b) = (0$^{\circ}$, 75$^{\circ}$). A larger field of view detector in the multi-TeV energy range with an improved sensitivity and resolution could improve the structure seen in this source.

An added benefit of a wide field of view, improved sensitivity and improved resolution is the possible discovery of other weak $\gamma$-ray sources. Observations of PWN J1825-137 (Figure~\ref{fig:hesssource2} right) shows a weak multi-TeV signal appearing in an unidentified EGRET source. The discovery of new unexpected sources with a large field of view detector will also apply to a Galactic plane survey. A large field of view plus an improved sensitivity will shorten the time required for a full Galactic plane survey. With less time required for a Galactic plane survey, more of the detector's annual observation time can be spent observing other regions of extreme interest. An exciting new picture of the Galactic plane at multi-TeV energies awaits the arrival of a new multi-TeV image atmospheric Cherenkov telescope.

A wide field of view also benefits the detection of large core distance events, thereby achieving a large collection area. A.V. Plyasheshnikov \cite{Plyasheshnikov} conducted Monte Carlo simulations to investigate the E $>$ 10 TeV regime with a larger detection area. A way to achieve this is by increasing the separation between telescopes in the array, creating a sparse array of telescopes. Since multi-TeV showers are larger and brighter than GeV showers, only small to moderately sized telescopes are required with small to moderately sized mirrors to collect all the light. Plyasheshnikov simulations utilised a new design consisting of relatively small mirrors of 5 to 10 \rm{m$^{2}$}, large cameras with a diameter of 8$^{\circ}$ and moderately sized pixels of 0.3 - 0.5$^{\circ}$. This design combined with a 300 to 500 \rm{m} separation between 4 telescopes can provide a 1 \rm{km$^{2}$} collecting area \cite{Plyasheshnikov}. The large collecting area combined with a large field of view allows the detection of events up 500 \rm{m} distances or more. The array design provided 20$\%$ energy resolution, good angular resolution ($\leq$ 0.1$^{\circ}$) and effective separation of $\gamma$-ray and proton events. Therefore, the studies conducted by A.V. Plyasheshnikov have shown the importance of large core distance events in the detection of multi-TeV $\gamma$-rays and have provided encouraging results above 10 TeV. These ideas provide strong motivation for the design of new multi-TeV image atmospheric Cherenkov telescope. The focus of this work will be on the detection of large core distance events to gain the largest possible collecting area combined with good angular resolution, good energy resolution and effective separation of $\gamma$-ray and cosmic ray events.

	
\begin{figure}[h]
\begin{centering}
\includegraphics[scale=1.1]{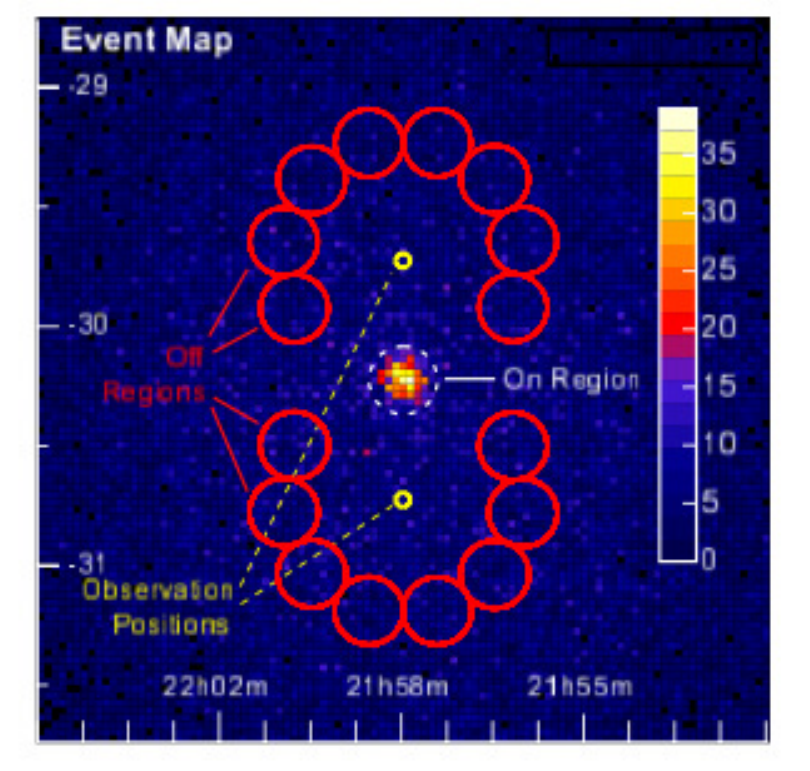}
\captionsetup{width=13cm}  
\caption{An example of on and off-source regions for a H.E.S.S. observation \cite{backmodel} with a 5$^{\circ}$ field of view. The pointing direction of the telescope is shown as yellow rings, the red rings represent the off-source regions while the dotted white ring shows the on-source region. The off-source regions start a reasonable distance from the on-source region to prevent any contamination. The on and off-source regions are similar distances away from the pointing direction.}
 \label{fig:background_2}
\end{centering}
\end{figure}

\subsubsection{Time Development in $\gamma$-ray showers}

Recent observations and simulations \cite{Stamatescu} with TeV $\gamma$-rays have shown the time development in Cherenkov images can be exploited. This is an interesting and exciting new aspect that could prove to be useful for multi-TeV observations. The time profile for $\gamma$-ray and cosmic ray showers can be used to improve shower reconstruction and event separation. Stamatescu \cite{Stamatescu} demonstrated the use of the time gradient along the major axis of Cherenkov images to improve the reconstruction algorithm \cite{Stamatescu}. 


Work done in \cite{Stamatescu} has shown such promise from the time development of the shower, and other uses for time measurements within the image can be investigated. The time development of a $\gamma$-ray shower shows a small time difference at ground level between Cherenkov photons that are emitted at the top of the shower compared to the bottom of the shower. The time difference is around 25 \rm{ns} \cite{Stamatescu}. Such a small time development can be utilised to apply additional cuts to the data which passes the cleaning algorithm. An extra cut has been investigated in a later chapter in this thesis. The results have shown some significant improvements when the night sky background is brightest.


For multi-TeV showers, the images produced in a camera will be larger than GeV showers. The current image parameters have to be re-optimised for multi-TeV energies. This work is investigated within this thesis and the best results are summarised for varying levels of night sky background, varying image parameters and varying observational sites. During this study a time cut will be investigated and applied.\\


\subsection{Proposed concept : TenTen / PeX}
	In this thesis, I will discuss the concept of a new multi-TeV detector. TenTen is a proposed array concept of Imaging Atmospheric Cherenkov Telescopes (IACTs) which aims for a collecting area of 10 \rm{km$^{2}$} above 10 TeV or TenTen \cite{Tentenrowell}. At multi-TeV energies the fluxes from observed sources decrease, so any IACTs operating in this region must have a large collecting area ($\sim$ few \rm{km$^{2}$}) to obtain significant statistics \cite{tenten2007}. To achieve a large collecting area, up to 30 - 50 moderately sized telescopes could be used in a stereoscopic mode. The arrangement of telescopes is still under investigation. However, a smaller $cell$ of 5 telescopes, known as PeX (Pevatron eXplorer), could be the first step towards a multi-TeV array like TenTen of the high energy component of CTA (to be discussed shortly). The idea behind PeX comes from the design studies of Plyasheshnikov et. al. \cite{Plyasheshnikov}, who looked at a cell of 4 telescopes with large 300 to 500 \rm{m} spacing for energies from several TeV to 100 TeV.

	Since the extensive air showers are large and bright at high energies compared to GeV energies, IACTs for multi-TeV $\gamma$-ray astronomy need only have moderately sized mirrors, eg. mirror area of 23.8\rm{m$^{2}$}, and pixels, eg. $0.24^{\circ}$ pixel side length, compared to the lower energy counterparts, eg. H.E.S.S. mirror areas of 108 \rm{m$^{2}$} and pixel side length of $0.16^{\circ}$.

	Based on this the field of view for PeX will approximately be $8.2^{\circ}$ by $8.2^{\circ}$ compared to $5^{\circ}$ by $5^{\circ}$ for H.E.S.S. which has the largest field of view of the current IACTs. New information for existing sources or new sources could be discovered through a survey of the Galactic plane with PeX and then with the full array. The large field of view will allow a Galactic plane survey to be conducted within a reasonable time period (2 - 3 years).


	The major future IACT array in the design stage is CTA, the Cherenkov Telescope Array \cite{CTAdesign}, which is a growing consortium of over 25 countries. The aim of CTA is to provide an order of magnitude improvement in sensitivity compared to current IACTs at 1 TeV; to increase the effective area and detection rates to benefit transient and high energy sources; to improve the angular resolution and provide improved resolution for extended source morphologies; to provide energy coverage from 10 GeV up to and beyond 100 TeV; and to enhance the all sky survey capability of IACTs \cite{CTAdesign}. For an all sky survey, two CTA arrays have been proposed, for northern and southern hemisphere sites \cite{CTAdesign}.

The CTA concept can be broken down into three energy ranges: the low energy range $\leq$ 100 GeV, the core energy range from 100 GeV to 10 TeV and the high energy range $\geq$ 10 TeV. CTA design studies \cite{CTAdesign}, have shown a possible layout of different size telescopes arranged into a compact array for the low energies with a sparse array for high energies having a total ground coverage area of $\approx$ 3\rm{km$^{2}$}. A possible layout consists of 3 different telescope types: 
\begin{itemize}
\item for low energies: 4 telescopes with 24\rm{m} mirror diameters and a 5$^{\circ}$ field of view with 0.09$^{\circ}$ pixels, which is known as the Large Size Telescopes (LST)
\item for core energies: 23 telescopes with 12\rm{m} mirror diameters and a 8$^{\circ}$ field of view with 0.18$^{\circ}$ pixels, which is known as the Medium Size Telescopes (MST)
\item for high energies: 32 telescopes with 7\rm{m} mirror diameters and a 10$^{\circ}$ field of view with 0.25$^{\circ}$ pixels, which is known as the Small Size Telescopes (SST)
\end{itemize}
With the above mentioned layout, the calculated sensitivity using standard analysis techniques is an order of magnitude  better than H.E.S.S. for most energies \cite{CTAdesign}. 

We could consider the PeX cell as a sub-array of the CTA SST. The results from PeX can provide an indication of how well the full CTA SST could perform. With staged funding, the SST could be built in PeX sized sub-arrays which motivates the development and investigation of a PeX sized cell. 

With PeX, the spacing between telescopes can be investigated and the results could be utilised by CTA especially for the SST. The PeX cell can determine the best telescope spacing for different energy ranges. Results in Chapter 4 suggest that at 10 TeV the optimum spacing could be 200 to 300 \rm{m} and at 100 TeV the optimum spacing could be 500 to 800 \rm{m}. Therefore, PeX could be considered a pathfinder detector to the high energy component of CTA.


The results presented in this thesis focus on the design of PeX. The standard reconstruction parameters are discussed and optimised for the multi-TeV energy range. Discussion is presented for varying levels of night sky background, varying cell configurations and two different observational levels to prove the benefit of a low altitude observational site, such as those found in Australia.

A new time constraint has been added to the Cherenkov image processing algorithm to provide improved performance for the reconstruction algorithm. The new time constraint has been tested for a low and high altitude observational site and with varying levels of night sky background. A key result is the improvement the new time constraint provides when varying or high levels of night sky background are present. The results show a strong robustness to different night sky background levels.

Finally, future upgrades to PeX will be discussed in the final chapter.

\begin{figure}[h]
\includegraphics[scale=0.5]{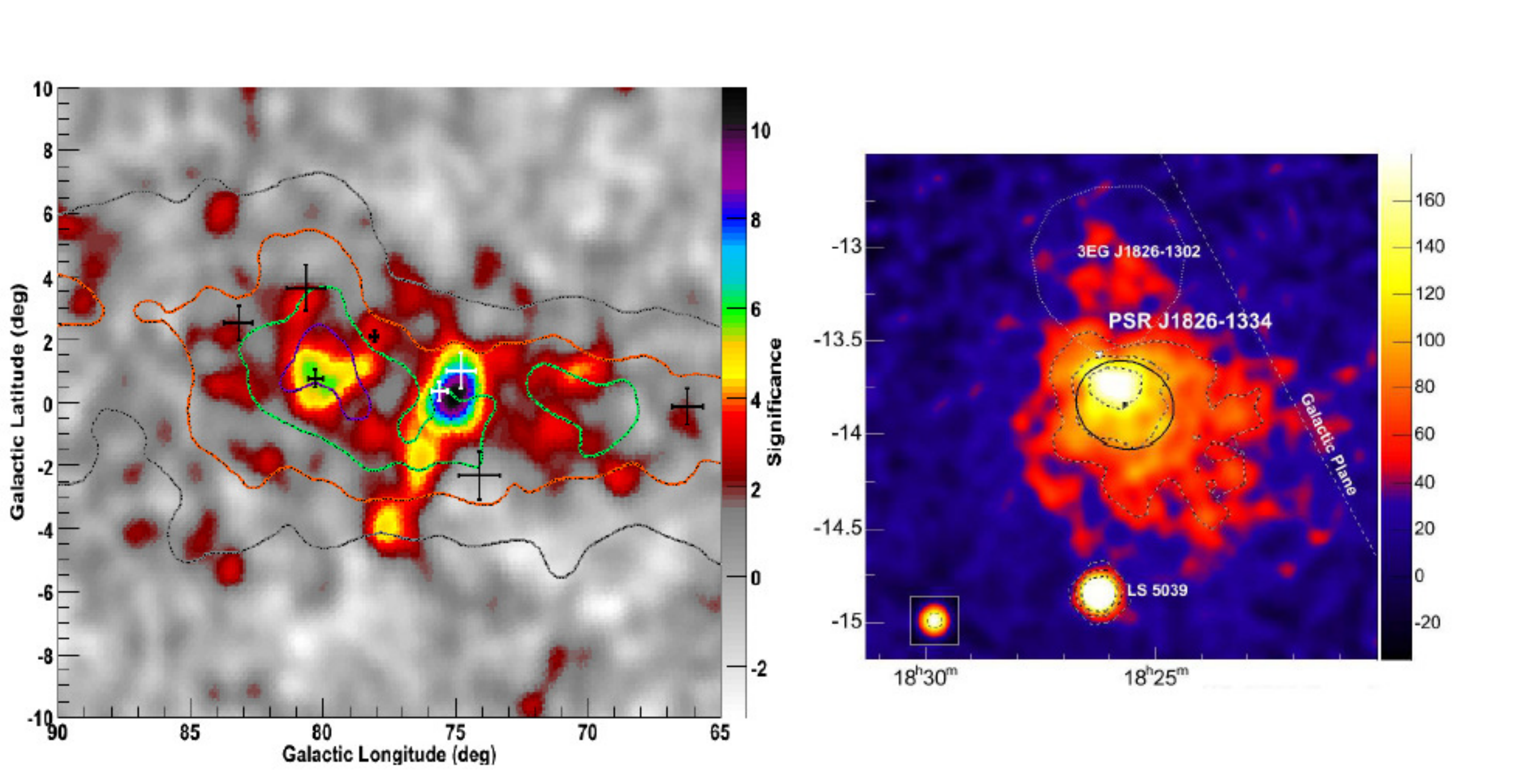}
\captionsetup{width=13cm}  
\caption{Left: Image of the diffuse TeV emission seen from the Cygnus region observed by Milagro \cite{Lagage}. Milagro is the large field of view Cherenkov extensive air shower array. The crosses represent corresponding EGRET GeV sources in the region. The contours represent matter density \cite{Lagage}. The energy threshold of the detection is $\sim$ 12 TeV. This shows the rough size of extended morphology seen in multi-TeV sources. Right: H.E.S.S. observation of PWN J1825-137. A total observing time of 53 hours has revealed that the unidentified EGRET source 3EG J1826-1302 to the north of the PWN emits weakly in the TeV energy regime, E $\approx$ 10 TeV. The image highlights the importance of a large field of view.}
 \label{fig:hesssource2}
\end{figure}


\chapter{Ground-Based Detection of Gamma-rays}

\section{Brief overview of $\gamma$-ray detectors}
	
	In the previous chapter it was mentioned that $\gamma$-ray detection provides tracers for particle acceleration. The detection of $\gamma$-rays was first undertaken by a number of early balloon-borne and satellite detectors in the 1950's and 60's. Satellite detectors produced the first significant $\gamma$-ray detection with the third Orbiting Solar Observatory (OSO-3) in 1967, which detected 621 events above 50 MeV \cite{OSO}. Other influential early space satellites were NASA's second Small Astronomy Satellite (SAS-II) in 1972 \cite{SAS} and the European collaboration COS-B in 1975 \cite{COS}. 

	The next generation of satellite detectors included the Energetic Gamma Ray Experiment Telescope (EGRET) \cite{EGRET} launched in 1991, which was based on SAS-II and COS-B. EGRET searched for $\gamma$-rays with energies from 20 MeV to 20 GeV. EGRET was on board the Compton Gamma Ray Observatory (CGRO), which contained three other detection instruments. The Burst And Transient Source Experiment (BATSE), was used to detect solar flares and cosmic $\gamma$-ray bursts occurring between 10 keV and 10 MeV. It also alerted the other instruments on-board the CGRO of incoming solar flares which activated a special burst mode to avoid unwanted interference from solar activity. The Oriented Scintillation Spectrometer Experiment (OSSE), had an objective to conduct spectroscopy of the cosmic $\gamma$-ray sources and solar flares between 100 keV and 10 MeV. Lastly, the Imaging Compton Telescope (COMPTEL), was used to detect $\gamma$-rays at energies from 1 to 30 MeV. 

	The Fermi large area telescope (Fermi-LAT) is a satellite detector, which has been operational since 2008 \cite{FERMIwebsite}. Fermi-LAT covers an energy range from 20 MeV to above 300 GeV with a field of view which covers approximately 20$\%$ of the sky at any time. It continuously scans the sky and can provide a full scan of the sky within three hours. Fermi-LAT is continuously providing new and exciting results to fill the gap in energy range between previous-generation satellite and ground-based detectors.

	For ground-based detectors, P. Blackett in 1949 \cite{Blackett, Lidvansky} suggested that part of the night sky background contribution could be produced by Cherenkov radiation from cosmic rays. He introduced tha fact that the air showers could generate Cherenkov radiation in air and not just in dense media. In 1952, Galbraith and Jelley conducted many experiments using a small Cherenkov detector. They discovered very short flashes of light on the background of the night sky glow \cite{Jelley}. They later established that the flashes were associated with the extensive air showers of cosmic rays and the production of Cherenkov radiation. 

	In 1960, Chudakov constructed an array of 12 detectors in Catsiveli, Crimea. Each detector used a single photomultiplier tube and a 1.5\rm{m} diameter reflecting mirror. This was the first detector to study extensive air shower Cherenkov radiation in detail. Chudakov provided an upper limit to the flux of high energy $\gamma$-rays from the Crab Nebula, which gave an indication of the acceleration of electrons \cite{Lidvansky}.

	The first larger mirror ground-based detector was the 10\rm{m} single telescope detector at the Whipple observatory in Arizona with an energy range from 300 GeV up to 10 TeV. The Whipple telescope in 1989 provided the first detection of the Crab Nebula in TeV $\gamma$-rays \cite{whipple}. The strong 9$\sigma$ detection demonstrated the power of ground-based detectors, which utilised the imaging atmospheric Cherenkov technique. Since Whipple, several IACTs have employed multiple telescopes in a stereoscopic technique, such as the High Energy Gamma Ray Array (HEGRA) \cite{HEGRAsite}. The design and results from HEGRA and Whipple, encouraged similar designs for the current IACTs, which are continually providing new results (MAGIC, VERITAS, H.E.S.S., CANGAROO and CAT) in the field of $\gamma$-ray astronomy. 

	 The main limitation of satellite detectors is their small collecting area. For EGRET the collecting area was 1500 \rm{cm$^2$} while Fermi-LAT has a collecting area of approximately 8000 \rm{cm$^2$}. The collecting areas of satellite detectors are limited by the equipment used to launch and deliver the detectors into space. The limit to detector size is usually 1 \rm{m$^2$} \cite{ground-based}. The $\gamma$-ray flux decreases as energy increases, so a larger collecting area is required for higher energy detection. Therefore to detect $\gamma$-rays in the higher energy regime, from 100 GeV to multi-TeV energies, a collecting area of approximately 0.1 to 10 \rm{km$^2$} is required, which is clearly not feasible with a satellite detector.

	The ground based detectors provide the required collecting area for the desired  TeV energy regime. The difference with ground based detectors is that the detection of $\gamma$-rays is not direct like satellite detectors, instead the observations are due to the interaction of the $\gamma$-rays with the atmosphere.

	An extensive air shower (EAS) is produced when a high energy particle enters the atmosphere, travels a short distance then interacts with atmospheric nuclei. The type of EAS depends on the primary particle. If the primary particle is a electron or photon, then an electromagnetic EAS is formed. If the primary particle is a nucleon or a nucleus, then a hadronic EAS is formed which contains an electromagnetic component. 

\section{Electromagnetic EAS}
 
	Once the $\gamma$-ray\footnote[1]{I will only refer to primary photons as the very high energy particle that initiates the electromagnetic EAS. We will not consider electrons as the primary particle, although electrons do initiate electromagnetic showers.} reaches our atmosphere, pair production initiates the electromagnetic EAS. The first interaction is between the primary photon and the Coulomb field of an atmospheric nucleus creating an electron/positron pair, known as pair production. The electrons and positrons are then deflected by the presence of other atmospheric nuclei, emitting photons, in a process known as bremsstrahlung. Each photon continues to travel through the atmosphere until it passes through the Coulomb field of another atmospheric nucleus which causes the photon to produce another electron/positron pair, and so on. The main difference between the interactions of the EAS particles in an electromagnetic shower is that, in the bremsstrahlung process only part of the initial electron energy is transferred to the photon, while in pair production the initial photon energy is completely transferred to the resultant electron/positron pair. Therefore, the initial photon which causes pair production ceases to exist, while the electron in bremsstrahlung continues to travel through the atmosphere causing more interactions.
	
	Bremsstrahlung interactions continue to occur until the energy of the electron drops below some critical energy whereby ionisation losses begin to dominate.\\ 

	The distance in the atmosphere travelled by the electron, positron or photon before an interaction is characterised by the radiation length, for the electron or positron, or the mean free path, for the photon. A radiation length is interpreted as the average column density of material required to reduce the energy of the electron to 1/e of its initial energy. In other words the energy loss rate of electrons, $dE/dx$ in \rm{eV (g\,cm$^{-2}$)$^{-1}$}, via bremsstrahlung is directly proportional to the initial energy of the electron, $E_{o}$, with proportionality constant $X_{L}$, known as the radiation length in \rm{g\,cm$^{-2}$} \cite{Bezak}:

\begin{eqnarray}
-\frac{dE}{dx} = \frac{E_{o}}{X_{L}} \;\; \rm{[eV\,(g\,cm^{-2})^{-1}]}
\end{eqnarray}
Integrating this gives:

\begin{eqnarray}
E(x)=E_{o} \exp{\frac{-x}{X_{L}}}\;\;\rm{[eV]}
\end{eqnarray}
where $x$ is the column density traversed, in \rm{g\,cm$^{-2}$}, by the electron, $E_{o}$ is the initial electron energy and $E(x)$ is the energy remaining after the electron has travelled length $x$.
	
	The mean free path for pair production, $\lambda$, happens to be very similar to the radiation length for bremsstrahlung and is given in terms of the radiation length, $\lambda=(9/7) X_{L}$. This fact implies that the EAS at its maximum size contains roughly equal parts of $\gamma$-rays, electrons and positrons, and the shower's growth may be understood in simple terms. 
	

A simple cascade model was established by Heitler \cite{Heitler} using certain assumptions. The model simulates the growth of the EAS, calculates the maximum number of particles produced in the EAS and determines the peak in EAS production. It includes a critical energy threshold, E$_{c}$, where energy loss processes switch from bremsstrahlung and pair production to ionisation and collisional losses. The critical energy occurs when the rate of energy loss by electrons via bremsstrahlung is equal to the rate of energy loss by ionisation. When the particle's energy is lower than the critical energy, no new particles are produced. Therefore when the energy of the shower particles reaches the critical energy, the maximum size of the shower is reached. The restrictions used in the model are:
\begin{itemize}
\item The initial energy of the electron is much greater than the critical energy, E$_{c}$. Once the energy of the electron drops below the critical energy, ionisation dominates over bremsstrahlung.
\item The model assumes that an electron with $E > E_{c}$ radiates half of its energy to a photon via bremsstrahlung after one radiation length.
\item The model assumes that a photon with $E > E_{c}$ is converted to an electron/positron pair via pair production after one radiation length, with both particles having half the energy of the photon.
\end{itemize}
	Using the above assumptions, we are approximating that the radiation length is equal to the mean free path, $\lambda \approx X_{L}$. With each interaction the energy is divided equally between the secondary particles.  

	The electromagnetic EAS starts with the initial photon with energy $E_{i}$ (Figure~\ref{fig:Heitler2}). After the photon travels one radiation length, so at X = $X_{L}$, pair production produces an electron/positron pair with each particle gaining energy $E_{i}/2$. The electron/positron pair travel another radiation length, so X = 2 $X_{L}$ since the initial interaction, before undergoing bremsstrahlung and each forming a $\gamma$-ray with all particles having roughly equal energy $E_{i}/2$. At X = 3 $X_{L}$ in Figure~\ref{fig:Heitler2}, the particles undergo either pair production or bremsstrahlung. At this stage in the shower development we have roughly equal parts electrons, positrons and photons with a total particle count of 8 particles. This procedure continues until $E < E_{c}$ at which point no new particles are produced.

\begin{figure}[here]
\begin{centering}
\includegraphics[scale=0.7]{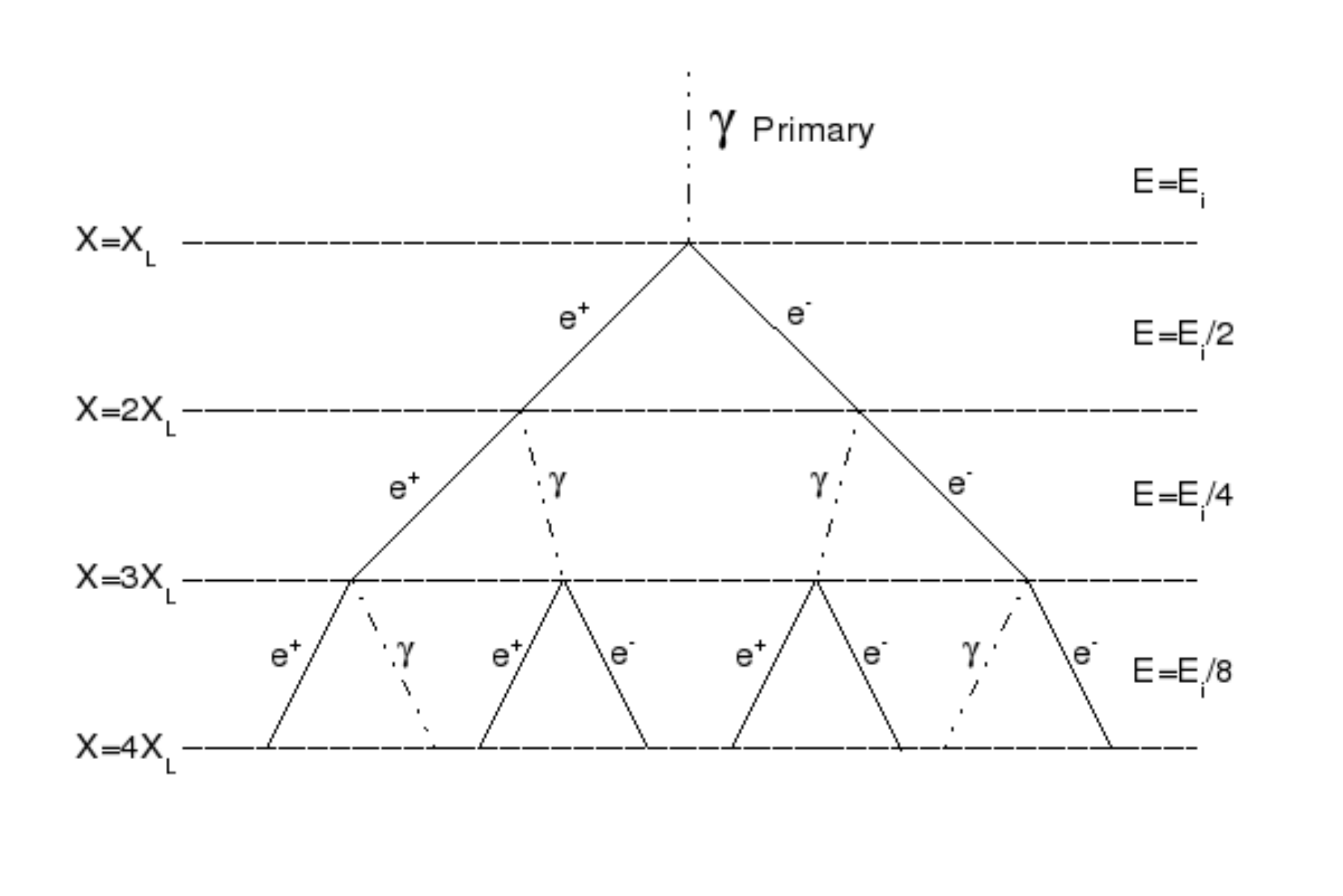}
\captionsetup{width=13cm}  
 \caption{The growth of the electromagnetic EAS after each radiation length. On the left is the depth in the atmosphere, X, represented by radiation lengths, X$_{L}$. The energy is divided equally between the particles with approximately equal components of each particle type. The figure is based on the Heitler model \cite{Heitler}.}
 \label{fig:Heitler2}
\end{centering}
\end{figure}

	The branching model indicates that after $n$ radiation lengths the shower size would have grown to incorporate $2^{n}$ particles with roughly equal numbers of electrons, positrons and photons, each having energy $E_{i} / 2^{n}$. At the critical energy, the shower particle count should be at a maximum since no new particles are produced after this point. The maximum particle count, $N_{max}$, occurs when all particles have the critical energy with $n$ being the number of radiation lengths such that:

\begin{eqnarray}
N_{max} = 2^{n} = \frac{E_{i}}{E_{c}}
\end{eqnarray}
The shower reaches the maximum particle count at a depth in the atmosphere X$_{max}$ where the particle energies are $E_{c}$ and no new particles are produced. With $n\,=\, X_{max}/X_{L}$, the depth of shower maximum, $X_{max}$ can be calculated:

\begin{eqnarray}
E_{c} = \frac {E_{i}}{2^{X_{max}/X_{L}}}
X_{max} \approx  X_{L} \frac {\ln E_{i}/{E_{c}}}{\ln 2}
\end{eqnarray}

The Heitler model \cite{Heitler} indicates that the depth of shower maximum, $X_{max}$, is proportional to the logarithm of the initial energy of the particle while the maximum number of particles, $N_{max}$, is proportional to initial energy of the particle. The model shows that the growth of the shower is approximately exponential until the maximum is reached \cite{Berge,Bezak,Hiller}. After the shower maximum, the previously neglected ionisation losses reduce the size of the shower. 

The lateral extent of the electromagnetic shower depends on photon scattering in the atmosphere and multiple Coulomb scattering of the electrons and positrons. The direction of secondary particles emitted from pair production and bremsstrahlung deviate from the direction of the primary particle. The transverse momentum gained from these interactions for photon scattering is roughly equivalent to the order of a single electron mass \cite{Stanev}. The angular deviation, $\delta \theta$, between the primary particle track and the secondary particle track is minimal. 
The major contribution to the shower lateral distribution comes from multiple Coulomb scattering for electrons/positrons. The angular deviation between primary and secondary particle directions for multiple Coulomb scattering is represented by \cite{Gaisser, Stanev}:

\begin{eqnarray}
 \delta \theta  = (\frac{E_{s}}{E_{}})^{2} \delta X 
\end{eqnarray}
where $\delta$X is in units of radiation length, E and E$_{s}$ are defined as above.

The lateral distribution width is usually quoted in terms of the Moli\`{e}re radius, $r_{1}$,  which gives the unit of lateral spread due to multiple Coulomb scattering. The Moli\`{e}re radius gives the $90 \%$ containment radius for electrons in a given medium. The Moli\`{e}re radius at different points in a shower is given by
\begin{eqnarray}
 r_{1} = (\frac{E_{s}}{E_{}}) X\;\; \rm{[g\,cm^{-2}]}
\end{eqnarray}
where X is the path length in \rm{g\,cm$^{-2}$}, E is the energy of the particle and $E_{s} = m_{e}c^{2} \sqrt{4 \pi / \alpha}$, which is the effective energy for Coulomb scattering $\approx$ 21 MeV \cite{Stanev}.

If the Moli\'{e}re radius is calculated for the critical energy for electrons, $E_{c}$, then $r_{1}$ = 9.3 \rm{g\,cm$^{-2}$}. At sea level the density of air is 1.2 \rm{kg\,m$^{-3}$}, so the Moli\`{e}re radius equates to $\approx$ 85\rm{m}. The value of $r_{1}$ is dependent on the energy of the particle. At the early stages of shower development the electrons have high energies so the Moli\`{e}re radius is smaller but this is partially compensated by a lower atmospheric density. The effect of higher energies dominates over the lower density. Therefore, the Moli\`{e}re radius increases as the shower progresses through the atmosphere until shower maximum. So high energy particles, on average, have small multiple Coulomb scattering deviations per path length. These deviations are almost negligible compared to the deviations seen at low energies.

\section{Electromagnetic and Hadronic EAS comparisons}
	
	A hadronic shower forms when a cosmic ray nucleon or nucleus, most commonly a proton, enters the atmosphere. At TeV energies the interaction length and/or mean free path for a proton is larger than an equivalent electron or photon, so the proton penetrates deeper into the atmosphere before the first interaction. On average, the proton shower has a deeper shower maximum, $X_{max}$ compared to a $\gamma$-ray of the same energy.

	The nucleon or nucleus interacts with the atmosphere via inelastic collisions with an atmospheric nuclei initiating the hadronic EAS. Strong interactions produce a number of disintegration products, protons and neutrons, and pions, charged and neutral. The higher energy nucleons and the primary particle continue colliding with atmospheric nuclei producing more disintegration products forming the nucleonic component of the hadronic EAS. 

	The charged pions, $\pi^{-}$ $\pi^{+}$, decay with a lifetime of $t = (2.6033\pm0.0005) \times 10^{-8}$\rm{s} \cite{pion} to muons and their neutrino/anti-neutrino counterparts. The charged pion decay products form the muonic component of the hadronic EAS. The neutrinos and long-lived muons retain a significant amount of the energy from the initial particle, which limits the energy available for further particle production. This reduces the number of particles at shower maximum for a hadronic EAS which causes hadronic EAS to possesses fewer particles than an equivalent-energy electromagnetic EAS. 

	The neutral pion, $\pi^{o}$, rapidly decays, $t = (8.4\pm0.6) \times 10^{-17}$ \rm{s} \cite{pion}, into two $\gamma$-ray photons of equal energy. The decay initiates the electromagnetic component of the hadronic EAS, with the electromagnetic component growth explained in the previous section. Roughly one third of the initial primary hadron energy ends up in the electromagnetic component of the proton EAS \cite{rowellhon}. So the electromagnetic component of a 1 TeV proton shower typically produces the same number of particles as a 300 GeV $\gamma$-ray shower. 

The lateral extent of electromagnetic and hadronic EAS is dominated by the types of interactions the particles undergo within the atmosphere. The electromagnetic EAS is produced by elastic interactions and processes which eject the secondary particles in the forward (downward) direction with small deviations from the primary particle trajectory. A hadronic EAS is produced by inelastic collisions with atmospheric nuclei, which transfers a large amount of transverse momentum causing the collisional products to deviate significantly from the forward (downward) direction. This extends the lateral extent of the hadronic EAS. The pions in the shower receive the largest transverse momenta, typically p$_{t}$ $\approx$ 350 to 400 MeV \cite{Reviewpt}, which then decay forming muons and photons.

\section{Cherenkov Light Production}

	The secondary and subsequent generation particles in EAS are high energy relativistic particles that can surpass the local phase velocity of light, $c/n$, where $n$ is the refractive index of the medium. As a charged particle propagates through the medium at a velocity, $v$, it polarises the atoms within the local medium. The locally polarised medium then emits an electromagnetic pulse. The constructive and destructive interference is best illustrated by Huygen's construction of wavelets \cite{Giancoli}. If  $v < c/n$, the electromagnetic pulse will not constructively interfere. The pulses are usually out of phase and can interfere in a destructive way (Figure~\ref{fig:cherenkov}). The blue circles in Figure~\ref{fig:cherenkov} left indicate no interference from the electromagnetic pulse, for this case.

	If $v > c/n$, the pulses interfere constructively re-enforcing the emission. This is represented by the green lines in front of the constructively interfering blue circles in Figure~\ref{fig:cherenkov} right. 


\begin{figure}[h]
\begin{centering}
\includegraphics[scale=0.6]{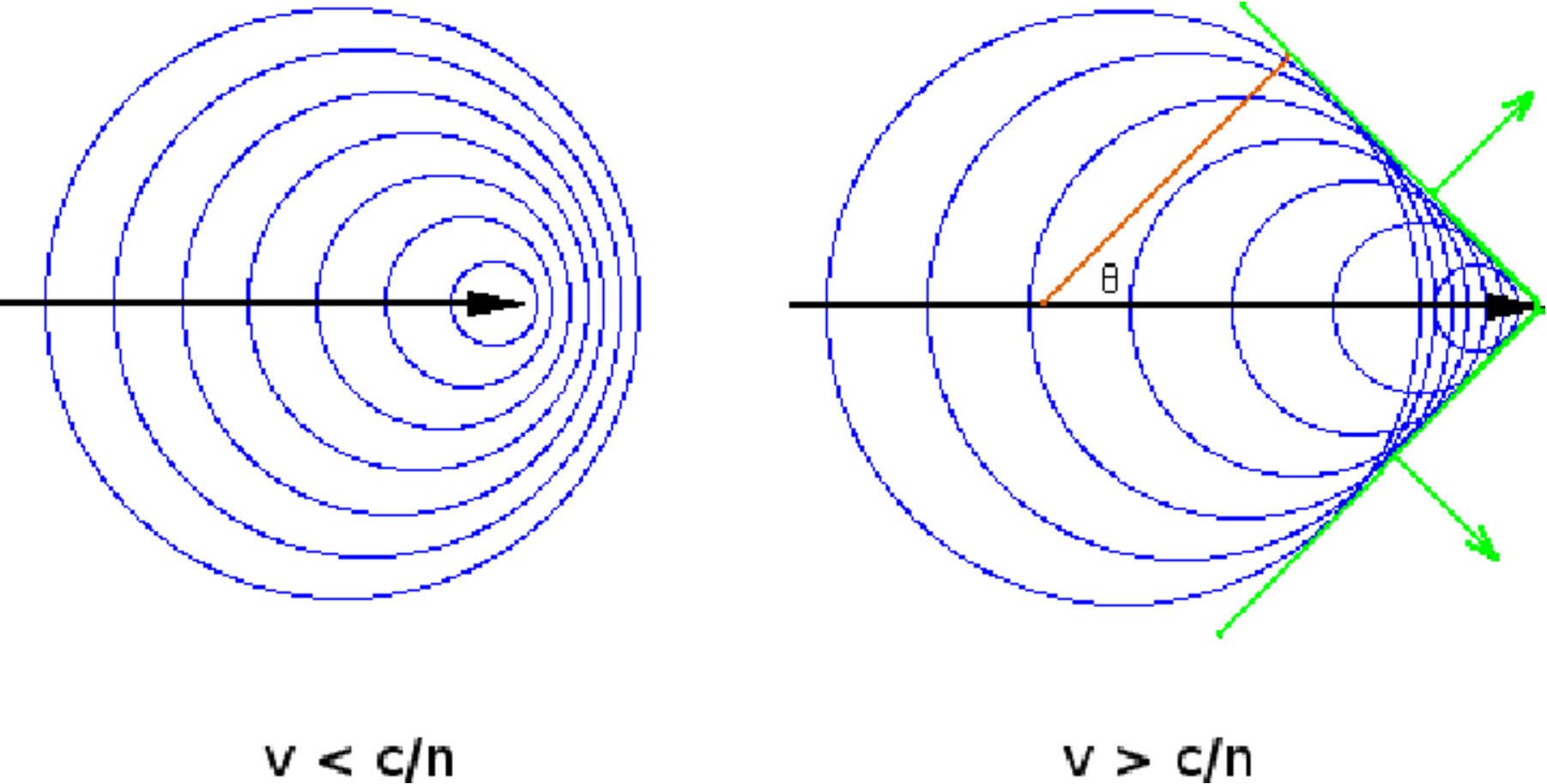}
\captionsetup{width=13cm}  
 \caption{The constructive interference of wavelets for light travelling though a medium. The black line represents the relativistic particle and its track. The blue circles indicate the electromagnetic pulses emitted from the locally polarised medium. The v $<$ c/n case represents no interference that prevents the emission of Cherenkov radiation. The v $>$ c/n case represents the emission of Cherenkov radiation from a particle travelling faster than the local phase velocity of light in the medium. So as the particle continues travelling, the wave-front continues to grow. The angle, $\theta$, is the Cherenkov angle. 
}
 \label{fig:cherenkov}
\end{centering}
\end{figure}

The phenomenon is analogous to a jet travelling at speeds greater than the local speed of sound. A shockwave is created which is the result of constructive interference. The constructive interference is similar to the wavelets re-enforcing along the wave-front at an angle $\theta$ with respect to the direction of motion the angle the black line particle trajectory makes with the green line wavefront in Figure~\ref{fig:cherenkov}. Cherenkov radiation propagates perpendicular to this wavefront. The angle between the charged particle track and wave-front is known as the Cherenkov angle, $\theta_{c}$:

\begin{eqnarray}
\theta_{c} = \arccos(\frac{1}{n \beta})\;\; 
 \label{eqn:cherenang}
\end{eqnarray}
where $\beta$ = \textit{v}/c.

	The maximum Cherenkov angle is achieved when $\beta$ approaches 1. The angle is also dependent on the refractive index, $n$, of the medium. The atmospheric refractive index has an exponential dependence on the altitude, h (m), such that: 

\begin{eqnarray}
n(h) = n_{o} \exp \frac{-h}{h_{o}}
 \label{eqn:index_atm}
\end{eqnarray}
with the sea level refractive index $n_{0} = 1.00029$, a scale height of $h_{0} = 7500$ \rm{m}. Therefore, as the shower develops and progresses through the atmosphere towards the observational level, the Cherenkov angle increases as the refractive index increases. So the spread of Cherenkov light is larger towards the shower maximum compared with earlier in the shower development.

The particle energy threshold for Cherenkov light production, $E_{thr}$, is also dependent on refractive index of the medium:

\begin{eqnarray}
E_{thr} = \frac{m_{o} c^{2}}{\sqrt{1 - {n^{-2}}}}
\end{eqnarray}
where $m_{o}$ is the mass of the particle. As the refractive index decreases the threshold energy increases. Cherenkov shower maximum typically occurs around 8 to 10 \rm{km} above sea level for 1 TeV primaries where the threshold energy is approximately 83 MeV for electrons. At sea level the Cherenkov threshold energy for electrons becomes $\approx$ 21 MeV. The refractive index is 1.33 in water, giving a threshold energy of $\approx$ 1 MeV for electrons in that medium \cite{Hiller}.

	The number of photons emitted from Cherenkov radiation is proportional to $ \lambda^{-2} $ from the calculation of Frank $\&$ Tamm \cite{FrankTamm}, 
\begin{eqnarray}
\frac{d^{2}N}{dzd\lambda} = \frac{2 \pi \alpha Z^{2}}{\lambda^{2}} (1-\frac{1}{\beta^{2} n(\lambda ,z)^{2}})
 \label{eqn:frank-tamm}
\end{eqnarray}
where \textit{dN} is the number of Cherenkov photons, \textit{dz} is the position along the z-axis, \textit{d$\lambda$} is the wavelength, \textit{Z} is the atomic number of the particle that emits the Cherenkov photons, \textit{n} is the refractive index and \textit{$\alpha$} is the fine structure constant. The bulk of photons will come from the UV range, due to the 1/$\lambda^{2}$ dependency. This dependency indicates that there are many more Cherenkov photons emitted between 10 and 300 \rm{nm}.

	For an electromagnetic EAS, the lateral extent of the shower is generally smaller compared to a hadronic EAS. Figure~\ref{fig:cherenkovshower} shows a diagram of the Cherenkov light pool development for a 1 TeV $\gamma$-ray and 1 TeV proton shower. The electromagnetic shower has a generally predictable structure which produces Cherenkov radiation around the shower axis. For a hadronic EAS, the lateral extent of the shower is larger and the shower has a more random structure compared to the electromagnetic shower. The more random structure comes from the interactions in hadronic showers that are not present in electromagnetic showers. These interactions provide the secondary particles with large transverse momenta, which shifts the particles away from the main shower. Since particles are shifted away from the main shower for hadronic EAS, the radiation is emitted away from the shower. Therefore, the resultant image of the Cherenkov radiation from a hadronic shower is more scattered around the core of the shower than for an electromagnetic shower.


As a shower at zenith progresses downward through the atmosphere, the Cherenkov angle increases according to Eq~\ref{eqn:cherenang} and~\ref{eqn:index_atm}. This is represented in Figure~\ref{fig:cherenkovshower} by the solid black lines emitted from the shower axis. The solid lined box in Figure~\ref{fig:cherenkovshower} represents the main emission region for the 1 TeV $\gamma$-ray shower. It has a radial spread of $\approx$ 20 \rm{m} and is roughly at an 8 \rm{km} altitude. The larger dotted lined box in Figure~\ref{fig:cherenkovshower} is the main emission region for the 1 TeV proton shower. It has a radial spread of $\approx$ 70 \rm{m} and is roughly at a 7.5 \rm{km} altitude.

The resultant effect of the Cherenkov emission in the atmosphere with increasing Cherenkov angle is seen at sea level. The black lines from the small rectangular region produces a bunch up at 120 \rm{m} from the shower axis and produce a peak in the photon intensity (Figure~\ref{fig:cherenkovshower} intensity plot). This feature is known as the Cherenkov shoulder and is around 120 \rm{m} for $\approx$ 0.1 to 100 TeV showers at zenith. A lot of the Cherenkov light is deposited into the Cherenkov shoulder, which leaves few particles to deposit light into the region between the shower axis and the shoulder. The photon intensity in this region is lower than the shoulder and is usually constant (Figure~\ref{fig:cherenkovshower} intensity plot). Some Cherenkov light is scattered beyond the Cherenkov shoulder at 120 \rm{m}, which produces the tail of the light distribution. With incresing energy, more particles are available to deposit light into the all regions of the shower at ground level since the number of particles increases proportionally with energy. 



These features provide a photon intensity plot at ground level which is also known as the lateral distribution of Cherenkov light for the shower. To further illustrate the lateral distribution for a $\gamma$-ray shower, we performed Monte Carlo simulation for multi-TeV $\gamma$-ray showers at 30$^{\circ}$ zenith angles. The energies ranged from 1 to 500 TeV and the showers are simulated at multiple distances from the detectors. The detector used in simulations was the PeX cell. The PeX cell consists of 5 telescopes in a sparse array, which use mirrors to reflect the light onto cameras of 804 photomultiplier tubes. The photomultiplier tubes convert the photon light into an electronic signal in photoelectrons (\textit{pe}), which provide an image of the shower in the camera. A more detailed discussion of the PeX cell is in Chapter 3.

The lateral distribution profile describes how the photon intensity varies with distance from the shower axis. Figure~\ref{fig:distance} shows the Cherenkov light intensity at ground level measured as the total \textit{size} of number or photoelectrons in the Cherenkov image as a function of core distance, distance from the telescope to the shower axis, for multi-TeV $\gamma$-ray showers at 30$^{\circ}$ zenith angles. All 3 features described above can be seen in Figure~\ref{fig:distance}: (i) the constant photon intensity between shower axis and Cherenkov shoulder from 0 to 150 \rm{m} core distance; (ii) the Cherenkov shoulder at 170 \rm{m} and, (iii) the tail of the light distribution from 170 \rm{m} to 1000 \rm{m}. The different coloured points represent different energy showers, while the blue line represents a 60\textit{pe} limit on the size of an image in the camera (total image size in photoelectrons). This value is a typical cut applied in the work of this thesis. As the energy increases, the photon intensity and size of the lateral distribution increases.

The important thing to note is the large tail to the lateral distribution in Figure~\ref{fig:distance}. For low energies, E $<$ 5 TeV, the tail of the lateral distribution is small. The photons are detectable up to and including 400 \rm{m} away from the shower location. As the energy increases, the number of photons in the shower increases. With an increased number of photons, the shower can be detected at distances further away from the shower position. For the 500 TeV events, the showers can be detected up to and including 900 \rm{m} away from the shower location. The Cherenkov shoulder at 170 \rm{m} starts to disappear due to the overwhelming number of photons in the showers at high energy $>$ 100 TeV. This tail in the lateral distribution provides considerable motivation for multi-TeV studies as it potentially allows detection of EAS at large distances, hence offering a large collection area. A large collecting area is required due to the decreasing $\gamma$-ray flux as energy increases.

\begin{figure}[p]
\begin{centering}
\includegraphics[scale=0.7]{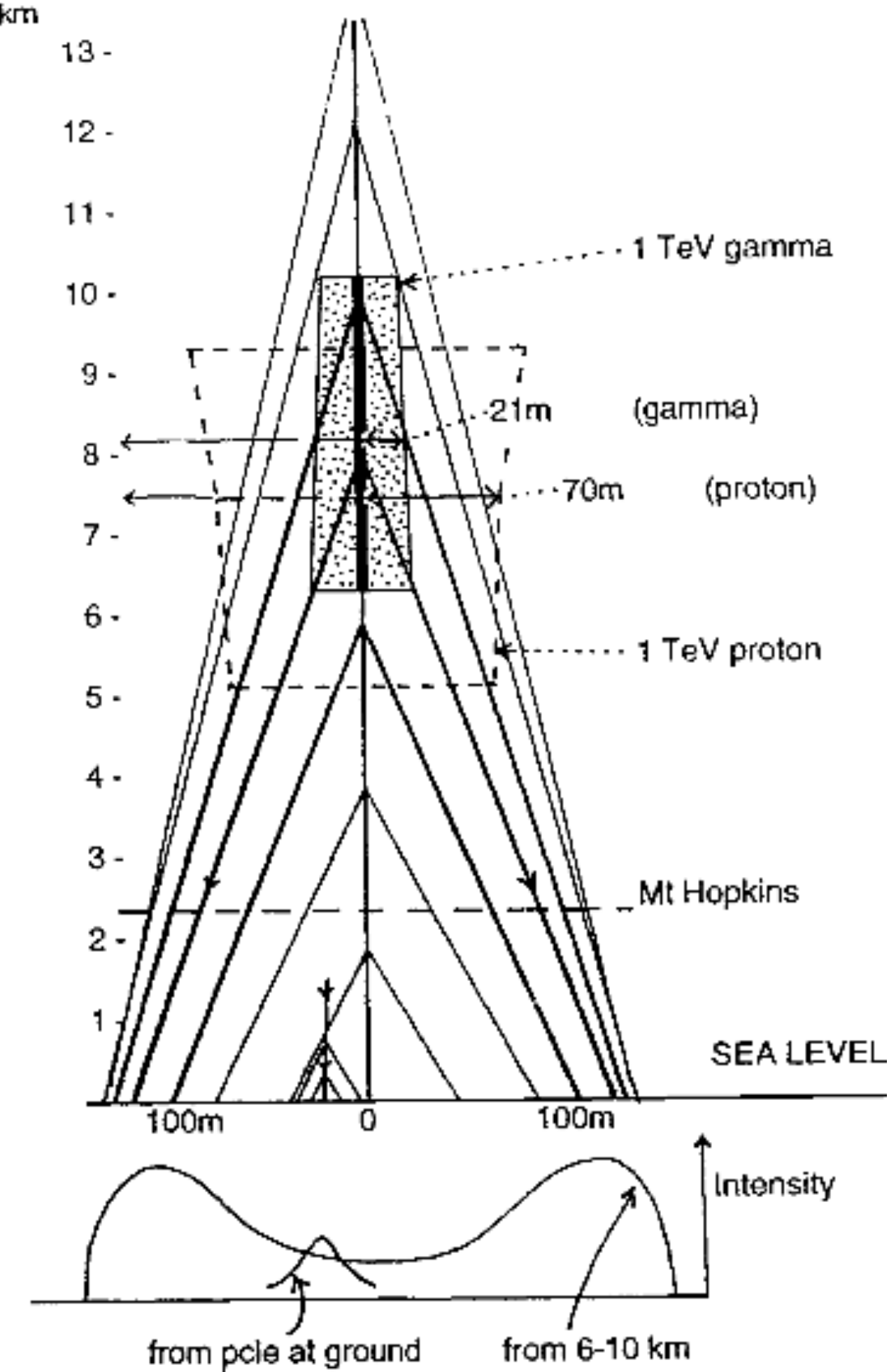} 
\captionsetup{width=13cm}  
 \caption{Cherenkov light pool development for a 1 TeV $\gamma$-ray and proton shower. For the $\gamma$-ray shower the small solid-lined rectangular box on the shower axis represent the main emission region for particles from the shower. The radial spread of the emission from the shower axis is approximately 21 \rm{m}. For the proton shower the large dashed line box on the shower axis represents the main emission region. The radial spread of the emission from the shower axis is 70 \rm{m}. The diagram shows the Cherenkov emission angles for the light emitted at different emission heights along the shower axis (the horizontal axis has been exaggerated). The intensity at ground level is shown and the peaks at $\approx$ 120 \rm{m} represent the Cherenkov shoulder. The small sub-shower to the left of the shower axis represent the intensity seen from penetrating muons and other charged particles. Figure taken from \cite{hillas1996}.}
 \label{fig:cherenkovshower}
\end{centering}
\end{figure}

The lateral distributions in Figure~\ref{fig:distance} have larger Cherenkov shoulder radii than that indicated in Figure~\ref{fig:cherenkovshower}. The difference is caused by the shower zenith angle $\theta_{z}$, the angle the shower axis makes with the zenith plus higher energies than shown in Figure~\ref{fig:cherenkovshower}. The lateral distributions in Figure~\ref{fig:distance} are produced by showers with a zenith angle of 30$^{\circ}$. Consider a shower with zenith angle $\theta_{z}$. It produces the bulk of its Cherenkov light at height h and the width of the Cherenkov light cone is 2$\theta_{c}$. The radius of the Cherenkov shoulder at the observation level can be approximately by 
\begin{eqnarray}
R_{c} = (\frac{h-h_{obs}}{\cos(\theta_{z})}).\tan(\theta_{c}) \;\; \rm{[m]}
\end{eqnarray}
where $h_{obs}$ is the observational height above sea level. If the shower enters the atmosphere at zenith, $\theta_{z}$ = 0, the Cherenkov shoulder will typically appear at about 120 \rm{m} from the shower axis. For larger zenith angles, the path length of the light from the source to the ground will increase, and the shoulder will move further out from the shower axis, as seen in Figure~\ref{fig:distance}. The Cherenkov shoulder is at 170 \rm{m} for showers which arrive with a zenith angle of 30$^{\circ}$. As the zenith angle increases, the Cherenkov shoulder shifts away from the shower position and the tail of the lateral distribution becomes larger.

\begin{figure}[h]
\begin{centering}
\includegraphics[scale=0.8]{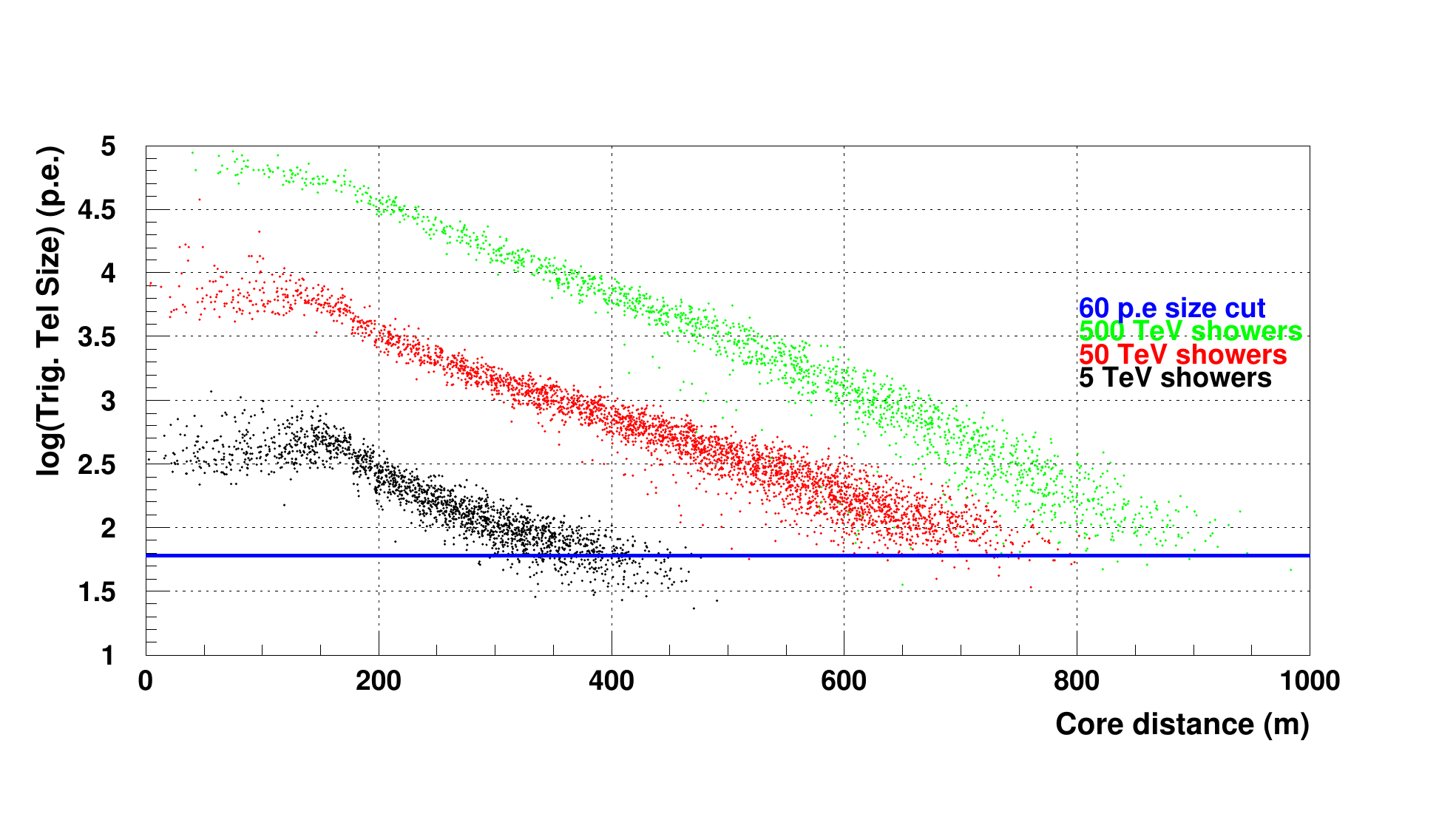} 
\captionsetup{width=13cm}  
 \caption{Cherenkov light intensity at ground level as a function of core distance for multi-TeV $\gamma$-ray showers with 30$^{\circ}$ zenith angles from simulation code. The y-axis represents the light intensity (total image size) or size of the image in the camera in photoelectrons (\textit{pe}) and the x-axis represents the core distance, distance from the telescope to the shower axis. The showers have been split into the energy bands to show the effect of increasing energy on the lateral distribution. The Cherenkov shoulder is  at a radius around 170 \rm{m}, which is further out than the shoulder ($\approx$120 \rm{m}) in vertical showers.}
 \label{fig:distance}
\end{centering}
\end{figure}

\section{Imaging Atmospheric Cherenkov Technique}

As the Cherenkov light reaches the observational level, it can be detected by a single telescope or multiple telescopes. For an IACT, the basic telescope design includes a large mirror and a camera of photomultiplier tubes that represents the focal plane for the mirror. The reflective mirror is made up of small mirror facets that are shaped into a spherical, parabolic or elliptical dish. A previous study \cite{Stamatescu} investigated the shape of the reflective mirror for PeX and TenTen. It was concluded that an elliptically curved dish would provide the best focusing at the focal plane over a wide off-axis range up to 4$^{\circ}$ \cite{Plyasheshnikov, Stamatescu}. The light is reflected from the mirror onto the camera, which contains a large number of photomultiplier tubes placed in a camera for the work described in this thesis. 

The mapping of light onto the camera depends on the angle the light makes with the telescope and the reflective mirrors. The light hits the elliptically curved mirror, which reflects the light back towards the camera, as illustrated in Figure~\ref{fig:core_distance3}. The light is mapped from a certain position in the atmosphere (\textit{x,y,z}) to the focal plane coordinates (\textit{u,v}) through the approximate expression \cite{Berge}:

\begin{eqnarray}
(\begin{array}{c}
u \\
v \\
\end{array}) = \frac{180}{\pi} \frac{1}{z} (\begin{array}{c}
x \\
y \\
\end{array})       
 \label{equation:optics}
\end{eqnarray}
where the units of (\textit{u}, \textit{v}) are degrees.

The image axis in the camera points towards the shower axis direction and the shower impact position on the ground. Light from the shower axis at height B (z = B) corresponds to the section of the image closest to the centre of the camera for on-axis observations (Figure~\ref{fig:core_distance3}, Point B). The light from the shower axis at height A (z = A) corresponds to the section of the image furthest from the centre of the camera, (Figure~\ref{fig:core_distance3}, Point A). The light from point B and point A make angles $\phi_{B}$ and $\phi_{A}$ with respect to the optical axis of the mirror. These angles determine the angular length of the image in the camera which is defined as $\Delta \phi = \phi_{A} - \phi_{B}$. 

	This angular length of the image in the camera depends on the distance of the shower core, R, with respect to the telescope. The image that forms on the camera is roughly elliptical in shape for R $>$ 0. Practically, the ellipse appears slightly wider towards the edge of the camera due to optical aberrations. The roughly elliptical nature of $\gamma$-ray shower images is due to the nature of the particle cascade in the atmosphere. The image in the camera represents the angular distribution of the particle cascade and the width of the cascade is generally shorter than its length. 


Figure~\ref{fig:core_distance3} shows the light from an image as it is reflected to the focal plane for both small and large core distance events, R. For small R the image appears elliptical and close to the centre of the camera for on-axis observations, where on-axis indicates that the telescope is pointing in a direction parallel to the shower axis. As R increases and the shower moves away from the telescope, the angles $\phi_{B}$ and $\phi_{A}$ become larger. However, the angle $\phi_{A}$ increases faster than $\phi_{B}$ since point A is closer to the ground. Therefore as R increases, the image becomes more elongated and moves towards the edge of the camera, seen in Figure~\ref{fig:core_distance3} top image. For large R there is a chance that the image could be truncated by the edge of the camera, which can cause problems for the reconstruction of the shower characteristics.

\begin{figure}[h]
\begin{centering}
\includegraphics[width=15cm]{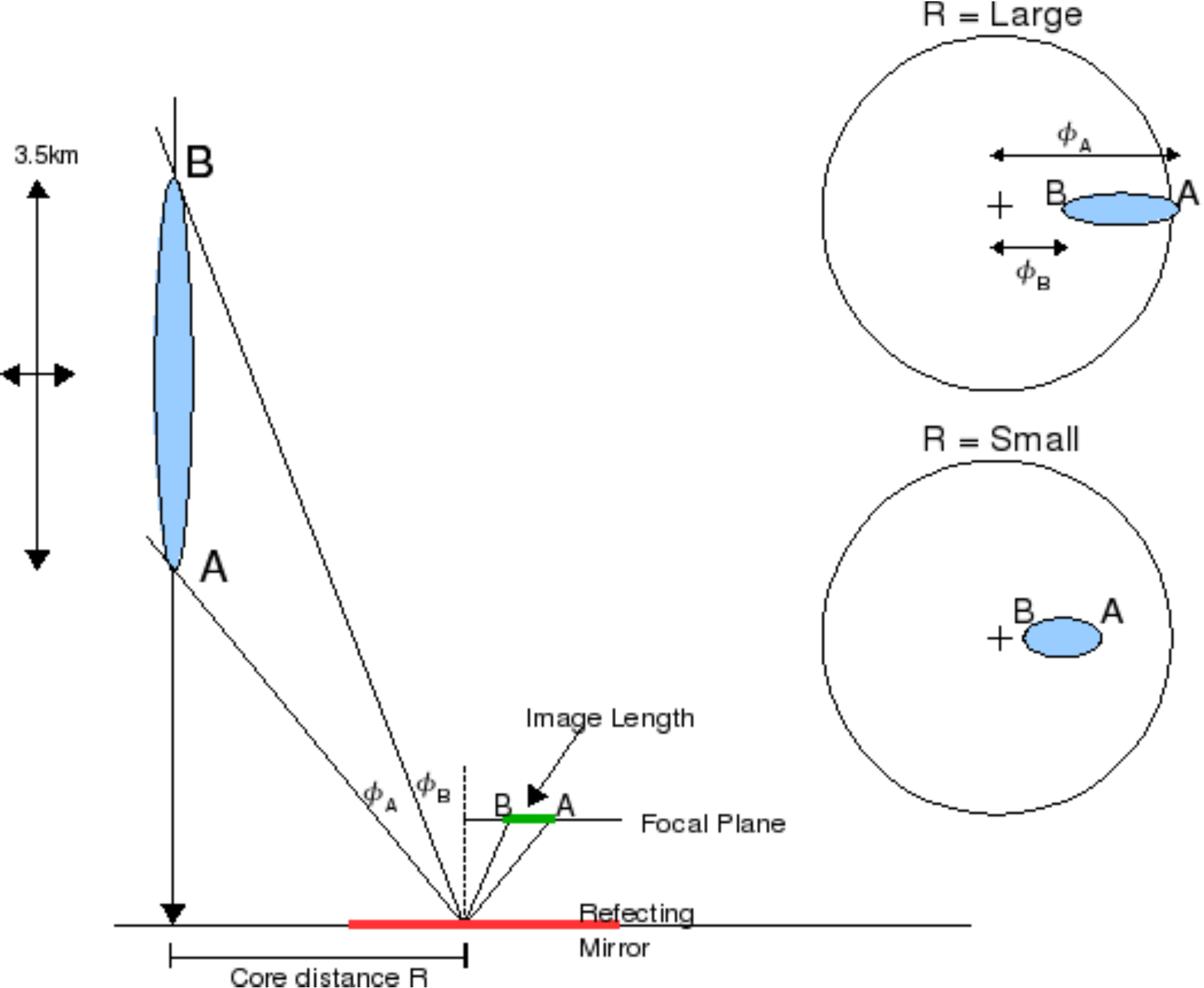} 
\captionsetup{width=13cm}  
 \caption{Shower mapping onto the camera plane via a reflecting mirror for an on-axis source. The angle, $\phi_{B}$ represents the angle between the light from the top of the shower, point B, and the optical axis of the reflective mirror and the angle, $\phi_{A}$, represents the angle between the light from the bottom, point A, of the shower and the optical axis of the reflective mirror. A camera image is shown for two values of R.}
 \label{fig:core_distance3}
\end{centering}
\end{figure}

The next chapter will discuss the concept of PeX and what happens to the Cherenkov light after it triggers the cameras in the array. This includes a discussion of telescope triggering, how the images are cleaned to reduce effects of night sky background light and how the images are parameterised so they can be used to reconstruct a shower.

\chapter{The PeV eXplorer (PeX) Five Telescope Cell: Standard Configuration}

	Current ground-based $\gamma$-ray observatories utilise the imaging atmospheric Cherenkov technique through the use of single or multiple telescopes. The MAGIC-II \cite{MAGICwebsite} telescopes comprise two large 17 \rm{m} telescopes and CAT used a single telescope \cite{CATdetector}, while H.E.S.S. \cite{Hesswebsite}, CANGAROO-III \cite{Cangaroowebsite} and VERITAS \cite{Veritaswebsite} use arrays of small to mid-sized telescopes. Table~\ref{table:IACT} represents the IACT telescope specifics.

Other $\gamma$-ray detectors include Milagro \cite{1994AAS...185.1604S}, the large field of view Cherenkov extensive air shower array. It was the first large area water Cherenkov detector use to provide a 24 hour survey of the sky. It was sensitive to electrons, photons, hadrons and muons and studied extensive air showers in the 0.1 to 100 TeV range. The detector consisted of 723 photomultipler tubes placed in a 60 $\times$ 80 $\times$ 8 \rm{m$^{3}$} pool of water and was located in New Mexico at an altitude of 2650 \rm{m}. The next generation in all sky $\gamma$-ray detectors is HAWC \cite{HAWCconcept}. HAWC will re-use the Milagro photomultipliers and re-deploy them in a different arrangement but at a 4000 \rm{m} altitude. It will have about 15 times better sensitivity compared with Milagro and will operate in the 10 to 100 TeV range.

The Tibet air shower (AS) + muon detector (MD) \cite{2010cosp382333T} experiment uses a combination of scintillator counters and water Cherenkov detectors to observe cosmic rays and $\gamma$-rays. The AS component of the array consists of 697 plastic scintillators placed on a lattice with 7.5 \rm{m} spacing and 36 scintillators placed on a lattice with 15 \rm{m} spacing to provide a 37,000 \rm{m$^{2}$} area. The MD component will consist of 192 muon detectors placed in waterproof concrete pools under the AS array and will provide a 10,000 \rm{m$^{2}$} effective area with a 1 GeV energy threshold. The muon detectors will help discriminate between $\gamma$-ray and hadron based showers. The combination of both arrays, will allow the Tibet array to effectively detect $\gamma$-rays up to approximately 200 TeV with an improved sensitivity compared to existing detectors \cite{2007ApSS309435A}. The MD component is still under construction \cite{tibetnew}.

\begin{table}[h]
\centering
\begin{tabular}{lrrrrrrrrr}
\hline
System & Lat. & Long. & Alt. & Tel. & Mirror Area & Pixels & FoV \\
 & ($\circ$) & ($\circ$) & (m) &  & (m$^{2}$) & (per tel.) & ($\circ$)  \\
\hline
H.E.S.S. & -23 & 16 & 1800 & 4 & 428 & 960 & 5 \\
VERITAS & 32 & -111 & 1275 & 4 & 424 & 499 & 3.5 \\
MAGIC-II & 29 & 18 & 2225 & 2 & 568 & 576 & 3.5\\
CANGAROO-III & -31 & 137 & 160 & 4 & 228 & 427 & 4 \\
Whipple & 32 & -111 & 2300 & 1 & 10 & 379 & 2.3 \\
HEGRA & 29 & 18 & 2200 & 5 & 8.5 & 271 & 4.3 \\
CAT & 42 & 2 & 1650 & 1 & 17.8 & 600 & 4.8 \\
\hline
\end{tabular}
\captionsetup{width=13cm}  
\caption{Table representing IACT telescope specifics. The table displays the IACT system, the coordinates, the altitude, the number of telescopes, the total mirror area, the number of pixels per telescope and the Field of View (FoV). The operating energy range for each telescope is displayed in Table~\ref{table:IACT2}. Adapted from \cite{teraelectron}}
 \label{table:IACT}
\end{table}

Here, we discuss the details of the concept of the Pevatron eXplorer or PeX. In section 1.3, the astrophysical motivations for building a multi-TeV detector were discussed. With these motivations in mind, a new approach to studying $\gamma$-rays was considered for the multi-TeV regime. The design for H.E.S.S., CANGAROO, MAGIC-II and VERITAS are optimised for 10's of GeV up to 10's of TeV. The flux for E $>$ 10 TeV $\gamma$-rays decreases with increasing energy and current detectors do not yield enough statistics due to a limited detection area $<$ 0.1 km$^{2}$. For good statistics in the E $>$ 10 TeV regime, a larger detection area $>$ 1 km$^{2}$ is required. A potential way to achieve this is by increasing the separation between telescopes in the cell/array. This creates a sparse array of telescopes. Since showers are larger and brighter, only small to moderately sized telescopes are required to collect the Cherenkov light. To investigate these claims, A.V. Plyasheshnikov et al. conducted Monte Carlo simulations with a new IACT design concept \cite{Plyasheshnikov}. The new design consisted of relatively small mirrors of 5 to 10 \rm{m$^{2}$}, large cameras with a diameter of 8$^{\circ}$ and moderately sized pixels of 0.3 - 0.5$^{\circ}$. They showed that this design combined with a 300 to 500 \rm{m} separation between 4 telescopes can provide a 1 \rm{km$^{2}$} collecting area, an angular resolution of 0.1$^{\circ}$ to 0.2$^{\circ}$ and a good energy resolution of $\approx$ 20$\%$ \cite{Plyasheshnikov}. 


The `Pevatron Explorer' or PeX, is a concept consisting of 5 small sized IACTs arranged into a square with a central telescope (Figure~\ref{fig:cell-1}) similar to the HEGRA system. The design was based on results from \cite{Plyasheshnikov}. The design incorporates 5 telescopes each with a 6 \rm{m} mirror providing a 23.8 \rm{m$^{2}$} mirror area. Each camera consists of 804 pixels arranged into a square grid with a pixel gap of 0.3 \rm{cm} between each 0.24$^{\circ}$ pixel (Figure~\ref{fig:cell-1}). PeX will provide an 8.2$^{\circ}$ by 8.2$^{\circ}$ field of view. The desired operational energy range is a few TeV to 500 TeV, which will allow energy overlap with current IACTs. PeX may be a pathfinder for a larger array known as TenTen, with an effective area of 10\rm{km$^{2}$} at 10 TeV. The TenTen detector could consist of 30 - 50 telescopes or 6 - 10 PeX cells. With multiple PeX cells combined into one array, the sensitivity and operational capabilities will improve. Another future IACT in design is CTA, which will have multiple telescopes ranging in size to cover the largest possible energy range. The high energy component may realise the goals outlined by the TenTen concept.

	In this Chapter, I will discuss the Monte Carlo simulations used in the design of PeX and the important parameters that will be studied further in later chapters.

\section{Monte Carlo EAS Simulations}	
	
	For the EAS and the emission of Cherenkov light, a Monte Carlo simulation package, CORSIKA v6.204 \cite{CORSIKA}, has been used. It models the interactions of each particle with the use of the mini-jet modelling and the interaction package SYBYLL \cite{Fletcher}. The SYBYLL package allows particle types that can be recognised by CORSIKA and deals with particles with energies as low as 12 GeV \cite{Fletcher}. Between CORSIKA and SYBYLL, the simulation packages provide the growth and development for an EAS for a variety of initial particles, up to and including iron nuclei. All particles are tracked until they interact producing new particles or reach ground level. The simulation of particle showers at multi-TeV energies with CORSIKA is time consuming and produces large output files containing shower information. To reduce the simulation time and file sizes generated, a \textit{bunchsize} parameter is used. The \textit{bunchsize} takes one Cherenkov photon and weights the photon by the \textit{bunchsize} value. If 20 Cherenkov photons are produced at one interaction length and the \textit{bunchsize} is 20, then only one Cherenkov photon is tracked and the photon is given a weighting of 20 to represent 20 Cherenkov photons. 

	The Cherenkov photon information from CORSIKA, such as the \textit{bunchsize}, position of the photon telescope level, the photon direction, the arrival time with respect to the extrapolated primary particle reaching the ground level and its emission height, are read into the telescope simulation program \textit{$sim_{-}telarray$} \cite{simtelarray}. The program \textit{$sim_{-}telarray$} is the telescope simulation program which places an array of telescopes at an observational level under the EAS from CORSIKA. The program includes detailed atmospheric models from MODTRAN \cite{Modtran}. This models a variety of atmospheres for different observational sites around the world. \textit{$Sim_{-}telarray$} deals with the transmission of the Cherenkov photons with the chosen atmosphere and follows the photons from emission height down to observational level to determine whether the photon is absorbed or scattered by the atmosphere. The Cherenkov photons that reach observational level interact with the array of telescope implemented by \textit{$sim_{-}telarray$}. To provide multiple events, one EAS from CORSIKA is used several times by \textit{$sim_{-}telarray$}. The parameter CSCAT determines the number of times an EAS is used. The events are randomly placed at different core positions. For PeX simulations, the \textit{bunchsize} and the CSCAT parameter is 20 and \cite{Stamatescu2} showed that this value is an adequate value to use. Therefore, the shower is thrown 20 times over a 1 \rm{km$^{2}$} area around the centre of the array. This results in fewer CORSIKA simulations being required which saves on output file size and simulation time. 

	For the simulations for PeX, we focus on energies between 1 and 500 TeV and we draw energies from a spectrum $dN/dE$ $\propto E^{0}$ ($\gamma$-rays) and $dN/dE$ $\propto E^{-1}$ (protons). The non-zero slope on the simulated protons provides higher proton event numbers at low energies, where statistics are low after the events have passed through the telescope simulation and selection cuts. Only protons are included in the simulations since after reconstruction and applying cuts the rate of helium is only 5$\%$ of the rate of protons \cite{Denman}. Cuts are designed to reject protons and accept $\gamma$-rays based on the shape of the images. We apply two different sets of cuts to the results: shape cuts which act on the width and length of the images, and selection cuts which include shape cuts but also a point source cut. Cuts are discussed in more detail in section~\ref{sec:mean_scaled}. Table~\ref{table:events} shows the number of events thrown, the number of events triggered, the number of events that pass shape cuts and the number of event which pass selection or all cuts. The numbers presented are for a standard PeX cell simulated at a 0.22 \rm{km} altitude site (Figure~\ref{fig:cell-1} and Table~\ref{table:stand_config_original}). The values show the total number of events that trigger the array and pass cuts. The same table is displayed in Chapter 5 for a 1.8 \rm{km} altitude site.

\begin{table}[h]
\centering
\begin{tabular}{lrrrr}
\hline
& Thrown & Triggered & Post-shape Cuts & Post-selection Cuts \\
\hline
$\gamma$-rays & 118440 & 41196 & 29061 & 23572 \\
Protons & 349200 & 39986 & 1007 & 139 \\
\hline
\end{tabular}
\captionsetup{width=13cm}  
\caption{The number of events thrown, number of events that trigger the PeX cell, number of events which pass shape cuts and number of events which pass selection or all cuts in the 1 to 500 TeV energy range. The simulations are done with a standard PeX cell at a 0.22 \rm{km} altitude site. }
 \label{table:events}
\end{table}

\begin{figure}[h]
\begin{centering}
\includegraphics[scale=0.3]{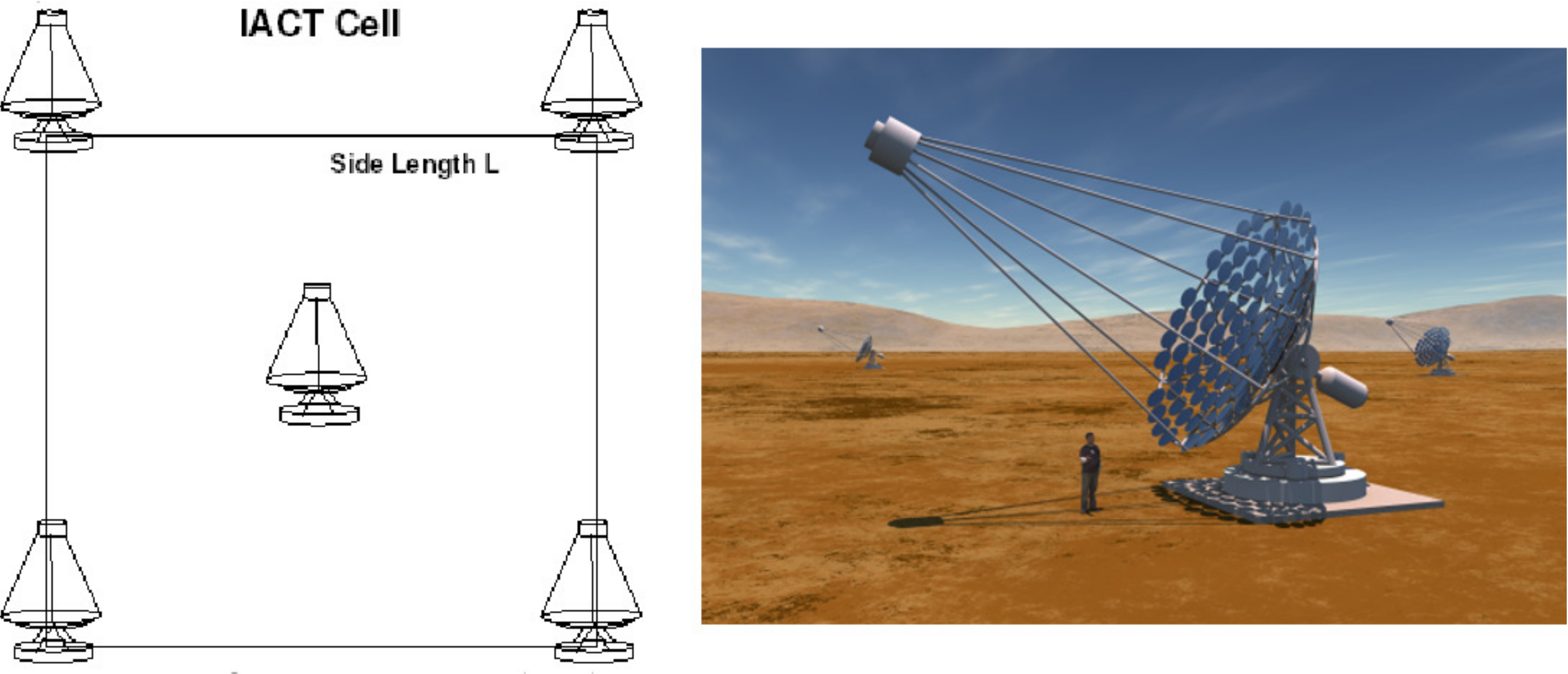}
\begin{minipage}[bottom]{11cm}
\begin{tabular}{lrrrrrrr}
\hline
Telescope Specifics\\
\hline
Parameters & Value\\
\hline
No. of Telescopes & 5 \\
Mirror diameter & 6 \rm{m} \\
Mirror area & 23.8 \rm{m$^{2}$}\\
Pixel length & 0.24$^{\circ}$\\
Pixel gap & 0.3 \rm{cm}\\
Total pixels & 804 \\
Field of view & 8.2$^{\circ}$ \\
Side Length (L) & 500 \rm{m} \\
\hline
\end{tabular}
\begin{minipage}[right]{12cm}
\includegraphics[scale=0.55]{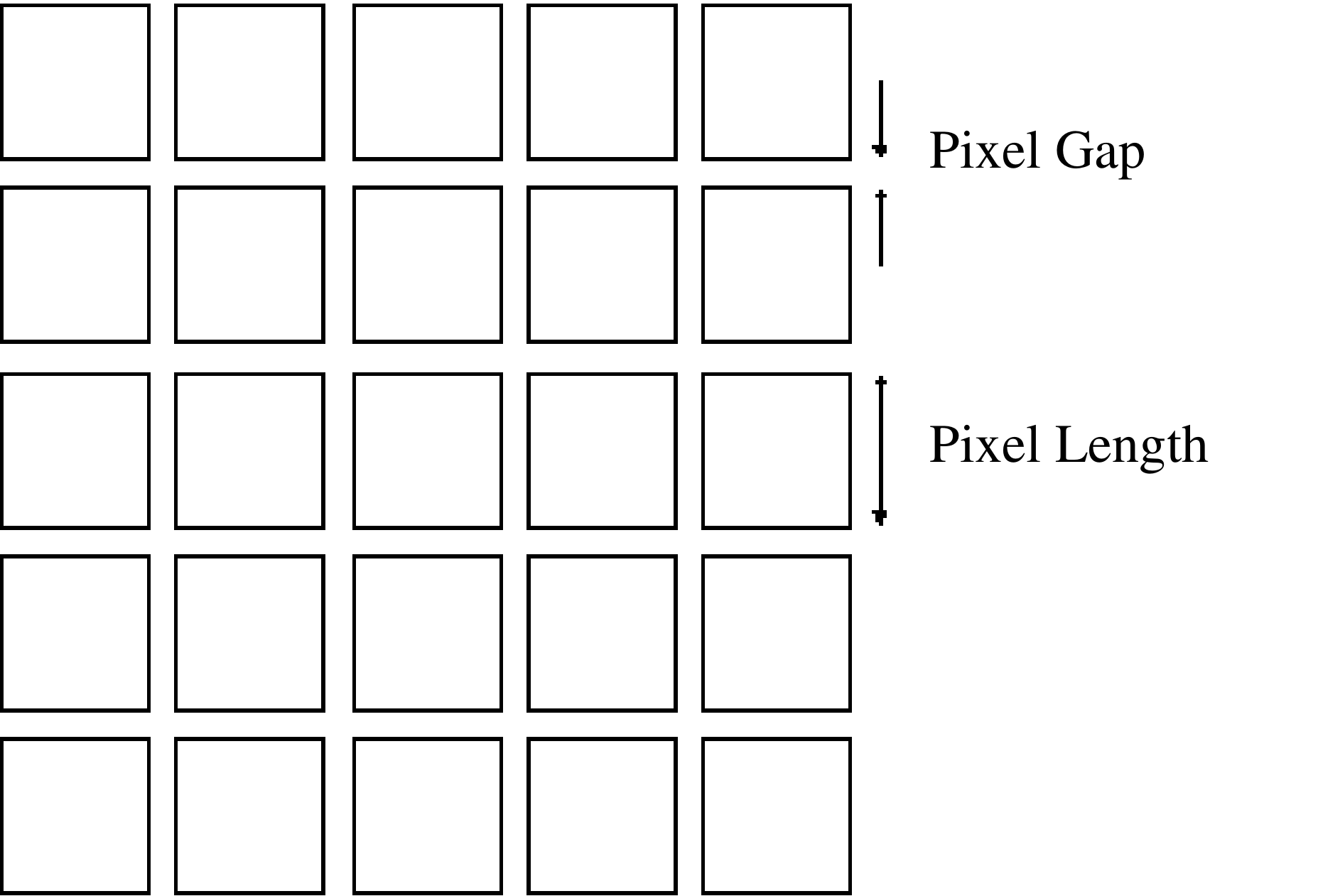}
\end{minipage}
\end{minipage}
\begin{minipage}[right]{7cm}

\end{minipage}

 \caption{Top Left: PeX or a single cell layout with all telescopes being the same size. The telescope separation or side length is represented by L. Top Right:  Scaled illustration of telescopes within the PeX cell, with a person for size comparison \cite{andrewsmith}. Bottom Left: A table displaying the telescope specifics e.g. mirror area. Bottom Right: A section of the camera illustrating the arrangement of pixels in the camera. The pixel length and the pixel gap is indicated on the diagram.}
 \label{fig:cell-1}
\end{centering}
\end{figure}

\section{Night sky background (NSB) contributions}
 \label{sec:NSB_cont}

The NSB consists of contributions from many sources, which depend on the pointing direction and the physical location of the detector. The various contributions are \cite{NSBbook}:
\begin{itemize}
\item Starlight - the light from stars within our galaxy, where the contribution to NSB increases over a wide area as the observation region approaches the Galactic plane.
\item Zodiacal light - the sunlight which has been scattered by interplanetary dust and is generally visible near the western horizon just after sunset and eastern horizon just before sunrise.
\item Air-glow - caused by photochemical reactions between the neutral and ionized parts of the upper atmosphere which results in the emission of light.
\item Aurorae - caused by energetic solar wind particles transferring energy to particles in the upper atmosphere, which produces a high level of excitation. Larger contributions occur at high latitudes.
\item Diffuse Galactic light - the light from stars scattered by the interplanetary dust. This contribution increases as the observation region moves towards the Galactic plane.
\end{itemize}

	For a low altitude site, 0.22  \rm{km} above sea level, auroral light is negligible unless at the poles and often transitory thus the other components will dominantly contribute to the NSB. Preu$\beta$ et al \cite{NSBthesis} measured the NSB level for the H.E.S.S. observational site, 1.8 \rm{km} above sea level. The NSB was averaged over multiple regions of the sky, using zenith angles $>$ 60$^{o}$ and regions outside the Galactic plane, $|$b$|$ $>$ 20$^{o}$. Any star that possessed a magnitude brighter than 6 was excluded from the calculation. The average NSB flux for the region at the H.E.S.S. site in Namibia is 2.21 $\times$ 10$^{12}$ \rm{photons (sr s m$^{2}$)$^{-1}$} for 300 \rm{nm} $<$ $\lambda$ $<$ 650 \rm{nm} \cite{NSB}. 

	To convert the average NSB flux to a rate per pixel for H.E.S.S. the following H.E.S.S. values were used: a mirror area of 94 \rm{m$^{2}$}, a pixel size of 0.16$^{o}$, an 80$\%$ mirror reflectivity, an 80$\%$ net collection of the light by the Winston cones in front of the photomultiplier tubes and an average quantum efficiency of 9.8$\%$ for 300 \rm{nm} $<$ $\lambda$ $<$ 650 \rm{nm} \cite{NSB}. The resultant NSB pixel rate becomes 0.1\rm{pe (ns pixel)$^{-1}$}. The current NSB used within the simulation code is 0.045 \rm{pe (ns pixel)$^{-1}$} since it was scaled from the H.E.S.S. value.
	
	To estimate an NSB value for a low altitude site, the NSB value from \cite{NSB} for a site at altitude 1.8 \rm{km} can be converted to a value appropriate for a 0.22 \rm{km} site. The convolution between the quantum efficiency, QE($\lambda$), of the photomultiplier tube and the input spectrum, S($\lambda$), will provide the average quantum efficiency Eq~\ref{eqn:qe},

\begin{eqnarray}
<QE> =  \int_{\lambda_{1}}^{\lambda_{2}} S(\lambda)QE(\lambda)\,d\lambda. \qquad \rm{for} \qquad 300\rm{nm} < \lambda < 650\rm{nm}
 \label{eqn:qe}
\end{eqnarray}

	The convolution between the optical depth, $\sigma$($\lambda$), from 1.8 \rm{km} to 0.22 \rm{km}, and the input spectrum will provide the average optical depth. The input spectrum will be taken from Figure~\ref{fig:input_spectrum} \cite{NSB}.

\begin{figure}[h]
\begin{centering}
\includegraphics[scale=0.6]{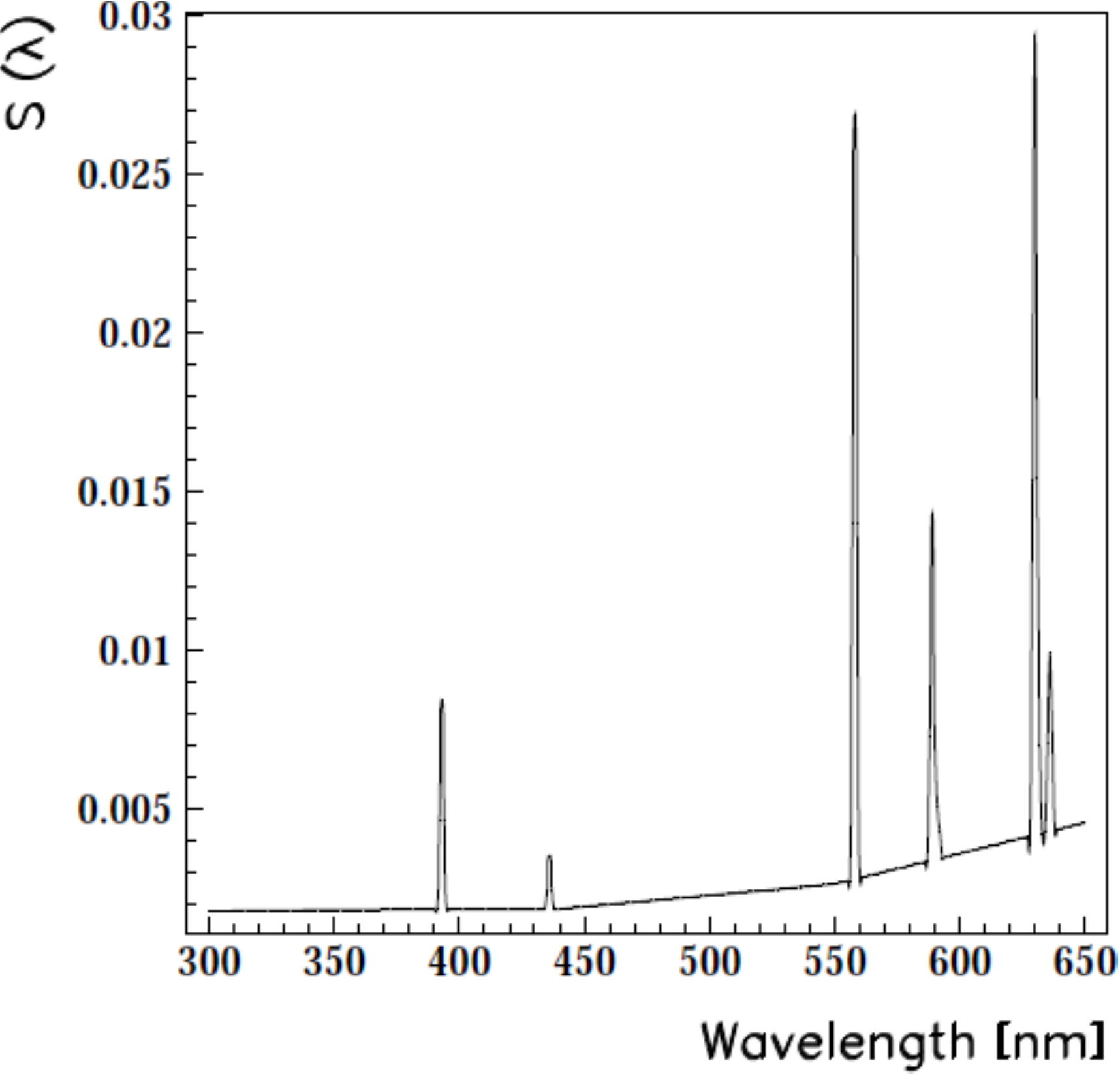}
\captionsetup{width=13cm}  
 \caption{Input spectrum used in the calculation of the average quantum efficiency of the photomultiplier and the average transmission. Adapted from \cite{NSB}}
 \label{fig:input_spectrum}
\end{centering}
\end{figure}

	
	The convolution between the input spectrum in \cite{NSB} and the quantum efficiency of the pixels used in $sim_{-}telarray$ \cite{simtelarray} provides the cell with an average quantum efficiency of 9.8$\%$ for the entire wavelength range, 300 \rm{nm} $<$ $\lambda$ $<$ 650 \rm{nm}. Using these simplifications, we can estimate values over the entire wavelength range.

The average optical depth, $<\sigma>$, between 1.8 \rm{km} and 0.22 \rm{km} is calculated, in the same way as the average quantum efficiency, to be 0.16. So the transmission of light from 1.8 \rm{km} to 0.22 \rm{km} is on average only 85$\%$. Therefore, the NSB flux down at an 0.22 \rm{km} altitude becomes 1.87 $\times$ 10$^{12}$ \rm{photons (sr s m$^{2}$)$^{-1}$}. 
	
This NSB flux can be converted to a rate per pixel for the PeX cell using a mirror area of 23.8 \rm{m$^{2}$}, a mirror reflectivity of 80$\%$, a pixel diameter of 0.24$^{o}$ and an average quantum efficiency of 9.8$\%$ for the entire wavelength range. The NSB pixel rate becomes 0.052 \rm{pe (ns pixel)$^{-1}$}. 
This values represents the average NSB away from the Galactic plane for the 0.22 \rm{km} altitude site. There is a small difference between the NSB value scaled from the H.E.S.S. NSB and the NSB from Preu$\beta$ et al \cite{NSBthesis}. The difference is so small that either value can be used in the simulation. For the PeX cell, 0.045 \rm{pe (ns pixel)$^{-1}$} has been included to represent an off-Galactic plane level of NSB.

	

\section{Telescope and Camera Specifications}
 \label{sec:trigger_def}

	For the investigation of PeX performance, 5 telescopes have been arranged into a design similar to HEGRA, the High Energy Gamma-Ray Array \cite{1997APh}. The HEGRA detector employed a 5 telescope system with 4 telescopes arranged into a square with one central telescope. For PeX, the telescopes also form a square with the 5th telescope situated in the centre of the square (Figure~\ref{fig:cell-1}). This arrangement allows telescopes to record multiple images of the shower at different positions which provides improved reconstruction compared to a system with fewer telescopes.

	The distance between the outer telescopes is known as the side length, $L$, or telescope separation. We define the standard telescope separation as 500 \rm{m}. 

	Each telescope dish within the cell contains 84 mirror facets. The combined area of the facets gives a total mirror area of 23.8 \rm{m$^{2}$} with a 6 \rm{m} diameter dish. Three dish shapes were considered: parabolic, Davies-Cotton and elliptical. Earlier studies concluded that an elliptical dish provided an improved performance for off-axis observations \cite{Schliesser,Stamatescu}. The shower image in a camera can appear slightly blurred towards the edge of the camera due to optical aberrations. In this case, coma aberrations can dominate and the effects are shown in Figure~\ref{fig:abberation} for a point source of light vs off-axis angle. At 4$^{\circ}$ the point spread function has reached the size of the PeX pixel (0.24$^{\circ}$).

\begin{figure}[h]
\begin{centering}
\includegraphics[scale=0.6]{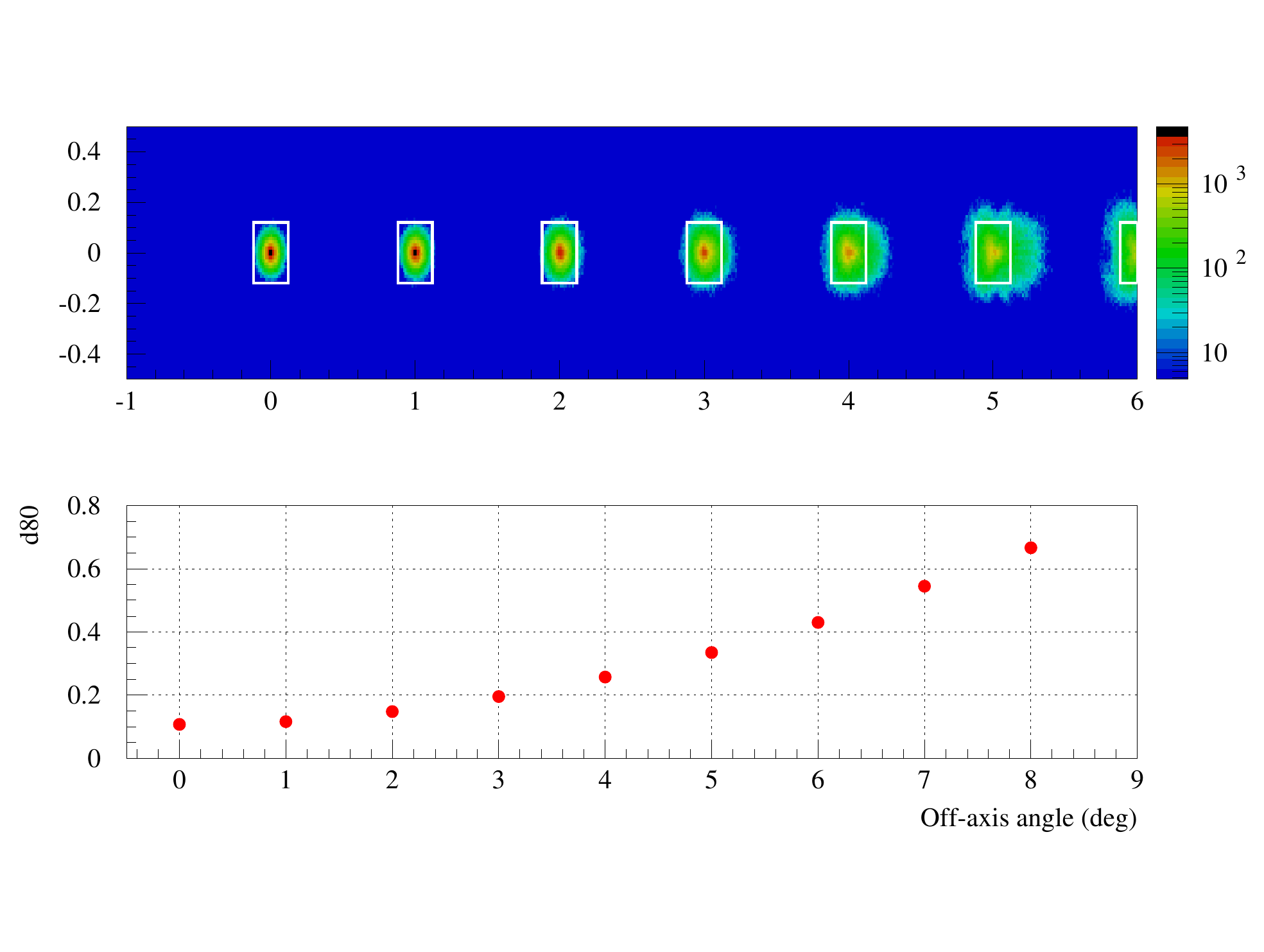}

\captionsetup{width=13cm}    
\caption{Top: The point spread function (photons per bin) compared to the pixel size (white box side length 0.24$^{\circ}$) for off-axis angles for the standard telescope in Figure~\ref{fig:cell-1}. The simulation includes the standard H.E.S.S. offsets in mirror alignment and mirror surface error. As the incoming light comes from angles further off-axis the aberrations become larger and the point spread function becomes larger than the pixel size. Bottom: The d80 (deg) curve for the point spread function. The d80 curve shows the diameter of a circle that contains 80$\%$ of the light.}
 \label{fig:abberation}
\end{centering}
\end{figure}

	We define the standard camera as being circular in shape with 804 pixels each having a side length of $0.24^{\circ}$. The square pixels are arranged into a square grid of 32 by 32 pixels with a 0.3 \rm{cm} gap between pixels. The resultant camera provides a 8.2$^{\circ}$ by 8.2$^{\circ}$ field of view. In Chapter 8, we consider the performance using alternative camera layouts and pixel sizes. Each pixel consists of a photomultiplier tube used to collect light which has been reflected from the elliptical mirror. The photons collected for each event enter the photomultiplier tubes which convert them to an electronic signal. In this case when the photon energy is absorbed by the photocathode in the photomultipler tube, it emits a photoelectron, via the photoelectric effect. The signals from the photomultiplier tubes are digitised by the Flash Analogue to Digital converters (FADCs). As the signal passes through the photomultiplier tubes, each photoelectron produces a single pulse in the FADC, which is designed to have a 100 \rm{ns} buffer. The Cherenkov pulses can pile up since they have similar arrival times in the buffer, while NSB photons on the other hand arrive randomly and so contribute to fluctuations in the combined pulse. These pile ups provide a total signal pulse in the buffer. The maximum value or peak in the FADC is used to find the signal in each pixel and to determine if an image triggers a telescope. There is usually one main peak in the FADC buffer unless the FADC only contains NSB then there could be multiple peaks in the buffer. The signal is continuously sampled at 1 GHz for all pixels.
	

	For an event to trigger a telescope it must satisfy certain conditions. First consider a 3 $\times$ 3 square grid or group of 9 pixels in the camera. The central pixel in the square grid is adjacent to the surrounding 8 pixels (Figure~\ref{fig:pixelnames2}). For a telescope trigger, the signal in $n$ adjacent pixels must surpass the threshold value. The trigger condition is usually denoted in the form \textit{threshold} value and $n$ pixels or (\textit{threshold}, $n$). The standard trigger condition for PeX is (6\textit{pe}, 3) i.e. 6\textit{pe} threshold per pixel with $\geq$ 3 pixels reaching this level, where \textit{pe} is the photoelectron count.

The triggering conditions help lower the accidental trigger rates from the night sky background. As discussed in section~\ref{sec:NSB_cont}, the assumed night sky background for PeX simulations is 1.87 $\times$ 10$^{12}$ \rm{photons (sr s m$^{2}$)$^{-1}$} or 0.045\rm{\textit{pe} (ns pixel)$^{-1}$}. 

The single pixel trigger rate comes from the probability density function for a Poisson distribution using the number of pulses observed and the mean number of pulses in a certain time interval. The probability can then be converted to the probability for a pixel discriminator surpassing a threshold value of Q in \textit{pe} in a certain time interval. The final step is to convert the probability to a trigger rate in Hz by dividing by the time interval to arrive at the single pixel rate. To find an adequate trigger \textit{threshold} and \textit{n} pixel value, the following equations can be used. The single pixel trigger rate given below, comes from \cite{PMTnightskybackground} and \cite{Ricky}:
\begin{equation}
R_{pixel}(Q) = R_{pe} \frac{(R_{pe}\tau)^{Q-1} \exp{-R_{pe}\tau}}{(Q-1)!} \; \rm{Hz}
\end{equation}
and with a certain \textit{n} pixel value:
\begin{equation}
R_{n} = n \;{^m}C_{n} R_{pixel} (R_{pixel} T)^{n-1} \; \rm{Hz}
\end{equation}
where R$_{pixel}$ is the pixel rate in Hz, T is the typical coincidence window and is around 5 \rm{ns}, $\tau$ is the pulse response and is around 2.8 \rm{ns} ,  ${^m}C_{n}$ is the number of unique pixel combinations in the camera where m $=$ 804, Q is the threshold value, R$_{pe}$ is the rate of NSB per pixel, n is the number of pixels in the triggering combination and m is the total number of pixels in the camera.

This level of night sky background provides a single pixel trigger rate and an accidental telescope trigger rate of about $10^{3}$ \rm{Hz} and $10^{-2}$ \rm{Hz} respectively for our current trigger conditions \cite{Ricky} with PeX Figure~\ref{fig:telescope_trigger} in Appendix~\ref{sec:appendix_plot} highlights the telescope trigger rate vs Q and n as part of an earlier study \cite{Ricky}.

A number of different triggering conditions were considered for the investigation and (6\textit{pe}, 3 pixels) was found to provide adequate rejection of night sky background.

	When an event triggers the telescope, the 100 \rm{ns} traces in all the FADCs are recorded. The electronic signal in each buffer has a contribution from 3 different sources $\colon$ the EAS Cherenkov light, night sky background and electronic noise. The buffers are scanned for the signal peak with a 20 \rm{ns} sliding integration window \cite{Stamatescu}. The centre of the integration window is placed on the peak in the buffer and the signal within the window is integrated to give the total signal in each pixel. The total signal is converted from an electronic signal to photoelectron counts. The signal is converted via a calibration function which converts the total signal in the intergration window in \rm{mV} to photoelectrons. Further processing to reduce the effect of night sky background is then applied as explained in the next section.

\section{Image Cleaning}
 \label{sec:cleaning_algorithm}
	A majority of the pixels contain pure night sky background and electronic noise. These pixels disrupt the shower reconstruction and hence must be removed from the image. To achieve this, the image in the camera undergoes an image cleaning algorithm which aims to reduce the effect of the pure noise pixels while leaving a majority of the image. 

	The cleaning algorithm uses a two-level tail cut system introduced by Whipple used in H.E.S.S. \cite{HESSclean} and HEGRA \cite{1997APh}. The two level tail cuts are known as picture and boundary threshold values denoted as \textit{picture} and \textit{boundary} values or (i,j). The picture pixel is taken as any pixel in the camera. Once a picture pixel is chosen (e.g the red pixel in Figure~\ref{fig:pixelnames2}), the 8 surrounding or adjacent pixels become the boundary pixels (orange pixels in Figure~\ref{fig:pixelnames2}). 

\begin{figure}[h]
\includegraphics[scale=0.48]{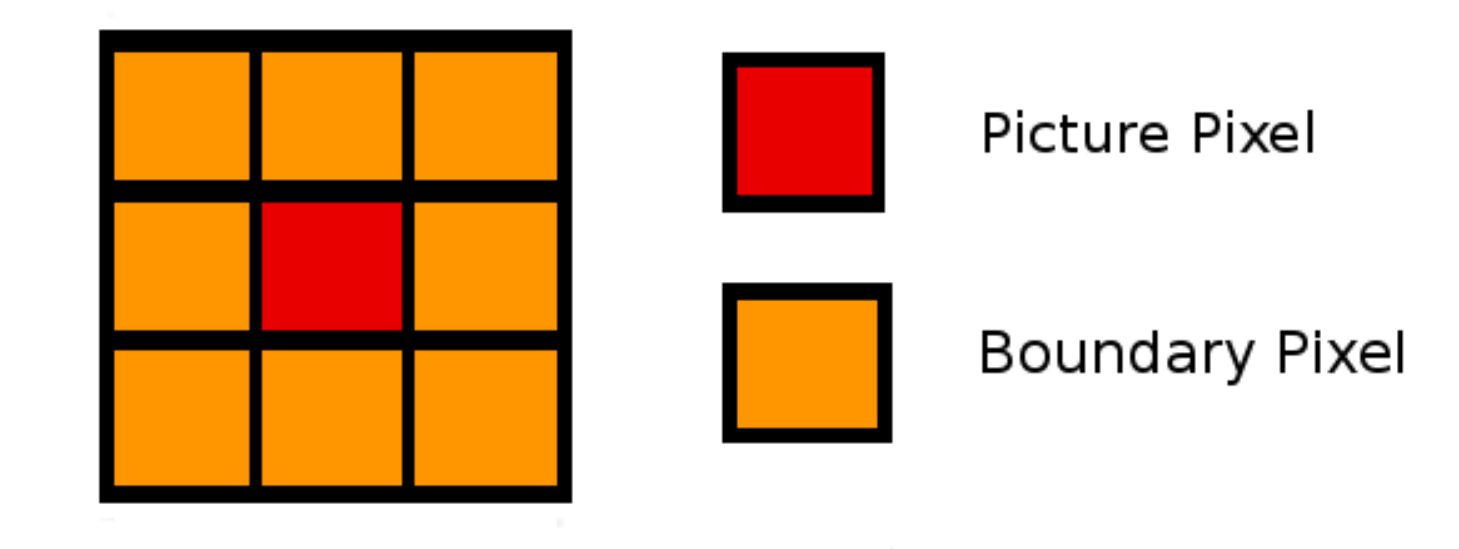}
\begin{minipage}[right]{7cm}
 \caption{A grid of 3 $\times$ 3 pixels from a section of the camera. This geometry is also used in camera triggering. The central pixel, red, represents the picture pixel while the surrounding or adjacent pixels to the central pixel, orange, represent the boundary pixels.}
 \label{fig:pixelnames2}
\end{minipage}

\end{figure}

The cleaning algorithm is used on every pixel within the camera. The algorithm methodically goes through all pixels one by one. The pixel \textit{pe} count is compared to the \textit{picture} value. If the \textit{pe} count surpasses the \textit{picture} value then the pixel is used in the final image regardless of its location in the camera (Figure~\ref{fig:cleaning_alg3} Image A). If the pixel \textit{pe} count is less than the \textit{picture} value but greater than the \textit{boundary} value, then the pixel is included in the final image if it is adjacent to a \textit{picture} pixel (Figure~\ref{fig:cleaning_alg3} Image B). Image C and Image D show two cases where the initial central pixel is not included in the final image.

To summarise the cleaning conditions:
\begin{itemize}
\item Pixels that contain more photoelectrons than the \textit{picture} value are kept in the final image
\item Pixels that contain more photoelectrons than the \textit{boundary} value but less than the \textit{picture} value are kept if an adjacent pixel meets the criteria for a picture pixel (above)
\end{itemize}

Any pixel which does not pass the cleaning algorithm has its signal set to zero while pixels which pass are used for the subsequent event reconstruction process. 

\begin{figure}[h]
\begin{centering}
\includegraphics[scale=0.5]{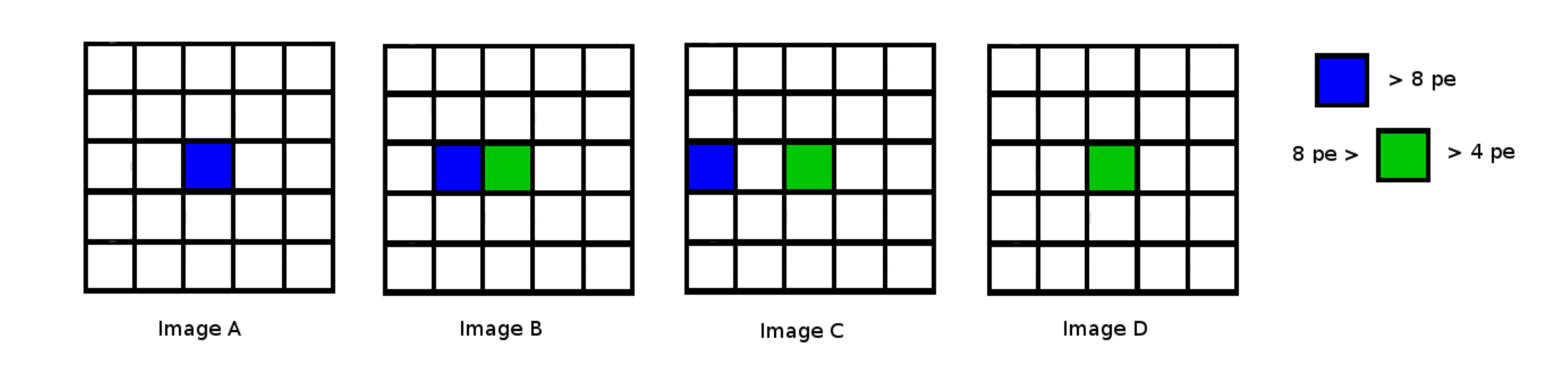}
\captionsetup{width=13cm}  
 \caption{Four different situations illustrating the cleaning algorithm. The blue pixel is $>$ 8\textit{pe} and represents a picture pixel, the green pixel $<$ 8\textit{pe} but $>$ 4\textit{pe} and represents a boundary pixel. Image A shows a picture pixel that would pass the cleaning algorithm. Image B represents a case where a boundary pixel is next to a picture pixel so it passes the cleaning algorithm. Image C and Image D represent cases where the central pixel (green) would not pass the cleaning algorithm.}
 \label{fig:cleaning_alg3}
\end{centering}
\end{figure}

The standard configuration values have been based on the H.E.S.S. values scaled to values appropriate for PeX. The scaling is done using the ratio between the pixel diameters (d$_{PeX}$/d$_{HESS}$), and the mirror diameters (D$_{PeX}$/D$_{HESS}$).  For example, the H.E.S.S. cleaning combination is (10\textit{pe}, 5\textit{pe}). To convert this cleaning combination to a value appropriate for PeX: 
\begin{eqnarray}
\rm{picture}_{\rm{PeX}} = \frac{6 \rm{m}}{12 \rm{m}} \times \frac{0.24^{\circ}}{0.16^{\circ}} \times 10\textit{pe} = 8\textit{pe}\\
\rm{boundary}_{\rm{PeX}} = \frac{6 \rm{m}}{12 \rm{m}} \times \frac{0.24^{\circ}}{0.16^{\circ}} \times 5\textit{pe} = 4\textit{pe}
\end{eqnarray}

Now that the relevant parameters have been introduced, the standard configuration (telescope specifications and image quality selection) for the PeX cell can be defined (Table~\ref{table:stand_config_original}). 

\begin{table}[h]
\centering
\begin{tabular}{lrrrrrrrrrr}
\hline
Parameters & Standard configuration\\
\hline
Number of Telescopes & 5 \\
Pixels & 804 \\
Telescope Separation & 500\rm{m} \\
Triggering Combination & (6\textit{pe}, 3) \\
Cleaning Combination & (8\textit{pe}, 4\textit{pe}) \\
\textit{Size} Cut & 60\textit{pe} \\
\hline
\end{tabular}
\captionsetup{width=13cm}  
\caption{The configuration for the standard PeX cell. The \textit{size} cut is the size of the image post cleaning.}
 \label{table:stand_config_original}
\end{table}

\section{Image Parameter Reconstruction}

	After the image cleaning algorithm, the images undergo Hillas parameterisation \cite{Hillas1985} which provides the necessary information to reconstruct images from the EAS. The reconstruction is based on the fact that the $\gamma$-ray images are approximately ellipses when viewed off-axis from the EAS core. The Hillas parameterisation appears to provide an accurate reconstruction of the shower, which can be broken down into different moments as follows (Figure~\ref{fig:hillasplot}):

\begin{itemize}
\item The zeroth order moment of the light distribution, size, is the sum over all pixel intensities (Eq~\ref{eqn:size} in Appendix~\ref{sec:hillas_a}).

\item The first order moments provide the centre of gravity, C.O.G, of the light distribution with coordinates $<$x$>$ and $<$y$>$ in degrees (Eq~\ref{eqn:first_order} in Appendix~\ref{sec:hillas_a}). The C.O.G represents the centre of the ellipse and the rough position of the shower maximum (red dot in Figure~\ref{fig:hillasplot}).

\item The second order moments $<$x$^{2}>$, $<$y$^{2}>$ and $<$xy$>$ (Eq~\ref{eqn:second_order} in Appendix~\ref{sec:hillas_a}) provide the width, W (Eq~\ref{eqn:width} in Appendix~\ref{sec:hillas_a}), and length, L (Eq~\ref{eqn:length} in Appendix~\ref{sec:hillas_a}), of the image. In Figure~\ref{fig:hillasplot}, the width corresponds to the RMS of the light distribution along the minor axis while the length corresponds to the RMS of the light distribution along the major axis. Relating these parameters to the shower, the width is somewhat correlated to the lateral distribution of the shower and the length is somewhat correlated to the longitudinal distribution of the shower.

\item The first and second order moments are combined to provide other shower information. The nominal distance, d (Eq~\ref{eqn:nomial_dis} in Appendix~\ref{sec:hillas_a}), is the distance between the telescope pointing direction and the C.O.G of the image. 

\item The major axis represents the shower axis of the EAS. The major axis is given by the vector $\vec{u}$ (Eq~\ref{eqn:major_axis} in Appendix~\ref{sec:hillas_a}). The major axis is the line which runs through the centre of the ellipse in Figure~\ref{fig:hillasplot}. 

\item The direction, $\phi$, and orientation, $\gamma$, are given by the second order moments (Eq~\ref{eqn:direction} and ~\ref{eqn:direction} in Appendix~\ref{sec:hillas_a}). The angle $\phi$ is the angle between the major axis and the x-axis of the camera while the angle $\gamma$ is the angle between the nominal distance vector and the x-axis of the camera both shown in Figure~\ref{fig:hillasplot}. 

\item For a $\gamma$-ray simulated on-axis as a point source, the centre of the camera represents the simulated source direction. The \textit{miss} parameter provides the perpendicular distance between the major axis and the centre of the camera which gives a good indication of the accuracy of the major axis and is expressed in Eq~\ref{eqn:miss} in Appendix~\ref{sec:hillas_a}.
\end{itemize}

\begin{figure}[h]
\begin{centering}
\includegraphics[scale=0.65]{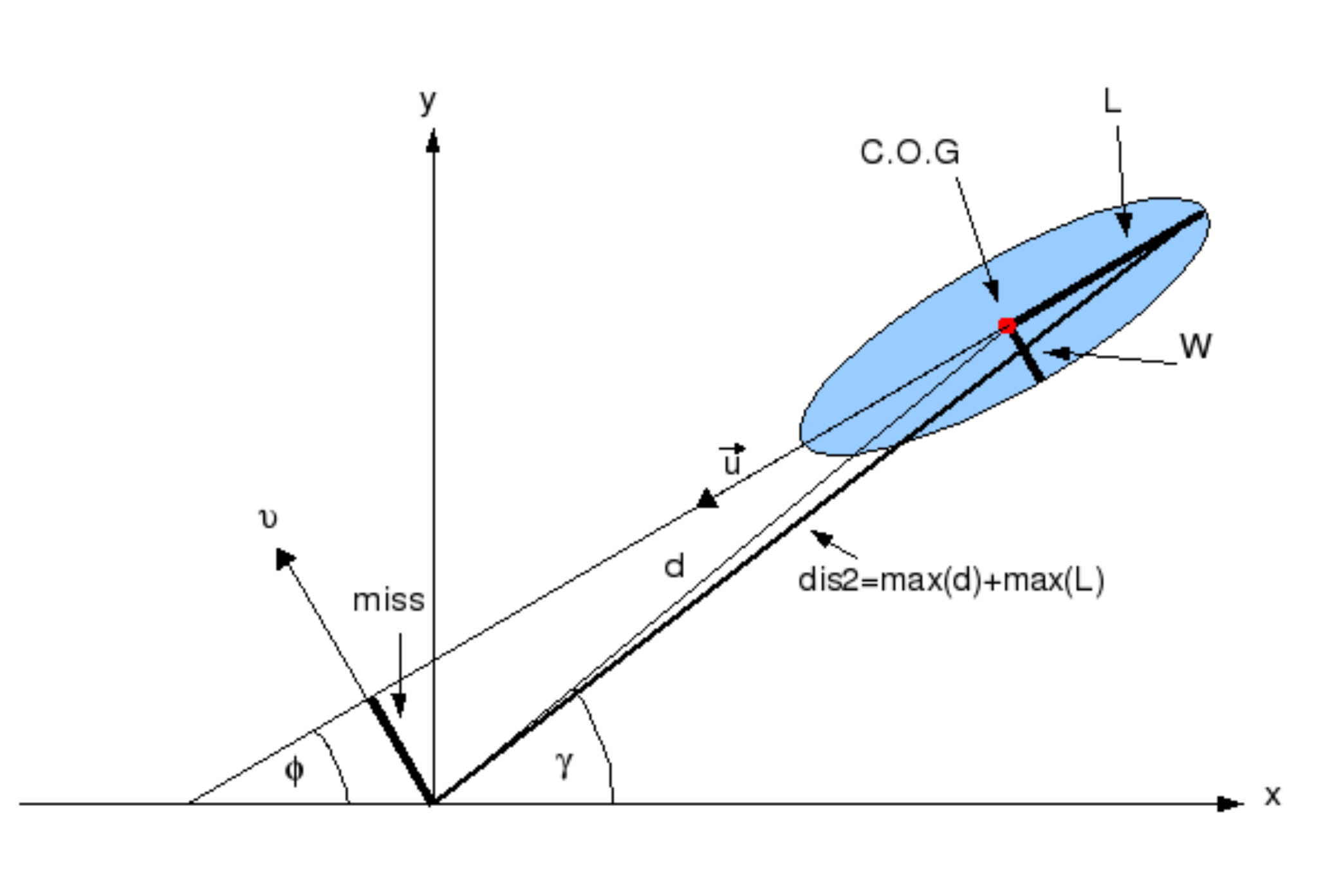}
\captionsetup{width=13cm}  
 \caption{The Hillas parameters for a single image, where (0,0) is the centre of the camera. The blue ellipse represents the image in the camera. The centre of gravity or C.O.G (red dot), image width, W, and length, L, are shown on the blue ellipse. The major axis runs through the middle of the ellipse along the image length. The angle the major axis makes with the x-axis of the camera is the direction of the image, $\phi$, and angle between the x-axis of the camera and the line between the origin and the C.O.G provides the orientation, $\gamma$, of the image with respect to the camera axis. Lastly the perpendicular distance from the major axis to the centre of the camera represents the \textit{miss} parameter, which is equivalent to $\theta$ the angular distance between the true source position and the reconstructed source position.}
 \label{fig:hillasplot}
\end{centering}
\end{figure}

The Hillas parameterisation includes some higher order moment terms (third and fourth order) that provide the skewness and kurtosis of the image respectively. For the parameterisation of the events for the cell, skewness and kurtosis will not be considered. Next the reconstruction of event directions will be discussed for several reconstruction techniques.

\section{Event Direction Reconstruction}
 \label{sec:event_reconstruction}

	For an image to be included in the reconstruction it must pass stereoscopic cuts. The stereoscopic cuts include an image \textit{size} cut, distance2 cut, \textit{dis2}, and telescope multiplicity cut, n$_{tel}$. The \textit{dis2} value is the vector sum of the maximum angular distance from the centre of the camera to the C.O.G of the ellipse plus the image length. The \textit{dis2} cut guarantees that the C.O.G is not at the very edge of the camera. It ensures that the images suffering from significant edge effects are removed. The \textit{dis2} cut also helps off-axis observations since it takes into account the orientation of the image via the image length vector \cite{Stamatescu}. The respective stereoscopic cuts are: image \textit{size} $>$ 60pe, \textit{dis2} $<$ 4.0$^{\circ}$ and n$_{tel}$ $\geq$ 2. 

\subparagraph{Algorithm 1}:\\

	After $\gamma$-ray selection cuts have been applied the parameterised images are used to calculate the core position and shower direction using a geometrical method Algorithm 1 \cite{hofmann}.

So-called `Algorithm 1' takes the major axes, provided by the Hillas parameterisation, for each pair of images and calculates the intersection point between the major axes. Then each intersection point is weighted according to the combined size of the images and the sine of the angle between the major axes of a pair of images. The weighting function, \textit{w$_{ij}$}, is given by:
\begin{eqnarray}
w_{\textit{ij}} = (size_{\textit{i}} + size_{\textit{j}})^{2}  \sin{a_{\textit{ij}}} \;\; 
 \label{eqn:weighting}
\end{eqnarray}
where $size_{\textit{i}}$ is the size of image \textit{i} in \textit{pe}, $size_{\textit{j}}$ is the size of image \textit{j} in \textit{pe} and $a_{\textit{ij}}$ is the angle between the major axes of images \textit{i} and \textit{j}. Figure~\ref{fig:gamma_image_stereo} represents multiple images from three different telescopes for a single event. Image 1 has a core distance of 362 \rm{m} and the size of the image is 1300\textit{pe}, Image 2 has a core distance of 471 \rm{m} and the size of the image is 690\textit{pe}, and Image 3 has a core distance of 310 \rm{m} and the size of the image is 2390\textit{pe}. For this example, Image 1 and Image 3 would have more weighting than Image 2 based on the size of the image. For pairs of major axes, Image 1 and Image 3 have a larger total size and larger angle between major axes and therefore would provide a larger weighting factor, \textit{w$_{13}$}, compared to the other pairs of major axes, \textit{w$_{12}$} and \textit{w$_{23}$}.

\begin{figure}[p]
\begin{centering}
\includegraphics[width=15cm]{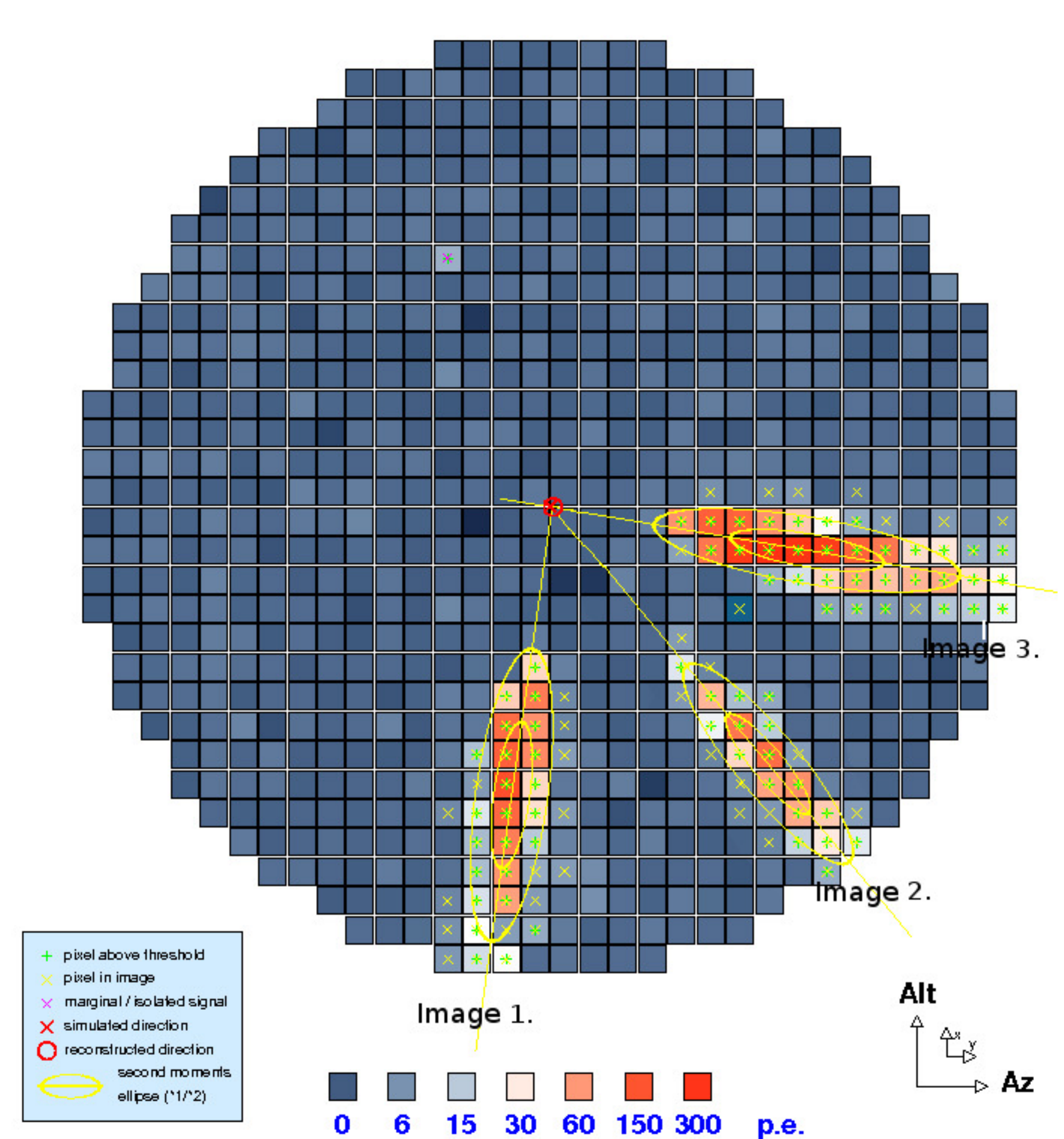}
\captionsetup{width=13cm}  
 \caption{Combined images for a single simulated 76 TeV $\gamma$-ray event to illustrate stereoscopic imaging. The event was seen by three telescopes. The image sizes and core distances are: 1300\textit{pe} at 362 \rm{m}, 690\textit{pe} at 471 \rm{m}, and 2390\textit{pe} at 310 \rm{m} for Images 1, 2 and 3 respectively. Therefore, we have three calculations of an intersection point between major axes. The weighting scheme, Eq~\ref{eqn:weighting}, would give increased weighting to the intersection point between Image 1 and Image 3 based on the angle between the major axis and the image \textit{size} of the images. The red ring represent the reconstructed shower direction while the red cross shows the true simulated shower direction. }
 \label{fig:gamma_image_stereo}
\end{centering}
\end{figure}

The size of the image and the angle between the major axes helps provide extra weighting to images that will provide an improved event reconstruction. The larger the combined image size, the larger the weighting since smaller images provide poor reconstruction due to low \textit{pe} counts and they are more susceptible to night sky background fluctuations. If the angle between axes is small, $<$ 10$^{\circ}$, then the major axes are almost parallel to each other. A small error in the major axes and C.O.G will increase the error on the intersection point particularly in the longitudinal direction which can affect the reconstruction. If the angle between the major axes is $>$ 30$^{\circ}$ then the images are further apart and the major axes will provide a more accurate intersection point. If the angle is large, $\approx$ 180$^{\circ}$, then the images are opposite each other and again the major axes are almost parallel. Taking the sine of the angle between major axes give more weighting to image pairs with angles between 45$^{\circ}$ and 135$^{\circ}$, with an optimal angle being 90$^{\circ}$.

	The event in Figure~\ref{fig:gamma_image_stereo} will provide three estimates of the intersection point between image pairs; point$_{12}$, point$_{13}$ and point$_{23}$. The coordinates of these intersection points are then multiplied by the weighting factor for that pair of images. The coordinates of each weighted intersection point are summed up and the final values are divided by the sum of the weighting factors:

\begin{eqnarray}
X_{final} = \frac{\sum_l w_{ij} X_{ij}}{\sum_l w_{ij}}
 \label{eqn:meanreco}
\end{eqnarray}

 where \textit{X$_{ij}$} is the x coordiate for a pair of images and \textit{l} is the number of pairs of images. The results provide the weighted average of intersection points. This point is then converted to the equivalent sky co-ordinate frame to provide the final calculation of the reconstructed shower direction. 

A similar method is used to calculate the core position for the event. The intersection points between images are calculated for pairs of images in the reconstructed shower plane at ground level, by starting from the positions of each telescope. Each intersection coordinate is multiplied by the weighting factor, summed up and then divided by the sum of the weights. This value is used to provide the average weighted intersection point. This average weighted intersection point provides the core position in the reconstructed shower plane. The core distance or position is the distance between the telescope and the shower maximum, calculated in the shower plane.


	To highlight the implications of a small angle between major axes, the simulated true core position for $\gamma$-ray events are presented on a scatter plot (Figure~\ref{fig:core_recon}) using the standard PeX layout. The events have been split into 3 energy bands: 1 - 10 TeV, 10 - 100 TeV and 100 - 500 TeV. To show the quality of event reconstruction compared to the telescope positions, the true shower positions have been split into angular reconstruction distance, $\theta$, bands: $\theta$ $<$ 0.2$^{\circ}$ (red), 0.2$^{\circ}$ $\leq$ $\theta$ $<$ 0.6$^{\circ}$ (green) and 0.6$^{\circ}$ $\leq$ $\theta$ (blue). The angular distance comes from the difference between the true source position and the reconstructed source position. The angular reconstruction distance will later be used to characterise the angular resolution in section~\ref{sec:ang_res_def}, but for now it shows the quality of reconstruction for each event. By splitting the events into angular distance bands, we can identify regions around the cell that provide poor event reconstruction. If the shower had a true core position in one of these regions, then the reconstruction source position would be poorly reconstructed. These regions correspond to positions around the cell where the shower core would lie between adjacent telescopes or behind each telescope outside of the cell. In these regions, the angles between the telescopes are $\approx$ 180$^{o}$ or $<$ $\approx$ 10$^{o}$.



\begin{figure}
\begin{centering}
\includegraphics[scale=0.8]{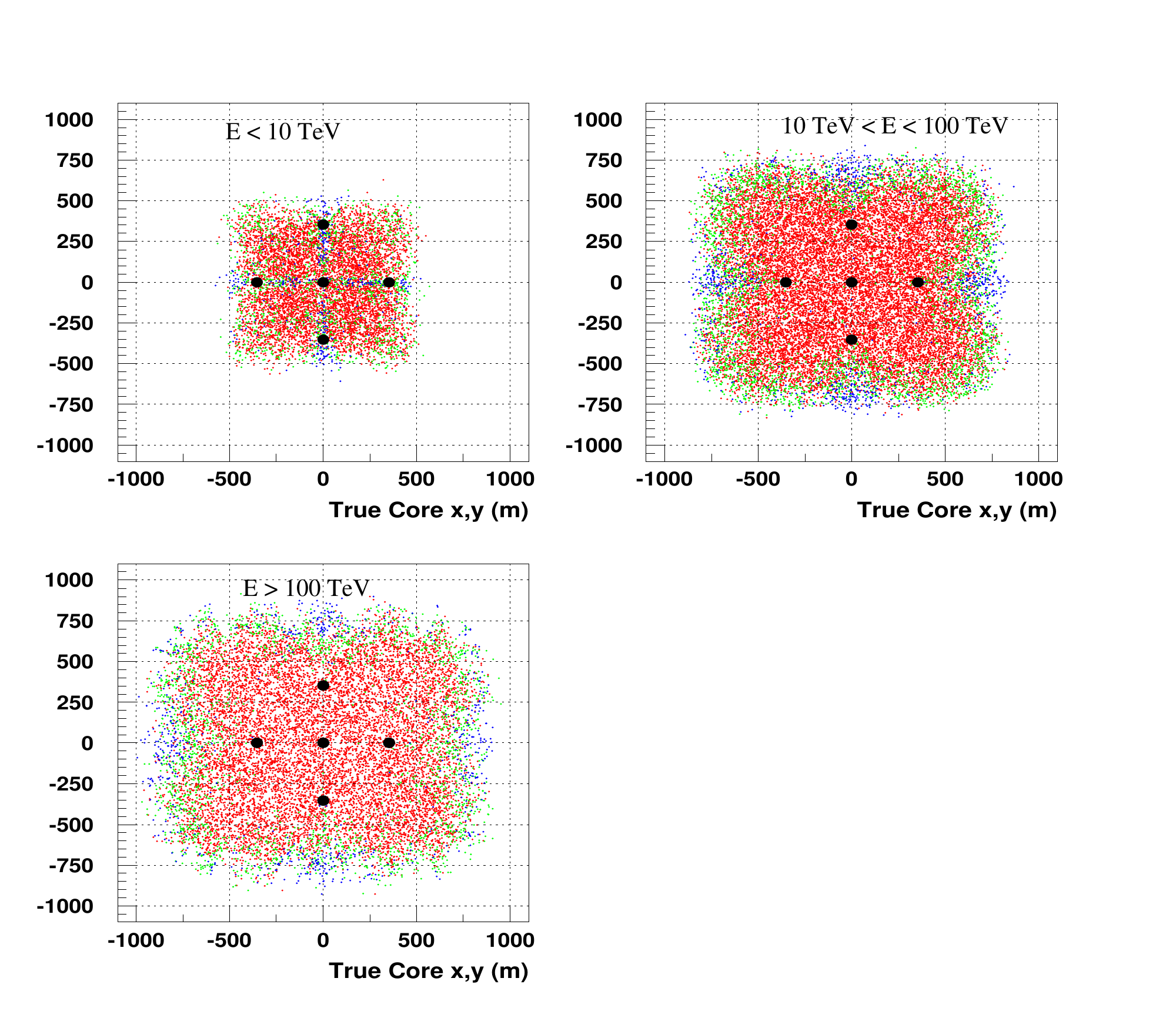}

\captionsetup{width=12cm}  
 \caption{Scatter plot for simulated true core locations for $\gamma$-ray events split into different energy bands for Algorithm 1. The colours represent different angular distance bands: $\theta$ $<$ 0.2$^{\circ}$ (red), 0.2$^{\circ}$ $\leq$ $\theta$ $<$ 0.6$^{\circ}$ (green) and 0.6$^{\circ}$ $\leq$ $\theta$ (blue). The blue sections indicate regions where the event reconstruction is poor. These regions correspond to positions between adjacent telescopes or behind each telescope outside of the cell.}

 \label{fig:core_recon}
\end{centering}
\end{figure}

\subparagraph{Algorithm 3}:\\

	Other reconstruction methods include a multi-parameter likelihood analysis, which is based on comparing the image on a pixel by pixel basis using a pre-calculated model of the shower image \cite{modelanalysis}. There is also a 3D model analysis, which uses a 3D version of the Hillas parameters to created a Gaussian photosphere in the atmosphere to predict the light expected in each pixel \cite{modelanalysis}. All three methods provide similar $\gamma$-ray efficiencies. More information can be found in \cite{modelanalysis}.

	Another improved reconstruction method under investigation is known as Algorithm 3 \cite{hofmann,Stamatescu} (Figure~\ref{fig:alg3_diagram}). Algorithm 3 uses image shape information to predict the distance, d$_{p}$, from the image C.O.G to the true shower direction and places this predicted distance to the shower along the major axis. The predicted distance along the major axis is actually feasible on both sides of the image since the predicted distance could be on either side of the image although `asymmetry' can help to resolve this. To calculate the predicted distance, a number of image parameters and parameter ratios can be used. The predicted distance is found to correlate well with ratios of \textit{W}, \textit{L} and log(\textit{size}) \cite{Stamatescu} for all core distances. Further discussion is presented in \cite{Stamatescu}. The image parameters are:
\begin{itemize}
\item $\frac{\textit{W}}{\textit{L}}$ and log(\textit{size})
\item  \textit{L} and log(\textit{size})
\end{itemize}

These image parameter ratios are compared against look-up tables of simulated $\gamma$-ray shower parameters. For each of the above mentioned combinations, the ratios are found using true shower parameters and a predicted distance is associated with each ratio. The image \textit{W}, image \textit{L} and log(\textit{size}) values are separated into small bin sizes to provide better resolution in look-up table values. 

\begin{figure}
\begin{centering}
\includegraphics[width=13cm]{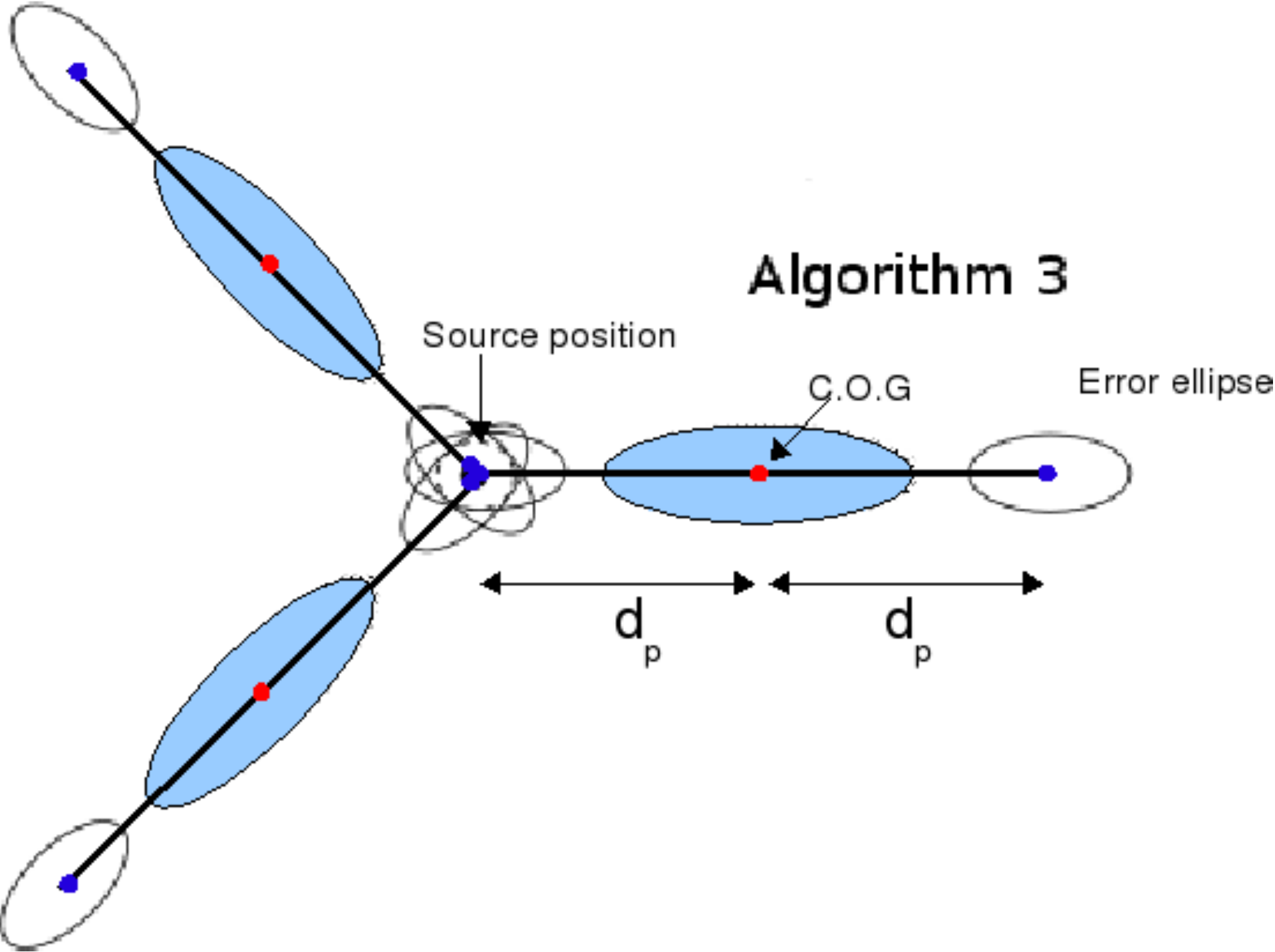}
\captionsetup{width=13cm}  
 \caption{A diagram of the Algorithm 3 reconstruction. The diagram contains 3 images for an event with the corresponding predicted distances, d$_{p}$, and error ellipses. The images are represented by the blue ellipses, the red dot indicates the C.O.G. and the black line indicates the major axis from Algorithm 1. The predicted distance is placed on both sides of the image using the major axis as a guide since it is not given a specific direction. The reconstructed source location is given by the blue dots for each image. The final image source position is calculated using the predicted source positions and the overlaid error ellipses for each image. Image adapted from \cite{hofmann}.}
 \label{fig:alg3_diagram}
\end{centering}
\end{figure}

Along with the predicted distance, its error ellipse is formed from look-up tables. The look-up tables for errors are constructed from simulations by imaging the true shower axis in the camera. These errors are based on the error in image C.O.G and major axis, and in the image \textit{size} and shape of the Hillas image ellipse. The size of the error ellipse reflects the error in the image parameterisation. The smaller the error ellipse the more accurate the W, L and \textit{size}.

	Stereoscopic imaging creates multiple predicted distances each with an error ellipse, which eliminates the predicted distance that appears on the wrong side of each image (Figure~\ref{fig:alg3_diagram}). The predicted distances and error ellipses are combined, see \cite{Stamatescu} for details, to provide an improved estimate for the true shower direction. An analogous method determines the core location. 

 
Algorithm 3 provides improved reconstruction over Algorithm 1 and this effect is seen in the true core position scatter plots for $\gamma$-ray events. Comparing Figure~\ref{fig:core_recon} for Algorithm 1 and Figure~\ref{fig:core_recon2} for Algorithm 3, the regions in the Algorithm 1 results are significantly reduced for Algorithm 3. 

Further refinements to Algorithm 3 have also been investigated \cite{Stamatescu} based on pixel timing. The image time gradient can be used to improve the prediction of d$_{p}$ for large core distance. The image time gradient, $c_{1}$, comes from the average slope of the time profile for each image, where the time profile is made up of the pixel times, t$_{i}$, with respect to a mean weighted pixel time. The linear equation, $t{i} = c_{o} + c_{1}.d_{i}$, is fitted to the time profiles. The slope of the time profile, c$_{1}$, can be found by minimising the following equation:

\begin{eqnarray}
\chi^{2}(c_{o}, c_{1}) = \sum_i^{N} (\frac{t{i} - c_{o} - c_{1}.d_{i}}{\sigma_{i}})^{2}
 \label{eqn:time_gradient}
\end{eqnarray}
\begin{eqnarray}
\frac{1}{\sigma_{i}} = \frac{pe_{i}}{\sum_i^{N} pe_{i}}
 \label{eqn:recon_light}
\end{eqnarray}
where $\chi^{2}$ is the goodness of fit parameter from the fit, t$_{i}$ is the pixel time with respect to the mean weighted pixel time of the image, d$_{\textit{i}}$ is the distance from pixel \textit{i} to the C.O.G along the major axis, $c_{o}$ is a constant, N is the number of pixels which pass cleaning and pe$_{\textit{i}}$ is the number of photoelectrons in the pixel.

The slope of the time profile or c$_{1}$ is otherwise known as the time gradient. The time gradient is correlated to the true source location so it makes a good predictor for the distance to the shower used in Algorithm 3 \cite{Stamatescu}. This parameter is used in a similar way to the image \textit{W}, image \textit{L} and \textit{size}. The effect of time gradient versus image \textit{W}, image \textit{L} and \textit{size} are presented in \cite{Stamatescu}. The only issues is that a shower that has a deeper depth of maximum in the atmosphere produces a larger predicted distance, while a shower higher in the atmosphere produces a smaller predicted distance but both can have similar c$_{1}$ values \cite{Stamatescu}. 

To improve the estimation of the predicted distance, the time gradient can be paired with the reconstructed light maximum parameter. The reconstructed light maximum is given by

\begin{eqnarray}
\rm{rec.}\;\rm{light}\;\rm{max} = \frac{r}{(\pi /180) \textit{d$_{r}$}} \cos(\textit{zen}) - r \sin(\textit{zen})
 \label{eqn:recon_light2}
\end{eqnarray}
where r is the core distance from Algorithm 1, \textit{d$_{r}$} is the angular distance from the C.O.G to the Algorithm 1 source position and \textit{zen} is the zenith angle.

The reconstructed light maximum represents the height of the maximum detected light emission above ground level. The time gradient and the reciprocal of the reconstructed light maximum show improved performance over using only the time gradient or the image \textit{L} and log(\textit{size}) combination to predict the distance to the shower \cite{Stamatescu,Heb}. Evidence of this is discussed in \cite{Stamatescu}. The time gradient and the reciprocal of the reconstructed light maximum also showed more robustness to large core distance showers and against fluctuations in night sky background. The improvement is seen over all energies and the improvement in angular resolution ranges from 5$\%$ to 10$\%$ compared to the standard \textit{W}/\textit{L} and log(\textit{size}) predictors.

The only issue is at small core distances, where the time gradient is not well defined for $\gamma$-ray events. This corresponds to the condition d$_{r}$ $<$ 0.85$^{\circ}$, and in this case, the standard \textit{W}/\textit{L} ratio is used to provide the predicted distance as discussed in \cite{Stamatescu}. 

Thus we will adapt the time gradient and reconstructed light maximum predictors here, as adopted in \cite{Stamatescu}.


\section{Stereoscopic Image Shape Reconstruction}
 \label{sec:mean_scaled}

	Now that the shower geometry has been reconstructed, cuts based on the shape of the image can be applied to suppress the cosmic ray events. As discussed in Chapter 2, cosmic ray showers have more irregular EAS development which can be exploited. The main parameters used to distinguish cosmic rays, in this case protons, from $\gamma$-rays are \textit{W} and \textit{L} of the image in the camera. 

	We will use three parameters for PeX; the \textit{W} of the image, the \textit{L} of the image and the number of pixels within an image, \textit{Npix} \cite{Tentenrowell}. 

The images from all telescopes which pass the stereoscopic cuts (section~\ref{sec:event_reconstruction}) are used to construct Mean Scaled Parameters (\textit{W}, \textit{L} or \textit{Npix}) \cite{MSWref}. The `mean scaled' approach has been successfully demonstrated by HEGRA, H.E.S.S., CANGAROO-III and VERITAS. The general expression for calculating any Mean Scaled Parameter is given by
\begin{eqnarray}
MS\textit{P} = \frac{\sum_i^{n_{tel}} w_{i} \textit{P}_{i}/\langle \textit{P} \rangle}{\sum_i^{n_{tel}} w_{i}}
 \label{eqn:meanscaled}
\end{eqnarray}
where \textit{P} = \textit{W}, \textit{L} or \textit{Npix} and represents the measured image value, $\langle$\textit{P}$\rangle$ is the expected image value from simulations, w$_{i}$ is the weighting factor explained shortly, and n$_{tel}$ is the number of accepted telescope images. The distributions for the mean scaled values for $\gamma$-rays and protons are displayed in Figure~\ref{fig:rejection}.

True \textit{W}, \textit{L} and \textit{Npix} distributions are binned based on \textit{size} and core distance for each telescope image to create look-up tables for $\gamma$-ray events. The look-up tables are used to compare \textit{W} of the image with the expected $\langle$\textit{W}$\rangle$ from simulated showers. By doing this, parameter \textit{P} can be expressed in terms of how much it deviates from its expected value $\langle$\textit{P}$\rangle$ for a $\gamma$-ray shower with the same \textit{size} and core distance.

The weighting factor \textit{w$_{i}$} is based on the \textit{size} for each image and the error in the expected $\langle$\textit{size}$\rangle$ from look-up tables. This is due to the fact that fainter images provide poor shower images and reconstruction. Figure~\ref{fig:rejection} top left panel shows the distribution of MSW values for $\gamma$-ray and proton showers over a 1 to 500 TeV energy range. There is a difference in peak positions in the MSW distributions for protons and $\gamma$-rays. The tails of the distributions over-lap. However, the fraction of $\gamma$-ray events left after a MSW cut can be maximised whilst the proton fraction is minimised. The MSW value that provides the optimum number of $\gamma$-ray events is then used to provide a quality factor.

The Q$_{fact}$ or quality factor is used to quantify the rejection power of analysis cuts. The Q$_{fact}$ is defined as,
\begin{eqnarray}
Q_{fact} = \frac{\kappa_{\gamma}}{\sqrt{\kappa_{p}}}  
 \label{eqn:qfactor}
\end{eqnarray}
using
\begin{eqnarray}
\kappa_{\gamma} = \frac{N_{\rm{{\gamma}_{cut}}}}{N_{\rm{{\gamma}_{total}}}}  \;\;\rm{and} \;\; \kappa_{p} = \frac{N_{\rm{p_{cut}}}}{N_{\rm{p_{total}}}}
 \label{eqn:kappa}
\end{eqnarray}
where $\kappa_{g}$ and $\kappa_{p}$ are fractions of $\gamma$-ray or proton events remaining after shape cuts.

\begin{figure}
\begin{centering}
\includegraphics[width=\textwidth]{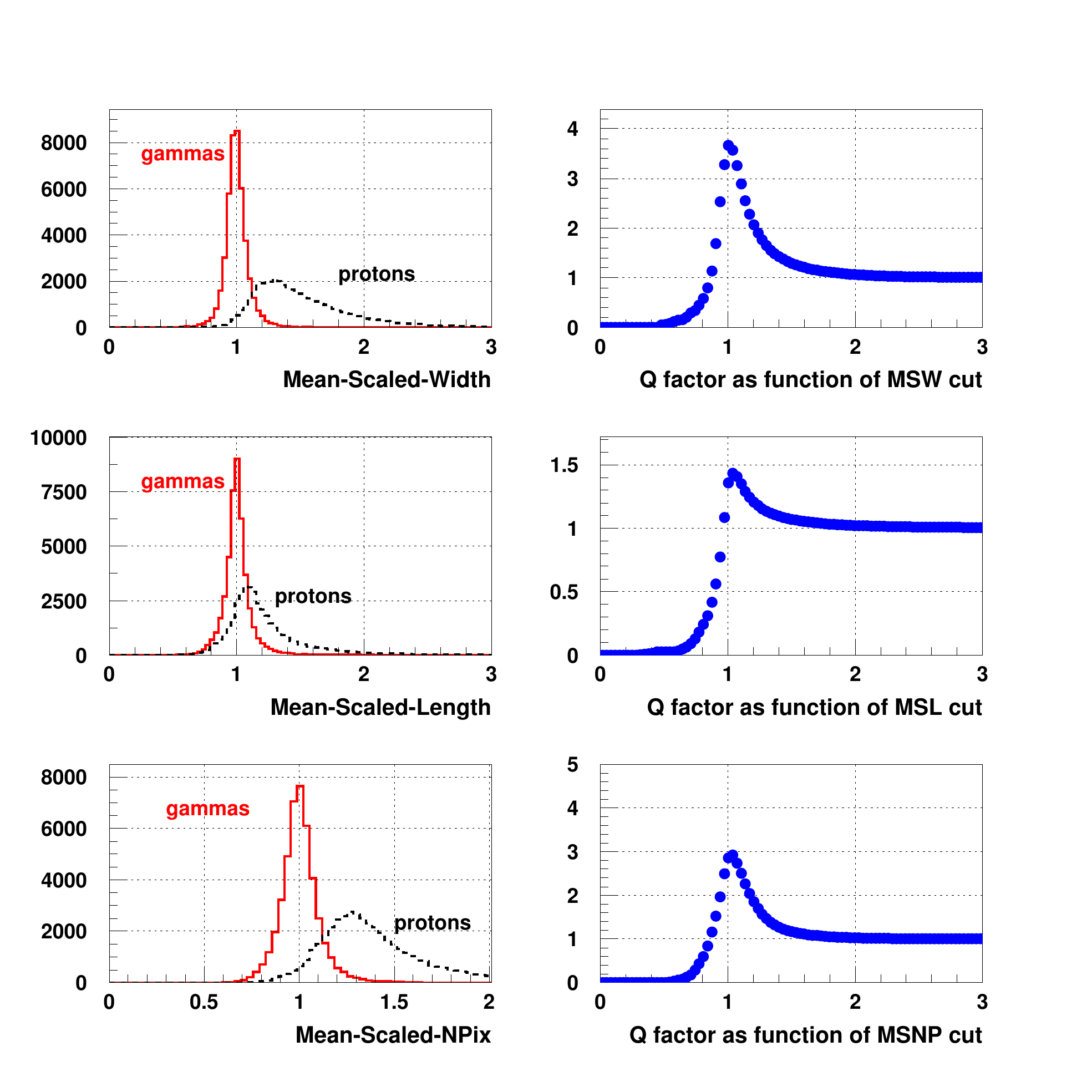}
\captionsetup{width=13cm}  
 \caption{The MSW, MSL and MSNpix distributions for $\gamma$-ray (red) and proton (black) events on the left and Q$_{fact}$ curves as a function of MSW, MSL and MSNpix on the right for the standard PeX configuration at all energies. The plots indicate no clear separation between $\gamma$-ray and proton events. However, the number of $\gamma$-ray events can be maximised while minimising the number of protons to find the optimum cuts. The right plots represent the Q$_{fact}$ values as a function of cut value. The optimum Q$_{fact}$ is provided by the cut values MSW $<$ 1.05, MSL $<$ 1.1 and MSNpix $<$ 1.05.}
 \label{fig:rejection}
\end{centering}
\end{figure}

	From Figure~\ref{fig:rejection} top panel, it can be seen that a MSW cut of $\approx$ 1.0 to 1.2 would provide the optimum acceptance of $\gamma$-ray showers and rejection of proton showers to maximise Q$_{fact}$. 

	A similar parameter, Mean Scaled Length (MSL), is obtained for the \textit{L} of the image and is calculated in the same way as MSW (Eq~\ref{eqn:meanscaled}). The middle panel in Figure~\ref{fig:rejection} shows the results for MSL. The left panel indicates that MSL provides a poorer separation between $\gamma$-ray and proton events. The Q$_{fact}$ as a function of MSL indicates that a MSL cut of 1.1 would provide the optimum Q$_{fact}$ (Figure~\ref{fig:rejection} middle right panel). 


	The last parameter for proton rejection comes from the number of pixels in the image, \textit{Npix}. The average number of pixels in a proton image should be greater than a $\gamma$-ray image since a protons image is usually wider. The Mean Scaled Npix (MSNpix) value is calculated the same way as MSW and MSL, Eq~\ref{eqn:meanscaled}. Figure~\ref{fig:rejection} bottom left panel shows that there is separation between $\gamma$-ray and proton pixel numbers. The Q$_{fact}$ as a function of MSNpix cut plot (Figure~\ref{fig:rejection} bottom right) indicates that a MSNpix cut of 1.1 provides the optimum cut. 


The optimum cuts for MSW, MSL and MSNpix are: MSW $<$ 1.05, MSL $<$ 1.2 and MSNpix $<$ 1.1. The cut values chosen are slightly larger than the peak values represented in Figure~\ref{fig:rejection} for the Q$_{fact}$ plots. This ensures that the maximum number of $\gamma$-ray events are retained while rejecting as many proton events as possible and still providing the best results. These cuts are defined as \textit{shape cuts} and will define what we term \textit{post-shape cut} results.

\begin{figure}[h]
\begin{centering}
\includegraphics[width=\textwidth]{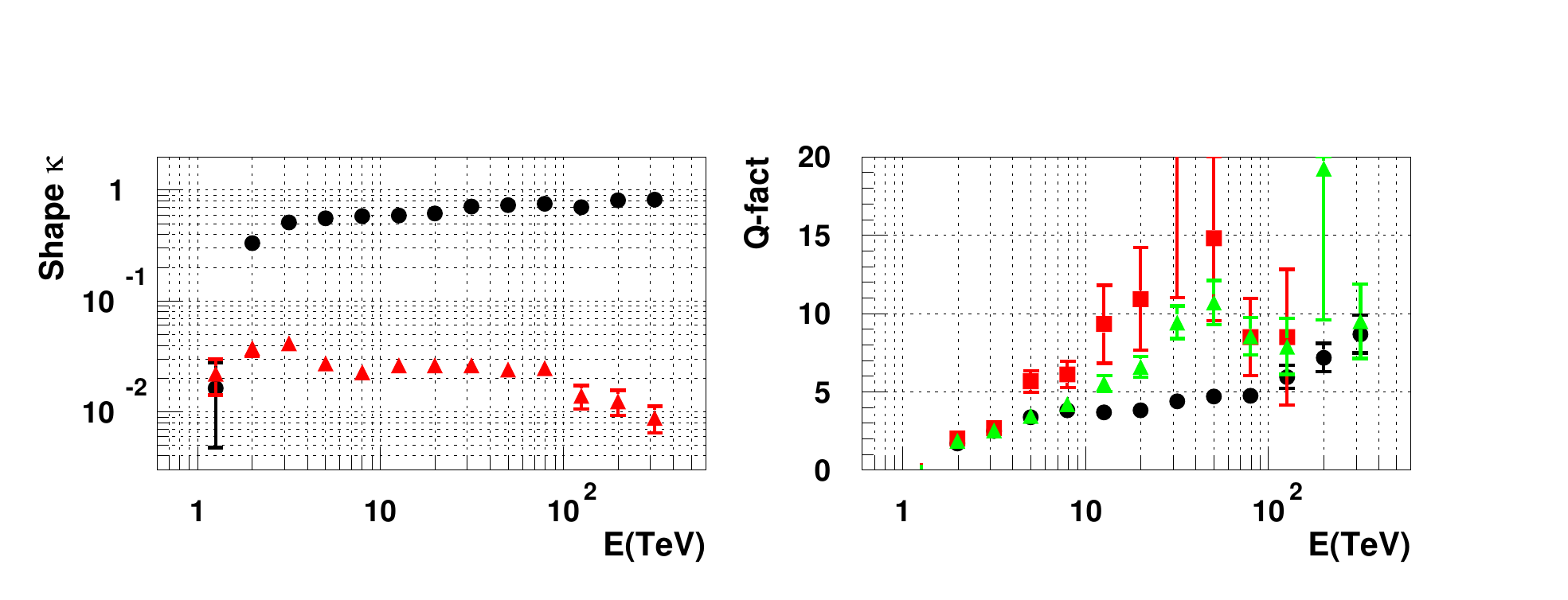}
\captionsetup{width=13cm}  
 \caption{Left: The shape acceptances for $\gamma$-rays (black circles) and protons (red triangles). After all shape-cuts are applied, the shape acceptances indicate that $\approx 50 \%$ of $\gamma$-rays events are accepted above 10 TeV while $\approx 2\%$ of protons event are accepted for the same energy range. Right: The Q-factor using the acceptances obtained from the left panel. The black circles indicate the Q-factor for all events, the red squares are for events which have a core distance $<$ 250\rm{m} and the green triangles are for events which have a core distance $<$ 500 \rm{m}.}
 \label{fig:shape_qfactor}
\end{centering}
\end{figure}

If the shape cuts are all applied to data then an overall Q$_{fact}$ value can be obtained. Figure~\ref{fig:shape_qfactor} left panel shows the $\kappa$ acceptances for $\gamma$-ray and proton events. We find that around 50$\%$ of $\gamma$-rays are accepted and around 97$\%$ of protons are rejected. Therefore, the shape cuts appear to reject a large majority of the proton events. The right panel in Figure~\ref{fig:shape_qfactor} shows the Q$_{fact}$ for all energies and core distances (black circles).



It is clear that the rejection power increases as energy increases. This effect is due to the increasing image \textit{size} and the increasing telescope multiplicity. With increasing image \textit{size}, the shape of the image is usually well defined which generally improves the parameterisation of the images allowing improved separation between $\gamma$-ray and proton events. 

\begin{figure}[h]
\begin{centering}
\includegraphics[width=13cm]{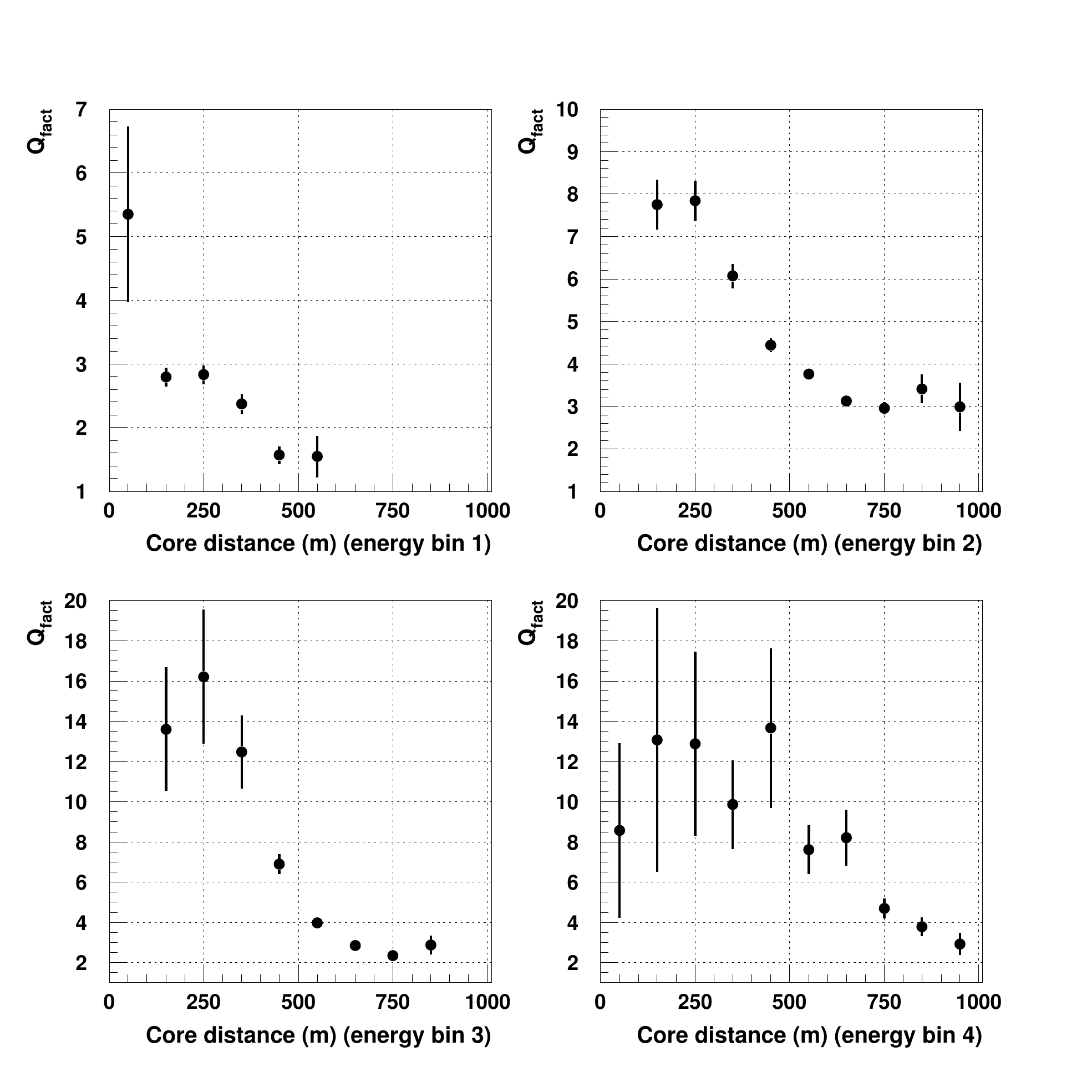}
\captionsetup{width=13cm}  
 \caption{Q-factor vs core distance results for Algorithm 3. The curves indicate the expected decrease in the Q-factor with increasing core distance. These curves show how dramatic the core distance effect is on Q$_{fact}$. The results have been split into four logarthmic energy bands: 1 - 4.2 TeV, 4.2 - 22.3 TeV, 22.3 - 105.7 TeV and 105.7 - 500 TeV.}
 \label{fig:q-fact_core}
\end{centering}
\end{figure}

We will now examine more closely the effect of core distance on the shape rejection power. Figure~\ref{fig:shape_qfactor} shows the shape acceptance for $\gamma$-ray and protons (right) and the Q$_{fact}$ (left). The Q$_{fact}$ plot (Figure~\ref{fig:shape_qfactor} right) displays two additional curves. The red curve represents the Q$_{fact}$ using only events with core distance $<$ 250 \rm{m}, while the green curve represents the Q$_{fact}$ using events with core distance $<$ 500 \rm{m}. The results show a steady increase in Q$_{fact}$ with energy for both the red and green curves. Therefore, the large core distance events affect the event reconstruction and hence the Q$_{fact}$. Figure~\ref{fig:q-fact_core} shows the Q$_{fact}$ as a function of core distance.


\begin{figure}[h]
\begin{centering}
\includegraphics[width=12cm]{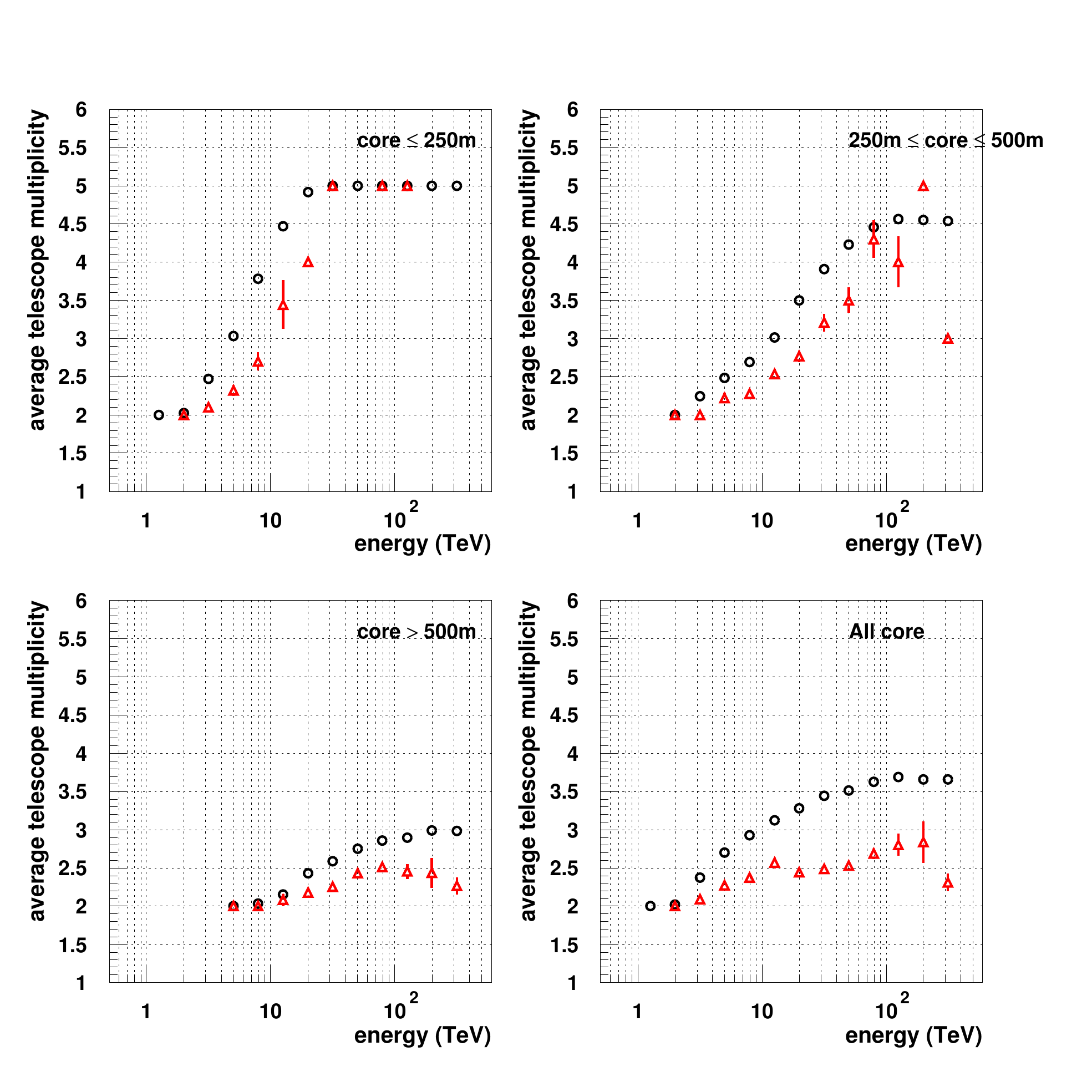}
\captionsetup{width=12cm}  
 \caption{The average telescope multiplicity vs energy for various core distances indicated on the top right of each plot. The open black circles represent $\gamma$-rays post-shape cuts with error bars which represent 1$\sigma$. The open red triangles represent the protons post-cuts, again with 1$\sigma$ error bars. These plots represent the effect of core distance on the telescope multiplicity.}
 \label{fig:multi_tel}
\end{centering}
\end{figure}

Figure~\ref{fig:q-fact_core} clearly shows that Q$_{fact}$ is reduced by large core distance events. If we consider the average telescope multiplicity (Figure~\ref{fig:multi_tel}), we see that the average telescope multiplicity quickly approaches the maximum telescope multiplicity with increasing energy especially for events with small core distances (Figure~\ref{fig:multi_tel} top left). The maximum telescope multiplicity is reached quickly for events with core distance $<$ 250 \rm{m}, due to larger sizes at high energies. By removing core distances less than 500 \rm{m}, we are left with a number of small and low multiplicity events (Figure~\ref{fig:multi_tel} bottom left). To show that the overall effect of core distance on telescope multiplicity, Figure~\ref{fig:multi_tel} bottom right provides the average telescope multiplicity for all core distance events over the whole energy range. The rise in average telescope multiplicity is slow as energy increases, which is the same as the effect seen in the Q$_{fact}$. Therefore, the Q$_{fact}$ is dependent on the telescope multiplicity, image \textit{size} and core distances of the events. 

PeX will ultilise the large core distance events to increase the collecting area of the cell. If the reconstruction of large core distance events can be improved then the Q$_{fact}$ will also improve. The aim is to improve the Q$_{fact}$ whilst maximising the number of events. A possible cut on core distance could be applied to PeX if the source is strong or there is an abundance of $\gamma$-ray statistics. A weak source will have a smaller number of events so it would best to not include a cut on core. However, if the Q$_{fact}$ can be improved by altering the standard configuration or reconstruction algorithm then the results should improve.

\section{Angular and Core Resolution}
 \label{sec:ang_res_def}

	A good angular resolution needs to be obtained for the PeX cell. To consider the precision of the reconstruction, we look at the angular distance, $\theta$, between the true shower direction and the reconstructed shower direction (Figure~\ref{fig:hillas2}).

\begin{figure}
\begin{centering}
\includegraphics[width=12cm]{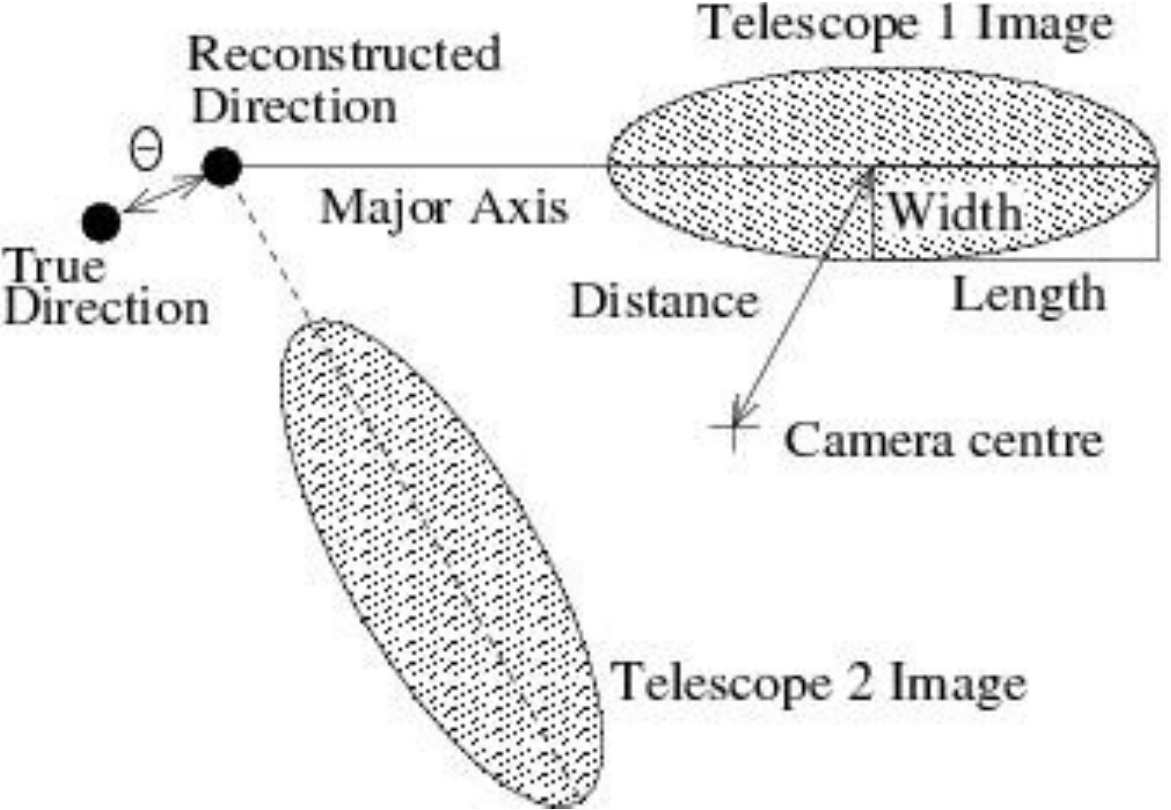}
\captionsetup{width=13cm}  
 \caption{The $\theta$ value for a pair of images. The $\theta$ value is the angular distance between the true shower direction and the reconstructed shower direction.}
 \label{fig:hillas2}
\end{centering}
\end{figure}

To calculate the angular resolution for PeX, we use the angular distance and put the values into a histogram. The histogram is fitted to a radial Gaussian curve $\approx$ $exp{-r^{2}/(\sigma^{2})}$. The r68 value is the radius that contains 68$\%$ of the events in the angular distance distribution. The low energy events provide an angular resolution of $\approx$ 0.15$^{\circ}$ for E $<$ 10 TeV. As the energy increases, the angular resolution improves to $\approx$ 0.1$^{\circ}$ for E $>$ 10 TeV. The improvement in angular resolution with energy for both Algorithm 1 and Algorithm 3 is significant. The improvement is larger for Algorithm 3 since it provides an enhanced reconstruction method. This improvement in angular resolution comes from multiple aspects.

\begin{figure}[h]
\begin{centering}
\includegraphics[width=13cm]{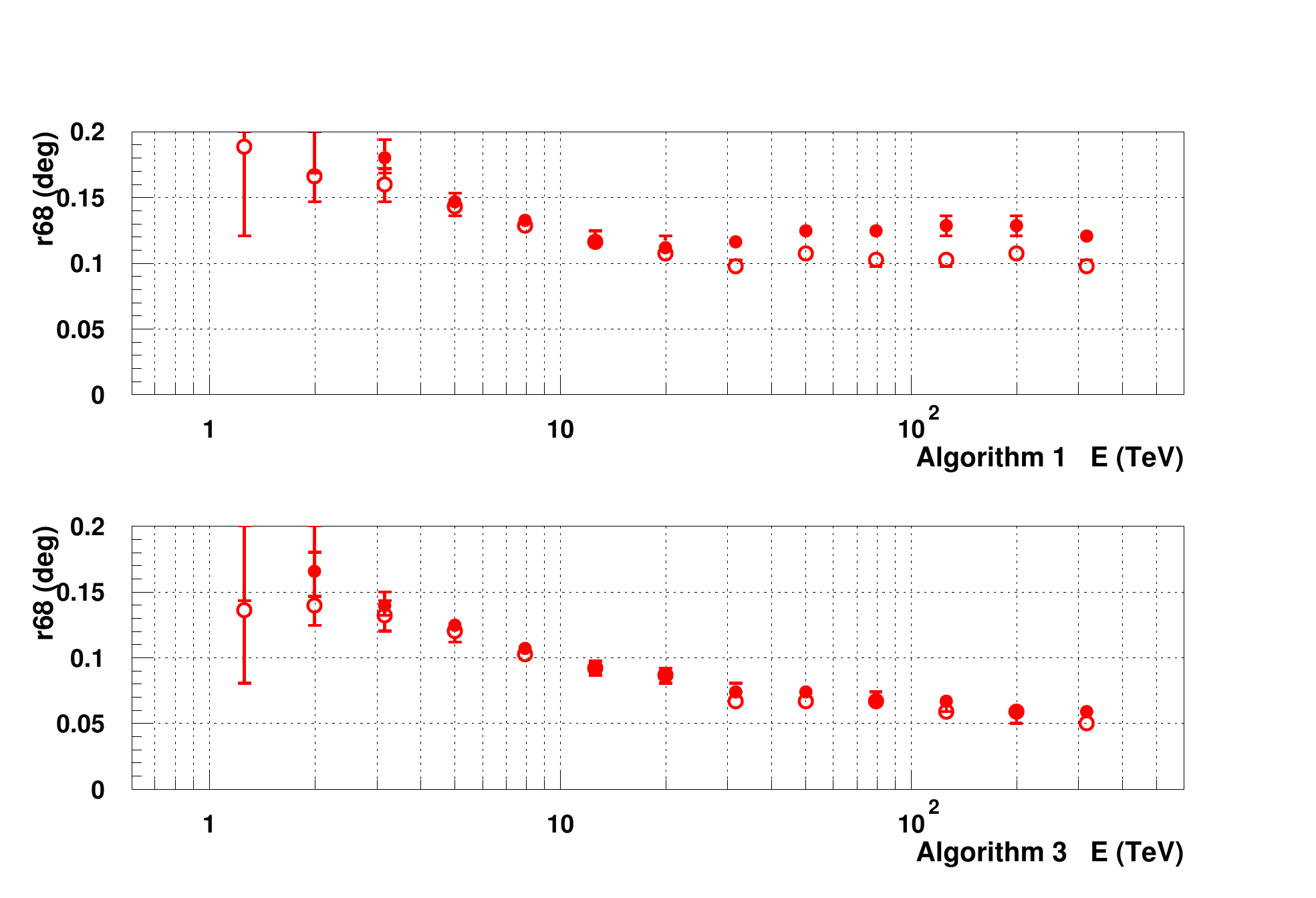}
\captionsetup{width=13cm}  
 \caption{The angular resolution (r68) for $\gamma$-ray events for the standard PeX cell. The top panel represents the Algorithm 1 angular resolution (deg) values pre- and post-shape cuts, while the bottom panel represents the Algorithm 3 angular resolution (deg) values pre- and post-shape cuts. The filled circles are for $\gamma$-rays pre-cuts and the open circles are for $\gamma$-rays post-shape cuts.}
 \label{fig:ang}
\end{centering}
\end{figure}
\begin{figure}[h]
\begin{centering}
\includegraphics[width=13cm]{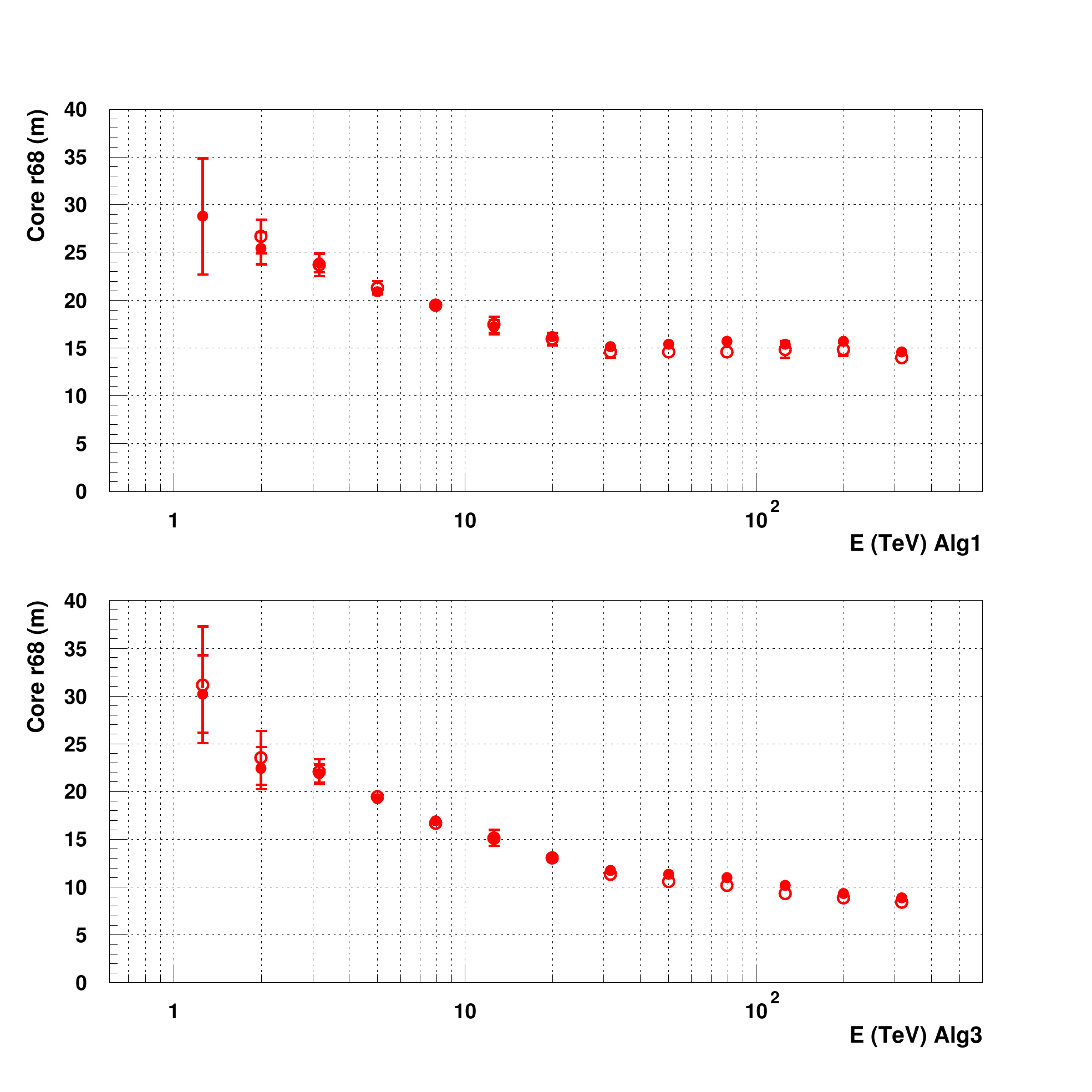}
\captionsetup{width=13cm}  
 \caption{The core resolution (r68) for $\gamma$-ray events for the standard PeX cell. The top panel shows the Algorithm 1 results and the bottom panel shows the Algorithm 3 results The circles represent the pre-cut core resolution and the open circles represents the post-shape cut core resolution.}
 \label{fig:core_res}
\end{centering}
\end{figure}

	The angular resolution improves due to an increasing size of the shower and increasing telescope multiplicity. As the size of the shower increases, the Hillas parameterisation provides an improved ellipse which gives a more accurate major axis. On the other hand, the angular resolution scales roughly with 1/$\sqrt{n_{tel}}$ as demonstrated by \cite{Puhlhofer2003267}. Since the angular resolution roughly scales with telescope multiplicity, the levelling off in the average telescope multiplicity (Figure~\ref{fig:multi_tel} bottom right) above 30 TeV is also seen in the angular resolution curve. The post-shape cut angular resolution reaches 0.1$^{o}$ (Figure~\ref{fig:ang} top panel open circles). A base angular or point source cut of 0.1$^{o}$ will be used for point source observations based on results presented in \cite{Buckley}.

The core resolution shows how well the core of the shower has been reconstructed. The core resolution is calculated in a similar way to the angular resolution. Figure~\ref{fig:core_res} shows the core resolution for Algorithm 1 and 3. As the energy increases, the size of the image increases and the reconstruction of the core position improves. Therefore, the core location of the shower improves and proves the same trend as the angular resolution results.

	After the shower direction and location have been reconstructed via Algorithm 1 or Algorithm 3, the $\gamma$-ray primary energy can be estimated.

\section{Energy Resolution}
 \label{sec:energy_res}
 
The energy resolution indicates how well the primary energy of the $\gamma$-ray can be predicted from shower reconstruction. To do this, the size and core distance of individual images are compared to simulated data stored in look-up tables. This technique is similar to that used to find the mean scaled parameters for the shower. The look-up tables return the expected energy $<$\textit{$E_{expected}$}$>$ value for each image:

\begin{eqnarray}
E_{recon} = \frac{\sum_i^{n_{tel}} w_{i} \langle E_{expected} \rangle}{\sum_i^{n_{tel}} w_{i}}
 \label{eqn:expect_e}
\end{eqnarray}

The estimated energy calculated from each triggered image is combined to create a weighted mean for the energy of the event (Eq~\ref{eqn:expect_e}). The images are weighted based on the image \textit{size} over the square of the uncertainty in $<$\textit{$E_{expected}$}$>$. This implies that larger images provide more weighting to the energy estimation. However, if the uncertainty is large then the image has less weight in the final energy calculation. The image \textit{size} is used as a weighting factor since larger images provide better reconstruction. 

The telescope energy resolution indicates how well the reconstruction algorithm can reconstruct the initial energy of the shower. The reconstructed energy is compared with the true energy of the modelled showers to create a distribution which represents the spread in energy reconstruction. The energy resolution is taken as the RMS of the distribution of $\Delta{E}/E$,
\begin{eqnarray}
\frac{\Delta{E}}{E} = \frac{E_{recon} - E_{true}}{E_{true}}
\end{eqnarray}
where E$_{recon}$ is the reconstructed energy and E$_{true}$ is the true energy.

To compare the energy resolution, the RMS of the $\Delta{E}/E$ distribution and the mean of the $\Delta{E}/E$ distribution are used. The RMS of the $\Delta{E}/E$ distribution represents the value which contains $68\%$ of the distribution. The pre-cut RMS in Figure~\ref{fig:energy_resolution}, increases as the energy increases above 10 TeV, which suggests that the reconstruction worsens at higher energies. A possible reason is that the large core distance events are truncated by the edge of the camera so only a partial image is collected which disrupts the reconstruction. The post-shape cut RMS shows that the distribution of $\Delta{E}/E$ becomes tighter and, on average, the RMS is around 15$\%$ for energies $>$ few TeV. 

The mean in the $\Delta{E}/E$ distribution pre-cuts shows that including the poorly reconstructed events produces a bias in energy. At low energies, the reconstructed energy has been overestimated. The post-shape cut values of the mean $\Delta{E}/E$ for E $>$ few TeV are $<$ 0.1. \\



\begin{figure}[h]
\begin{centering}
\includegraphics[width=12cm]{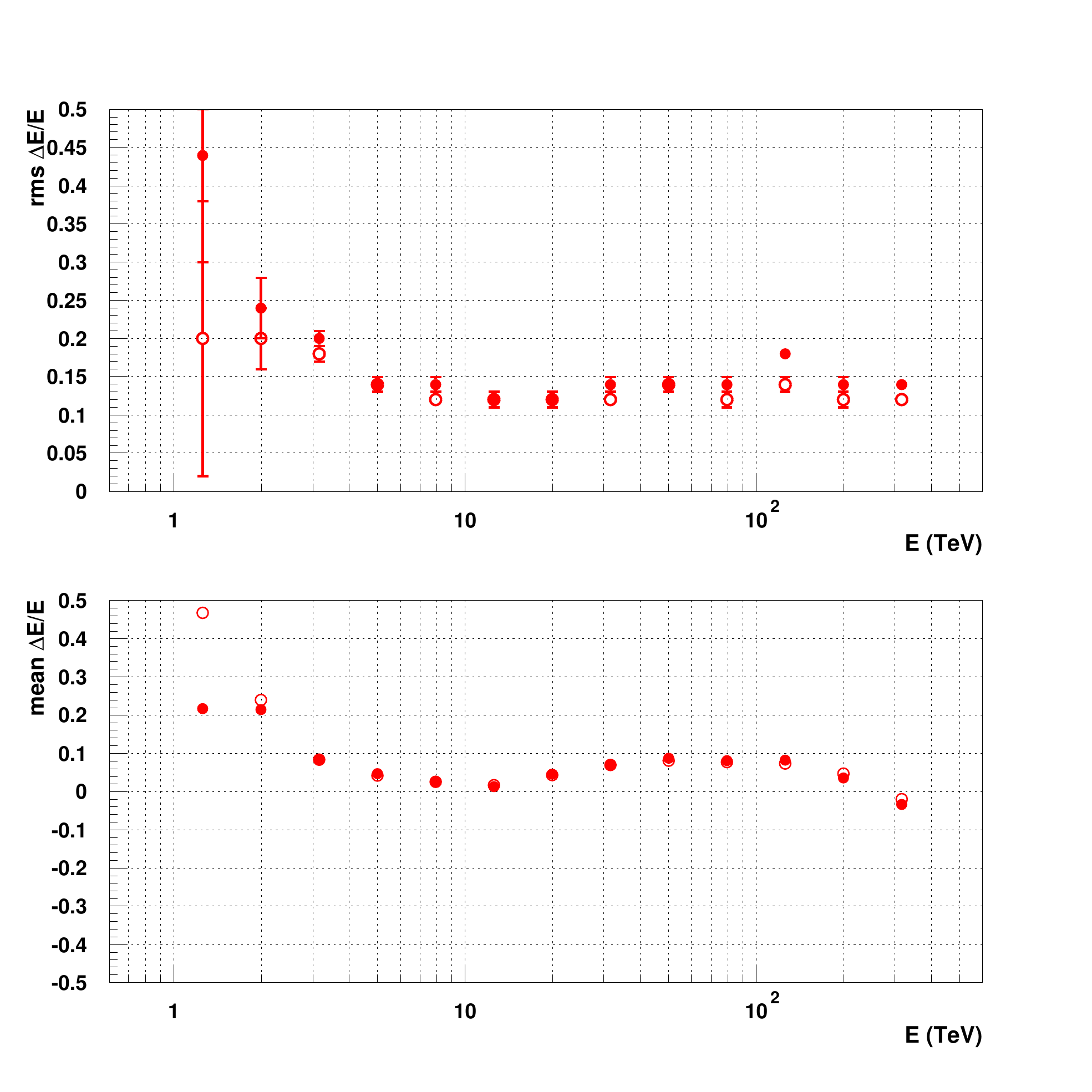}
\captionsetup{width=13cm}  
 \caption{Top: The RMS of the $\Delta{E}/E$ distribution for Algorithm 3. The red circles represent pre-cut results, while the red open circles represent the post-shape cut results. The smaller the RMS, the better the overall energy reconstruction for the events. Bottom: The mean in $\Delta{E}/E$ represents the accuracy of the energy reconstruction for Algorithm 3.}
 \label{fig:energy_resolution}
\end{centering}
\end{figure}

\section{Effective Area}

One of the aims of the PeX design is to optimise the effective area for improved multi-TeV sensitivity. While maximising the effective area, the telescope performance in other areas such as angular resolution and Q$_{fact}$ need to be maintained. To calculate the effective area, the ratio of the number of events detected, $N_{cut}$, over the number of events thrown, $N_{thrown}$, is used together with the CORSIKA area over which the events have been thrown, A$_{thrown}$. The expression for effective area is

\begin{eqnarray}
A_{eff} = A_{thrown} \frac{N_{cut}}{N_{thrown}}
 \label{area_eqn}
\end{eqnarray}

The effective area is calculated for a point source so an angular cut on reconstructed direction is applied as well as the shape cuts. Figure~\ref{fig:effective_area} shows the effective area for PeX using the standard configuration (Table~\ref{table:stand_config_original}). Comparing the PeX effective area against the H.E.S.S. effective area, a large improvement is seen due to a larger field of view (camera size) and telescope separation. A larger field of view allows the camera to trigger on events with larger core distances, which increases the effective area. The H.E.S.S. telescope separation is $\approx$ 120 \rm{m}, while the PeX telescope separation is 500\rm{m}.

\begin{figure}[h]
\begin{centering}
\includegraphics[width=\textwidth]{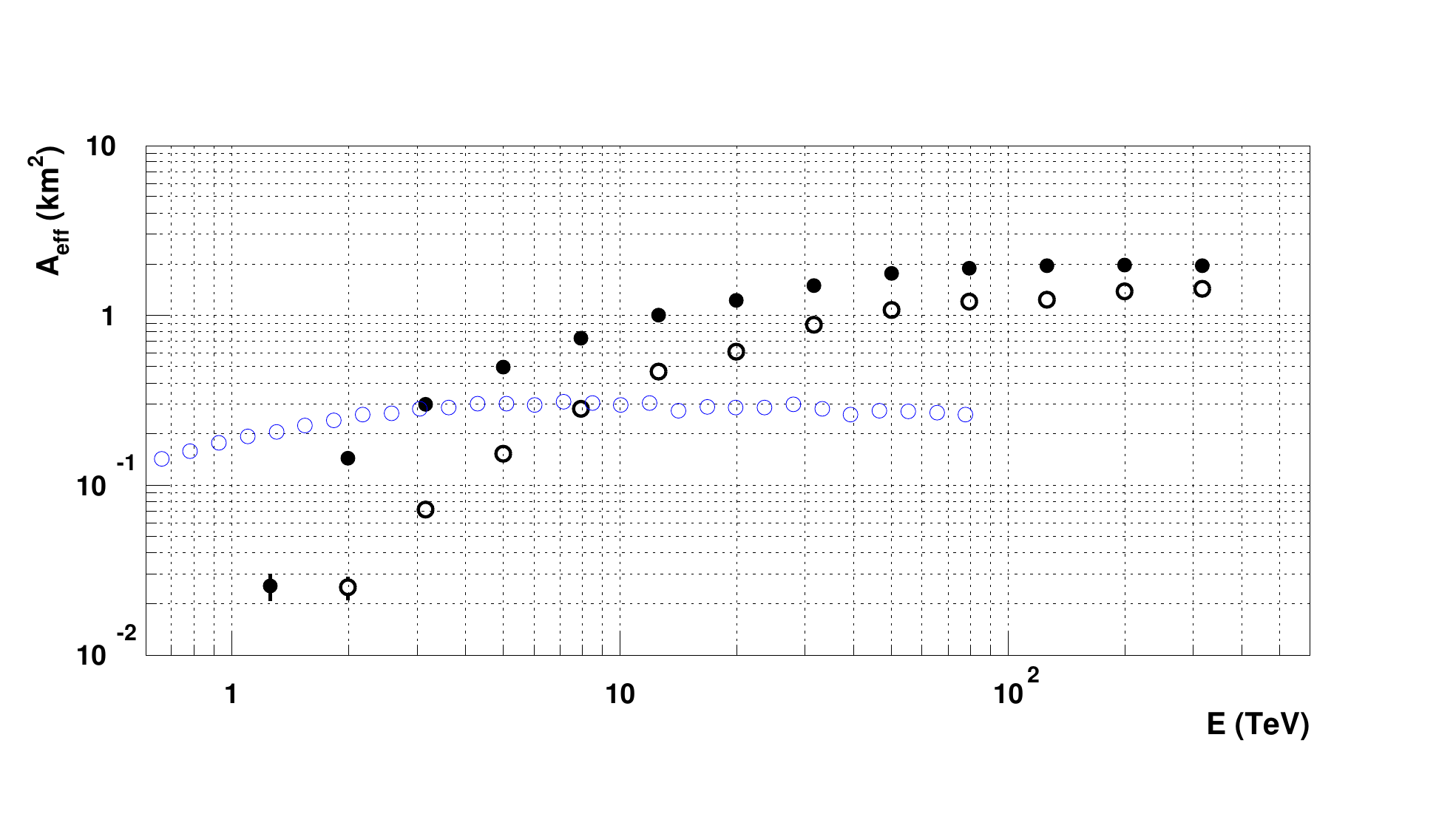}
\captionsetup{width=13cm}  
 \caption{The effective collecting area for PeX using the standard configuration with Algorithm 3. The black circles represent the $\gamma$-ray pre-cut effective area. The open black circles represent the $\gamma$-ray post-cut effective area. The open blue circles represent the H.E.S.S. $\gamma$-ray post-cut effective area. Post-selection cut results have passed an angular cut, and a MSW, MSL, and MSNpix cut.}
 \label{fig:effective_area}
\end{centering}
\end{figure}

\section{Flux Sensitivity}
 \label{sec:flux_sens}

	The flux sensitivity is defined as the minimum flux required to detect a $\gamma$-ray signal at a given significance level. The significance level, $S$, is expressed in terms of standard deviations of the background cosmic rays, $\sigma_{B}$. The signal significance is denoted, $S=N_{\gamma}/\sigma_{B}$, where $N_{\gamma}$ is the number of $\gamma$-rays. If the background follows Poisson statistics then the background deviation can be expressed as $\sigma_{B} = \sqrt{N_{B}}$.

	The $\gamma$-ray signal flux is required to pass a minimum significance of 5$\sigma$, where the signal is five times the level of the background fluctuations. The main contribution to the background fluctuations come from cosmic-rays. Therefore, for the calculation of significance the background contribution comes purely from cosmic ray protons, $N_{p}$. 
\begin{eqnarray}
S=\frac{N_{\gamma}}{\sqrt{N_{p}}}
 \label{eqn:signif}
\end{eqnarray}

The significance can be rearranged to calculate the minimum detectable flux above the current background, $N_{\gamma\:min} = S \sqrt{N_{p}}$. After cuts the number of $\gamma$-rays from a point source, $N_{\gamma}$, and the number of protons, $N_{p}$, are presented by
\begin{eqnarray}
N_{\gamma} \: = \: F_{\gamma}\: A_{eff}\: \lambda_{\gamma}\: t \: dE
 \label{eqn:ngam}
\end{eqnarray}
\begin{eqnarray}
N_{p} \: = \: F_{p}\: A_{eff_{p}} \: \lambda_{p} \: \Omega \: t \: dE  
 \label{eqn:nprot}
\end{eqnarray}
where $F_{p}$ is the $\gamma$-ray flux in \rm{ph (cm$^{2}$ s TeV)$^{-1}$}, $A_{eff}$ is the pre-cut effective area in \rm{cm$^{2}$} and we assume that $A_{eff}$ $\approx$ $A_{eff_{p}}$, $\lambda_{\gamma}$ is the $\gamma$-ray shape and angular cut acceptance, $t$ is the observational time, usually 50 hrs, $dE$ is the energy bin width, $\lambda_{p}$ is the proton shape and angular cut acceptance, $\Omega$ is the solid angle of the angular cut in \rm{sr} and $F_{p}$ is the proton flux in \rm{particles (cm$^{2}$ s sr TeV)$^{-1}$}. Only protons are considered since after shape cuts, the Helium rate is 5$\%$ of the proton rate \cite{Denman}. Inserting $N_{\gamma}$ and $N_{p}$ into eqn~\ref{eqn:signif} and rearranging, we find the minimum $\gamma$-ray flux 

\begin{eqnarray}
F_{\gamma} \: = \: 5 \: \frac{1}{Q_{total}} \: \sqrt{\frac{F_{p} \: \Omega}{A_{eff} \: t \: dE}} 
 \label{eqn:flux}
\end{eqnarray}

where Q$_{total}$ is the quality factor with shape and angular cut.
	From this equation, the factors which improve the flux sensitivity are seen. Each factor can be considered:
\begin{itemize}
\item The flux sensitivity scales with $\sqrt{A_{eff}}$. Increasing the effective area $A_{eff}$ will improve the flux sensitivity. To increase the effective area, a larger telescope separation may be considered in conjunction with a large field of view. Placing telescopes further apart will provide detection of events at larger core distances, which in turn increases the effective area at higher energies. Changing the pixel size and their arrangement can allow for improved image shapes, which provides better event reconstruction with larger telescope separation and field of view.
	
\item Improving the angular resolution will improve sensitivity. An improved angular resolution or $\Omega$, implies that a smaller on-source region is required for point sources. A way to improve the angular resolution is to obtain improved parameterisation of images. This will give a more accurate major axis for the images, improve the reconstruction of the shower direction and hence the angular resolution. 
	
\item Increasing the Q$_{total}$ will provide a better sensitivity. To achieve an improved Q$_{total}$ a number of changes can be made: varying the cleaning threshold, increasing the stereoscopic size cut demanding stronger images, or varying the triggering conditions. These changes can improve the shape of the image and parameterisation. Smaller pixels or different pixel arrangements, such as in a hexagonal grid, could provide better event images within the cameras. Some of these aspects will be investigated in Chapter~\ref{sec:future_work}.
	
\item Increasing the observational time will provide a better flux sensitivity. A standard observational time of 50 \rm{hrs}, currently used by other IACTs, will be used for the cell. This allows for a direct comparison between detector sensitivities.
The flux sensitivity improve as 1/$\sqrt(t)$.
\end{itemize}

For the final calculation of flux sensitivity, an alternative representation of significance will be used. Eq~\ref{eqn:signif} provides a simple estimation for significance that does not take into account errors in the true measurement of $\gamma$-ray and background events. However for PeX, the Li and Ma statistic \cite{liandma} (Eq~\ref{eqn:sig_li_ma}) will be used as it is based on the maximum likelihood/Poisson method. 

We use the Li and Ma equation, eqn~\ref{eqn:sig_li_ma}, since it is more accurate than the simplified $N_{\gamma}/\sqrt{N_{B}}$ from Poisson statistics when considering multiple background or off-source regions. We use the $N_{on}$ and $N_{off}$ regions which represent the number of event counts towards the source and the number of event counts away from the source respectively.

The final flux sensitivity results are calculated for 5 $\sigma$ signal, 50 hour observations and 5 energy bins per energy decade as per commonly adopted in TeV $\gamma$-ray astronomy. The final results will be displayed in Chapter 8.

Plotting the proton events in terms of true simulated energy creates a bias in the results. For $\gamma$-rays, the whole shower can be detected whereas for proton events only the electromagnetic component of the shower is detected. Therefore, the initial proton energy has been underestimated. The true proton shower energy needs to be shifted down to match the energy which represents the reconstructed energy. The way to account for this is to plot the results with reconstructed energy. The issue is that the energy reconstruction for protons uses the $\gamma$-ray look-up tables and it can produce additional fluctuations in the results. 

\begin{figure}[h]
\begin{centering}
\includegraphics[width=10cm]{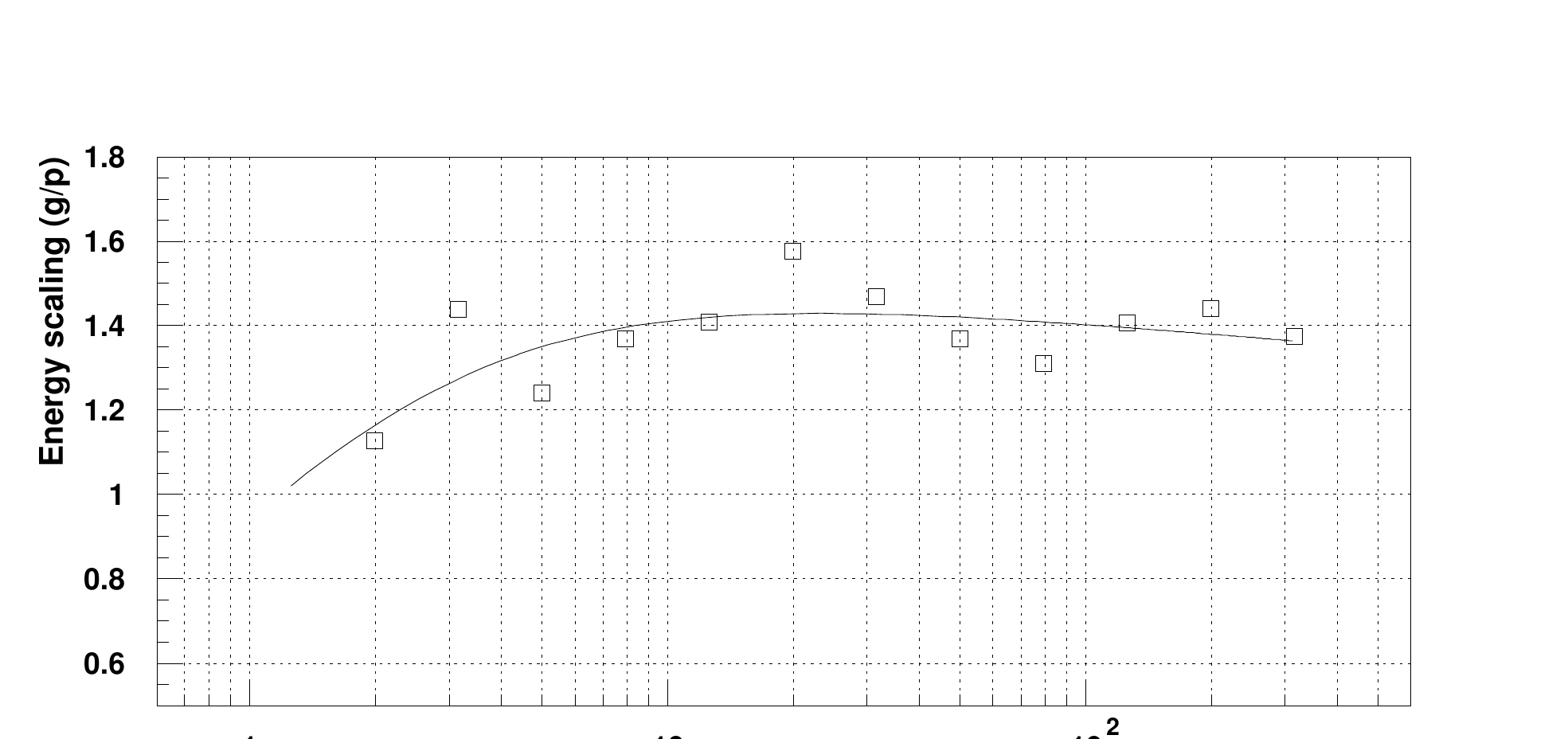}
\captionsetup{width=13cm}  
 \caption{The scaling factor for the energy mapping function. The squares represent the ratios of the means in the $\Delta{E}/E$ distributions. The line represents a fit to the ratio of mean values and produces the smoothing function.}
 \label{fig:scaling_factor}
\end{centering}
\end{figure}

An energy mapping function (Eq~\ref{eqn:scaling_f}) was created to correct the mismatched proton Cherenkov effect using reconstructed energy.
\begin{eqnarray}
E_{a}= s_{p} E_{b}
 \label{eqn:scaling_f}
\end{eqnarray}
\begin{eqnarray}
\rm{s_{p}} = \frac{\frac{\Delta{E_{g}}}{E_{g}} + 1}{\frac{\Delta{E_{p}}}{E_{p}} + 1}
 \label{eqn:scaling_factor}
\end{eqnarray}
where E$_{a}$ is the proton energy after scaling, s$_{p}$ is the scaling factor, E$_{b}$ is the proton energy before scaling, $\Delta{E_{g}}/E_{g}$ is the $\gamma$-ray energy resolution and $\Delta{E_{p}}/E_{p}$ is the proton energy resolution for each energy bin. Both energy resolutions assume a $\gamma$-ray model. The mapping function uses the ratio of means from the $\Delta{E}/E$ distribution to provide a scaling for each energy bin. The scaling is calculated by taking the mean in the $\gamma$-ray distribution over the mean in the proton distribution Eq~\ref{eqn:scaling_f}. The scaling is shown in Figure~\ref{fig:scaling_factor} by the open squares. The figure illustrates the fluctuations in the ratios produced, which creates artificial fluctuations in the results. To remove these fluctuations, a fit is applied to the square points in Figure~\ref{fig:scaling_factor}. The fit smooths out the large fluctuations between energy bins. The scaling is then applied to the proton energies to smooth out any the fluctuations via Eq~\ref{eqn:scaling_f}. By applying this scaling factor to the proton energies, the result can be displayed in reconstructed energy which best represents the true results from observational data.

The first step towards improving the PeX cell, will be to investigate the standard parameters such as telescope separation, cleaning combinations, triggering combinations and image size cut. Altering these values could improve the performance of the cell and will indicate which parameters can provide the biggest improvements.

\chapter{Low Altitude PeX cell optimisation} 
 \label{sec:optimise_low}
 
 In this Chapter, the PeX cell will be simulated for a site which is 0.22 \rm{km} above sea level. All simulations have been done using the Algorithm 1 analysis technique. In later chapters, the second reconstruction algorithm, Algorithm 3, will be considered.
 
To recap (Table~\ref{table:stand_config_original}), the standard configuration for PeX uses: a triggering threshold of (6\textit{pe}, 3), a cleaning threshold of (8\textit{pe}, 4\textit{pe}), an image \textit{size} cut of 60\textit{pe} and a telescope separation of 500 \rm{m}.

Although the first three parameters are scaled according to the optimal H.E.S.S. parameters it is worthwhile investigating if this combination is also optimal for PeX. Thus we will look at the effect of changing these four parameters on PeX performance. Each parameter will be investigated separately and optimised while the other values remain constant. A number of parameters will be investigated:

\begin{itemize}
\item The telescope separation for PeX, with values ranging from 200 \rm{m} to 800 \rm{m}
\item The triggering combination for images to trigger the camera, with values ranging from (\textit{threshold}, \textit{n}) = (4\textit{pe}, 2) to (12\textit{pe}, 5)
\item The cleaning combination for the image, with values ranging from (\textit{picture}, \textit{boundary}) = (5\textit{pe}, 0\textit{pe}) to (10\textit{pe}, 5\textit{pe})
\item The image size cut, which also determines the images used for reconstruction, with values ranging from 60\textit{pe} to 300\textit{pe}
\item The site altitude, to determine which site would provide the optimum results, either a 0.22 \rm{km} or 1.8 \rm{km} site based on real locations (Chapter~\ref{sec:optimise_high})
\end{itemize}

Goals for PeX are to provide the largest possible collecting area and to achieve the lowest detectable flux. The investigation will focus on the factors which improve the flux sensitivity and the cosmic ray rejection power such as:
\begin{itemize} 
\item The effective area of the cell
\item The angular resolution (r68)
\item The quality factor (Q$_{fact}$)
\item The energy resolution and precision of energy resolution
\end{itemize}



Initial work will be conducted using a low altitude site, which is 0.22 \rm{km} above sea level. Later, a second altitude site similar to H.E.S.S., which is 1.8 \rm{km} above sea level, will be investigated. All results presented here are optimised for point source observations. The results will be presented for both pre- and post-cuts. The cuts that will be applied to the results include:
\begin{itemize} 
\item The shape cuts: MSW $<$ 1.05, MSL $<$ 1.2 and MSNpix $<$ 1.1 (section~\ref{sec:mean_scaled})
\item The angular cut: $\theta$ $<$ 0.1$^{\circ}$ 
\item The selection cuts: shape cuts + angular cut
\item The quality cuts: size $>$ 60 \textit{pe}+ dis2 $<$ 4.0$^{\circ}$ and n$_{tel}$ $\leq$ 2
\end{itemize}
If only shape cuts are applied then the results represent post-shape cuts, if both shape cuts and an angular cut are applied then the results represent post-selection cuts. The terms post-shape cuts and post-selection cuts will be continually used throughout the rest of this thesis.

\section{Telescope Separation}
 \label{sec:tel_sep_op}

A square/diamond layout has been utilised by HEGRA, H.E.S.S., VERITAS and CANGAROO III. This layout provides good variation in image locations for a single event so that accurate event reconstruction can been achieved. For the current cell, a square layout will be used. However, an extra telescope at the centre of the square will be added. This design is similar to the design layout for HEGRA \cite{1997APh}.

	Recall that Figure~\ref{fig:distance} illustrates the lateral distributions of showers with different energies. The difference in lateral distribution comes from the number of particles produced for different energy showers. For the lower energy events, 100 - 500 GeV, the number of particles at shower maximum, N$_{max}$, ranges from 1250 to 6250 particles. For higher energy events, 1 - 100 TeV, N$_{max}$ ranges from 12500 to 1250000 particles. With a small N$_{max}$, the lateral extent of the shower is usually small with a smaller photon count, which implies that the shower must be closer to the telescopes for the events to trigger, i.e: smaller core distances (Figure~\ref{fig:distance} black points). A larger N$_{max}$ implies that the lateral extent of the shower can be larger with high photon counts which allows events to trigger at larger distances from the telescopes (Figure~\ref{fig:distance} blue points).

The implementation of a larger telescope separation will allow the events at approximately 800 \rm{m} core distances (Figure~\ref{fig:distance}) to trigger PeX. As more large core distance events trigger PeX, the effective area at the highest energies, E $>$ 100 TeV, will increase. The downside to increasing the telescope separation is the low energy event detection efficiency. As the core distance increases the photon intensity of low energy events decreases rapidly limiting the events which can trigger at large core distances (Figure~\ref{fig:distance}). It is best to achieve a region of overlap with current IACTs. Therefore, the optimum energy threshold of the cell would include 1 to 30 TeV energies. A desired energy range for the cell would start at a few TeV and go up to 500 TeV.

\begin{figure}
\begin{centering}
\includegraphics[scale=0.65]{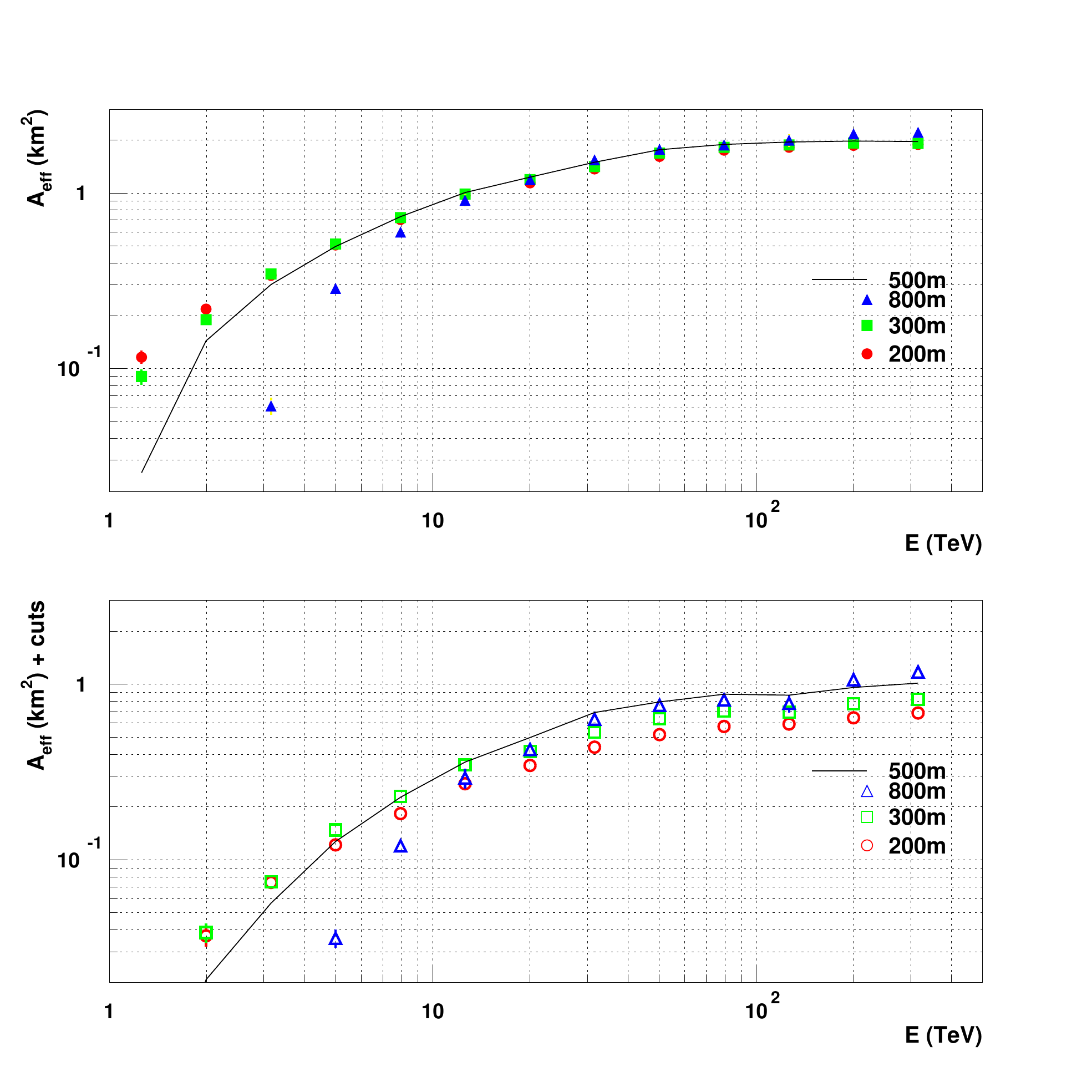}
\captionsetup{width=13cm}  
\caption{Effective areas for varying telescope separations with standard triggering combination, cleaning combination and image \textit{size} cut.
Top panel: The pre-cut effective area. For E $>$ 10 TeV, the effective areas are similar with $\approx$ 5 to 10$\%$ difference between all separations.
Bottom panel: The post-selection cut effective area. The differences are larger with 20$\%$ between the 200 \rm{m} and 300 \rm{m} separations, and 20$\%$ between the 300 \rm{m} and 500 \rm{m} separations.}

 \label{fig:area_sep_low_alt}
\end{centering}
\end{figure}

With this in mind, the effect of varying the telescope separation on the factors mentioned above can been investigated. By varying the telescope separation, the aim is to improve the collecting area of the cell whilst maintaining the event reconstruction and the rejection of proton showers. To illustrate the effect, results using four telescope separations will be shown; a 200 \rm{m}, 300 \rm{m}, 500 \rm{m} and 800 \rm{m}. These separations provide the smallest, mid-ranged, standard and largest separations considered in this investigation.\\

Figure~\ref{fig:area_sep_low_alt} shows the pre- and post-selection cut effective areas for varying telescope separations. The pre-cut effective area for an 800 \rm{m} separation displays a significant loss of events for E $<$ 10 TeV. The 800 \rm{m} separation thus increases the energy threshold for the cell which is undesired for PeX. The 200 \rm{m}, 300 \rm{m} and 500 \rm{m} separations provide similar effective areas for energies $<$ 10 TeV (Figure~\ref{fig:area_sep_low_alt} top panel). For E $>$ 100 TeV, the 500 \rm{m} and 800 \rm{m} separations provide slightly larger effective areas. The difference between all results is approximately 5 to 10$\%$ for E $>$ 10 TeV.
 
The post-selection cut effective area shows how the telescope separations affect the shape of the images in the camera and the reconstruction of shower direction (Figure~\ref{fig:area_sep_low_alt} bottom panel). For E $<$ 10 TeV, the smaller telescope separations provide the largest effective area while the 500 \rm{m} and 800 \rm{m} separations lose more events due to the telescopes being too far apart for these energies. For the E $>$ 10 TeV range, the 200 \rm{m} post-selection cut effective area loses more events via the shape and point source cut than the 500 \rm{m} and 800 \rm{m} separations. Therefore, the 200 \rm{m} separation provides poor shape parameterisation and shower reconstruction. The 300 \rm{m} effective area also suffers post-selection cuts compared to the 500 \rm{m} and 800 \rm{m} separations. However, not as many events are lost from the 300 \rm{m} separation as from the 200 \rm{m} separation.

Although that the pre-cut differences are small, the post-selection cut differences are 20$\%$ between the 200 \rm{m} and 300 \rm{m} separation results (Figure~\ref{fig:area_sep_low_alt} line and red circles) and 20$\%$ between the 300 \rm{m} and 500 \rm{m} separation results (Figure~\ref{fig:area_sep_low_alt} line and green squares). This suggests that increasing the separation between telescopes improves the event reconstruction and the calculation of parameters such as MSW, MSL and MSNpix. The angular resolution (r68) (Figure~\ref{fig:angres_sep_low_alt}) shows improvements with increasing telescope separation for most energies. 

\begin{figure}
\begin{centering}
\includegraphics[scale=0.65]{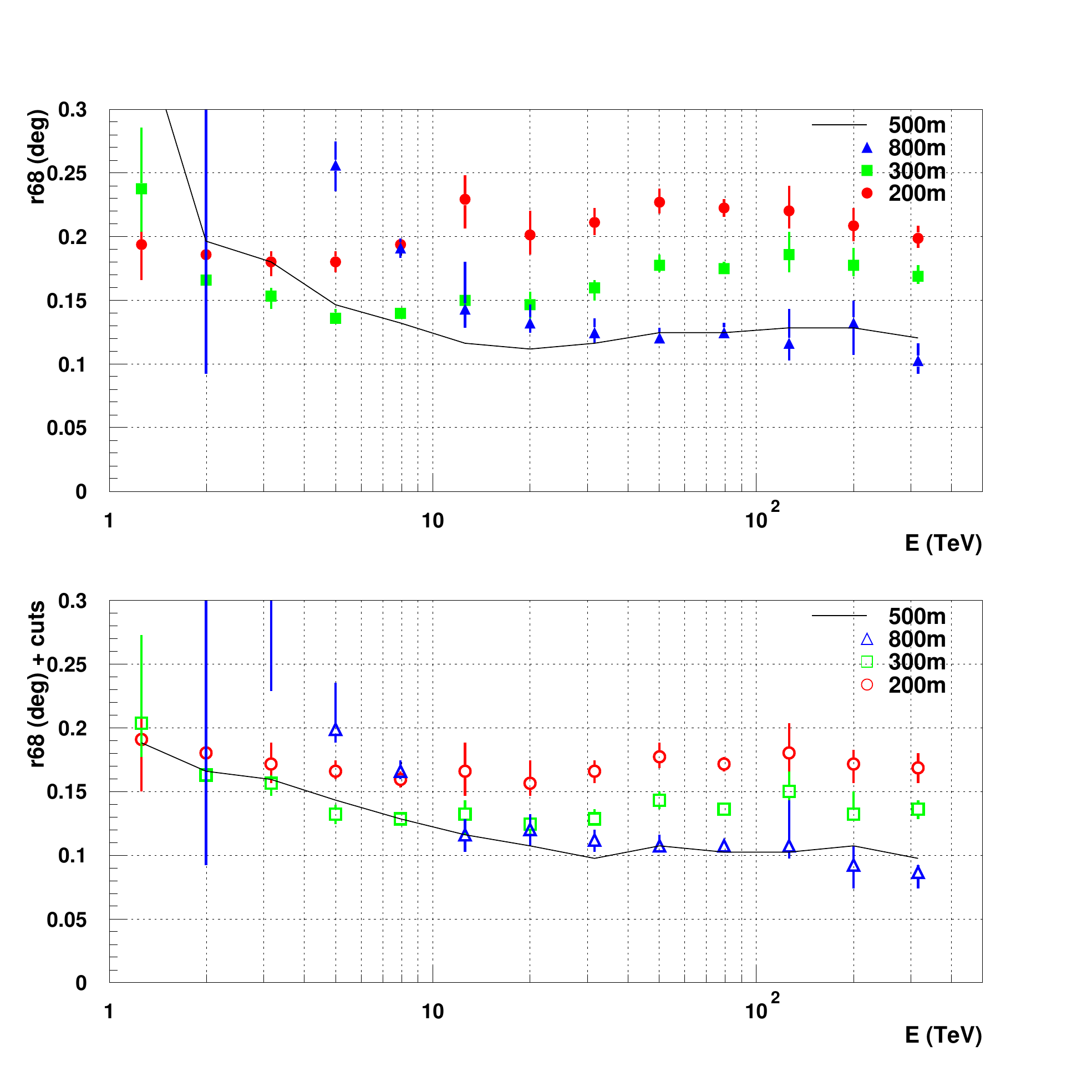}
\captionsetup{width=13cm}  
\caption{Angular resolution (r68) for varying telescope separations with standard triggering combination, cleaning combination and image \textit{size} cut.
Top panel: Pre-cut angular resolution. Bottom panel: Post-shape cut angular resolution. The trends seen in the pre- and post-shape cut angular resolution indicate that it improves with increasing telescope separation.}
 \label{fig:angres_sep_low_alt}
\end{centering}
\end{figure}


The 200 \rm{m} telescope separation overall provides poor event reconstruction compared to the larger separations except at the largest energies. The 500 \rm{m} separation provides a suitable post-shape cut angular resolution for all energies.

The post-shape cut energy resolution results are broken down into two components: the mean in the $\Delta{E}/E$ distribution and the RMS of the $\Delta{E}/E$ distribution (Figure~\ref{fig:energy_res_sep_low_alt}). The mean in the $\Delta{E}/E$ distribution appears similar for all telescope separations. The fact that the mean suffers minimal variation with varying telescope separation indicates that the telescope separation does not affect the estimation of the shower's true energy (Figure~\ref{fig:energy_res_sep_low_alt} top panel). All results appear to be biased towards positive $\Delta{E}/E$ mean values, except at the highest energies.


The RMS of the $\Delta{E}/E$ distribution clearly indicates that the 200 \rm{m} separation produces less precise energy reconstruction (Figure~\ref{fig:energy_res_sep_low_alt} red circles in bottom panel) than 500 \rm{m} and 800 \rm{m} separations.

\begin{figure}
\begin{centering}
\includegraphics[scale=0.65]{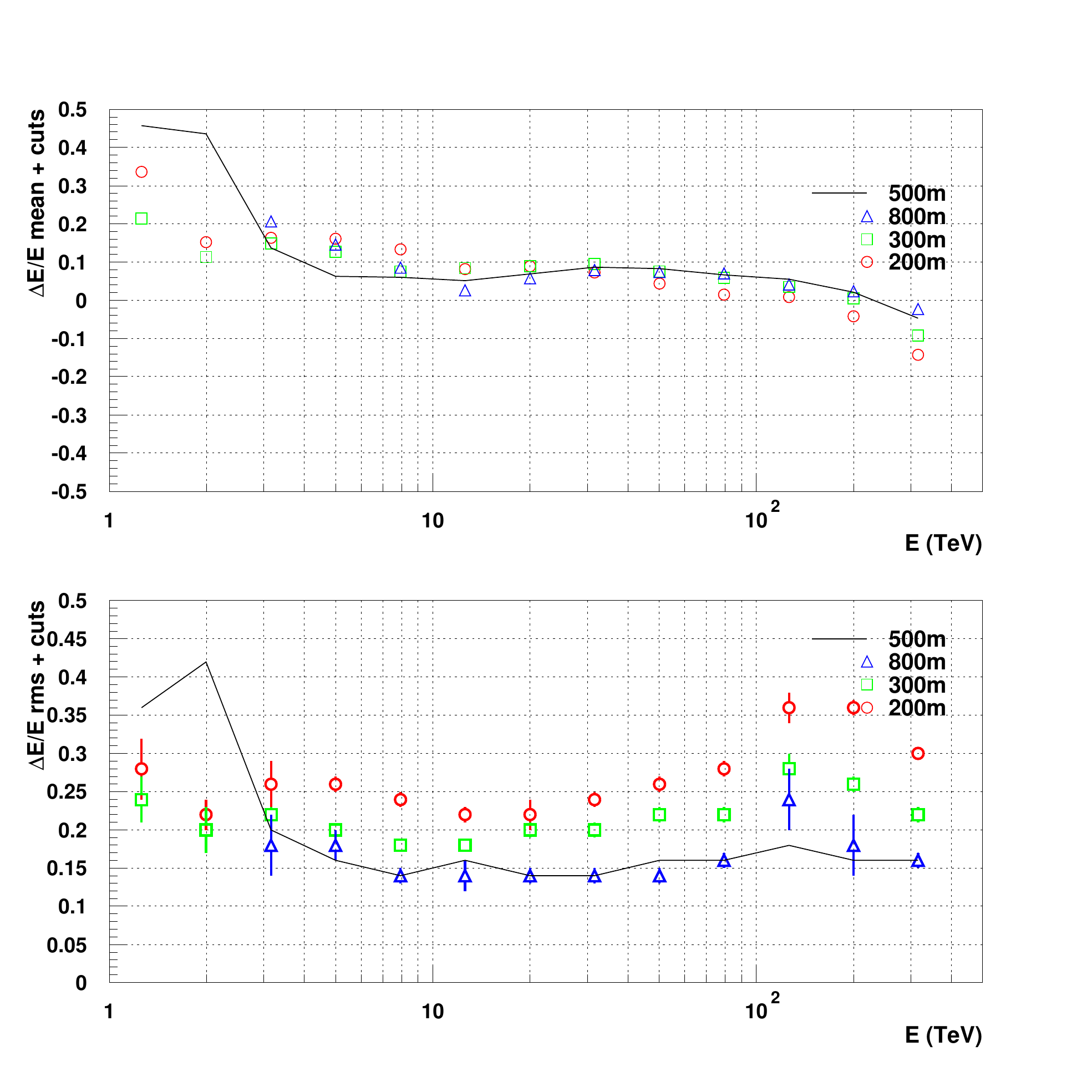}
\captionsetup{width=13cm}  
\caption{Post-shape cut energy resolution for varying telescope separation with standard triggering combination, cleaning combination and image \textit{size} cut. The top panel represents the mean in the $\Delta{E}/E$ distribution while the bottom panel represents the RMS of the $\Delta{E}/E$ distribution. A reasonable improvement is seen in RMS with large telescope separations compared to a small telescope separation, particularly for E $>$ 10 TeV.}

 \label{fig:energy_res_sep_low_alt}
\end{centering}
\end{figure}

Figure~\ref{fig:q-factor_low_sep_alt_low} represents the Q-factor for varying telescope separations. The protons for the 800 \rm{m} results have only been simulated up to 100 TeV. The results show that a 500 \rm{m} separation provides the best Q$_{fact}$ over the entire energy range. The 800 \rm{m} separation provides a poor quality factor for E $<$ 10 TeV, which is due to limited collection of low energy events. The events that have triggered the 800 \rm{m} separation appear to be of poor quality. As the energy increase, the Q$_{fact}$ increases which is similar to the angular resolution results.

\begin{figure}
\begin{centering}
\includegraphics[scale=0.65]{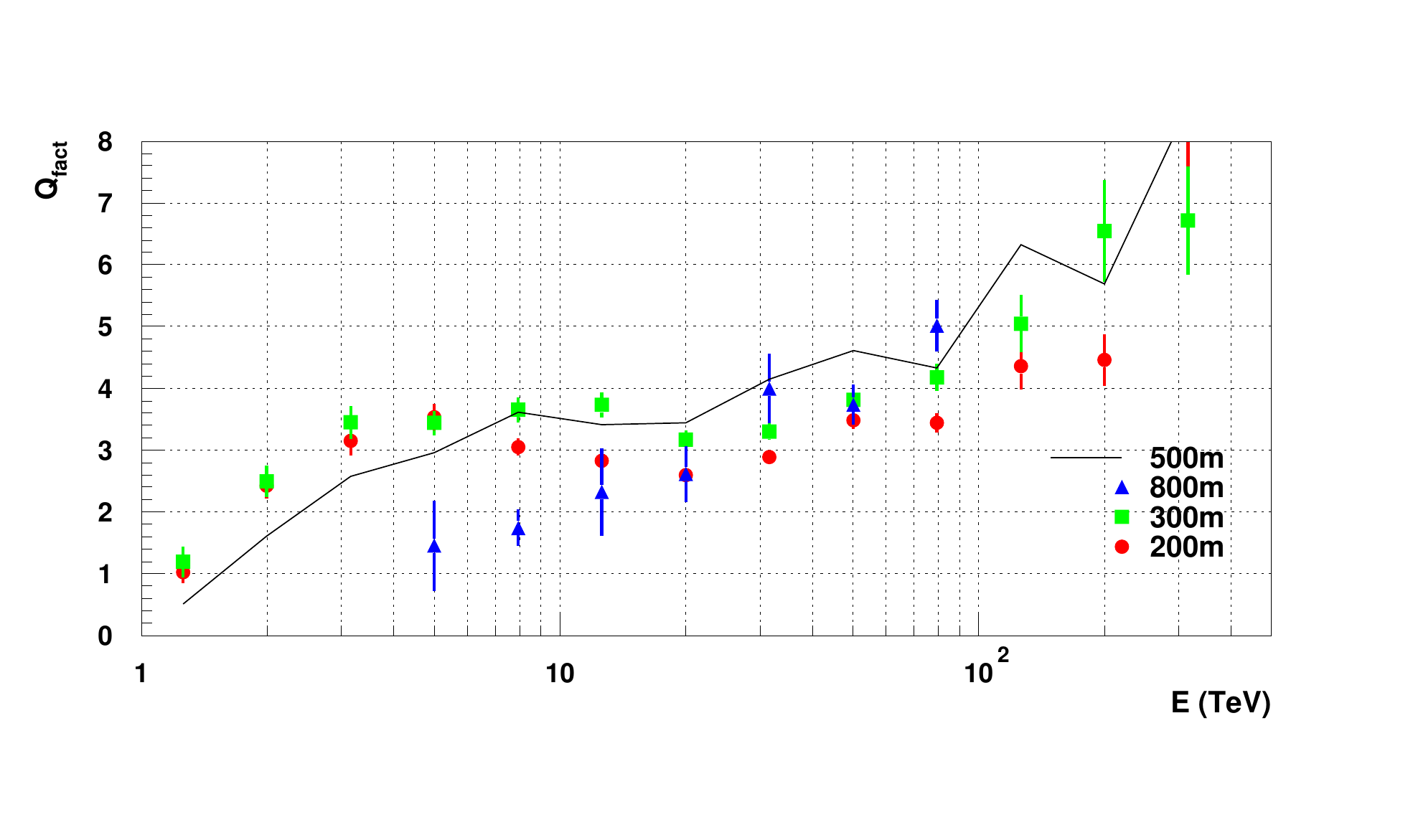}
\captionsetup{width=13cm}  
\caption{Q-factor for varying telescope separations with standard triggering combination, cleaning combination and image \textit{size} cut. The 800 \rm{m} results were only simulated from protons up to 100 TeV. The 500 \rm{m} separation provides the best Q$_{fact}$ value over the entire energy range.}

 \label{fig:q-factor_low_sep_alt_low}
\end{centering}
\end{figure}

The 800 \rm{m} separation has a significant loss of events at low energies, which is seen in the effective area curve and the large error bars associated with the angular resolution and the energy resolution RMS curves. The low energy showers cannot trigger multiple telescopes, so the events do not pass stereoscopic cuts.

The effective area curves suggest that similar $\gamma$-ray event statistics are collected for E $>$ 10 TeV for all separations (Figure~\ref{fig:area_sep_low_alt}). Figures~\ref{fig:area_sep_low_alt},~\ref{fig:angres_sep_low_alt} and~\ref{fig:energy_res_sep_low_alt} indicate that 200 \rm{m} or 300 \rm{m} separations offer a degraded event reconstruction performance. For such separations, stereo pairs of images will tend to be more parallel thereby reducing the direction and core reconstruction for Algorithm 1. The 800 \rm{m} separation appears suitable for the highest energies but has a high threshold of 10 TeV. The 500 \rm{m} separation appears to be a good compromise between telescope separations.

The 500 \rm{m} telescope separation provides the largest post-selection cut effective area suggesting that the separation allows for good collection of events, collects enough Cherenkov light from the showers to provide large sized images and gives a large opening angle between major axes for improved parameterisation and event reconstruction. 

The conclusion is that a 500 \rm{m} separation is a good choice to maintain sensitivity at high energies coupled with a few TeV threshold. For further studies, a 500 \rm{m} separation will be considered as the optimal value.

\section{Effect of Triggering Combinations}
 \label{sec:triggering_op}

	The triggering combination is used to minimise telescope triggers from the night sky background, also shown in \cite{Ricky}. However, by varying the triggering threshold, there could be a way to further optimise the trigger rate differences between $\gamma$-ray and proton events. By increasing the number of pixels required for a trigger, one limits the number of low energy or large core distance events due to the small size or elongated nature of images. These images usually provide poor reconstruction, which implies that a high \textit{threshold} value or a high \textit{n} pixel value, where \textit{n} is the adjacent number of pixels, should improve event reconstruction. 

	Varying the triggering combination could help reject proton events at this level since proton events have irregularly shaped images compared to equivalent $\gamma$-ray events. To test various triggering combinations, the \textit{threshold} value was varied from 4\textit{pe} up to and including 12\textit{pe} while the \textit{n} pixel value was varied from 2 pixels up to and including 5 pixels. These variations covered the main range of triggering combinations that are appropriate for PeX given the desire to maintain a few TeV trigger threshold. The total trigger size is defined as the number of \textit{pe} in a triggered image. For triggering combinations, the smallest total trigger size is 8\textit{pe} (4\textit{pe} $\times$ 2) while the largest total trigger size is 60\textit{pe} (12\textit{pe} $\times$ 5).

To best represent the angular resolution trend with varying triggering combinations, a 2D grey scale plot has been produced (Figure~\ref{fig:angres4_trigger}). The results have been broken into four logarithmic energy bands to better represent the angular resolution variation with different triggering combinations: 1 - 4.2 TeV, 4.2 - 22.3 TeV, 22.3 - 105.7 TeV and 105.7 - 500 TeV. The results show the angular resolution values vary by 0.05 degrees between the lowest and the highest angular resolution values for all energy brackets. The general trend indicates that a higher \textit{threshold} value and/or a higher \textit{n} pixel value improves the angular resolution. The improvement is larger when both \textit{threshold} value and \textit{n} pixel value are increased together. We note that the standard (6\textit{pe}, 3) combination does not provide the best angular resolution. However, other aspects need to be considered before this standard triggering combination is replaced. 

\begin{figure}
\begin{centering}
\includegraphics[scale=0.55, angle=270]{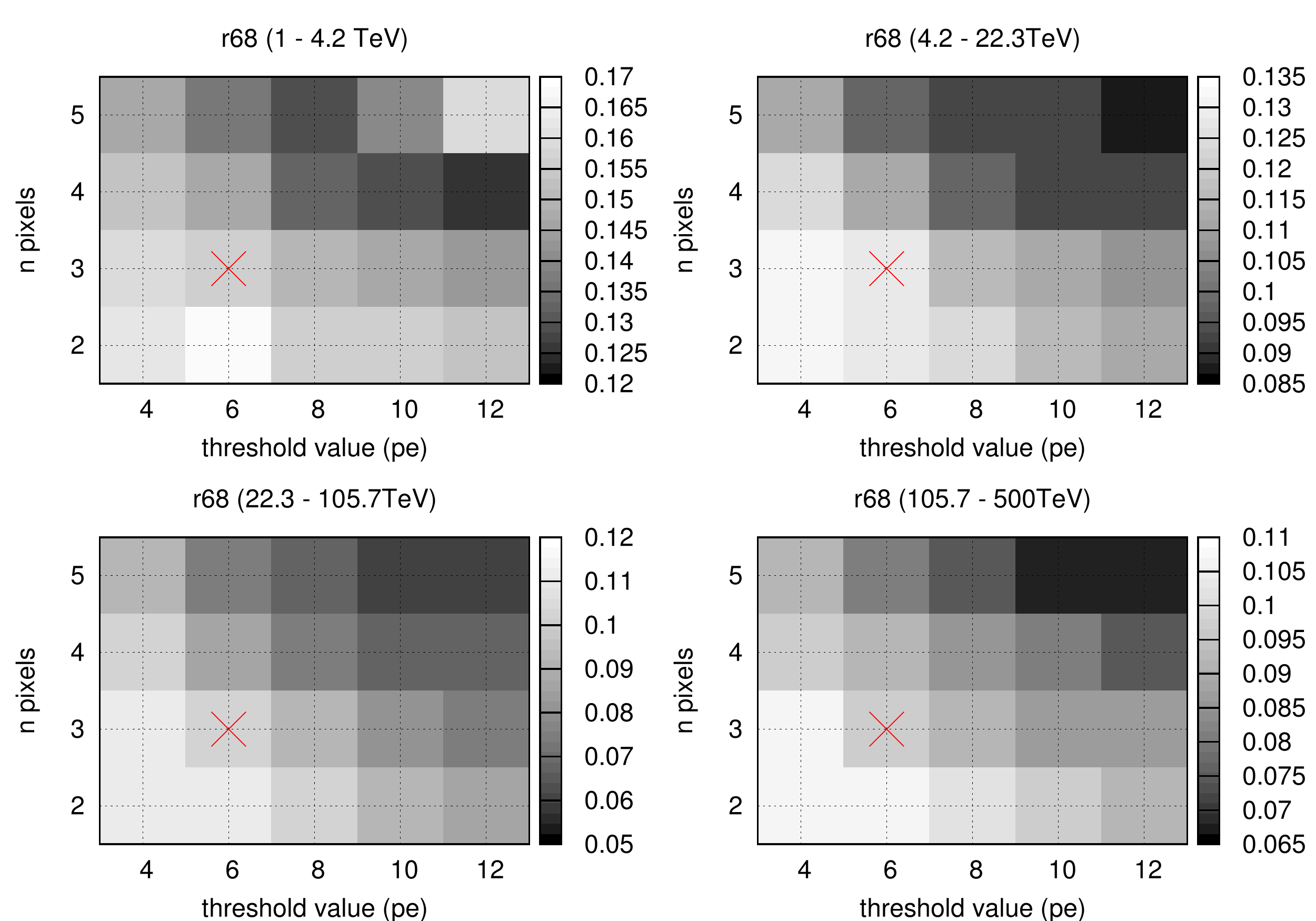}
\captionsetup{width=13cm}  
\caption{The angular resolution r68 (deg) for all triggering combinations with the standard cleaning combination, telescope separation and image \textit{size} cut. The x-axis represents the \textit{threshold} value in \textit{pe} and the y-axis represents the number of pixels \textit{n}. The 2D grey scale plots show the triggering combinations which provide the best angular resolution for each energy band. The general trend for all energy bands indicates that higher triggering \textit{threshold} values and higher \textit{n} pixel values provide improved angular resolution. The red cross represent the standard value.}

 \label{fig:angres4_trigger}
\end{centering}
\end{figure}

\begin{figure}
\begin{centering}
\includegraphics[scale=0.55, angle=270]{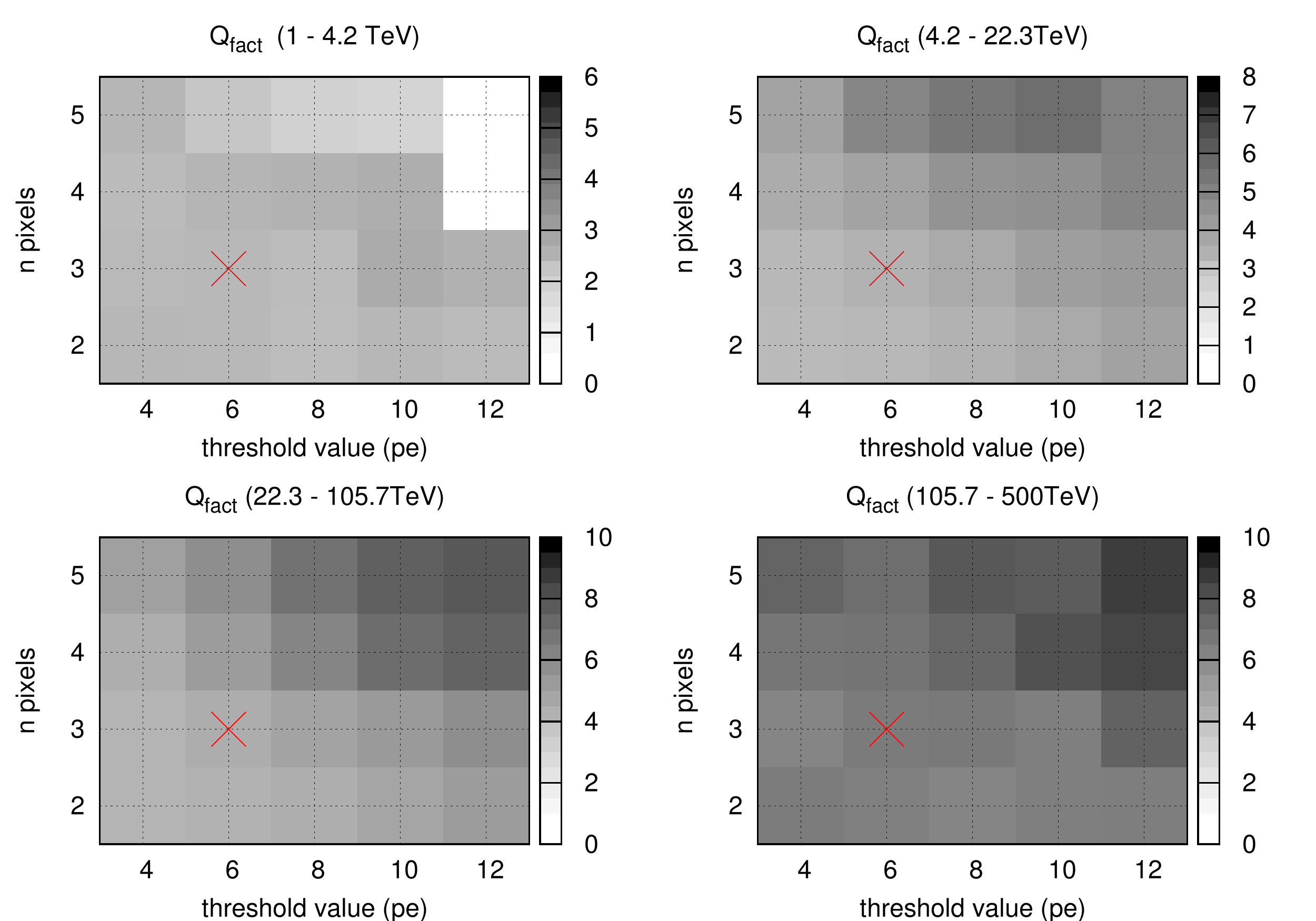}
\captionsetup{width=13cm}  
\caption{The Q$_{fact}$ values for all triggering combinations with the standard cleaning combination, telescope separation and image \textit{size} cut. The x-axis represents the \textit{threshold} value in \textit{pe} and the y-axis represents the number of pixels \textit{n}. The 2D grey scale plots display the Q$_{fact}$ as a function of triggering combinations. The results suggest that a higher triggering combination is preferred over a lower triggering combination. The white squares represent combinations that have remove all events. The red cross represent the standard value.}

 \label{fig:qfactor4_trigger}
\end{centering}
\end{figure}

The Q$_{fact}$ results have been displayed in the same way as the angular resolution (Figure~\ref{fig:qfactor4_trigger}). The same trend appears with the Q$_{fact}$. The Q$_{fact}$ improves with increasing \textit{threshold} value and \textit{n} pixel value. Again the standard triggering combination produces a slightly lower Q$_{fact}$ than some of the higher triggering combinations. The fraction of $\gamma$-rays lost due to increasing triggering combinations are displayed with the effective area plot (Figure~\ref{fig:area_triggering_low_alt}).

The angular resolution and Q$_{fact}$ results have shown that the standard triggering combination provides good but not necessarily best performance. The effect of varying triggering conditions on the effective area is also important to check. In Figure~\ref{fig:area_triggering_low_alt} we show the effective area for four specific triggering combinations. 



The weakest triggering combination, (4\textit{pe}, 2), provides the largest pre-cut effective area. The improvement in effective area compared to the standard triggering is around 10$\%$ for E $<$ 100 TeV. The pre-cut effective area above 100 TeV shows no difference. The strongest triggering combination, (12\textit{pe}, 5), shows a significant drop in the total number of events for the entire energy range.

\begin{figure}
\begin{centering}
\includegraphics[scale=0.65]{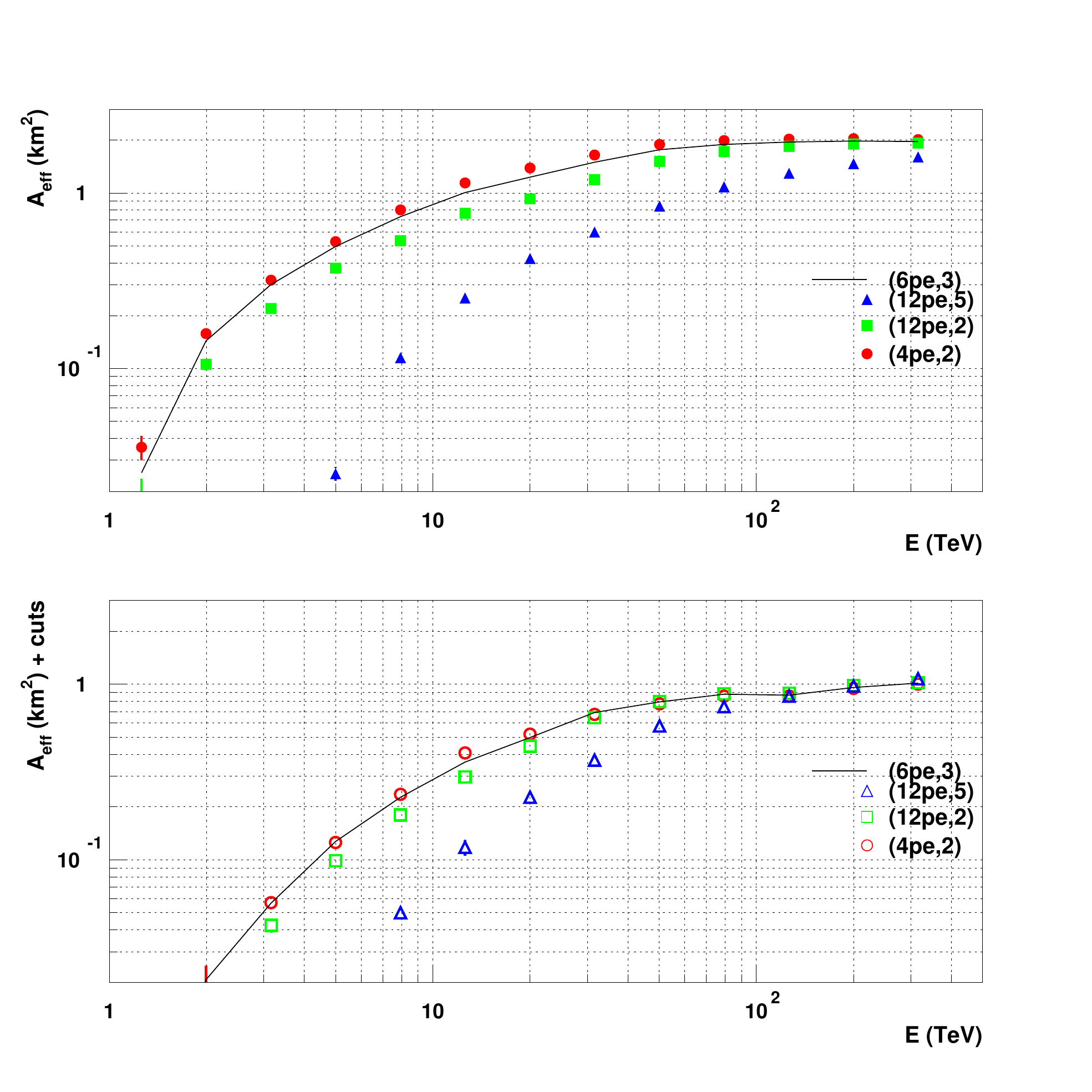}
\captionsetup{width=13cm}  
\caption{Effective area for varying triggering combinations with the standard cleaning combination, telescope separation and image \textit{size} cut. Top: The pre-cut effective area for the four selected triggering combinations. As the triggering \textit{threshold} value and \textit{n} pixel value increase, the effective area decreases. Bottom: The post-selection cut effective area for the same triggering combinations.}

 \label{fig:area_triggering_low_alt}
\end{centering}
\end{figure}

The post-selection cut effective area results show similar trends (Figure~\ref{fig:area_triggering_low_alt} bottom panel). The (4\textit{pe}, 2) combination provides the same post-selection cut effective area as the standard triggering combination. The (12\textit{pe}, 5) combination still causes a significant loss in the post-selection cut effective area. The triggering appears to be too restrictive for the low energy events, which pushes the energy threshold of the cell to above 10 TeV. 

\begin{figure}
\begin{centering}
\includegraphics[scale=0.65]{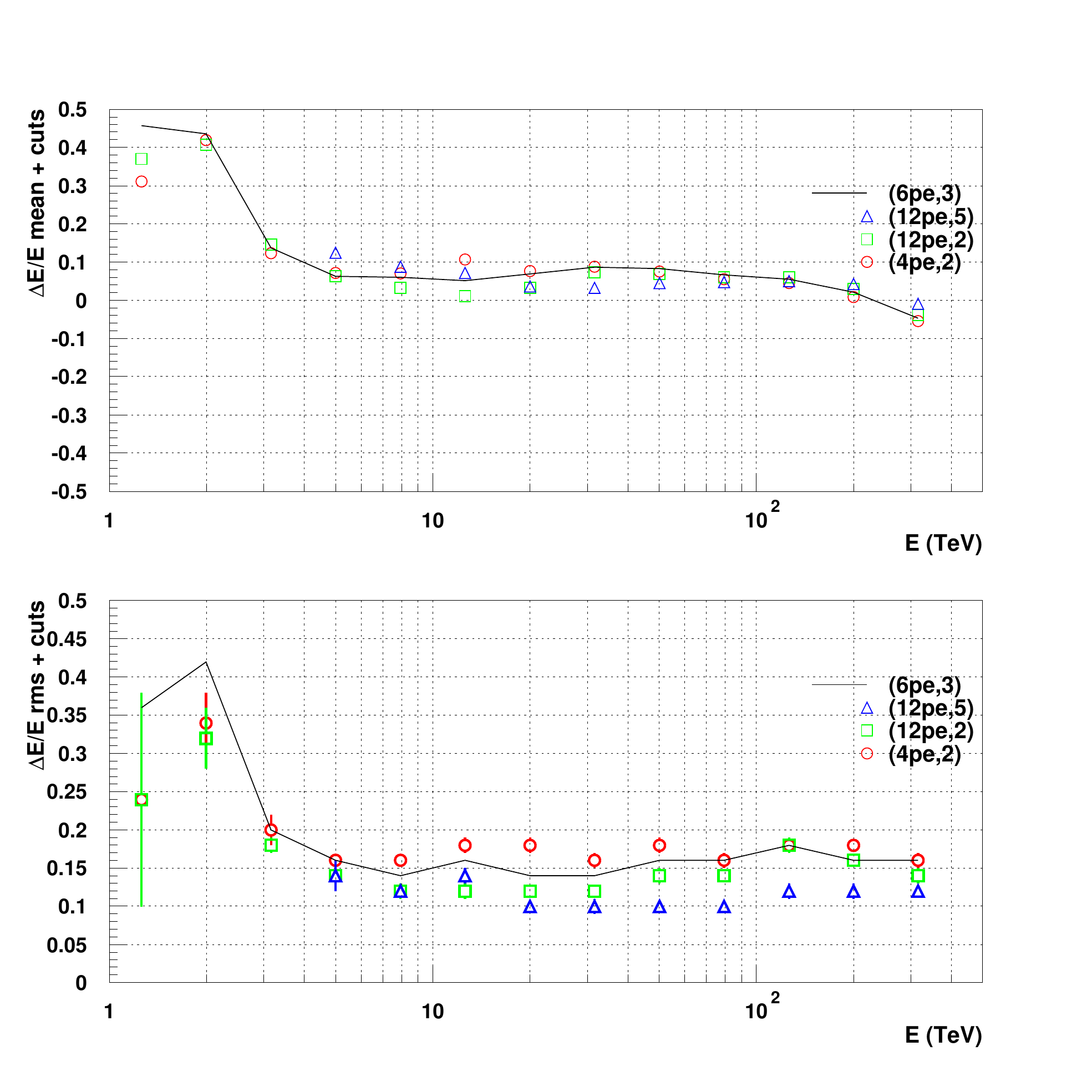}
\captionsetup{width=13cm}  
\caption{Post-shape cut energy resolution for the selected triggering combinations with the standard cleaning combinations, telescope separation and image \textit{size} cut. The top panel represents the mean in the $\Delta{E}/E$ distribution while the bottom plot represents the RMS of the $\Delta{E}/E$ distribution. The average RMS for the standard triggering combination is 15$\%$ for E $>$ 5 TeV.}

 \label{fig:energy_res_triggering_low_alt}
\end{centering}
\end{figure}

Finally, we compare the post-shape cut energy resolution for the triggering combinations in Figure~\ref{fig:energy_res_triggering_low_alt}. The mean in the $\Delta{E}/E$ points to more accurate energy reconstruction for the (12\textit{pe}, 5) combination. The RMS improves as the triggering \textit{threshold} value and \textit{n} pixel value increase. The standard triggering combination provides an average RMS of 15$\%$ for E $>$ 10 TeV, while a stronger triggering combination such as (12\textit{pe}, 5) provides an RMS on average of 10$\%$ for E $<$ 10 TeV (Figure~\ref{fig:energy_res_triggering_low_alt} bottom panel). The trend suggests that a stronger triggering combination does provide tighter $\Delta{E}/E$ distributions, but the selection of events is stricter and only the high quality events trigger the telescopes. \\

The trends in Figures~\ref{fig:angres4_trigger},~\ref{fig:qfactor4_trigger},~\ref{fig:area_triggering_low_alt} and~\ref{fig:energy_res_triggering_low_alt} are expected since, as the \textit{threshold} value and \textit{n} pixel value increase, so does the total trigger size. The low energy and/or large core distance events, which have smaller sized images and are elongated, fail to pass the triggering combination. Increasing the \textit{threshold} value and \textit{n} pixel value requires that the image is strong and spread over multiple pixels. The events which have small image sizes contain few pixels with weak Cherenkov signal. These small sized images occur at all energies from large core distance events, so a strong triggering combination will remove events at all energies.

Conversely, the weak triggering combination allows extremely small events to pass the triggering conditions. From Figure~\ref{fig:area_triggering_low_alt}, we see that the extra events from a (4\textit{pe}, 2) triggering combination offer no benefit since they are removed via cuts. This is seen when comparing the pre- and post-selection cut results. Therefore, any triggering combination weaker than the standard triggering seems to provide no improvement.

	The strong \textit{threshold} and strong \textit{n} pixel combination, (12\textit{pe}, 5), does provide improved angular resolution and Q$_{fact}$ results (Figures~\ref{fig:angres4_trigger} and Figure~\ref{fig:qfactor4_trigger}). However, the fraction of events lost compared to standard trigger values is significant. The strong triggering combination selects the best quality images, which is the reason for the significant improvement. The strong triggering combination improves the mean and RMS of the $\Delta{E}/E$ distribution. Again the selection of high quality images helps the energy reconstruction, provides a $\Delta{E}/E$ mean closer to zero, and gives a tighter distribution for reconstructed energies. The downside to a strong triggering combination is the loss in events, since a significant reduction in effective area both pre- and post-selection cuts is seen (Figure~\ref{fig:area_triggering_low_alt} green triangles). The other curves in Figure~\ref{fig:area_triggering_low_alt} suggest that a number of events which can produce good reconstruction are lost with a triggering combination that is too strong.\\

We can rule out the (4\textit{pe}, 2) triggering combination by considering the accidental night sky background trigger rates (Figure~\ref{fig:telescope_trigger} in Appendix~\ref{sec:appendix_plot}). The standard combination has an accidental night sky background rate of 10$^{-2}$ Hz \cite{Ricky}, while the (4\textit{pe}, 2) combination has an accidental night sky background rate of 80 Hz \cite{Ricky}, which is extremely high and is not acceptable for PeX. The other triggering combinations provide accidental NSB rates much lower than standard triggering combinations. Therefore, based on accidental trigger rates alone, the (4\textit{pe}, 2) triggering combination can be rejected as a possible combination for PeX.


The concluding remark is that a high \textit{threshold} value and high \textit{n} pixel value select higher quality images which improves reconstruction. However, the loss in event numbers is quite significant and not optimal for PeX given that once events are rejected at the trigger level, they are lost forever. It is therefore best to accept as many events as possible and to improve the cosmic ray rejection via software based methods. The lowest tolerable triggering combination for the cell appears to be the (6\textit{pe}, 3\textit{pe}) combination. Anything weaker offers no benefit to any parameter and the extra events are removed via software cuts later. 

\section{Effect of Image Cleaning Combinations}
 \label{sec:cleaning_op}

	The image cleaning (outlined in section~\ref{sec:cleaning_algorithm}) helps to mitigate the effect of the night sky background on the Cherenkov image. The cleaning algorithm sets pixels which contain low levels of photoelectrons to zero. The pixels that are dominated by NSB cause errors in the parameterisation of images. By varying the cleaning thresholds, more of the initial NSB is removed. The basic cleaning algorithm is usually denoted (\textit{picture}, \textit{boundary}) (see section~\ref{sec:cleaning_algorithm}). It is worthwhile considering if the standard image cleaning is optimal. Factors that need to be considered are the over- or under-cleaning of images and what is expected to occur in both scenarios:


\subsubsection{Over-cleaning}

	Over-cleaning an image implies that the cleaning thresholds are too high and start to remove Cherenkov signal from the image. The distinguishing features between $\gamma$-ray and proton events may be removed by over-cleaning, which makes it harder to separate them. With only the brightest part of the image remaining, event direction reconstruction may improve but the shape rejection power may suffer. Another effect of over-cleaning is that it might remove smaller sized images, these images are usually from low energy or large core distance events. This effect may lower the total effective area of PeX and raise its energy threshold.
	
\subsubsection{Under-cleaning}

	Under-cleaning of the image may leave too much night sky background pixels in the image. Pixels in the outer parts of the image (the boundary pixels) may contain a high fraction of night sky background compared to the Cherenkov signal. This disrupts the Hillas parameterisation hence the shower reconstruction. However, a weak cleaning threshold has an increased chance of improving the effective area by allowing smaller images to pass the cleaning algorithm. Optimisation of the cleaning thresholds will provide a balance between over- and under-cleaning. \\

\begin{figure}
\begin{centering}
\includegraphics[scale=0.55, angle=270]{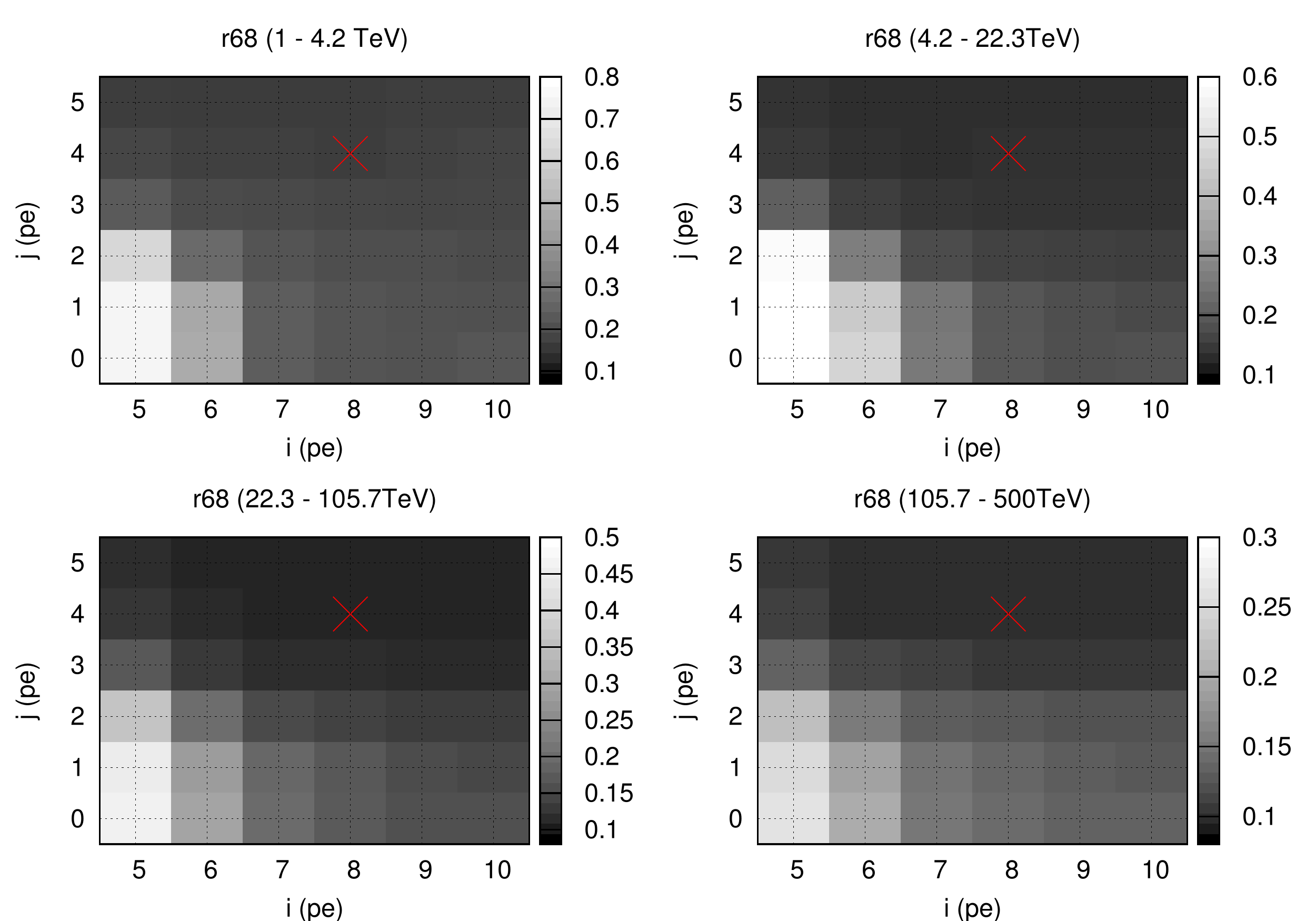}
\captionsetup{width=13cm}  
 \caption{Angular resolution r68 (deg) for a range of cleaning combinations with standard triggering combination, telescope separation and image \textit{size} cut. The x-axis represents the cleaning \textit{picture} value, \textit{i} and the y-axis represents the cleaning \textit{boundary} value, \textit{j}. The 2D grey scale plots show the cleaning combinations which provide the best angular resolution in each energy band. The general trend for all energies indicates that a stronger cleaning combination provides an improved angular resolution. The red cross represents the standard value.}
 \label{fig:angres4_clean}
\end{centering}
\end{figure}

	For this investigation, both the \textit{picture} and \textit{boundary} values have been varied within the definition of cleaning values (section~\ref{sec:cleaning_algorithm}). The triggering combination, telescope separation and image \textit{size} cut are all left at standard values and are not changed throughout the cleaning combination study.

	The \textit{picture} value has been varied from 5\textit{pe} to 10\textit{pe} while the \textit{boundary} value varies from 0\textit{pe} to 5\textit{pe}. \\

	The angular resolution (r68) changes have been displayed in Figure~\ref{fig:angres4_clean} as 2D grey scale plots and split into logarithmic energy bands: 1 - 4.2 TeV, 4.2 - 22.3 TeV, 22.3 - 105.7 TeV and 105.7 - 500 TeV . The angular resolution trend with cleaning combination indicates that a strong cleaning combination provides the improved angular resolution for all energies. A weak cleaning combination, \textit{picture} value $<$ 6 and \textit{boundary} value $<$ 2, leads to a considerable worsening of angular resolution.

	The angular resolution results for the stronger cleaning combinations show minimal to no variation, which implies that the pixels left within the image contain strong signal. Figure~\ref{fig:angres4_clean} clearly shows that a \textit{boundary} value is required when a weak \textit{picture} value is used. For a \textit{picture} value of 10 \textit{pe}, having 0 or 1 \textit{pe} \textit{boundary} value provides reasonable angular resolution. The standard cleaning combination provides good angular resolution compared to the other strong cleaning combinations. 
	The best angular resolution is provided by \textit{picture} value $\geq$ 7 and \textit{boundary} value $\geq$ 3. The next question to ask is; are these combinations removing distinguishing details from the images which affect the shape separation of $\gamma$-ray and proton events?

\begin{figure}
\begin{centering}
\includegraphics[scale=0.55, angle=270]{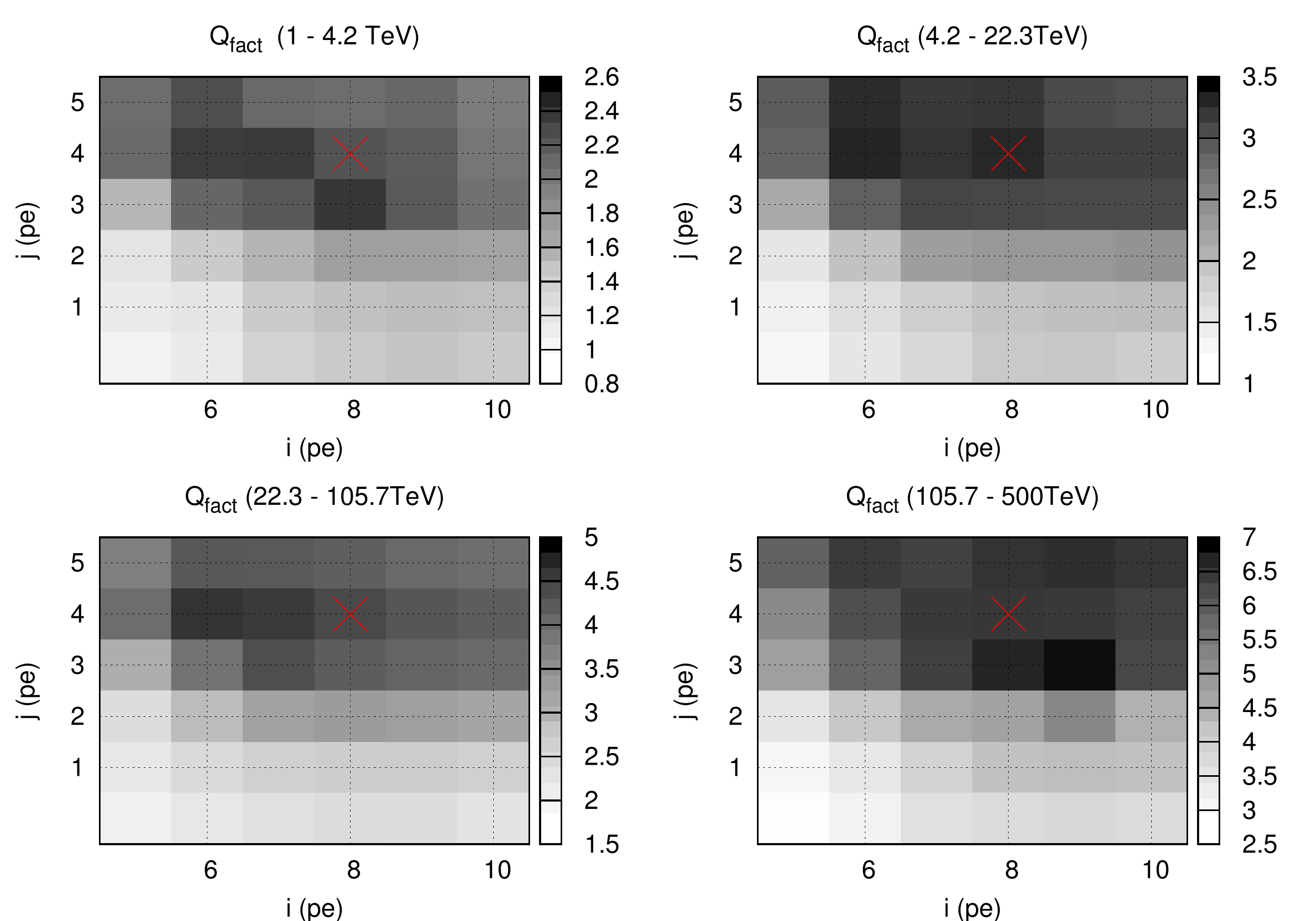}
\captionsetup{width=13cm}  
 \caption{The Q$_{fact}$ values for a range of cleaning combinations with standard triggering combination, telescope separation and image \textit{size} cut. The x-axis represents the \textit{picture} value, \textit{i}, and the y-axis represents the \textit{boundary} value, \textit{j}. The 2D grey scale plots display the Q$_{fact}$ as a function of cleaning combination. The Q$_{fact}$ shows variation over the four energy bands with no distinct region that provides the best shape separation between $\gamma$-ray and proton events. The red cross represents the standard value.}
 \label{fig:qfactor4_clean}
\end{centering}
\end{figure}

The Q$_{fact}$ results have been displayed in Figure~\ref{fig:qfactor4_clean} as 2D grey scale plots in four different energy bands as for Figure~\ref{fig:angres4_clean}. The Q$_{fact}$ results show an interesting trend across all energies. 
For the lowest energy band, 1 - 4.2 TeV, a weak \textit{picture} and strong \textit{boundary} combination, (5\textit{pe}, 5\textit{pe}), provides the best Q$_{fact}$. The other cleaning combinations are too strong for the energy range and remove a significant fraction of the image. For the next energy band, 4.2 - 22.3 TeV, a slightly stronger \textit{picture} value of 6\textit{pe} is preferred for the optimum Q$_{fact}$ value. For the last two energy bands, E $>$ 22.3 TeV, a slightly stronger \textit{picture} value can be used, 8 or 9\textit{pe}, since there is more Cherenkov light in each shower. The \textit{boundary} value must be $\geq$ 3 for all energies, which ensures that the night sky background, electronic noise and low level Cherenkov signal are removed via the cleaning algorithm. A \textit{boundary} value $<$ 3 results in the separation between $\gamma$-ray and proton events being lost and both shower images start to look similar. A (10\textit{pe}, 0 or 1\textit{pe}) cleaning combination is therefore not appropriate. The standard cleaning combination appears to provide the best or near best Q$_{fact}$ result for most energy bands. A strong combination, (10\textit{pe}, 5\textit{pe}), provides a slightly worse Q$_{fact}$ than the standard cleaning. \\

\begin{figure}
\begin{centering}
\includegraphics[scale=0.65]{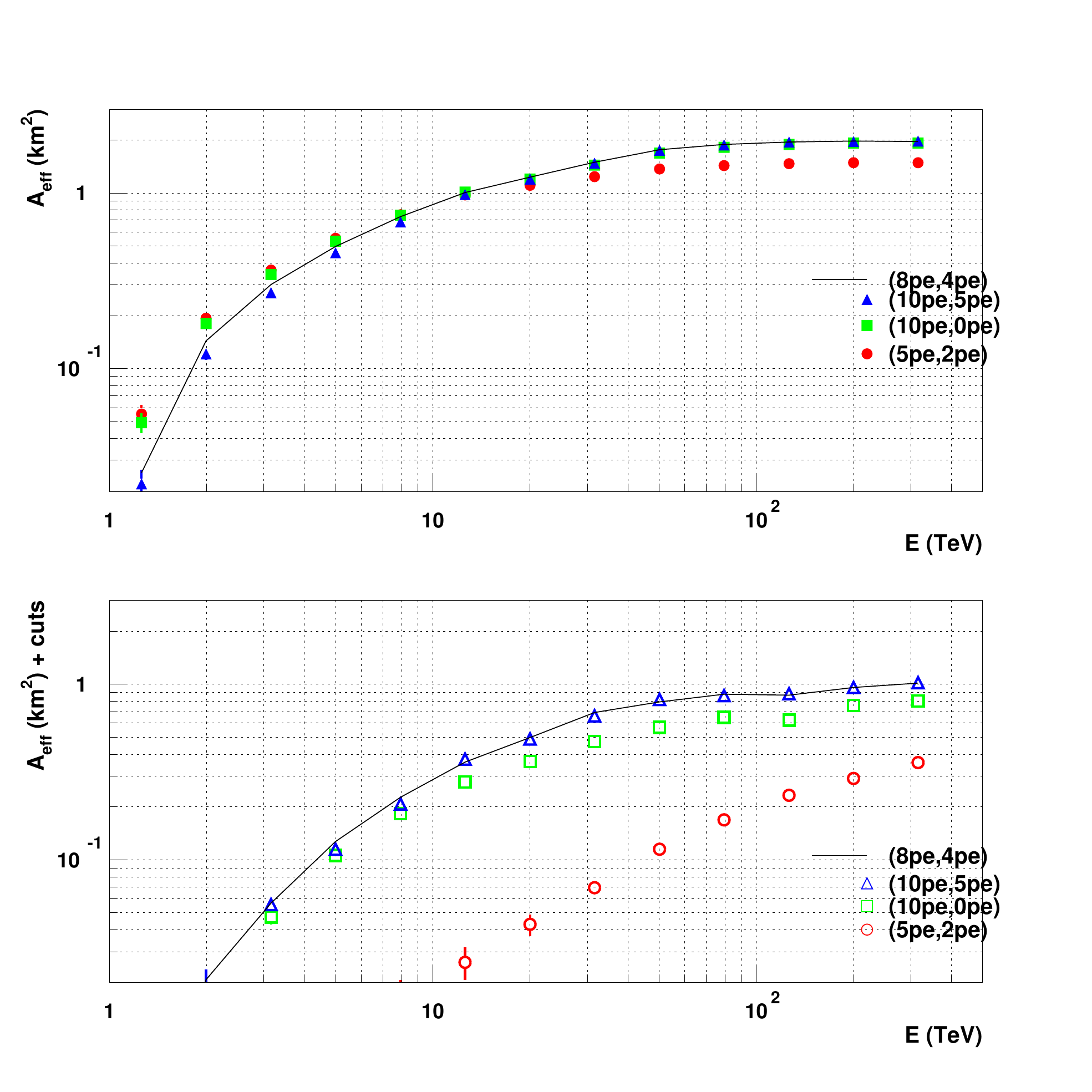}
\captionsetup{width=13cm}  
 \caption{Effective area for the four specific cleaning combinations with standard triggering combination, telescope separation and image \textit{size} cut. Top: The pre-cut effective area. For E $>$ 10 TeV, three of the cleaning combinations provide the same effective area while the (5\textit{pe}, 2\textit{pe}) combination has reached an upper limit. Bottom: The post-selection cut effective area. The standard cleaning produces the optimum post-selection cut effective area.}
 \label{fig:area_cleaning_low_alt}
\end{centering}
\end{figure}

We also looked at the effect on effective area and energy resolution, using four specific cleaning combinations.


	The effective area results in Figure~\ref{fig:area_cleaning_low_alt} provide a look at the total number of $\gamma$-ray events. The pre-cut effective area shows that the (5\textit{pe}, 2\textit{pe}) and (10\textit{pe}, 0\textit{pe}) combinations provide the largest area for E $<$ 10 TeV. The other cleaning combinations are too strong for the small image sizes produced at these energies. For E $>$ 10 TeV, the pre-cut effective area for the (5\textit{pe}, 2\textit{pe}) combination plateaus around 1.4\rm{km$^{2}$}, while the other combinations plateau at 2\rm{km$^{2}$}. 
	
	The post-selection cut effective area in Figure~\ref{fig:area_cleaning_low_alt} bottom panel shows a significant drop in $\gamma$-ray events for the (5\textit{pe}, 2\textit{pe}) combination. As seen in Q$_{fact}$ (Figure~\ref{fig:qfactor4_clean}), the \textit{boundary} value must be $\geq$ 3, otherwise the image shapes are affected by the extra pixels passing the cleaning algorithm. Therefore, the weak cleaning combination produces a loss in $\gamma$-ray events after cuts (Figure~\ref{fig:area_cleaning_low_alt}). Again the (10\textit{pe}, 0\textit{pe}) cleaning combination does not produce the same quality of results as the standard combination.
	
	The standard and the strong threshold combination appear to provide the generally optimal post-selection cut effective areas.

\begin{figure}
\begin{centering}
\includegraphics[scale=0.65]{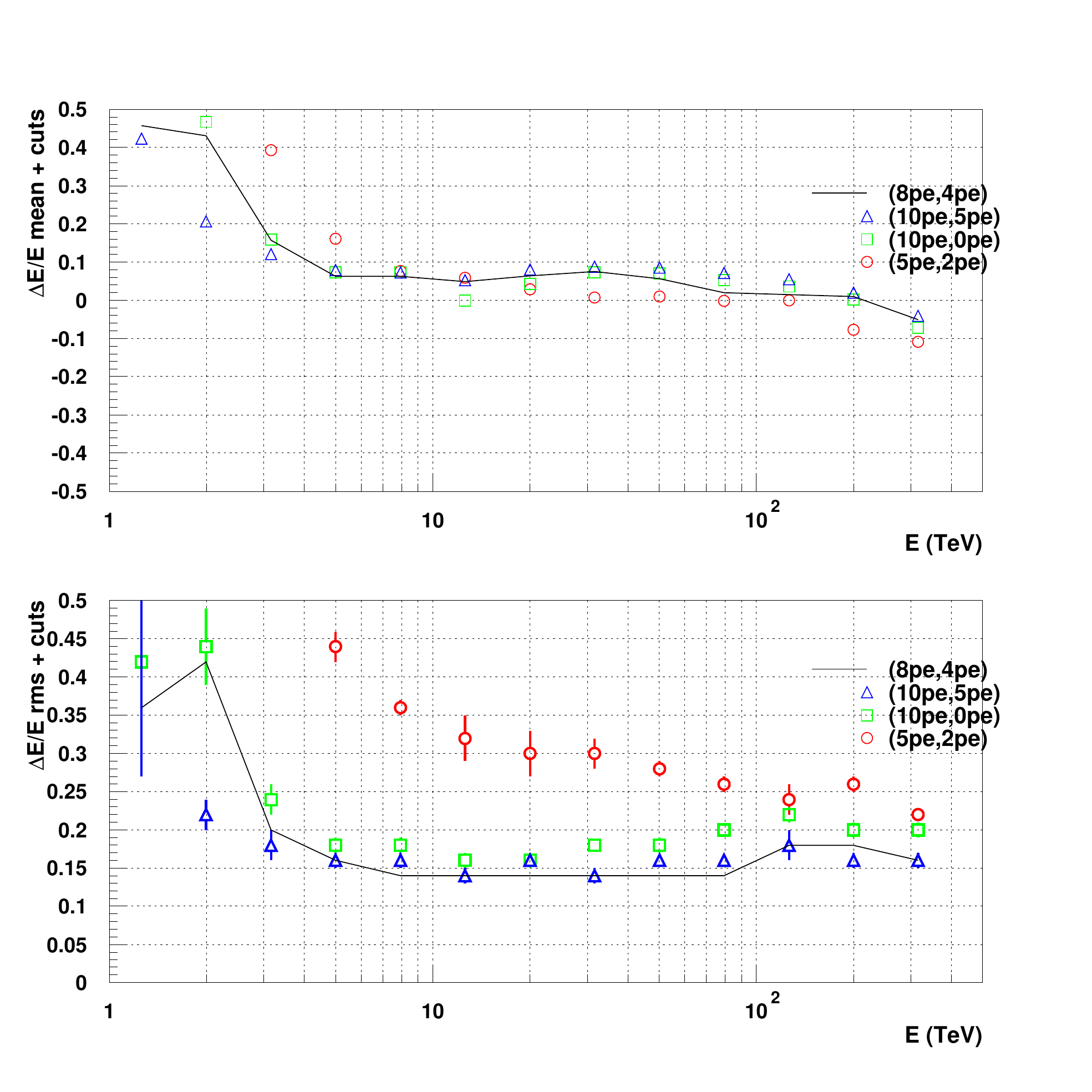}
\captionsetup{width=13cm}  
 \caption{Mean and RMS for the post-shape cut $\Delta{E}/E$ distribution using the standard triggering combination, telescope separation and image \textit{size} cut. Top: The mean for the (5\textit{pe}, 2\textit{pe}) combination shows most variation over the whole energy range, while the other cleaning combinations are more stable.
Bottom: The RMS for The standard cleaning combination and the highest combination, (10\textit{pe}, 5\textit{pe},) provide the tightest $\Delta{E}/E$ distribution.}
 \label{fig:energy_res_cleaning_low_alt}
\end{centering}
\end{figure}

	Figure~\ref{fig:energy_res_cleaning_low_alt} shows the post-shape cut energy resolution for varying cleaning combinations. The figure illustrates that the mean in the $\Delta{E}/E$ distribution is positive for E $<$ 100 TeV. For E $>$ 100 TeV, the $\Delta{E}/E$ mean approaches 0, which indicates that the over estimation of the reconstructed energy is reduced at high energies. The low energy events produce small images, which allow for larger errors in parameterisation and event reconstruction. The weak cleaning combination, (5\textit{pe}, 2\textit{pe}), shows variation in $\Delta{E}/E$ mean over the whole energy range compared to the other combinations. The improvement in mean for the (5\textit{pe}, 2\textit{pe}) combination is due to the selection of events passing shape cuts.
	
	The post-shape cut $\Delta{E}/E$ RMS is roughly 15$\%$ for E $>$ 10 TeV for the best cleaning combination. The weakest cleaning combination does show a significant worsening of the in RMS across all energies. The general trend shows that the RMS for the post-shape cut $\Delta{E}/E$ distribution improves with increasing cleaning threshold, but the standard cleaning appear quite adequate. \\
	
	The cleaning is used to mitigate NSB and unwanted pixels from the image. The unwanted pixels contain some mixture of Cherenkov photons from the tail of the shower plus night sky background (NSB). The tail of the shower usually consists of Cherenkov photons produced very low in the atmosphere and after the shower maximum. The signal from the tail of the shower is diffuse and generally weak compared to the main part of the shower and is more susceptible to NSB contamination. The evidence for under cleaning and unwanted photons from the tail of the shower is indicated in the angular resolution plots (Figure~\ref{fig:angres4_clean}), where the (5\textit{pe}, 2\textit{pe}) combination shows poor reconstructed directions. 

	These pixels can also shift the C.O.G of the image especially for large core distance events, which can be truncated by the edge of the camera. The extra pixels, which pass the low cleaning combination, usually appear towards the edge of the camera (Figure~\ref{fig:tail_of_shower} in Appendix~\ref{sec:appendix_plot}). This implies that the section of the image towards the edge of the camera has a higher \textit{pe} count, which shifts the C.O.G further away from the centre of the camera. With the C.O.G closer to the edge of the camera, there is a higher chance that more images will be removed by the \textit{dis2} cut. Figure~\ref{fig:dis2_plots_low_alt2} shows the \textit{dis2} values for a (5\textit{pe}, 2\textit{pe}) cleaning combination. We see that above 10 TeV, there is a number of images that would be removed by a \textit{dis2} cut. A majority come from small sized images and large core distance events. A shift in the C.O.G is seen in the pre-cut effective area (Figure~\ref{fig:area_cleaning_low_alt} top panel) since the weakest cleaning combination, (5\textit{pe}, 2\textit{pe}), shows a plateau in maximum effective area compared to the other cleaning combinations. This is due to the events being removed by a \textit{dis2} cut. 
The current \textit{dis2} cut provides good rejection of images that have a C.O.G at the edge of the camera. However, the \textit{dis2} cut has not been optimised and further work is required to refine this cut. As it stands, the edge effects are not completely removed but the effect has been reduced with a \textit{dis2} cut.

\begin{figure}
\begin{centering}
\includegraphics[scale=0.8]{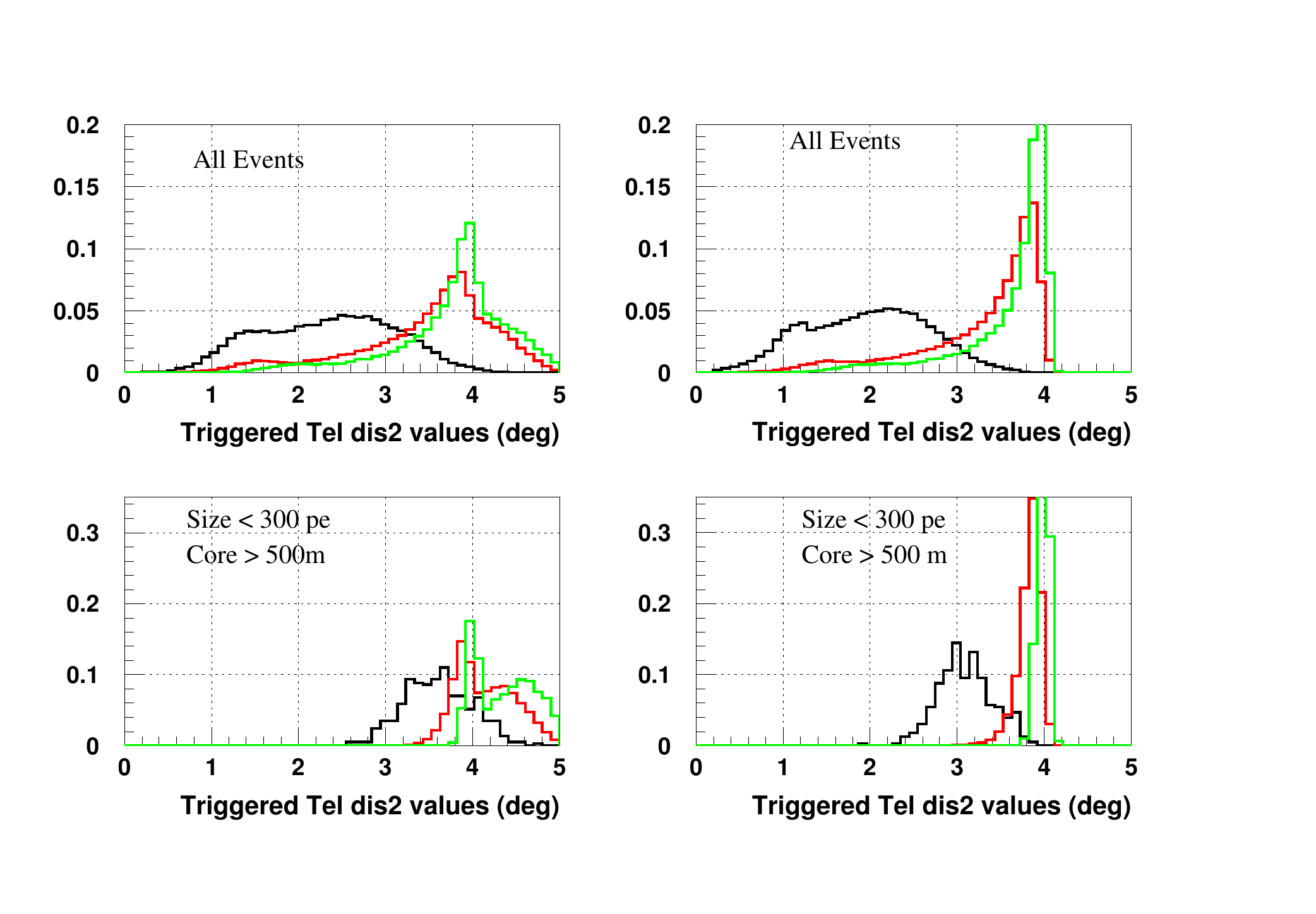}
\captionsetup{width=13cm}  
 \caption{The various \textit{dis2} value for events with (5\textit{pe}, 2\textit{pe}) cleaning combination (left) and (8\textit{pe}, 4\textit{pe}) cleaning combination (right). The events have been split into three different energy bands: 1 to 10 TeV (black), 10 to 100 TeV (red) and 100 to 500 TeV (green). The top panels shows all events and the bottom panels show events with an image size of 300\textit{pe} or less and a core distance greater than 500 \rm{m}. The edge of the physical camera is 4.1$^{\circ}$.}
 \label{fig:dis2_plots_low_alt2}
\end{centering}
\end{figure}

The strong cleaning combinations remove these extra pixels, which shifts the C.O.G back towards the centre of the camera and allows the images to pass the \textit{dis2} stereoscopic cut. Figure~\ref{fig:dis2_plots_low_alt2} shows the \textit{dis2} values for a (8\textit{pe}, 4\textit{pe}) cleaning combination. We can see that the extra pixels are removed and fewer events are removed by a \textit{dis2} cut.

\begin{figure}
\begin{centering}
\includegraphics[scale=0.5]{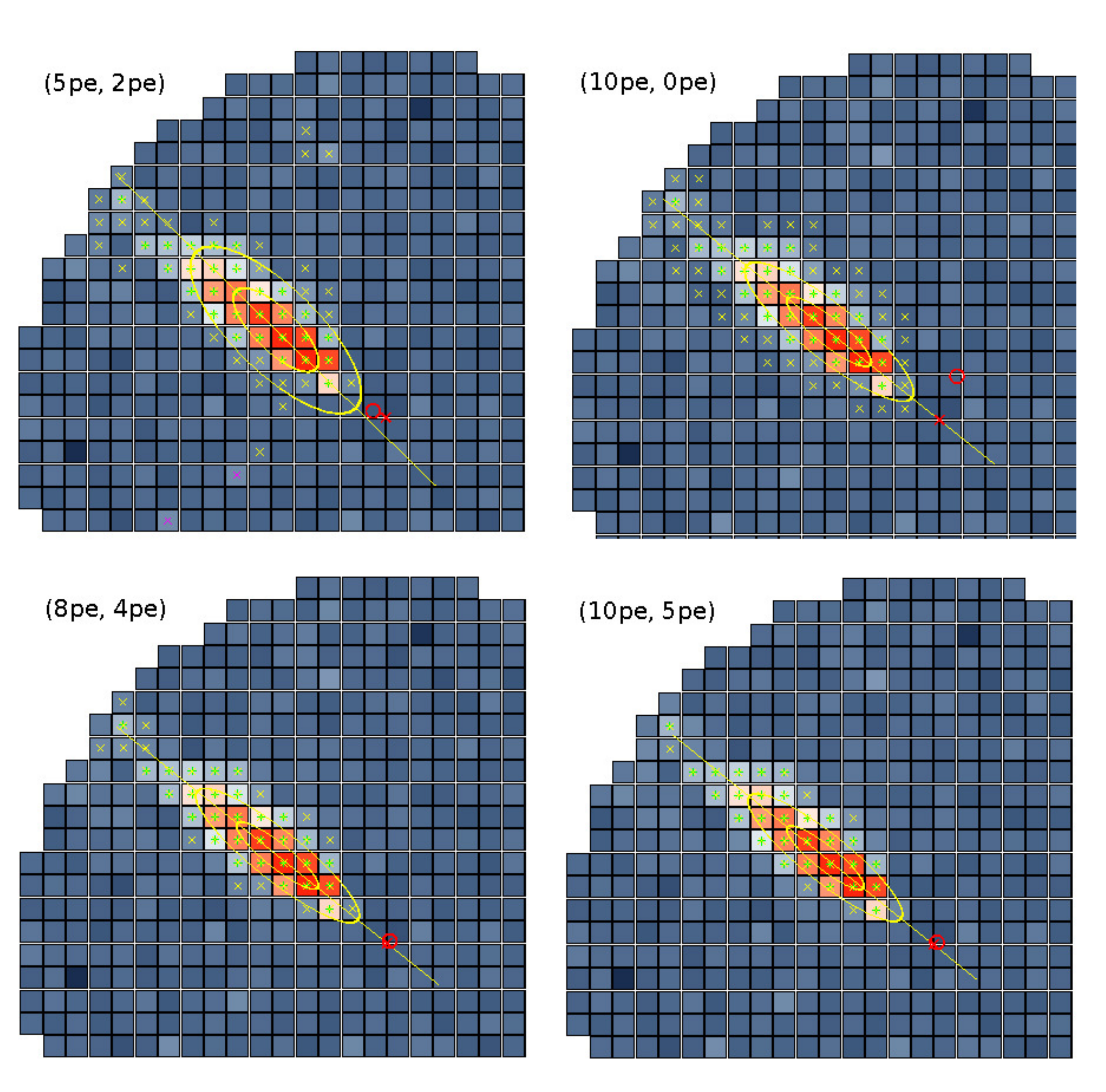}
\captionsetup{width=13cm}  
 \caption{A single 70 TeV image at a core distance of 350 \rm{m} that has been cleaned by the four different cleaning combinations. The image is from a shower that triggers two telescopes. The yellow crosses represent the pixels which pass the cleaning combinations, the red ring represents the reconstructed direction from Algorithm 1 using multiple images and the red cross represents the simulated true direction.}
 \label{fig:cleaning_images}
\end{centering}
\end{figure}
	
	If the tail of the shower is included, it also widens the image as it moves further from the centre of the camera (for an on-axis point source) (Figure~\ref{fig:tail_of_shower} in Appendix~\ref{sec:appendix_plot}). A widening of the image towards the edge of the camera can provide a larger variation or error on the major axis of the image. Using a major axis which has been affected by a widened image will disrupt the reconstruction of shower direction.  Figure~\ref{fig:cleaning_images} illustrates the reconstruction of a shower when the four cleaning combinations are applied. Comparing the (5\textit{pe}, 2\textit{pe}) and (8\textit{pe}, 4\textit{pe}) combinations, a small contribution from the the tail of the shower appears in the (5\textit{pe}, 2\textit{pe}) combination. There is more Cherenkov signal appearing at the outer edge of the image and around the sides of the image. By comparing the major axis, it can be seen that the (5\textit{pe}, 2\textit{pe}) and (10\textit{pe}, 0\textit{pe}) combinations do not pass through the simulated or true shower direction, whilst the (8\textit{pe}, 4\textit{pe}) and (10\textit{pe}, 5\textit{pe}) major axes run through the true simulated direction. 
	
	The image \textit{width} and the number of pixels in an image are also disrupted (Figure~\ref{fig:rejection_cleaning} in Appendix~\ref{sec:appendix_plot}). These are the two main factors which distinguish between $\gamma$-ray and proton showers. Figure~\ref{fig:rejection_cleaning} in Appendix~\ref{sec:appendix_plot} illustrates the effect of over- and under-cleaning on the shape parameters. Under-cleaning the image causes fewer $\gamma$-ray events to pass the shape cuts and it widens the distributions since the events have more variation in image parameters.

A widening in the mean scaled distribution lowers the performance of parameters like Q$_{fact}$ (Figure~\ref{fig:comparison_rej}). The peak in Q$_{fact}$ is at the same value for all curves. However, the maximum Q$_{fact}$ peak illustrates how well the events can be separated after each cleaning combination has been applied. The results indicate that the same MSW, MSL and MSNpix cuts can be used for all cleaning variations. The (5\textit{pe}, 2\textit{pe}) combination provides the lowest Q$_{fact}$ value while the standard and (10\textit{pe}, 5\textit{pe}) combination provide similar results. For (5\textit{pe}, 2\textit{pe}), a huge fraction of $\gamma$-ray events are removed via cuts for the post-selection cut effective area (Figure~\ref{fig:area_cleaning_low_alt} red circles). 

\begin{figure}
\begin{centering}
\includegraphics[scale=0.70]{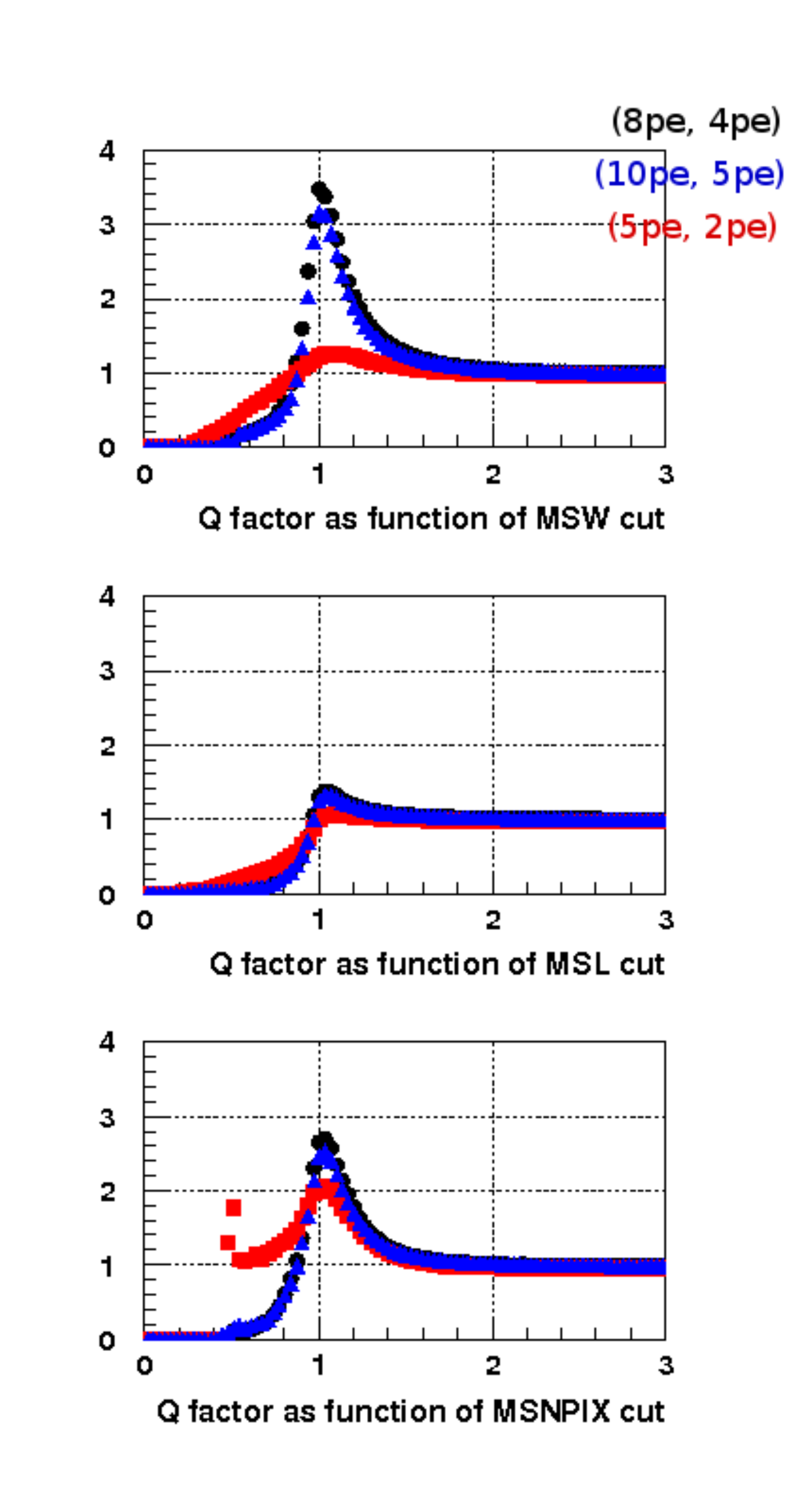}
\captionsetup{width=13cm}  
 \caption{The MSW, MSL and MSNpix values as a function of Q$_{fact}$. Three different cleaning combinations are represented; (8\textit{pe}, 4\textit{pe}) (black), (5\textit{pe}, 2\textit{pe}) (red) and (10\textit{pe}, 5\textit{pe}) (blue). The position of the peak in Q$_{fact}$ does not change but the height of the peak clearly depends on the level of cleaning.}
 \label{fig:comparison_rej}
\end{centering}
\end{figure}

	High energy showers are able to trigger the telescopes at large core distances. These events have small sized images so the higher cleaning combination removes a significant portion of the image. 


There are also one or two other cleaning combinations which provide results on par with the standard cleaning. These combinations are (10\textit{pe}, 5\textit{pe}) and (6\textit{pe}, 4\textit{pe}). Both combinations are within the best angular resolution and Q$_{fact}$ ranges.

\begin{figure}
\begin{centering}
\includegraphics[width=15cm]{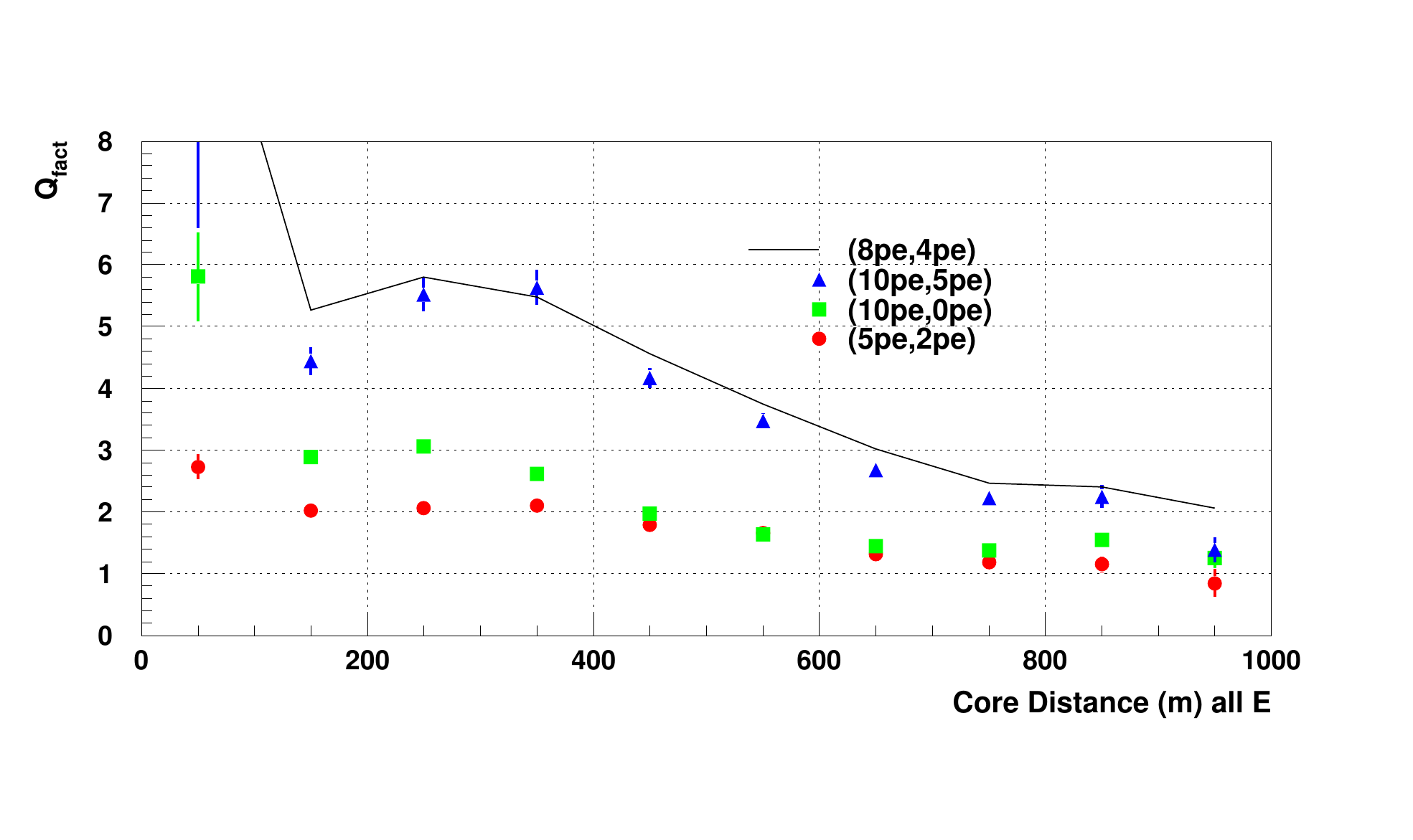}
\captionsetup{width=13cm}  
 \caption{Q$_{fact}$ vs core distance with varying cleaning combinations for algorithm 1 with standard telescope separation, triggering combination and image \textit{size} cut. }
 \label{fig:qfactor_v_core_clean}
\end{centering}
\end{figure}

Figure~\ref{fig:qfactor_v_core_clean} shows Q$_{fact}$ vs core distance which helps illustrate that Q$_{fact}$ in general decreases with increasing core distance. The Q$_{fact}$ vs core distance figure helps further separate the cleaning combinations since a number of combinations provide similar Q$_{fact}$ vs energy results. Varying the cleaning thresholds affects the Q$_{fact}$ at the smallest core distance. The (10\textit{pe}, 5\textit{pe}) combination appears to be too strong for the small sized images since the Q$_{fact}$ drops for all core distances compared with the standard cleaning combination. The standard (8\textit{pe}, 4\textit{pe}) appears to be the best combination.

\section{Effect of Image \textit{Size} Cut}
 \label{sec:image_size_op}

	The image \textit{size} is the total number of \textit{pe} of the image after the cleaning algorithm. The image \textit{size} cut is one of the stereoscopic cuts which determines the events used for shower reconstruction. The larger sized images will provide a more accurate Hillas parameterisation and increasing the image \textit{size} cut guarantees that the images are well defined. This should provide improved parameterisation, which will help the calculation of the major axis and should improve the event reconstruction. We investigate here the effect of varying the image \textit{size} cut from 60\textit{pe} to 300\textit{pe}.

	We can see that the angular resolution (r68) (Figure~\ref{fig:angres_size_low_alt}) improves with increasing image \textit{size} cut. As the image \textit{size} cut increases, the quality of images increases, which improves the parameterisation and hence the final reconstructed shower direction and location. The small images have larger errors in both parameterisation and reconstruction of the shower direction and location. 

\begin{figure}
\begin{centering}
\includegraphics[scale=0.65]{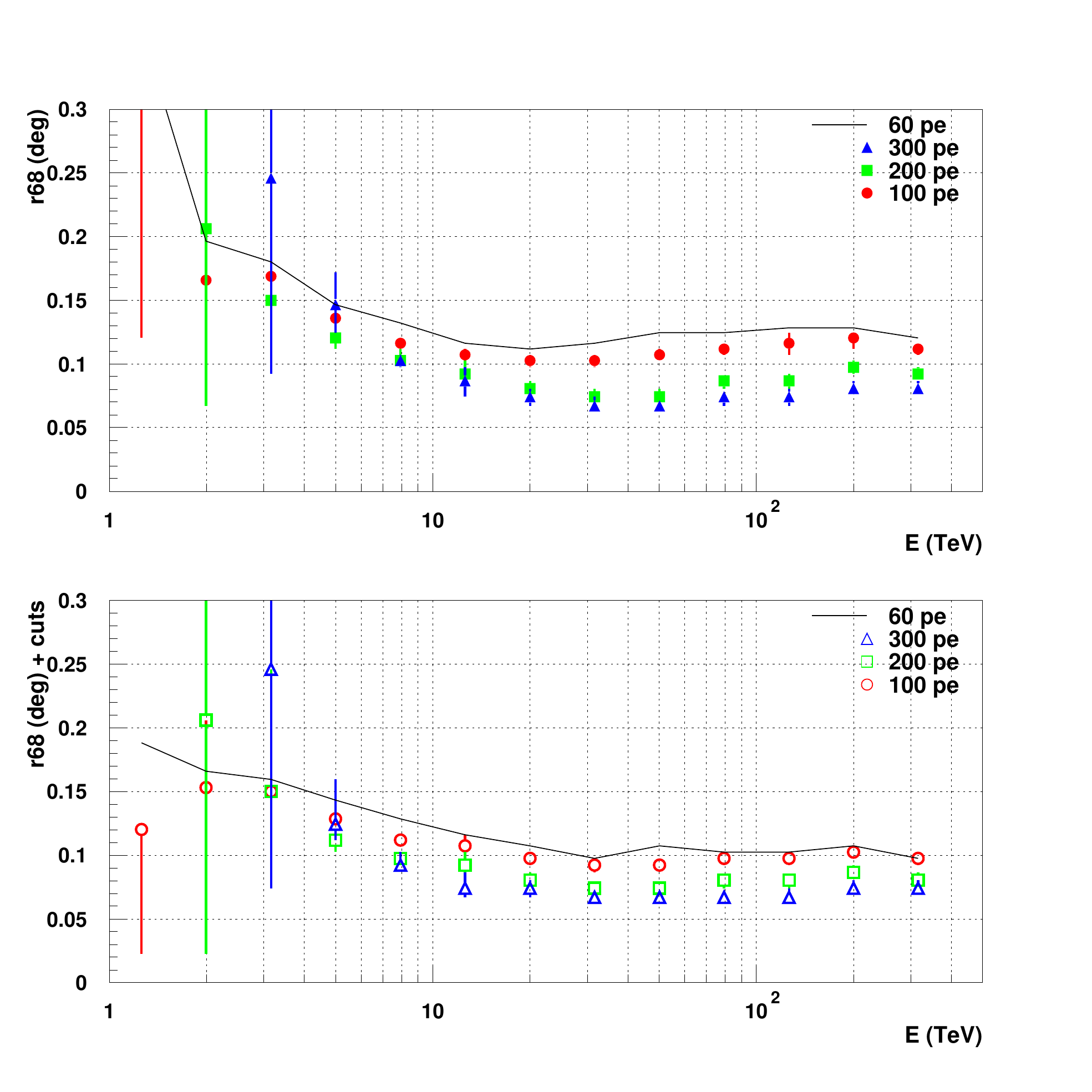}
\captionsetup{width=13cm}  
 \caption{Angular resolution (r68) for the four selected image \textit{size} cuts with standard triggering combination, cleaning combination and telescope separation. 
Top: Pre-cut angular resolution results. As the \textit{size} cut increases, angular resolution increases. The difference between the 200\textit{pe} and 300\textit{pe} results is minimal.
Bottom: The post-shape cut angular resolution results. }
 \label{fig:angres_size_low_alt}
\end{centering}
\end{figure}

	The Q$_{fact}$ results show an improvement for large image \textit{size} cuts (Figure~\ref{fig:qfactor_size_low_alt}). The event statistics for E $<$ 10 TeV are limited due to the large image \textit{size} cut. The error bars associated with each curve indicate that the Q$_{fact}$ has improved with a 300\textit{pe} cut. This size cut leads to a significantly lower total number of events in each energy bin. This effect is also evident in the effective area (Figure~\ref{fig:area_size_low_alt}). Little difference in Q$_{fact}$ is seen for 200\textit{pe} vs 300\textit{pe}.

\begin{figure}
\begin{centering}
\includegraphics[scale=0.65]{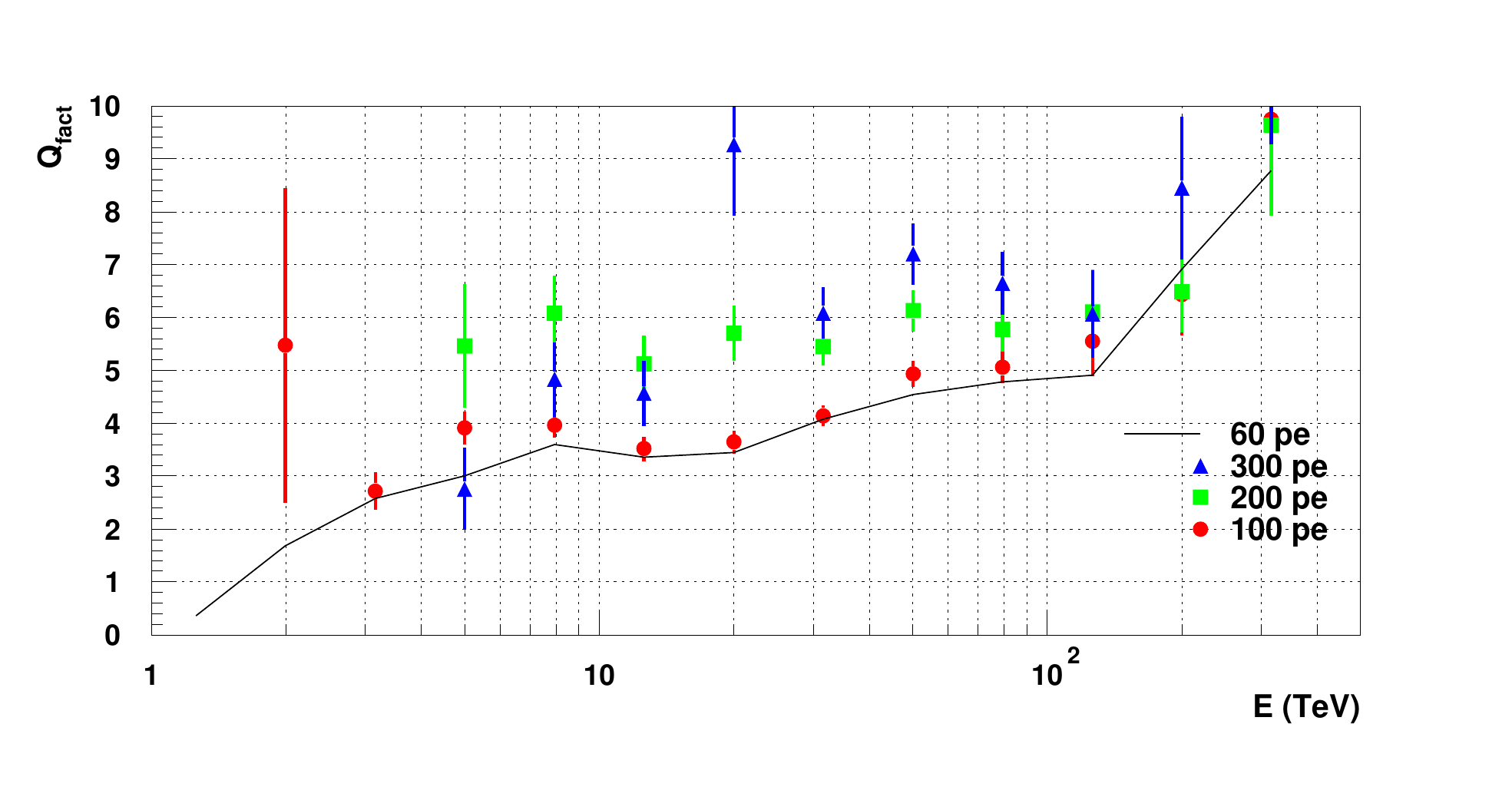}
\captionsetup{width=13cm}  
 \caption{Q$_{fact}$ for image \textit{size} cuts with standard triggering combination, cleaning combination and telescope sepatation. The Q$_{fact}$ for E $<$ 10 TeV does not exist for a high image \textit{size} cut since all protons events have been removed by cuts, which is due to limited Monte Carlo statistics. }
 \label{fig:qfactor_size_low_alt}
\end{centering}
\end{figure}

The effective area results (Figure~\ref{fig:area_size_low_alt}) show the loss in event numbers for a variety of image \textit{size} cuts. The 300\textit{pe} cut removes a large fraction of events for E $<$ 10 TeV, as indicated by the Q$_{fact}$ results. For E $>$ 10 TeV, the pre-cut effective area for 300\textit{pe} shows the number of small sized images in the data set. These results indicate that the energy threshold of the cell would increase to 10 TeV with an image \textit{size} cut of 300\textit{pe} since minimal $\gamma$-ray events are detected below 10 TeV.

The post-selection cut effective area for the standard image \textit{size} cut is largest for E $<$ 100 TeV. However, above 100 TeV the effective area is similar for all image \textit{size} cuts (Figure~\ref{fig:area_size_low_alt} bottom panel).

\begin{figure}
\begin{centering}
\includegraphics[scale=0.65]{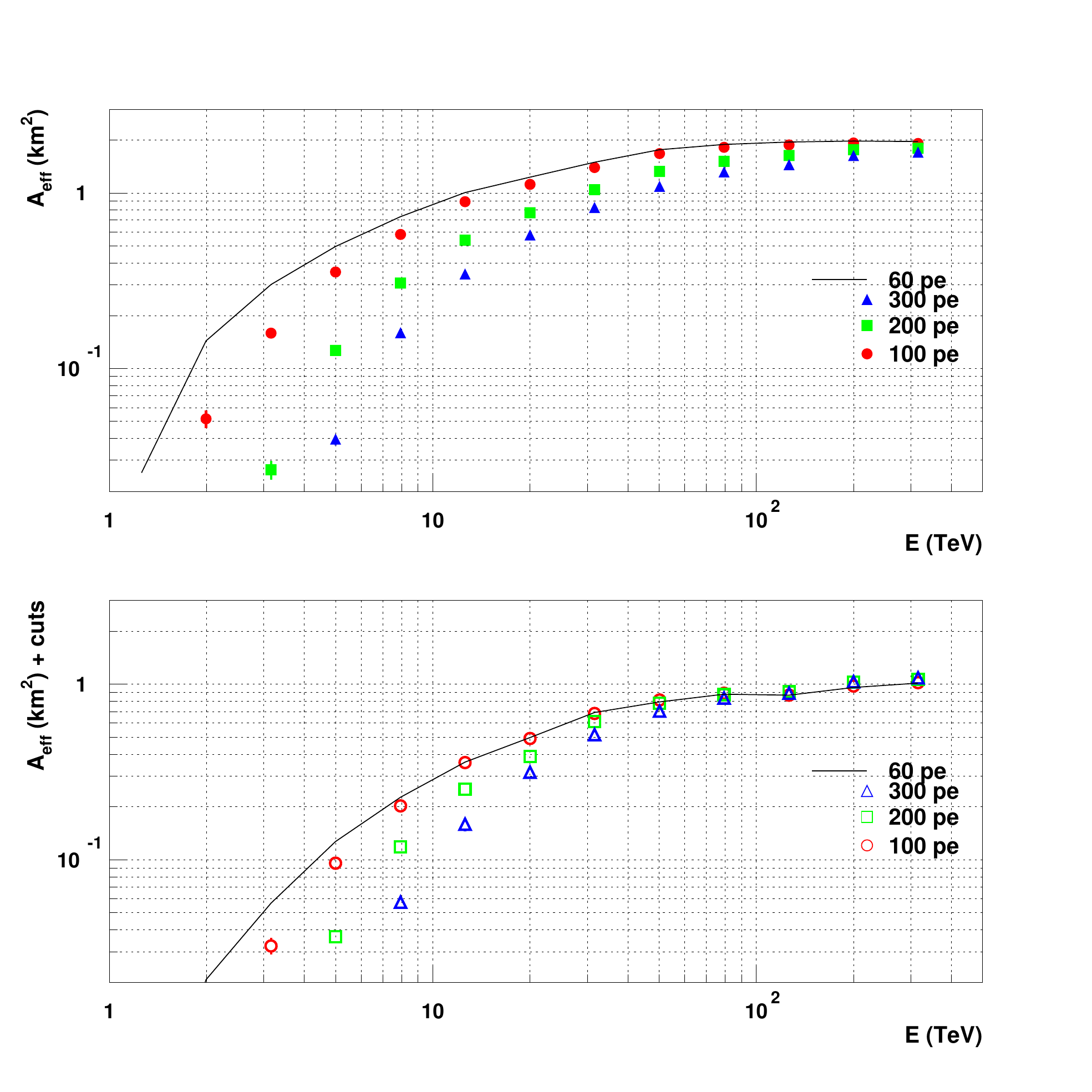}
\captionsetup{width=13cm}  
 \caption{Effective areas for a variety of image \textit{size} cuts with standard triggering combination, cleaning combination and telescope separation. 
Top: The general trend in the pre-cut effective area shows that as the image \textit{size} cut increases, the effective area decreases.
Bottom: The post-selection cut effective area still shows a loss in the number of events for E $<$ 50 TeV. For E $>$ 50 TeV, the post-selection cut effective areas are the same.}
 \label{fig:area_size_low_alt}
\end{centering}
\end{figure}

Figure~\ref{fig:energy_res_size_low_alt} in Appendix~\ref{sec:appendix_plot} illustrates the energy resolution for various image \textit{size} cuts. The mean in the post-shape cut $\Delta{E}/E$ distribution shows minimal variation with increasing image \textit{size} cut. The largest image \textit{size} cut, 300\textit{pe}, provides more stability in the distribution.

The RMS in the $\Delta{E}/E$ distribution appears to improve with the 300\textit{pe} cut. This is expected since the larger cut will select the strongest images. Therefore, the energy reconstruction improves with increasing cut values. The RMS for the 60\textit{pe} cut is 15$\%$ while the 300\textit{pe} cut produces an RMS of 12$\%$ for all energies. The improvement is mitigated by the fact that the 300\textit{pe} cut removes a significant faction of $\gamma$-ray events.\\

	The large error obtained from small image reconstruction is due to the number of pixels in the images. A small image of 60 - 70\textit{pe} usually has an average of 7 to 8 pixels. With a small number of pixels, calculating the C.O.G, \textit{width} and \textit{length} of the image is made more prone to fluctuations. Small sized images come from large core distance events or low energy events. For the low energy case, the showers are too faint to produce images with enough information to separate event types. For the large core distance case, only the brightest part of the shower reaches the telescope. Both cases affect the event reconstruction. Another effect of the large core distance shower, is the elongation of the image. Elongating an image with 7 or 8 pixels makes it difficult to calculate the \textit{width} of the image, especially when the image \textit{width} is one of the major proton rejection parameters.

 The effect of small sized images are seen in the Q$_{fact}$ and angular resolution (Figure~\ref{fig:angres_size_low_alt} and Figure~\ref{fig:qfactor_size_low_alt}). The 200\textit{pe} and 300\textit{pe} cuts improve the parameterisation and reconstruction compared to the standard image \textit{size} cut. There is a significant improvement in angular resolution between the standard image \textit{size} cut and the 300\textit{pe} cut. The improvement is 50$\%$ above 10 TeV for pre-cut angular resolution, while the improvement for post-shape cut angular resolution is 40$\%$ above 10 TeV. Below 10 TeV, the larger image \textit{size} cuts suffer from lack of statistics. The down side to this is the loss in events for both protons and $\gamma$-rays. The angular resolution improves for the 300\textit{pe} cut over the standard image \textit{size} cut but the events which are removed could still provide good reconstruction.\\


	The 300\textit{pe} cut removes too many events for observational use. The standard image \textit{size} cut of 60\textit{pe} is sufficient for the cell giving the largest possible effective area. However for more accurate results at high energies, a larger image \textit{size} cut of 200\textit{pe} can be used. One can have two sets of cuts, soft and hard cuts, which can be used depending on the observational source. A 60\textit{pe} size would be a soft cut and 200\textit{pe}size would be hard cuts. The current shape and stereoscopic cuts are classified as standard or soft cuts. This method of applying soft and hard cuts has already been applied to the H.E.S.S. analysis. A different set of shape and stereoscopic cuts can be set up for hard cuts, where hard implies that only the strong $\gamma$-ray like images are accepted.\\

\section{Concluding remarks on a low altitude PeX cell}

The aim of this Chapter was to find an optimal combination of PeX parameters with the limits of our Monte Carlo statistics. To best compare the results, we can plot a variation of the flux sensitivities. The flux sensitivities have been plotted as the ratio of varied flux over the standard flux, $F_{standard}$/$F_{varied}$, where the varied flux is the flux sensitivity for each parameter variation. Therefore, if the variation provides an improvement in flux sensitivity over the standard configuration then the plotted flux ratio will be $>$ 1. 



\begin{figure}
\begin{centering}
\includegraphics[scale=1]{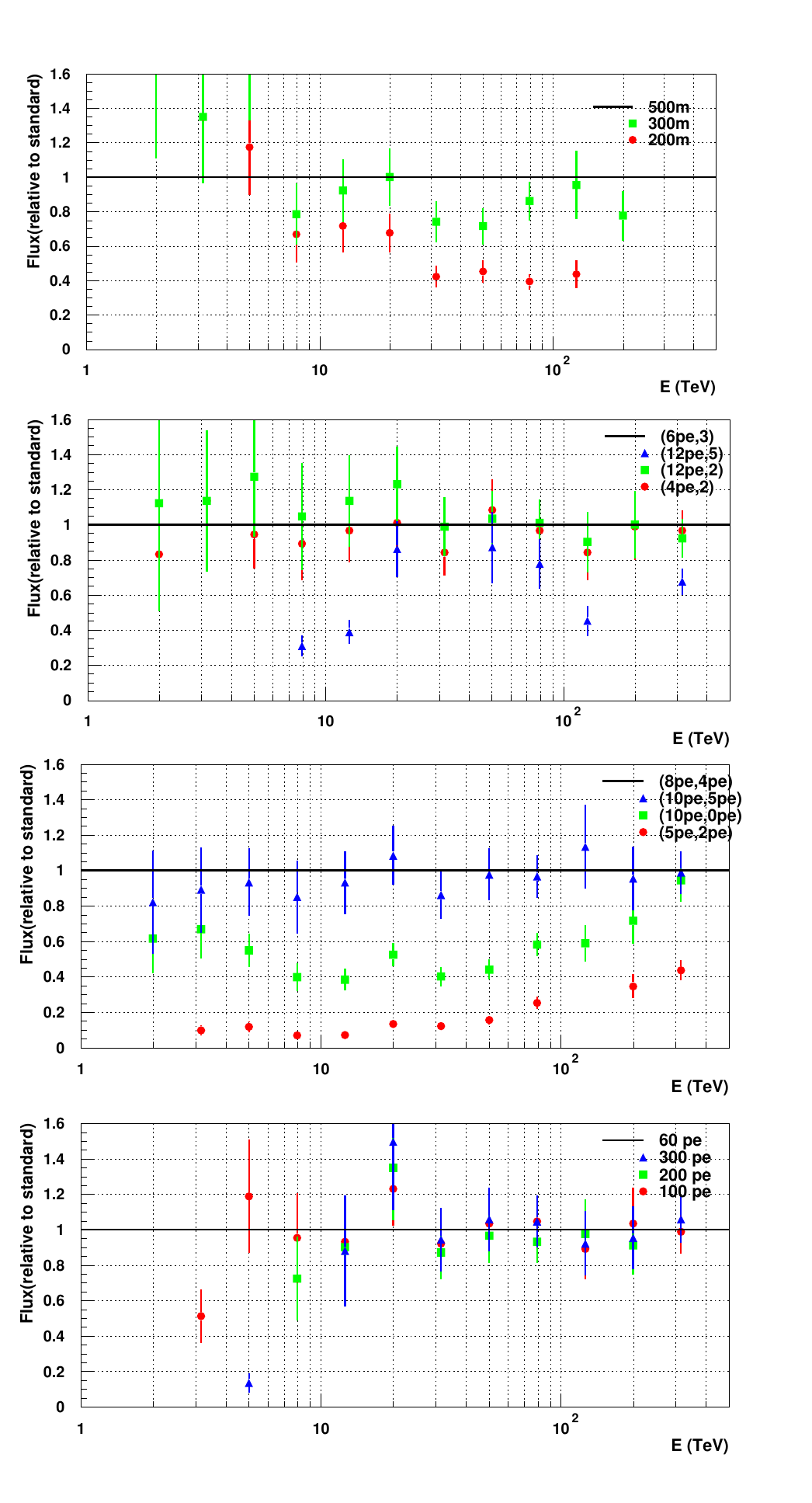}
\captionsetup{width=13cm}  
 \caption{The flux sensitivity for telescope separations, various triggering combinations, cleaning combinations, and image \textit{size} cuts normalised to the standard configuration. A ratio larger than 1 implies an improvement in the flux sensitivity, compared with the standard configuration.}
 \label{fig:flux_parameter_low}
\end{centering}
\end{figure}

The 500 \rm{m} telescope separation provides the best flux sensitivity (Figure~\ref{fig:flux_parameter_low} first panel). The 200 \rm{m} and 300 \rm{m} separations decrease the flux sensitivity due to worsened event reconstruction. The 800 \rm{m} separation has limited post-selection cut proton events and does not provide a full flux sensitivity curve.

In Figure~\ref{fig:flux_parameter_low} second panel, the standard triggering combination (6\textit{pe}, 3) provides the best flux sensitivity for all energies. The (12\textit{pe}, 5) combination does provide a significant improvement to the flux sensitivity for E $>$ 10 TeV, although it suffers a large loss in events below 10 TeV. The (12\textit{pe}, 5) combination could be used for strong point sources where an improved angular resolution is required. The improvement seen at high energies for the (12\textit{pe}, 5) combination could be gained by other means without a loss in events.

The standard cleaning combination (8\textit{pe}, 4\textit{pe}) provides the best flux sensitivity for all energies (Figure~\ref{fig:flux_parameter_low} third panel). The (10\textit{pe}, 5\textit{pe}) combination is the other combination that would provide a similar flux sensitivity. These combinations are the best choices for studying a point source since they also provide the best angular resolution for the cleaning combinations investigated.

In Figure~\ref{fig:flux_parameter_low} fourth panel, the standard image \textit{size} cut of 60\textit{pe} provides an adequate flux sensitivity. The stronger size values provide an improvement in the flux parameter for E $>$ 10 TeV. However, the low energy events are affected. The standard image \textit{size} cut provides good results for the entire energy range. Since the cut is applied to the results once the events have been recorded, it is easier to vary a size cut compared to the triggering combinations. Therefore, a 200 or 300\textit{pe} image \textit{size} cut can be used for hard-cuts analysis which may be suitable for strong and/or hard spectra sources as is used by H.E.S.S. \cite{HESShardcut}.

From this study of the individual optimisation of triggering combination, cleaning combination, telescope separation and image \textit{size} cuts for PeX at 0.22 \rm{km} observational site, we can conclude:
\begin{itemize}
\item The 500 \rm{m} telescope separation provides the adequate post-selection cut effective area for all energies. The improvement is 20$\%$ over the 300 \rm{m} separation for E $>$ 10 TeV. The large separation provides significant improvements in angular resolution, ranging from 20 to 30$\%$ for E $>$ 20 TeV.
\item The standard triggering combination (6\textit{pe}, 3) provides the largest post-selection cut effective area. A stronger triggering combination removes too many events. The angular resolution and Q$_{fact}$ are slightly worse compared to the (12\textit{pe}, 5) combination but the effective area is very important. The standard triggering combination produces good angular resolution and Q$_{fact}$. 
\item Several of the cleaning combinations provide adequate results. The standard cleaning is within the group of cleaning combinations which provide the largest effective area, best angular resolution and best Q$_{fact}$. Other cleaning combinations that provide near optimum results are (10\textit{pe}, 5\textit{pe}), (6\textit{pe}, 3\textit{pe}), (7\textit{pe}, 4\textit{pe}).
\item The standard image \textit{size} cut of 60\textit{pe} is the preferred cut. The other cuts remove too many events, which increases the energy threshold of the cell. The post-selection cut effective area showed a significant difference between image \textit{size} cuts suggesting that the events lost by a large cut could have been reconstructed accurately. So, vital events are lost with a large image size cut. The 200\textit{pe} cut could be used as a hard cut for the cell.
\end{itemize}

The useful configuration for a 0.22 \rm{km} altitude site appears to be a 500 \rm{m} telescope separation, with a (6\textit{pe}, 3) triggering combination, an (8\textit{pe}, 4\textit{pe}) cleaning combination and a 60\textit{pe} image size cut. Since many of these parameters are scaled from the optimal values for the H.E.S.S. telescope, we can also conclude that the scaling to PeX also provides a reasonably well optimised system. In the next chapter, we will repeat this study but with a PeX cell situated at a higher altitude site as is used for the H.E.S.S. telescope (1.8 \rm{km} above sea level).

\chapter{Mid altitude PeX cell optimisation}
 \label{sec:optimise_high}

	In Chapter 4 we optimised individual parameters for a PeX cell at a sea level or low altitude site. An interesting question is how the PeX cell would operate at a mid-level altitude of 1.8 \rm{km} above sea level which is used for H.E.S.S., VERITAS and MAGIC-II (Table~\ref{table:IACT}). As for Chapter 4, we define the standard configuration for the PeX cell at a mid-level altitude as: a triggering combination of (6\textit{pe}, 3), a cleaning combination of (8\textit{pe}, 4\textit{pe}), an image \textit{size} cut of 60\textit{pe}, a telescope separation of 500 \rm{m} and a \textit{dis2} cut of 4.0$^{\circ}$. The atmospheric absorption in the 300 \rm{nm} to 650 \rm{nm} wavelength range from a 1.8 \rm{km} altitude site to a 0.22 \rm{km} altitude site is roughly 17$\%$ as shown in section~\ref{sec:tel_sep_high}. The NSB at a 1.8 \rm{km} altitude site will therefore be around 0.053 \rm{pe (ns pixel)$^{-1}$} or 17$\%$ higher, which will not produce any significant effect on the results (see section~\ref{sec:timing_cut}). Therefore, the NSB for the 1.8 \rm{km} altitude site was left at 0.045 \rm{pe (ns pixel)$^{-1}$}. Hampf et al \cite{Hampf} showed that the night sky brightness measured at a low altitude Australian site (0.22 \rm{km}) was consistent with the night sky brightness measured at the H.E.S.S. site in Namibia (1.8 \rm{km}) and at La Palma (2.2 \rm{km}).

In this chapter each parameter will be varied individually so the best results can be obtained. Past and current IACTs have produced outstanding results and advances in the field of $\gamma$-ray astronomy. The MAGIC-II, VERITAS and H.E.S.S. collaborations have shown that a medium altitude or $\approx$ 2 \rm{km} a.s.l site provides good event reconstruction over the $\approx$ 0.1 to 10's of TeV energy range. Since the energies covered by PeX are higher than that of current IACTs, the aim of this chapter is to determine the performance of PeX placed at a medium altitude observational site.

The investigation will include varying the same parameters as in Chapter 4 over specified values. Then the factors which alter the flux sensitivity and the rejection power will be compared. 

The total number of events simulated for the 1 - 500 TeV range, the number of triggered events, the number of triggered events that passed shape cuts and the events that pass selection cuts for the 1.8 \rm{km} altitude site are displayed in Table~\ref{table:events_high} for $\gamma$-rays and protons. The same cuts for the 0.22 \rm{km} altitude site have been applied to the 1.8 \rm{km} altitude site. A similar table has been produced for the 0.22 \rm{km} altitude site (Table~\ref{table:events}).

\begin{table}[h]
\centering
\begin{tabular}{lrrrr}
\hline
&Thrown & Triggered & Post-shape Cuts & Post-selection Cuts \\
\hline
$\gamma$-rays & 71520 & 28182 & 20361 & 11025 \\
Protons & 136000 & 10567 & 175 & 25 \\
\hline
\end{tabular}
\captionsetup{width=13cm}  
\caption{The table represents the number of events thrown, number of events that trigger the PeX cell, number of events which pass shape cuts and number of events which pass selection cuts in the 1 to 500 TeV energy range. The simulations are done with a standard PeX cell at a 1.8\rm{km} altitude site. Only protons are included in the simulations since after reconstruction and applying cuts the rate of Helium is only 5$\%$ of the rate of protons \cite{Denman} at low altitudes. This rate should be similar for high altitudes.}
 \label{table:events_high}
\end{table}


\section{Telescope Separation}
 \label{sec:tel_sep_high}

\begin{figure}
\begin{centering}
\includegraphics[scale=0.65]{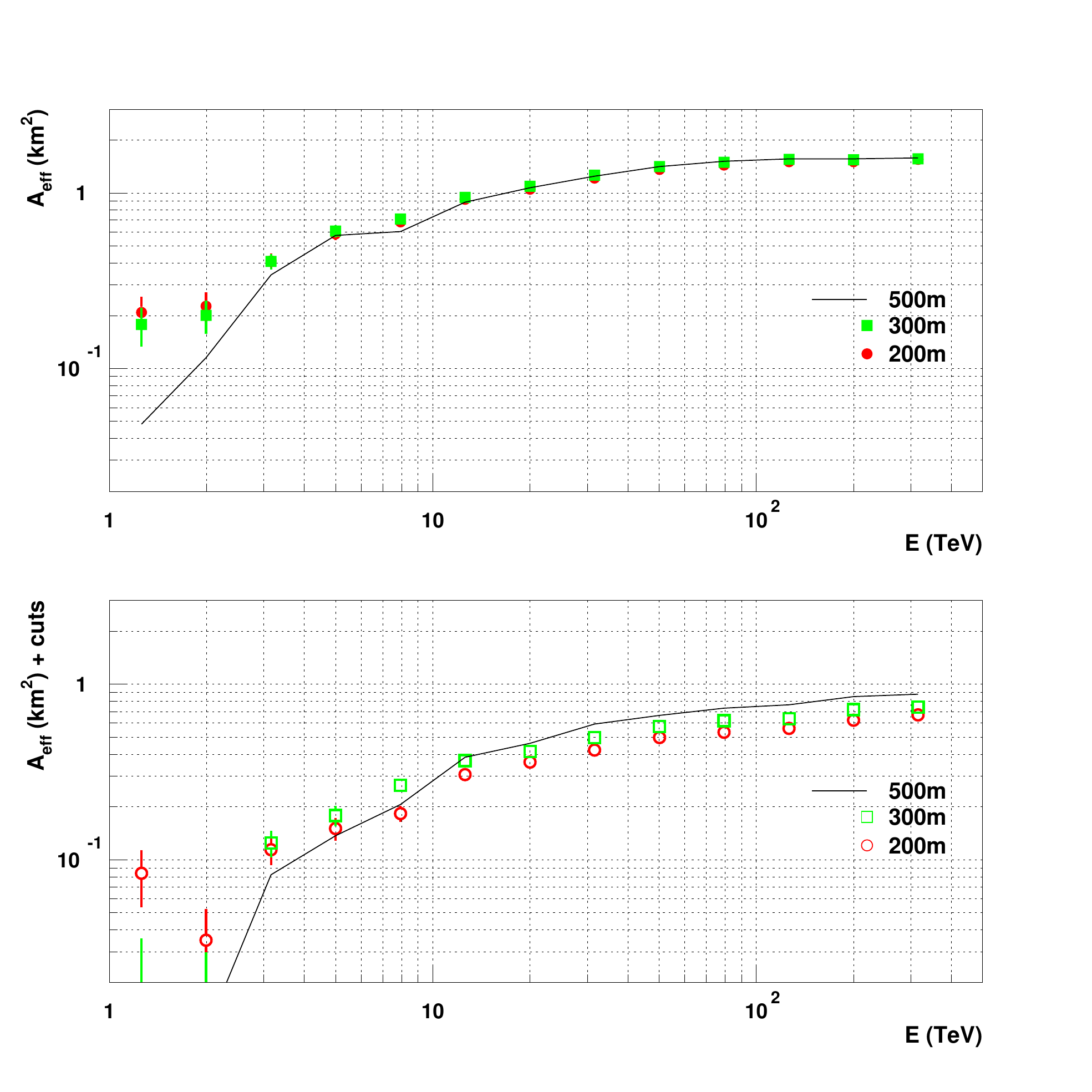}
\captionsetup{width=12cm}  
 \caption{Effective area for a variety of telescope separations at a 1.8 \rm{km} altitude observational site with standard triggering combination, cleaning combination and image \textit{size} cut. Top: Pre-cut effective area for telescope separations. The telescope separations provide no difference in pre-cut effective area. Bottom: Post-selection cut effective area for the same telescope separations. The 500 \rm{m} telescope separation provides the largest effective area.}
 \label{fig:area_sep_high_alt}
\end{centering}
\end{figure}

The telescope separation for the mid-altitude PeX cell was varied from 200 \rm{m} to 500 \rm{m}. The 800 \rm{m} separation considered in Chapter 4 was not considered for the mid-altitude site since the results for a 0.22 \rm{km} altitude site indicated that low energy events failed to trigger multiple telescopes. Therefore, with a mid-level altitude and smaller lateral distributions, an 800 \rm{m} telescope separation seemed an inappropriate choice.

Figure~\ref{fig:area_sep_high_alt} displays the effective area for the 1.8 \rm{km} altitude site. The 500 \rm{m} telescope separation for the 1.8 \rm{km} altitude site provides the largest pre-cut effective area for all energies. An improvement in the pre-cut effective area has been gained for the 200 \rm{m} telescope separation for E $<$ 2 TeV. There is no difference between the telescope separations for the E $>$ 10 TeV range. The post-selection cut effective area shows that the larger telescope separation, 500 \rm{m}, also offers the largest area above 10 TeV (Figure~\ref{fig:area_sep_high_alt} bottom panel). 

\begin{figure}
\begin{centering}
\includegraphics[width=15cm]{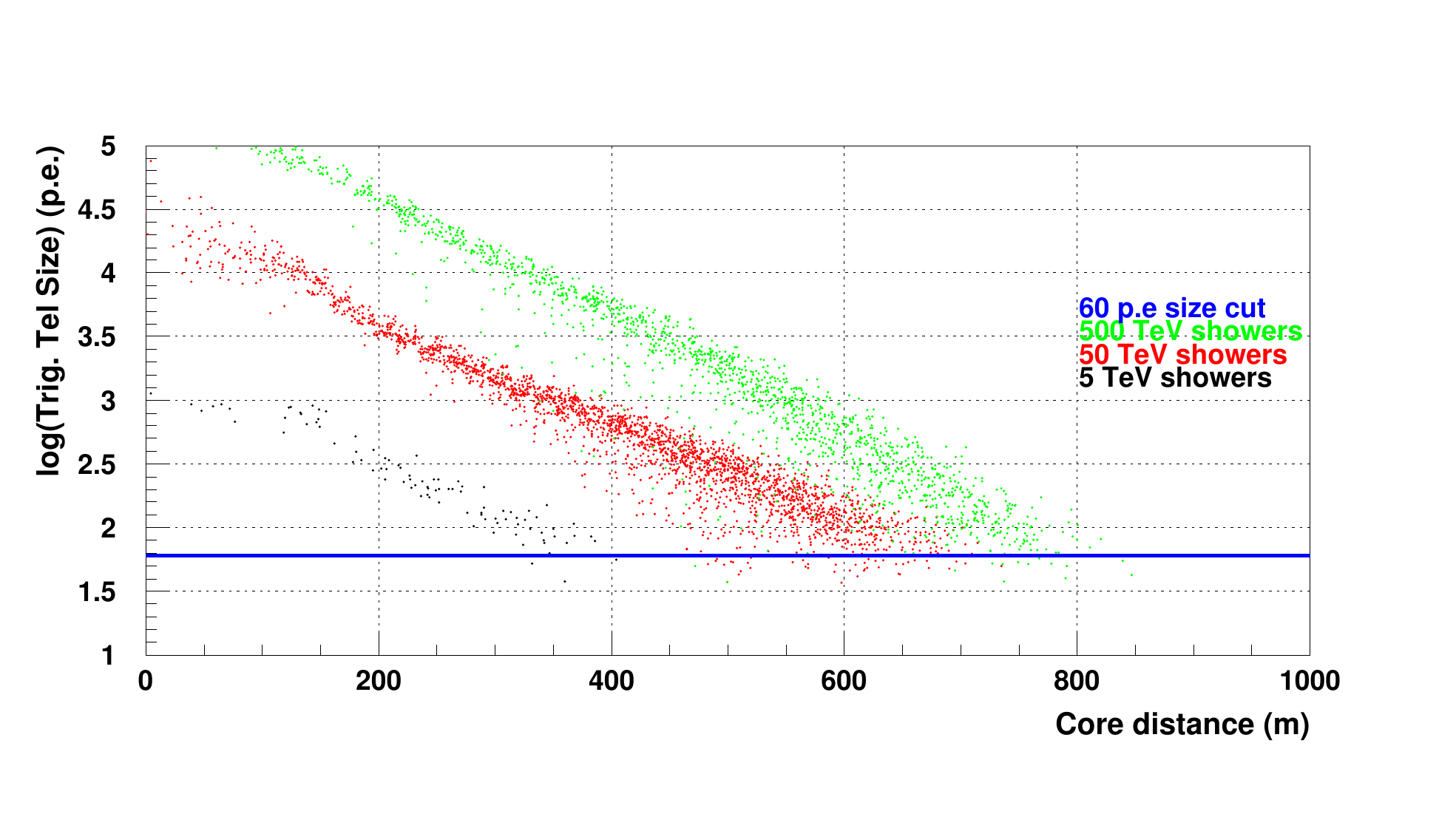}
\captionsetup{width=13cm}  
 \caption{Shower size vs core distance, which is equivalent to the lateral distribution, for multi-TeV $\gamma$-ray showers with 30$^{\circ}$ zenith angles from simulations code for a 1.8 \rm{km} altitude site. The y-axis represents the shower size in photoelectrons (\textit{pe}) and the x-axis represents the core distance from the telescope. The showers have been split into the energy bands to show the effect of increasing energy on the lateral distribution. Compare with Figure~\ref{fig:distance} at 220 \rm{m} a.s.l.}
 \label{fig:distance_high}
\end{centering}
\end{figure}

\begin{figure}
\begin{centering}
\includegraphics[scale=0.65]{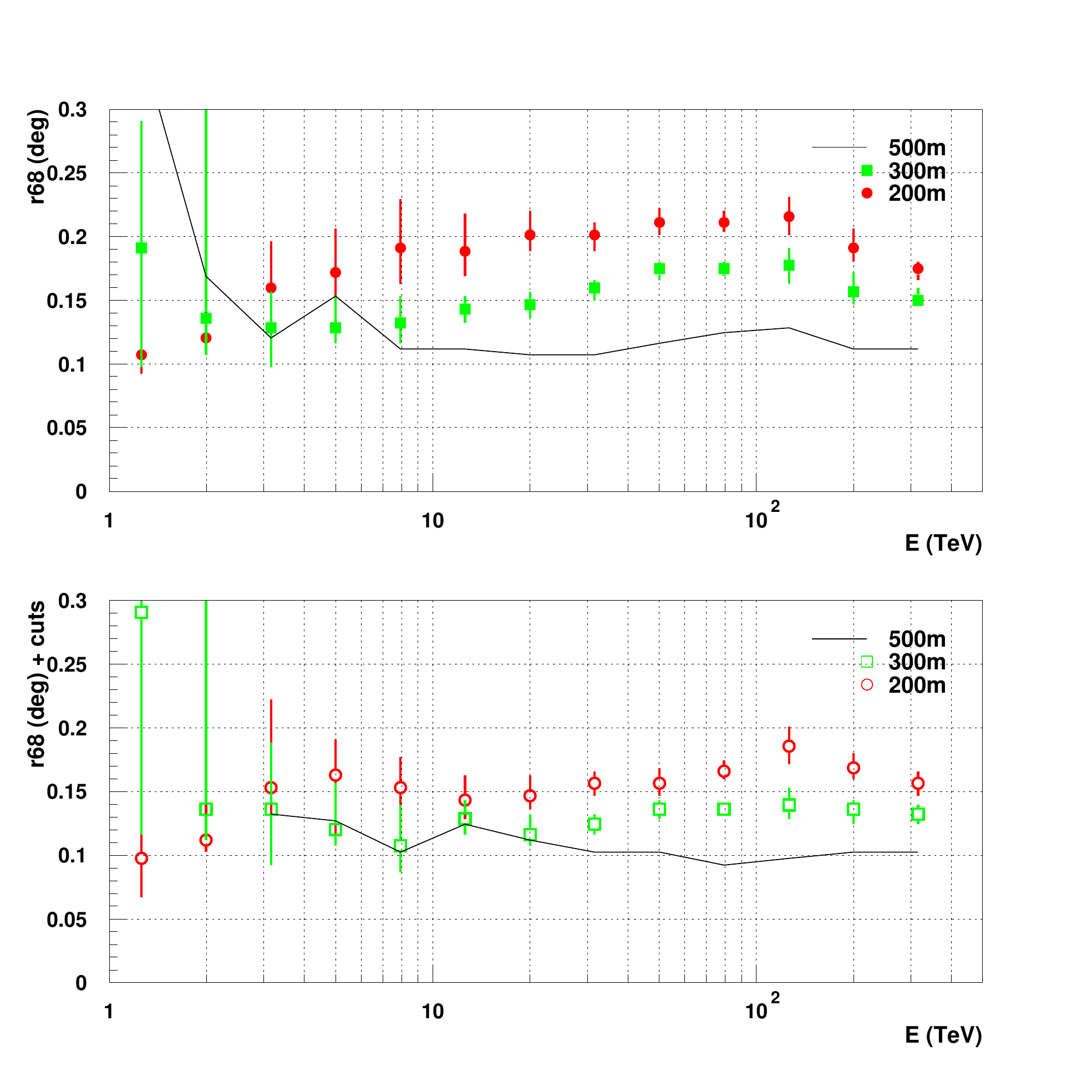}
\captionsetup{width=12cm} 
 \caption{Angular resolution (r68) for a 1.8 \rm{km} observational site for varying telescope separation with standard triggering combination, cleaning combination and image \textit{size} cut. Top: Pre-cut angular resolution for varying telescope separations. Bottom: Post-selection cut angular resolution for varying telescope separations. There is roughly 30$\%$ difference between the 300 \rm{m} and 500 \rm{m} curves for E $>$ 10 TeV.}
 \label{fig:angres_sep_high_alt}
\end{centering}
\end{figure}

The angular resolution (r68) is shown in Figure~\ref{fig:angres_sep_high_alt} for the 1.8 \rm{km} altitude site. The reconstructed shower direction improves with the larger 500 \rm{m} telescope separation. The event reconstructions for the 200 \rm{m} and 300 \rm{m} separations provide considerably poorer quality reconstruction compared to a 500 \rm{m} separation. The difference between the 500 \rm{m} and 300 \rm{m} separation results is roughly 25 to 30$\%$ for E $>$ 10 TeV.

\begin{figure}
\begin{centering}
\includegraphics[scale=0.65]{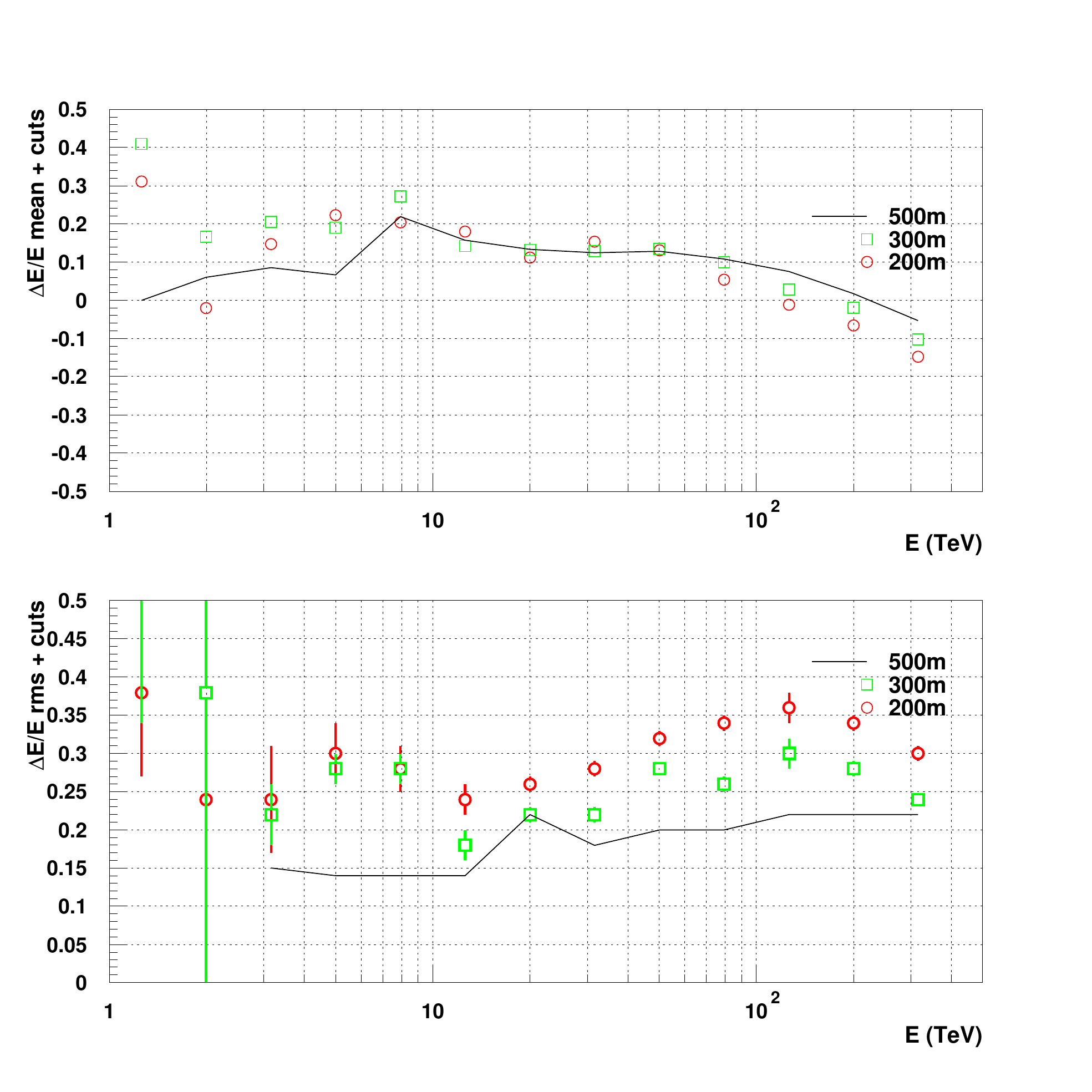}
\captionsetup{width=12cm}
 \caption{Post-shape cut energy resolution mean and RMS for a 1.8 \rm{km} altitude observational site with standard triggering combination, cleaning combination and image \textit{size} cut. 
 Top: The mean in $\Delta{E}/E$ distribution.
 Bottom: The RMS for the $\Delta{E}/E$ distribution.}
 \label{fig:energy_res_sep_high_alt}
\end{centering}
\end{figure}

The post-shape cut energy resolution for a variety of telescope separations is shown in Figure~\ref{fig:energy_res_sep_high_alt}. The results show minimal difference between the means in the $\Delta{E}/E$ distribution. The results indicate that E$_{recon}$ $>$ E$_{true}$ for a large majority of events and creates a positive bias for E $>$ 100 TeV. The RMS for the $\Delta{E}/E$ distribution (Figure~\ref{fig:energy_res_sep_high_alt} bottom panel) provides the same trend as angular resolution and effective area results. The telescope separation improves the performance for all results. The RMS for the 500 \rm{m} separation fluctuates around 20$\%$ for E $>$ 10 TeV, which is the best achievable RMS for all separations. A higher image \textit{size} cut of 70\textit{pe} could be more appropriate for the 1.8 \rm{km} PeX cell since the photon intensity has increased due to less absorption in the atmosphere. The slightly higher \textit{size} cut could improve the RMS in the energy resolution. We will see the effect of varying image \textit{size} cut in section~\ref{sec:image_size_high}\\

\begin{figure}
\begin{centering}
\includegraphics[scale=0.65]{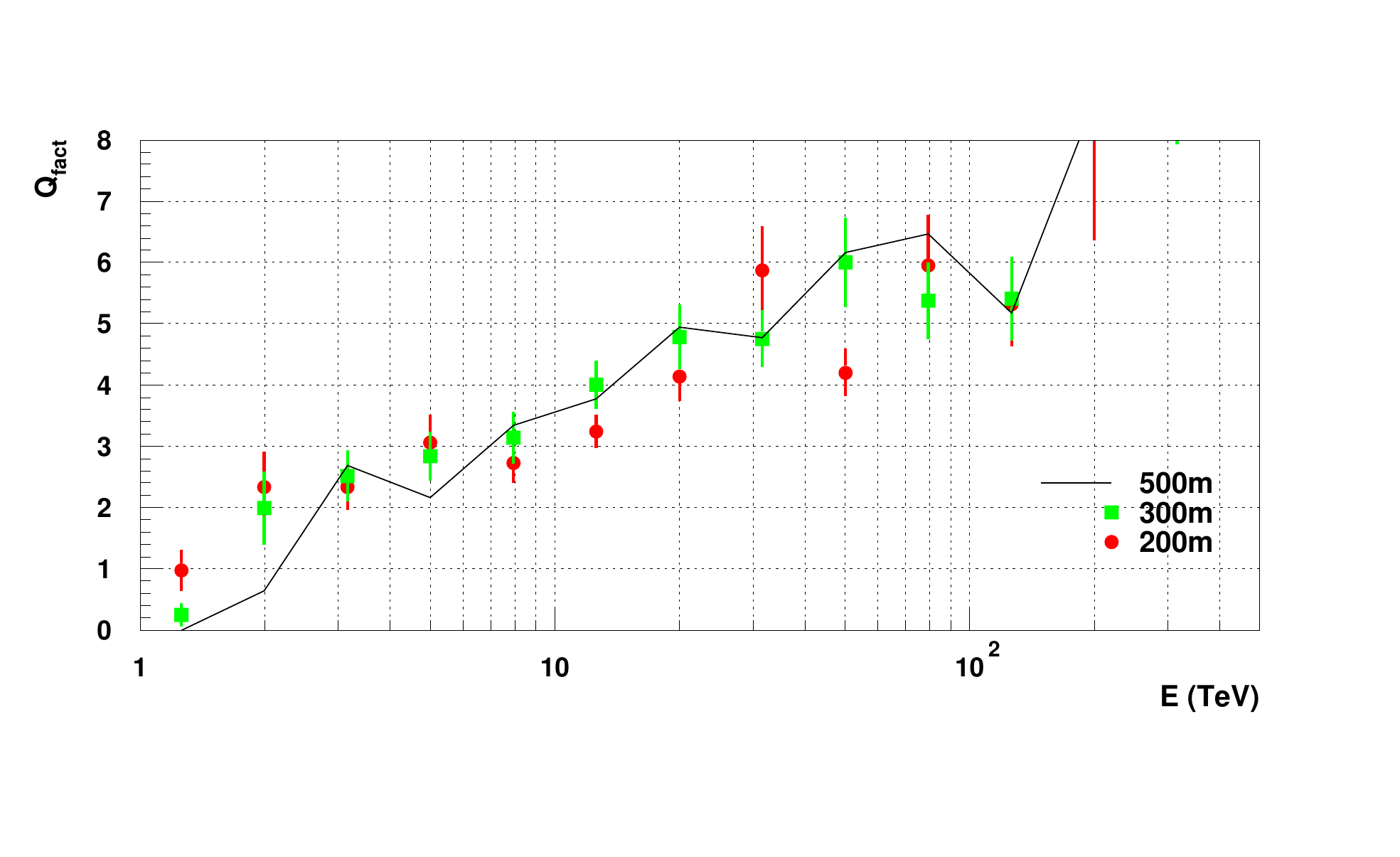}
\captionsetup{width=12cm}
 \caption{Q$_{fact}$ for various telescope separations at a 1.8 \rm{km} altitude observational site with standard triggering combination, cleaning combination and image \textit{size} cut. The Q$_{fact}$ is similar for all telescope separations.}
 \label{fig:q-factor_high_alt}
\end{centering}
\end{figure}

Figure~\ref{fig:q-factor_high_alt} shows the Q$_{fact}$ for various telescope separations. The telescope separations have minimal affect on the shape rejection for PeX. The 200 \rm{m}, 300 \rm{m} and 500 \rm{m} results are all within error bars of each other.

	For the 1.8 \rm{km} altitude site, the cell is at least 1.6 \rm{km} closer to the shower maximum compared with the low 0.22 \rm{km} altitude site. The Cherenkov light travels through less atmosphere and suffers less absorption and scattering. The transmission from a 1.8 \rm{km} altitude down to a 0.22 \rm{km} altitude is roughly 83$\%$ so 17$\%$ absorption occurs between the two altitude sites. The transmission between sites is calculated in more detail in Chapter~\ref{sec:timing_cleaning}. The Cherenkov angle for the light cone (Figure~\ref{fig:cherenkovshower}) is the same since the shower maximum above sea level does not change. The issue is that the shower does not spread out as far as it does for the low altitude site. This effect can been seen by comparing the lateral distribution plots in Figure~\ref{fig:distance} and Figure~\ref{fig:distance_high} for a 0.22 \rm{km} and 1.8 \rm{km} altitude site respectively. The figures also show the increase in photon intensity seen at the 1.8 \rm{km} altitude site.

The change in altitude also has an impact on the position of the images in the camera. For a 500 \rm{m} separation at a 0.22 \rm{km} altitude the angle between the light from the shower and the camera optical axis is small so the image is closer to the centre of the camera (Figure~\ref{fig:core_distance3} $\phi_{A}$ and $\phi_{B}$). The image in the camera would be equivalent to the image for R = small in Figure~\ref{fig:core_distance3}. For the same 500 \rm{m} separation placed at a 1.8 \rm{km} altitude, the cell would be closer to the shower maximum (Figure~\ref{fig:core_distance3} if the reflective mirror is shifted directly upwards). Therefore, the angles between the light from the shower and the camera optical axis are larger. This shifts the image further towards the edge of the camera. The angles $\phi_{A}$ and $\phi_{B}$ become larger, which is equivalent to R = large in Figure~\ref{fig:core_distance3}. This reduces the effective area and makes it harder for large core distance events to trigger a high altitude cell.



The 500 \rm{m} separation provides the largest average opening angle between major axes, which in turn improves the angular resolution (Figure~\ref{fig:angres_sep_high_alt} red circles). A similar situation was noticed for the 0.22 \rm{km} altitude PeX results (Figure~\ref{fig:angres_sep_low_alt}).


For the 1.8 \rm{km} altitude site, the 500 \rm{m} telescope separation appears to provide the best results for all parameters.

\section{Effect of Triggering Combinations}

	The triggering combinations were varied over the same \textit{threshold} values and \textit{n} pixel values as for the 0.22 \rm{km} altitude site optimisation (section~\ref{sec:triggering_op}).

\begin{figure}
\begin{centering}
\includegraphics[scale=0.55, angle=270]{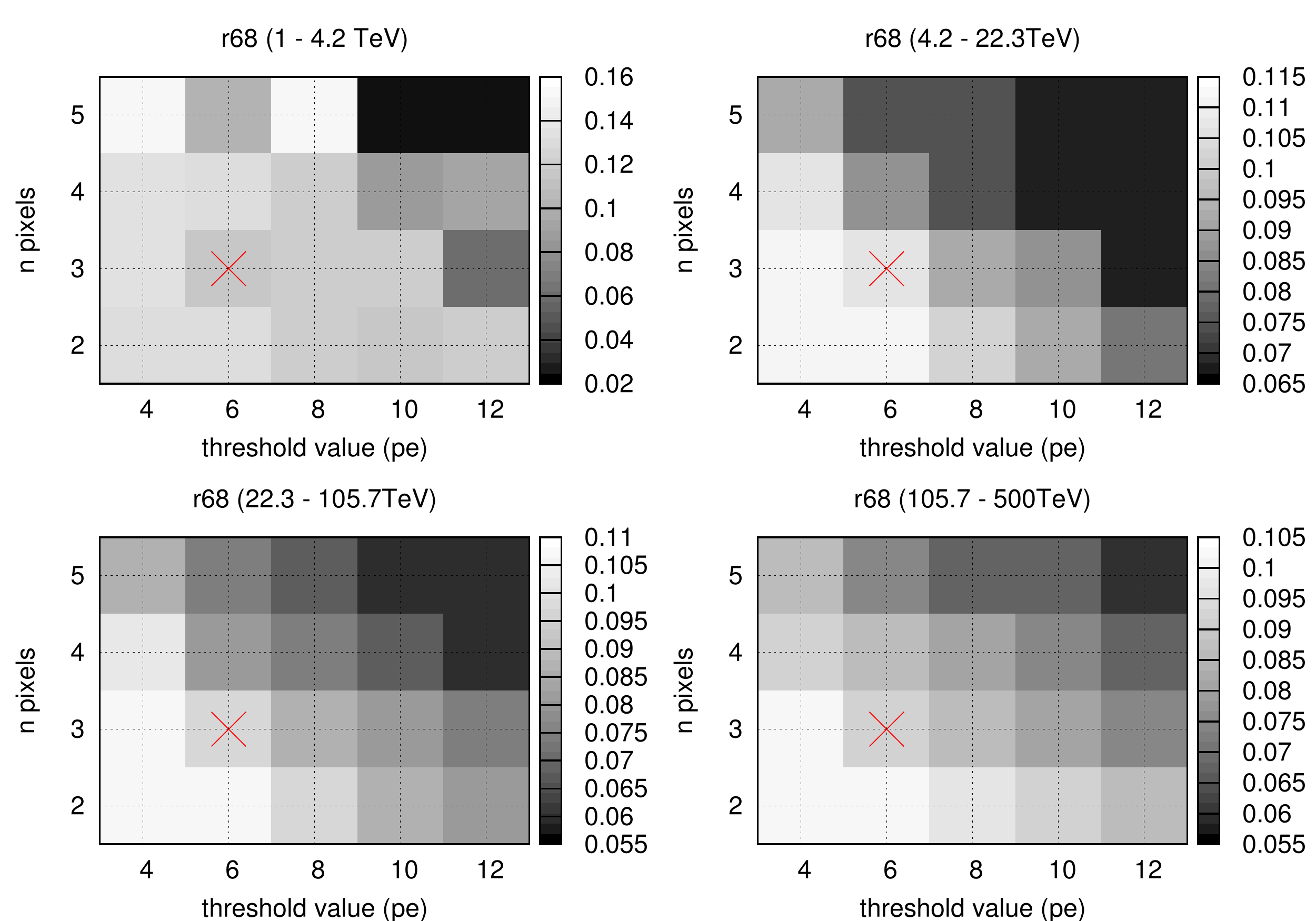}
\captionsetup{width=13cm}
 \caption{Angular resolution (r68), in degrees, for all triggering combinations for a 1.8 \rm{km} altitude observational site with standard cleaning combination, telescope separation and image \textit{size} cut. The x-axis represents the \textit{threshold} value in \textit{pe} and the y-axis represents the \textit{n} pixel value. The general trend indicates that the angular resolution improves with increasing triggering \textit{threshold} value and/or \textit{n} pixel value. The red cross represent the standard value. The equivalent 0.22 \rm{km} altitude results for angular resolution are shown in Figure~\ref{fig:angres4_trigger}.}
 \label{fig:angres4_trigger_high}
\end{centering}
\end{figure}

The angular resolution (r68) is displayed in Figure~\ref{fig:angres4_trigger_high}. As in Chapter 4, the results have been broken into four energy bands to better represent the angular resolution with varying triggering combinations: 1 - 4.2 TeV, 4.2 - 22.3 TeV, 22.3 - 105.7 TeV and 105.7 - 500 TeV. The angular resolution improves with increasing \textit{threshold} value and \textit{n} pixel value. However, if both values increase together then the improvement in angular resolution is larger. The same trend appeared for (towards the top right in all panels in Figure~\ref{fig:angres4_trigger}) the 0.22 \rm{km} altitude observational site (Figure~\ref{fig:angres4_trigger}). The (12\textit{pe}, 5) trigger combination provides the best angular resolution for all energy bands, whereas it is clear that the standard triggering combination, (6\textit{pe}, 3), does not provide the best angular resolution. However, all performance factors must be considered.

\begin{figure}
\begin{centering}
\includegraphics[scale=0.55, angle=270]{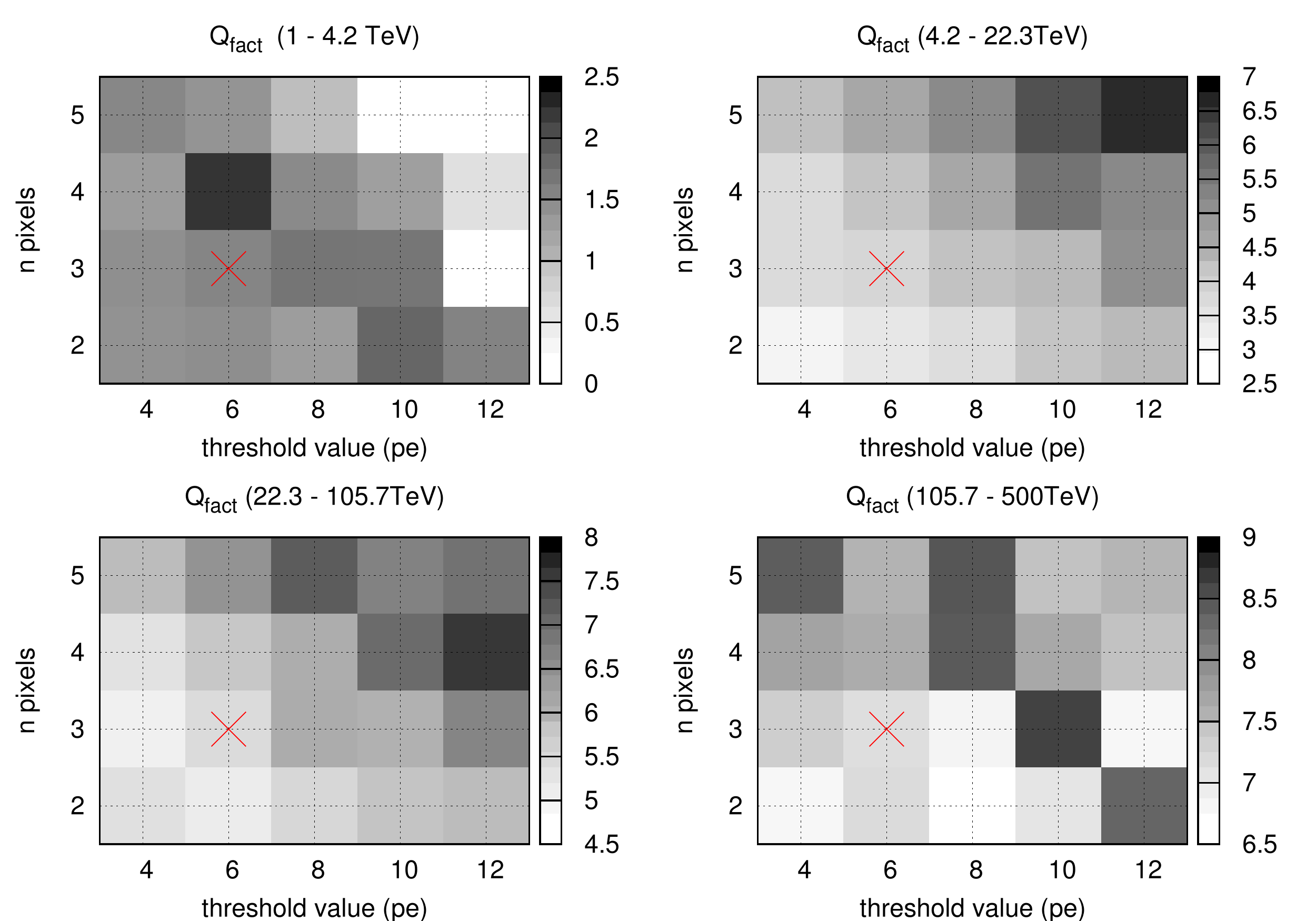}
\captionsetup{width=13cm}
 \caption{Q$_{fact}$ for all triggering combinations for a 1.8 \rm{km} altitude observational site with standard cleaning combination, telescope separation and image \textit{size} cut. The x-axis represents the \textit{threshold} value in \textit{pe} and the y-axis represents the \textit{n} pixel value. The standard triggering combination (6\textit{pe}, 3) provides an adequate Q$_{fact}$ in comparison to the other triggering combinations. The white squares represent combinations that have removed all proton events. The red cross represent the standard value. The equivalent 0.22 \rm{km} altitude results for Q$_{fact}$ are shown in Figure~\ref{fig:qfactor4_trigger}.}
 \label{fig:qfactor4_trigger_high}
\end{centering}
\end{figure}

The Q$_{fact}$ results (Figure~\ref{fig:qfactor4_trigger_high}) for 1.8 \rm{km} altitude observational site with varying triggering combinations. The optimum result varies with energy band. The same trend appeared with the 0.22 \rm{km} altitude Q$_{fact}$ results (Figure~\ref{fig:qfactor4_trigger}). 

The strong trigger \textit{threshold} value and \textit{n} pixel value do tend to provide an improved Q$_{fact}$ for the 2nd and 3rd energy bands, 5 TeV $<$ E $<$ 100 TeV. In the highest energy band E $>$ 100 TeV, all triggering combinations appear to provide similar Q$_{fact}$. Thus the standard triggering combination provides reasonable Q$_{fact}$ over all energy ranges.

For effective area (Figure~\ref{fig:area_trigger_high_alt}) and energy resolution (Figure~\ref{fig:energy_res_trigger_high_alt}) only selected cases are displayed, as in Chapter 4. The pre-cut effective area shows a significant loss in events as the triggering combination increases. The (12\textit{pe}, 5) combination shows a significant loss in events at low energies, E $<$ 10 TeV, compared to the standard triggering combination. 

	The post-selection cut effective area (Figure~\ref{fig:area_trigger_high_alt} bottom panel) shows that the (12\textit{pe}, 5) combination appears to be too strong for low energies, E $<$ 10 TeV.  Above 10 TeV, the difference between the (12\textit{pe}, 5) and (6\textit{pe}, 3) curves is reduced. However, the results show that a significant number of $\gamma$-ray events are lost with the (12\textit{pe}, 5) combination. The difference is not present in the post-selection cut effective areas for E $>$ 10 TeV. The extra events collected by the (4\textit{pe},2) and (6\textit{pe}, 3) combinations do not provide adequate parameterisation or event reconstruction. Overall, the standard triggering combination, (6\textit{pe}, 3), seems to provide adequate pre- and post-selection cut effective areas.

\begin{figure}
\begin{centering}
\includegraphics[scale=0.65]{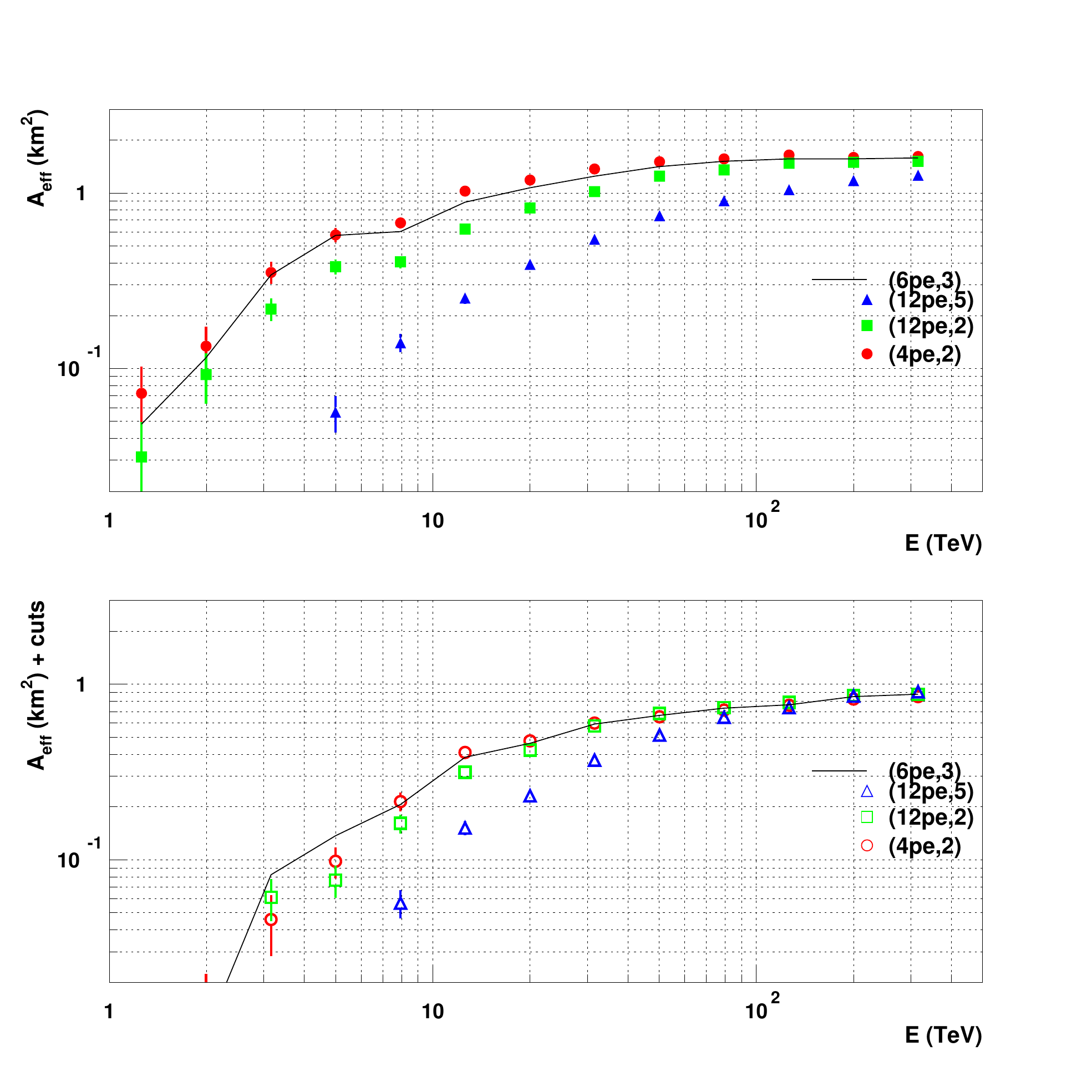}
\captionsetup{width=13cm}
 \caption{Effective area for different triggering combinations at a 1.8 \rm{km} altitude observational site with standard cleaning combination, telescope separation and image \textit{size} cut.
Top: Pre-cut effective area for varying triggering combination. Bottom: post-selection cut effective area for varying triggering combinations.}
 \label{fig:area_trigger_high_alt}
\end{centering}
\end{figure}

Figure~\ref{fig:energy_res_trigger_high_alt} shows the mean and RMS in the energy resolution. The mean in the $\Delta{E}/E$ distribution appears to be positively biased for E $<$ 100 TeV, suggesting that the reconstructed energy is over estimated slightly for all triggering combinations.

The RMS of the $\Delta{E}/E$ distribution for different triggering combinations shows more variation. The biggest improvement is seen above 10 TeV with the strong triggering combination, (12\textit{pe}, 5), selecting the highest quality images which should improve the reconstruction process. The weak triggering combination, (4\textit{pe}, 2), allows more events to trigger but the smaller sized images provide larger $\Delta{E}/E$ RMS. The average RMS for the (12\textit{pe}, 5) combination is 12$\%$ for E $>$ 10 TeV and 20$\%$ for the standard combination. \\

\begin{figure}
\begin{centering}
\includegraphics[scale=0.65]{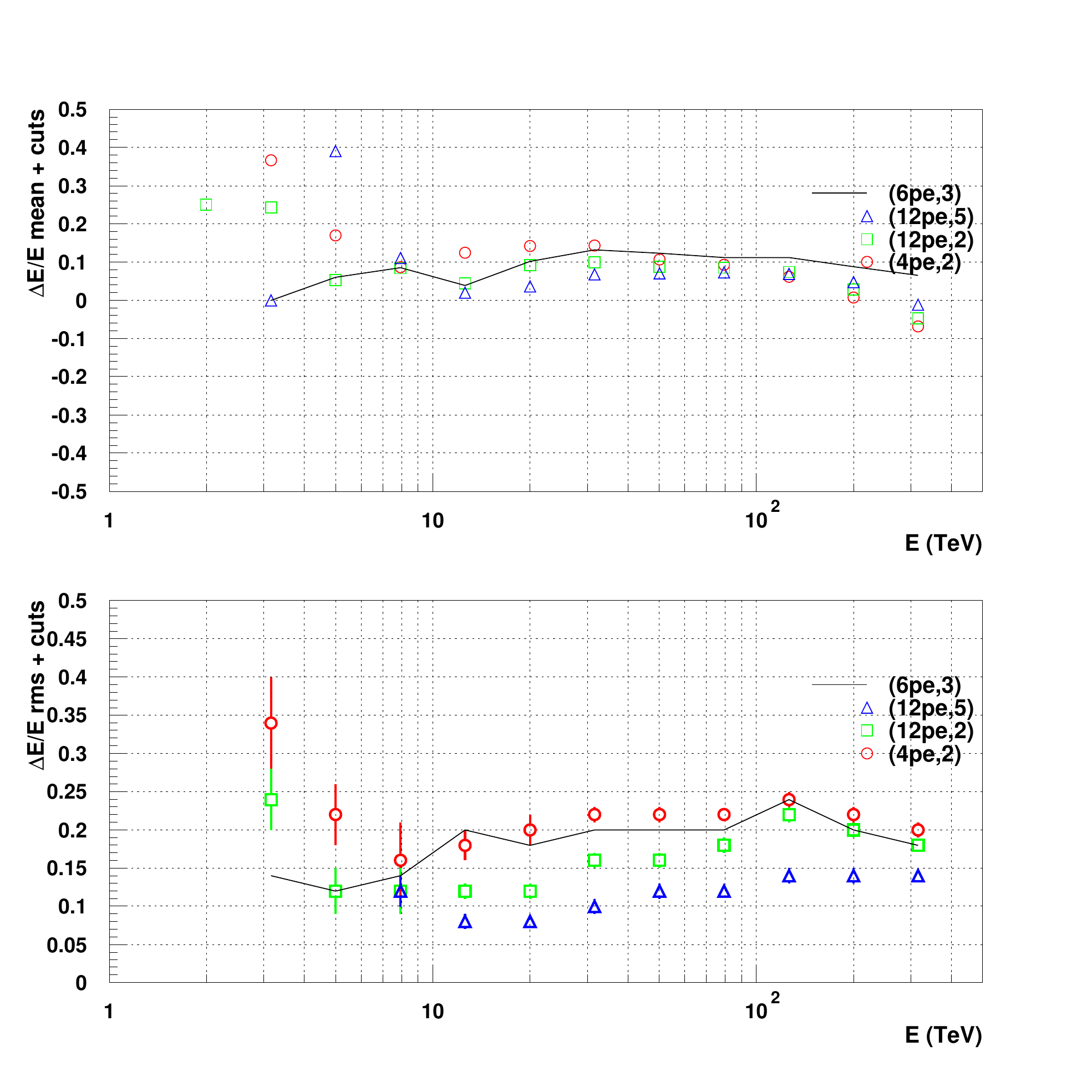}
\captionsetup{width=13cm}
 \caption{Post-shape cut energy resolution for different triggering combinations at a 1.8 \rm{km} altitude observational site with standard cleaning combination, telescope separation and image \textit{size} cut.
 Top: The mean in the $\Delta{E}/E$ distribution. Bottom: Post-shape cut RMS for the $\Delta{E}/E$ distribution. The strongest triggering combination, (12\textit{pe}, 5), provides a large improvement in RMS over the other combinations for E $>$ 10 TeV.}
 \label{fig:energy_res_trigger_high_alt}
\end{centering}
\end{figure}

	The main trend is that the strong triggering combinations, e.g (12\textit{pe}, 5), provide improved angular resolution (Figure~\ref{fig:angres4_trigger_high}) but show a significant loss in $\gamma$-ray events (Figure~\ref{fig:area_trigger_high_alt}) for E $<$ 50 TeV. This trend is expected since the images from the low energy events or large core distance events will fail to trigger a telescope, which leaves the higher quality images. The same trends are seen in the 0.22 \rm{km} altitude study for varying triggering combinations (section~\ref{sec:triggering_op}). A weak triggering combination produces poor shower reconstruction and weaker shape separation. However, more events are obtained due to a lower total trigger size. 



The concluding remark is the same as for the 0.22 \rm{km} altitude results. The high \textit{threshold} value and high \textit{n} pixel value select higher quality images which improves reconstruction. However, the loss in event numbers is quite significant and not optimal for PeX given that once events are rejected at the trigger level, they are lost forever. It is therefore best to accept as many events as possible and to improve the cosmic ray rejection via software based methods. The lower limit to the triggering combination for the cell appears to be the (6\textit{pe}, 3\textit{pe}) combination. Anything weaker offers no benefit to any parameter and the extra events are removed via software cuts later. The standard triggering combination is optimum for the 1.8 \rm{km} altitude site cell.


\section{Effect of cleaning combinations}

	Here, cleaning combinations, using different picture and boundary pixel thresholds values, will be tested at the 1.8 \rm{km} altitude site. We expect that the same effects that were seen with the low altitude site (section~\ref{sec:cleaning_op}) will be observed at the 1.8 \rm{km} altitude. 
	
	The 2D grey scale plots shown in Figure~\ref{fig:angres4_clean_high} and Figure~\ref{fig:qfactor4_clean_high} have been produced for angular resolution and Q$_{fact}$ respectivitely. For both figures, the results are split into four energy bands; 1 - 4.2 TeV, 4.2 - 22.3 TeV, 22.3 - 105.7 TeV and 105.7 - 500 TeV.

	The angular resolution trend shows that a higher cleaning combination provides the optimum angular resolution. These results show how the shower reconstruction is affected by varying cleaning threshold values. A \textit{boundary} value must be implemented, otherwise the image suffers from night sky background and electronic noise interference. A \textit{boundary} value $\geq$ 2\textit{pe} is required with a \textit{picture} value $\geq$ 7\textit{pe}. As the \textit{picture} value increases, the importance of the \textit{boundary} value decreases but is still required. The \textit{boundary} value improves the mitigation of night sky background, which improves the reconstruction. The optimum angular resolution is obtained when the \textit{picture} value is $\geq$ 7\textit{pe} and \textit{boundary} value is $\geq$ 3\textit{pe}. The standard cleaning combination, (8\textit{pe}, 4\textit{pe}), is within the range of combinations which provide a good angular resolution for all energy ranges.
	
\begin{figure}
\begin{centering}
\includegraphics[scale=0.55, angle=270]{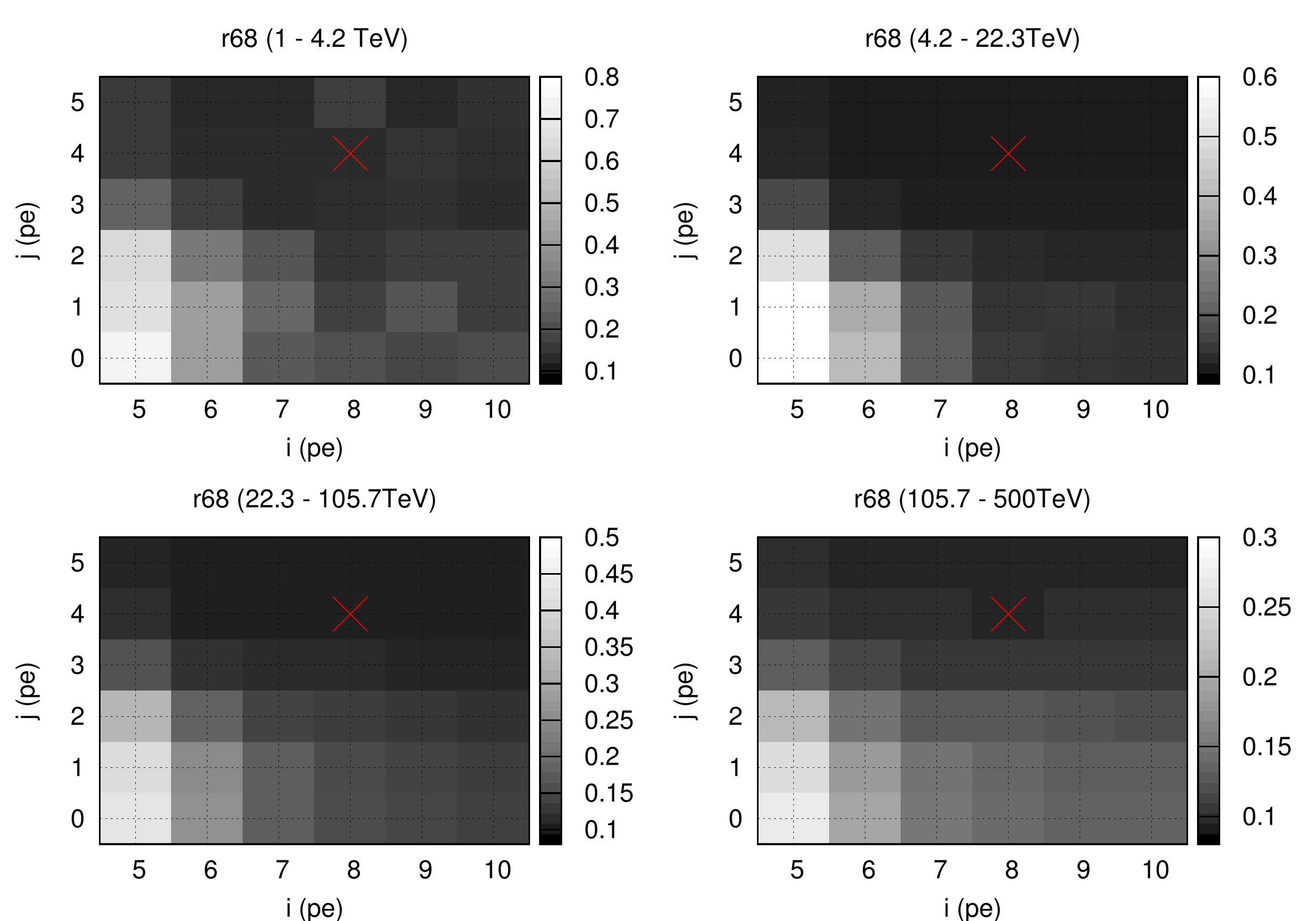}
\captionsetup{width=13cm}
 \caption{Angular resolution (r68) in degrees for all cleaning combinations for a 1.8 \rm{km} altitude observational site with standard triggering combination, telescope separation and image \textit{size} cut. The x-axis represents the \textit{picture} value, \textit{i}, and the y-axis represents the \textit{boundary} value, \textit{j}. The 2D grey scale plots display the r68 as a function of cleaning thresholds. A high \textit{picture} and \textit{boundary} value provide the most accurate angular resolution. The red cross represent the standard value. The equivalent 0.22 \rm{km} altitude results for angular resolution are show in Figure~\ref{fig:angres4_clean}.}
  \label{fig:angres4_clean_high}
\end{centering}
\end{figure}

	Figure~\ref{fig:qfactor4_clean_high} shows how Q$_{fact}$ varies with cleaning combination for the 1.8 \rm{km} altitude observational site. There is a clear trend, which shows that a \textit{boundary} value of $\geq$ 3 \textit{pe} must be used in the cleaning combination. Anything lower than 3\textit{pe} produces insufficient shape separation. For the other cleaning combinations, the standard appears to produce the optimum Q$_{fact}$. A few other cleaning combinations provide good results, such as any cleaning combination with a \textit{boundary} value of 4\textit{pe}. For a \textit{boundary} value of 5\textit{pe}, the cleaning appears to be too strong.

\begin{figure}
\begin{centering}
\includegraphics[scale=0.55, angle=270]{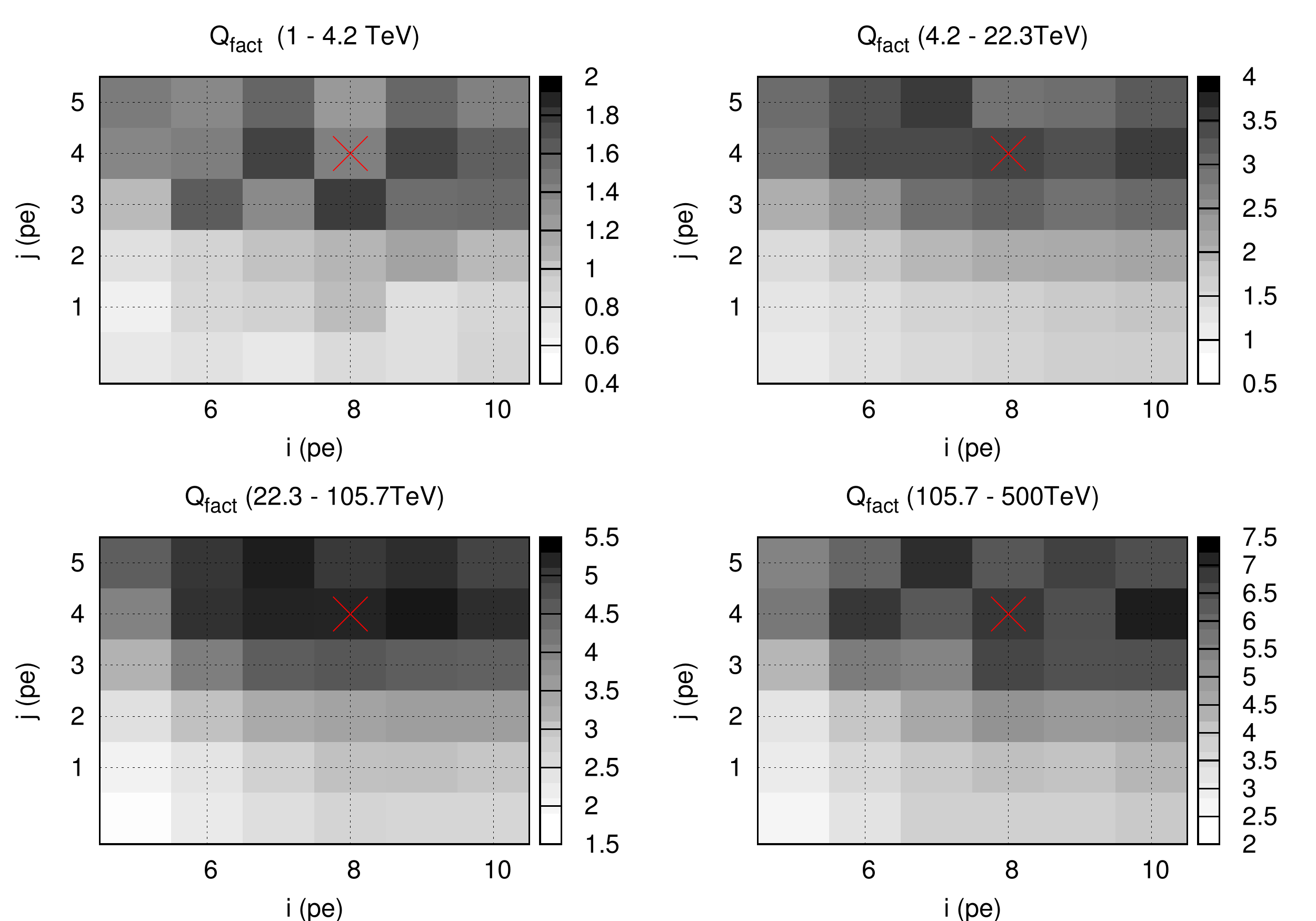}
\captionsetup{width=13cm}
 \caption{Q$_{fact}$ for all cleaning combinations for a 1.8 \rm{km} altitude observational site with standard triggering combination, telescope separation and image \textit{size} cut. The x-axis represents the \textit{picture} value, \textit{i}, and the y-axis represents the \textit{boundary} value, \textit{j}. The 2D grey scale plots display the Q$_{fact}$ as a function of cleaning thresholds. The Q$_{fact}$ shows variation over the four energy bands. The \textit{boundary} must be $>$ 3\textit{pe}. The red cross represent the standard value. The equivalent 0.22 \rm{km} altitude results for Q$_{fact}$ are show in Figure~\ref{fig:qfactor4_clean}.}
  \label{fig:qfactor4_clean_high}
\end{centering}
\end{figure}

\begin{figure}
\begin{centering}
\includegraphics[scale=0.65]{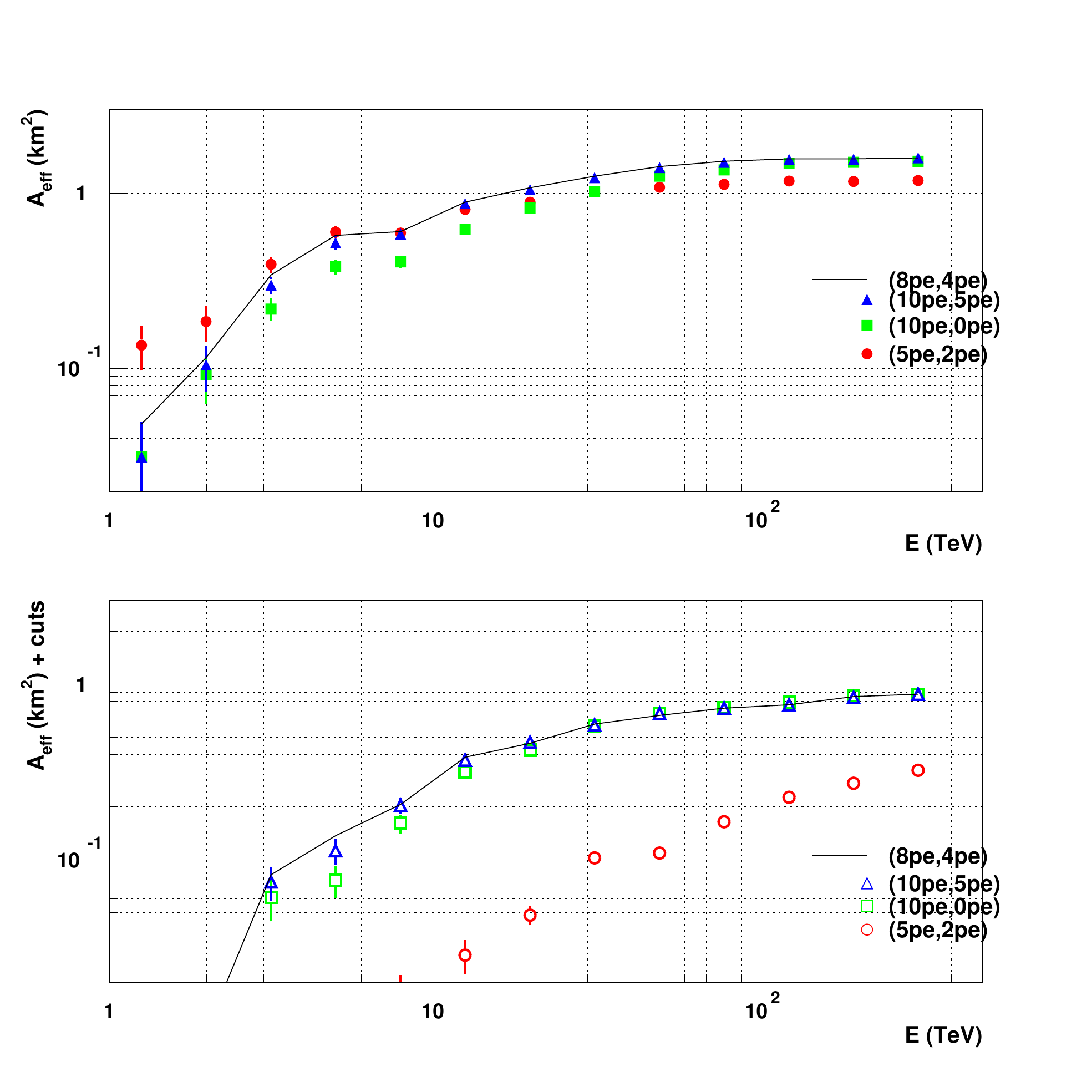}
\captionsetup{width=13cm}
 \caption{Effective area for different cleaning combinations at a 1.8 \rm{km} altitude observational site with standard triggering combination, telescope separation and image \textit{size} cut. 
Top: Pre-cut effective areas for the four cleaning combinations. Bottom: post-selection cut effective area for the four cleaning combinations. The optimum post-selection cut effective area appears to be the standard cleaning combination of (8\textit{pe}, 4\textit{pe}).}
 \label{fig:area_clean_high_alt}
\end{centering}
\end{figure}

	Figure~\ref{fig:area_clean_high_alt} shows the effective area for the 1.8 \rm{km} altitude observational site. We find a minimal difference in effective area is achieved for cleaning variations for E $<$ 10 TeV. The major difference appears for E $>$ 10 TeV, where the lowest cleaning combination, (5\textit{pe}, 2\textit{pe}), reaches an upper limit in the effective area of about 1.3\rm{km$^{2}$}.

	The other strong cleaning combinations provide larger pre-cut effective areas up to 1.8\rm{km$^{2}$}. The standard cleaning combination, (8\textit{pe}, 4\textit{pe}), and the strongest cleaning combination, (10\textit{pe}, 5\textit{pe}), both achieve similar large post-selection cut effective areas. The (5\textit{pe}, 2\textit{pe}) combination is under cleaning images which affects the reconstruction. Hence a larger fraction of events are removed, which is seen by the post-selection cut effective area.

\begin{figure}
\begin{centering}
\includegraphics[width=15cm]{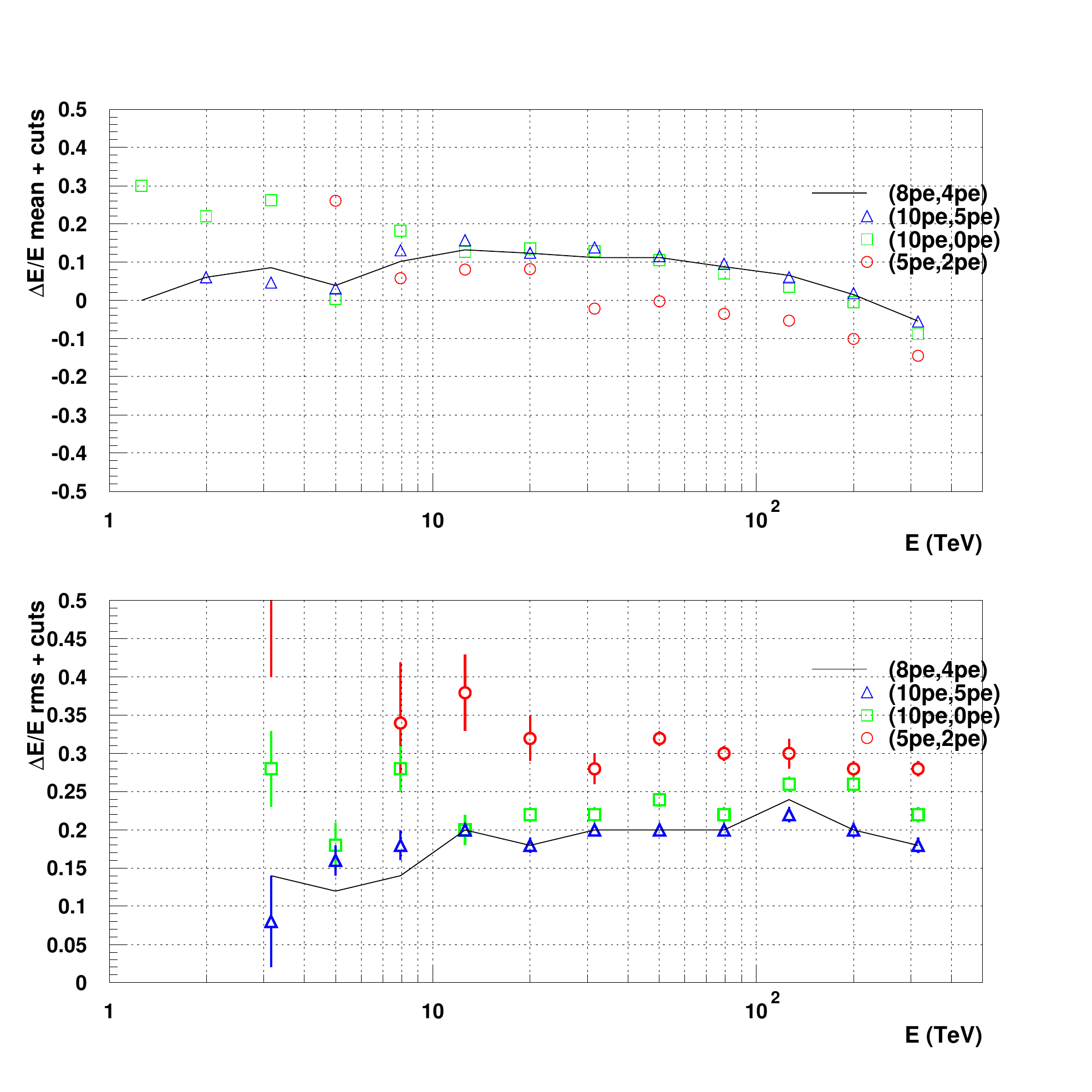}
\captionsetup{width=13cm}
 \caption{Post-shape cut energy resolution for different cleaning combinations at a 1.8 \rm{km} altitude observational site with standard triggering combination, telescope separation and image \textit{size} cut. Top: Mean in the $\Delta{E}/E$ distribution. Bottom: RMS of the $\Delta{E}/E$ distribution. The RMS is approximately 20 to 25$\%$ for three cleaning combinations. }
 \label{fig:energy_res_clean_high_alt}
\end{centering}
\end{figure}

	Figure~\ref{fig:energy_res_clean_high_alt} shows the post-shape cut energy resolution. The figure illustrate that the (5\textit{pe}, 2\textit{pe}) combination provides significant improvement for the mean in the $\Delta{E}/E$ distribution.
The general trend indicates that a stronger cleaning combination works better although the standard cleaning combination still provides adequate results (Figure~\ref{fig:energy_res_clean_high_alt}).

	 A similar effect was seen for the low altitude results (Figure~\ref{fig:area_cleaning_low_alt}) and the same argument can be used to explain it. With a weaker cleaning threshold, the images are affected by under cleaning and an overall decrease is seen. This includes an upper limit to the pre-cut effective area (Figure~\ref{fig:area_clean_high_alt}). More pixels are left in the final image, which can shift the position of the reconstructed C.O.G depending on where the extra pixels are situated in the image. The other factor is that images from showers with E $>$ 10 TeV can be truncated by the edge of the camera. This causes more pixels to remain at the camera edge. The reconstructed C.O.G. is pushed further towards the edge of the camera, which increases the \textit{dis2} parameter (Figure~\ref{fig:dis2_plot_high_alt2} left). Figure~\ref{fig:dis2_plot_high_alt2} illustrates that a (5\textit{pe}, 2\textit{pe}) cleaning combination affects the \textit{dis2} value for small sized images or large core distance events. Since the \textit{dis2} cut is 4.0, a number of these small sized images or large core distance events are removed. The (8pe, 4pe) cleaning combination fixes the problem that a (5\textit{pe}, 2\textit{pe}) combination causes (Figure~\ref{fig:dis2_plot_high_alt2} right). The large fraction of images removed by a \textit{dis2} cut have caused the upper limit in the (5\textit{pe}, 2\textit{pe}) pre-cut effective area (Figure~\ref{fig:area_clean_high_alt}).

\begin{figure}
\begin{centering}
\includegraphics[scale=0.8]{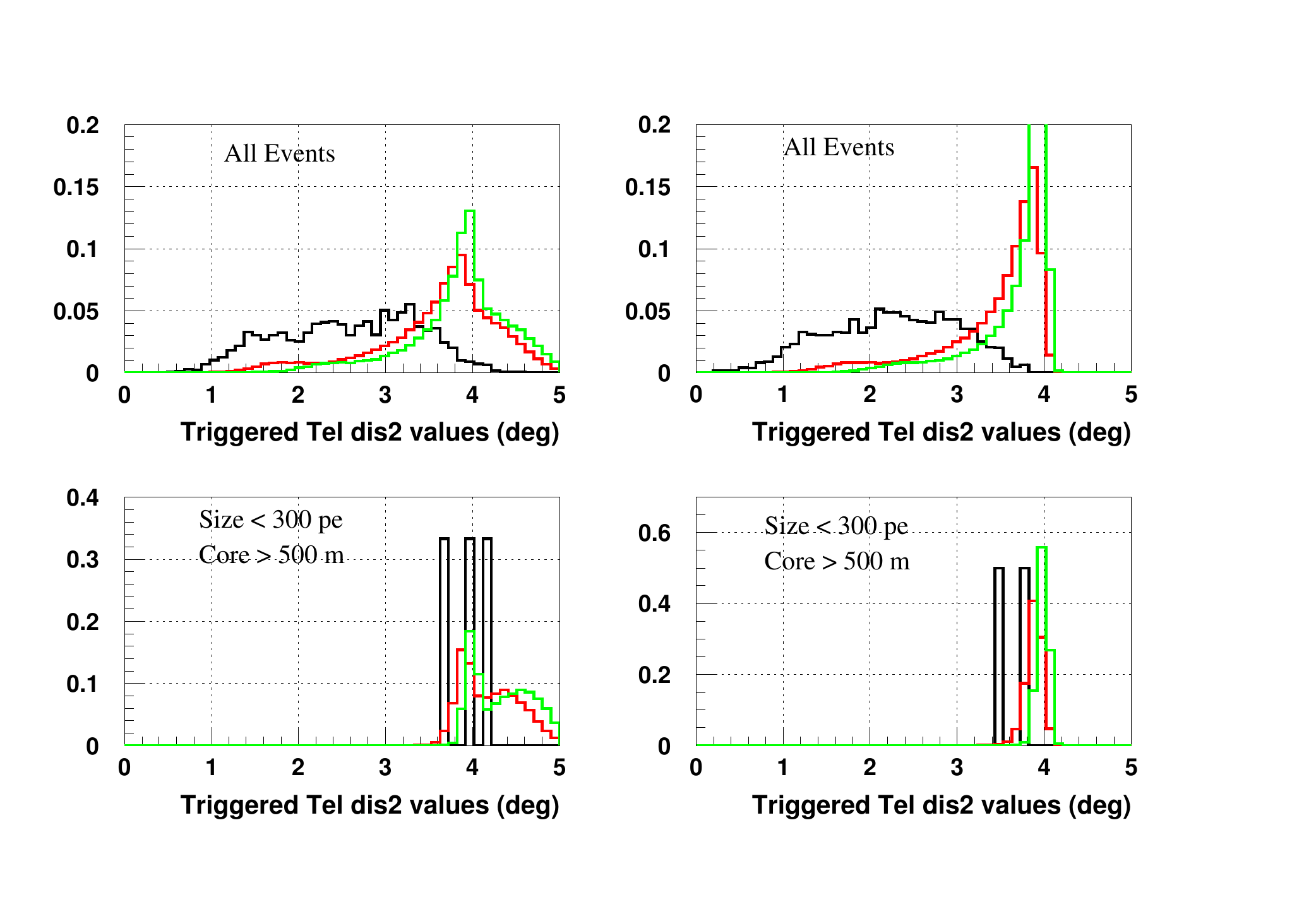}
\captionsetup{width=13cm}  
 \caption{The various \textit{dis2} value for events with (5\textit{pe}, 2\textit{pe}) cleaning combination (left) and (8\textit{pe}, 4\textit{pe}) cleaning combination (right). The events have been split into three different energy bands: 1 to 10 TeV (black), 10 to 100 TeV (red) and 100 to 500 TeV (green). The top panels shows all events and the bottom panels show events with an image size of 300\textit{pe} or less and a core distance greater than 500 \rm{m}. The edge of the physical camera is 4.1$^{\circ}$.}
 \label{fig:dis2_plot_high_alt2}
\end{centering}
\end{figure}


	 Using a high \textit{picture} value with no \textit{boundary} value does not provide the same quality of images as a combination with both threshold values (see (10\textit{pe}, 0\textit{pe}) and (10\textit{pe}, 5\textit{pe}) in Figure~\ref{fig:area_clean_high_alt}). Having no \textit{boundary} value allows noise to pass cleaning and disrupts the shape of the images in the camera. The disruption is not great but shows that a \textit{boundary} value must be used. Otherwise the parameterisation of images is off-set and shape rejection fails to separate $\gamma$-ray and proton events. Overall, the standard (8\textit{pe}, 4\textit{pe}) combination provides adequate results for the 1.8 \rm{km} altitude site.

\section{Effect of image \textit{size} cut}
 \label{sec:image_size_high}
	
	The image \textit{size} cut was varied over the same values as for the 0.22 \rm{km} altitude optimisation (section~\ref{sec:image_size_op}). The image \textit{size} cut is expected to have a similar result at a 1.8 \rm{km} altitude site as it did for the low altitude site. As the image \textit{size} cut increases, the rejection power of the cell will increase and the shower reconstruction will also improve.
The four image \textit{size} cuts used for comparison are the standard cut, 60\textit{pe}, and three larger cuts; 100\textit{pe}, 200\textit{pe} and 300\textit{pe}.

\begin{figure}
\begin{centering}
\includegraphics[scale=0.65]{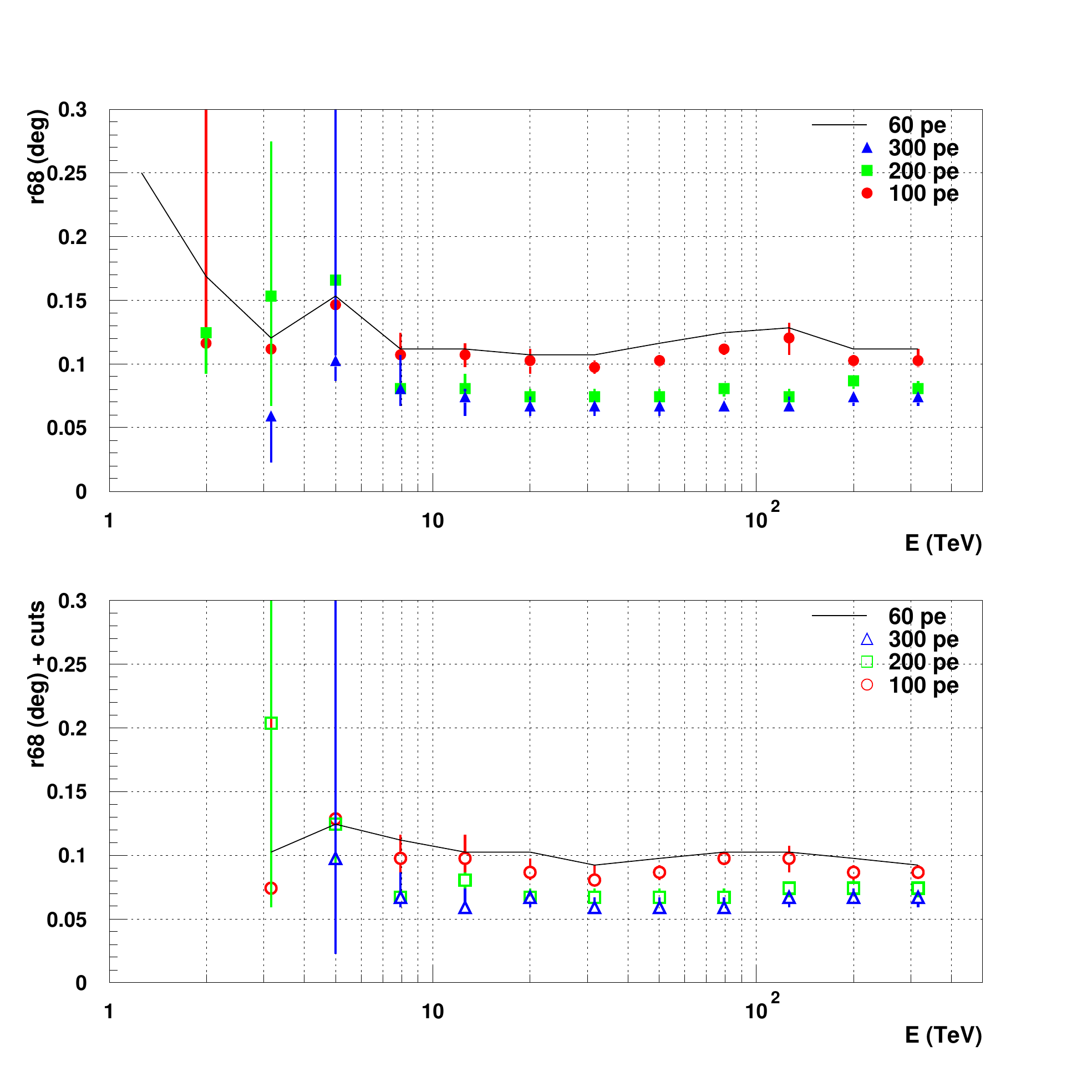}
\captionsetup{width=13cm}
 \caption{Angular resolution (r68) for varying image \textit{size} cuts at a 1.8 \rm{km} altitude observational site with standard triggering combination, cleaning combination and telescope separation. Top: Pre-cut angular resolution for varying image \textit{size} cuts. The improvement is approximately 50$\%$ for E $>$ 10 TeV between the standard and 300\textit{pe} image \textit{size} cut.
Bottom: Post-cut angular resolution for varying image \textit{size} cuts. The difference between the standard and 300\textit{pe} image \textit{size} cut is now 40$\%$.}
 \label{fig:angres_size_high_alt}
\end{centering}
\end{figure}

	Figure~\ref{fig:angres_size_high_alt} shows the angular resolution (r68) for a 1.8 \rm{km} altitude site.  The results show that the angular resolution improves as the image \textit{size} cut increases. This trend in angular resolution is expected, since increasing the lower limit on image size for all events improves the reconstruction. For E $<$ 10 TeV, the pre-cut angular resolution fluctuates due to limited events in all results (Figure~\ref{fig:angres_size_high_alt} top panel). For E $>$ 10 TeV, the pre-cut angular resolution is stable and the 300\textit{pe} image \textit{size} cut provide a significant improvement in angular resolution. The improvement is 50$\%$ for E $>$ 10 TeV between the standard and 300\textit{pe} image \textit{size} cuts (Figure~\ref{fig:angres_size_high_alt} black line and blue triangles).
	The post-cut angular resolutions show the same trend (Figure~\ref{fig:angres_size_high_alt} bottom panel). The 200\textit{pe} and 300\textit{pe} image \textit{size} cuts provide significant improvements in angular resolution over the standard image \textit{size} cut. The improvement between the standard and 300\textit{pe} image \textit{size} cut has decreased to about 40$\%$. There appears to be minimal difference between the 60\textit{pe} and 100\textit{pe} image \textit{size} cuts. Therefore, using a 60\textit{pe} image \textit{size} cut as the standard for the 1.8 \rm{km} altitude appears adequate. The small sized images in the reconstruction do provide good angular resolution. However, increasing the lower limit for the image size improves the angular resolution since the reconstruction is guaranteed to contain strong images.

\begin{figure}
\begin{centering}
\includegraphics[scale=0.65]{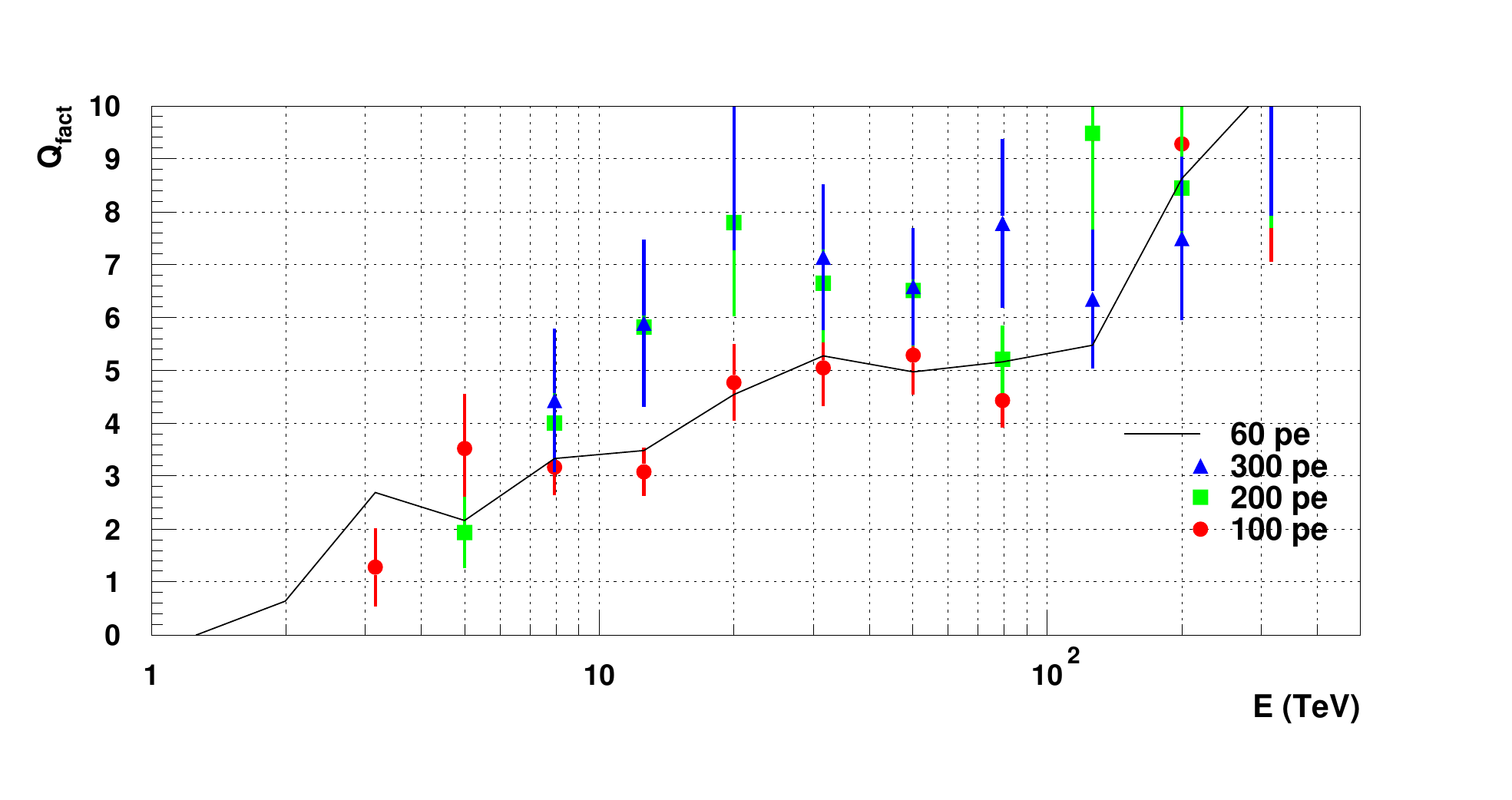}
\captionsetup{width=13cm}
 \caption{Q$_{fact}$ for varying image \textit{size} cuts at a 1.8 \rm{km} altitude site with standard triggering combination, cleaning combination and telescope separation. }
 \label{fig:qfactor_size_high_alt}
\end{centering}
\end{figure}

Figure~\ref{fig:qfactor_size_high_alt} shows the Q$_{fact}$ results. The figure illustrates that a larger image \textit{size} cut provides a small improvement to the rejection power of the cell. The curves for a 300\textit{pe} and 200\textit{pe} image \textit{size} cut show strong fluctuations but appear to provide a systematic improvement in the mid energy ranges, 10 TeV $<$ E $<$ 100 TeV. For E $<$ 10 TeV, all curves have a similar Q$_{fact}$. For E $>$ 100 TeV, the Q$_{fact}$ seems similar for all \textit{size} cuts. The results tend to suggest that a significant number of small events occur in the E $<$ 100 TeV energy range. Therefore, the largest improvement is gained in this range.

\begin{figure}
\begin{centering}
\includegraphics[scale=0.65]{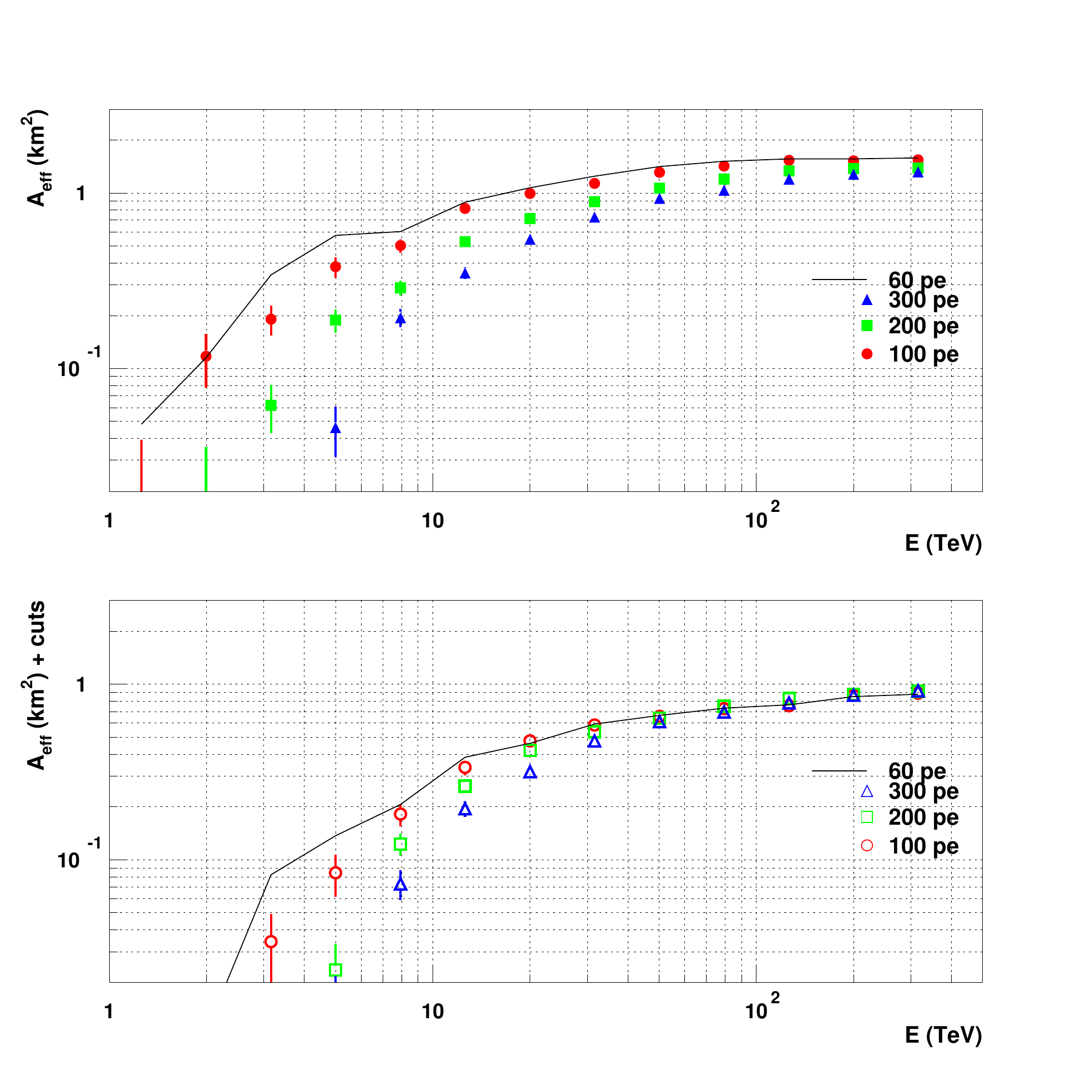}
\captionsetup{width=13cm}
 \caption{Effective area for a variety of image \textit{size} cuts at a 1.8 \rm{km} altitude observational site with standard triggering combination, cleaning combination and telescope separation. Top: Pre-cut effective area for the four different image \textit{size} cuts.
Bottom: post-selection cut effective area for the four different image \textit{size} cuts. The trend shows that as the image \textit{size} cut increases, the effective area decreases.}
 \label{fig:area_size_high_alt}
\end{centering}
\end{figure}

	The trend seen in the pre- and post-selection cut effective areas (Figure~\ref{fig:area_size_high_alt}) is somewhat expected. For low energies, E $<$ 10 TeV, a 300\textit{pe} image \textit{size} cut provides a reduced effective area compared to the standard image \textit{size} cut. The post-selection cut areas indicate that some of the events removed via a 200\textit{pe} or 300\textit{pe} cut provide good reconstruction and would pass shape and a point source cuts. However, there is a considerable loss of events caused by such a large cut.

	As the energy increases, the difference between the 60\textit{pe} and 300\textit{pe} effective area decreases. There is still an overall loss in events for 300\textit{pe} up to 500 TeV for the pre-cut effective area. However, the post-selection cut effective areas show that all results produce the same area for E $>$ 80 TeV. The extra events gained with a small image \textit{size} cut are removed via shape and point source cuts. These events are likely to be the small sized, large core distance, images.

	Comparing the post-shape cut energy resolution for different image \textit{size} cuts, the mean in the $\Delta{E}/E$ distribution is similar for all cuts. The RMS for the $\Delta{E}/E$ distribution for the 300\textit{pe} cut shows almost 40$\%$ improvement for E $>$ 10 TeV. The improvement is due to the lower number of events or the selection of higher quality events seen in Figure~\ref{fig:energy_res_size_high_alt} and Figure~\ref{fig:angres_size_high_alt}.\\ 

	A small image creates a larger error in reconstructed C.O.G, image \textit{width}, image \textit{length} and major axis. The larger error is due to minimal pixels being available for an ellipse fit. The larger errors create larger variations in the reconstructed direction. By removing the small images, the reconstructed direction and location improve (Figure~\ref{fig:angres_size_high_alt}). Sometimes the small images are paired with larger images at small core distances. This reduces the error in reconstruction. There are two scenarios:
\begin{itemize}
\item If the small image is part of a group of three images for a single event, then removing the small image will benefit the results. 
\item If the small image is part of a group of two images then removing the small image will remove the whole event due to stereoscopic or quality requirements ( $leq$ two images)
\end{itemize}
Therefore, removing these small images does improve the angular resolution but the number of events post-shape cuts are affected.

	The effective area results (Figure~\ref{fig:area_size_high_alt}) showed that a significant portion of events that trigger a telescope consists of small images. Including these small images is important for triggering off large core distance events. By limiting the image \textit{size}, one essentially limits the core distance at which events can trigger the cell. Hence, one limits the effective area of the cell. The improvement in angular resolution is not worth the loss in effective area. If the source flux is strong and no minimum energy is required, then a large image \textit{size} cut could be used to extract the best events. For a majority of observations, the largest possible effective area will be required, especially if PeX is used for an all sky survey of the Galactic plane since less observation time will be required. Therefore, a large cut of 200\textit{pe} or even 300\textit{pe} could be used in certain situations. We can classify the use of a large image \textit{size} cut as part of the hard cuts for the cell, as we did for the 0.22 \rm{km} altitude optimisation (Chapter 4). \\

\section{Concluding remarks on high altitude cell optimisation}

From this investigation, the aim was to find an optimised result for each individual parameter given that we are working with limited Monte Carlo statistics. To best compare the results, we can plot the flux sensitivity relative to the standard configuration for each parameter (Figure~\ref{fig:flux_parameter_high}). The flux sensitivities have been plotted as the ratio of varied flux over the standard flux, $F_{standard}$/$F_{varied}$, where the varied flux is the flux sensitivity for each parameter variation. Therefore, if the variation provides an improvement in flux sensitivity over the standard configuration then the plotted flux ratio will be $>$ 1.

\begin{figure}
\begin{centering}
\includegraphics[scale=0.75]{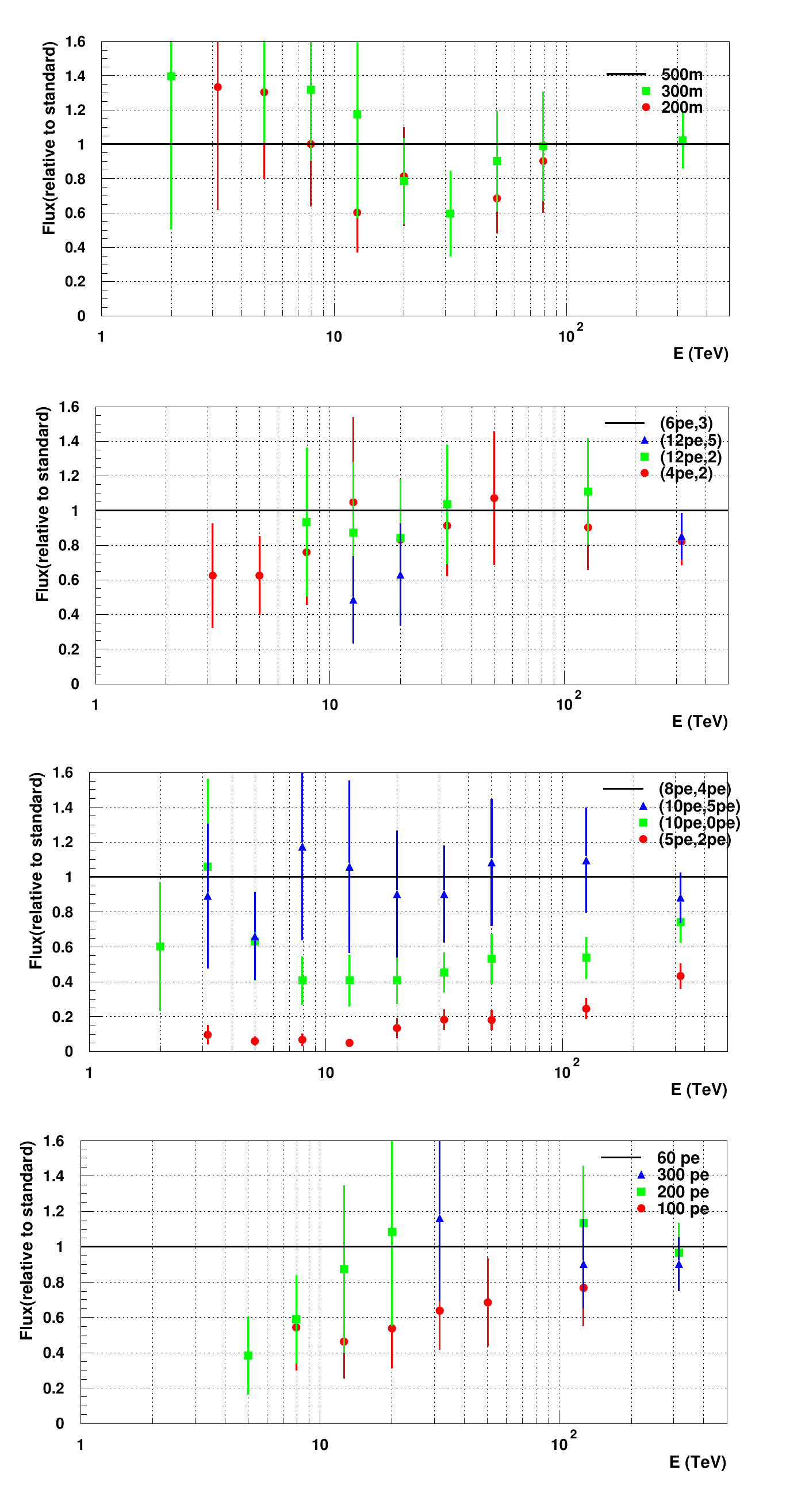}
\captionsetup{width=13cm}
 \caption{Flux sensitivity plots for various telescope separations (first panel), various triggering combinations (second panel), cleaning combinations (third panel) and image \textit{size} cuts (fourth panel) for a 1.8 \rm{km} altitude site. A ratio larger than 1 implies an improvement in the flux sensitivity, compared with the standard configuration. Compared to Figure~\ref{fig:flux_parameter_low} for 0.22 \rm{km}.}
 \label{fig:flux_parameter_high}
\end{centering}
\end{figure}

The standard telescope separation provides the best flux sensitivity for E $>$ 20 TeV (Figure~\ref{fig:flux_parameter_high} first panel). The smaller telescope separations detect more events at low energies. 

The standard triggering combination (6\textit{pe}, 3) provides the best flux sensitivity for all energies (Figure~\ref{fig:flux_parameter_high} second panel). The (12\textit{pe}, 5) combination does not provide an improvement to the flux sensitivity for E $>$ 10 TeV and there is a large loss in events below 10 TeV. The (12\textit{pe}, 5) combination could be used for strong point sources where an improved angular resolution is required since it provides a good angular resolution (Figure~\ref{fig:angres4_trigger_high}). 

The standard cleaning combination (8\textit{pe}, 4\textit{pe}) provides the best flux sensitivity for all energies (Figure~\ref{fig:flux_parameter_high} third panel). The (10\textit{pe}, 5\textit{pe}) is the other combination that would provide a similar flux sensitivity. These combinations would be the best choices for studying a point source since they also provide the best angular resolution for the cleaning combinations investigated. The other cleaning combinations shown in Figure~\ref{fig:flux_parameter_high} would provide significant loss in flux sensitivity.

The standard image \textit{size} cut of 60\textit{pe} provides an adequate flux sensitivity (Figure~\ref{fig:flux_parameter_high} fourth panel). All other image \textit{size} cut results are within error bars of the standard cut. Like previous parameters, the harder or stronger values in this case the 300\textit{pe} cut provides an improvement in the flux for E $>$ 10 TeV. The image \textit{size} cut is applied to the results once the events have been recorded, it is easier to vary compared to the triggering combinations. Therefore, a 200 or 300\textit{pe} image \textit{size} cut can be used for a strong source (extended or point) to help select only strong events. For all results regarding Figure~\ref{fig:flux_parameter_high}, the larger error bars indicate a lower total number of events used to provide each curve. Taking these results into account, the final conclusions can be made.

The optimisation of the triggering combination, the cleaning combination, the telescope separation and the image \textit{size} cut for a 1.8 \rm{km} observation site provides the following conclusions:
\begin{itemize}
\item A 500 \rm{m} separation provides a good post-selection cut effective area for E $>$ 10 TeV. For E $<$ 10 TeV, the post-selection cut effective areas produce the same results between telescope separations. A significant improvement of $35\%$ was seen for the post-shape cut angular resolution with the 500 \rm{m} telescope separation compared to the 300 \rm{m} separation.
\item The standard triggering combination (6\textit{pe}, 3) provides a high post-selection cut effective area. However, the strong triggering combinations do provide better angular resolution and Q$_{fact}$ but with a significant loss in effective area for E $<$ 80 TeV. Due to the large loss in events, the strong triggering combinations might not be optimal for PeX when detecting/discovering sources.
\item The standard cleaning combination, (8\textit{pe}, 4\textit{pe}), provides the optimum effective area, angular resolution and Q$_{fact}$. However, a number of other cleaning combinations provide similar results, (10\textit{pe}, 5\textit{pe}) and (6\textit{pe}, 4\textit{pe}).
\item The standard image \textit{size} cut of 60\textit{pe} appears to be the appropriate cut. The 200\textit{pe} or 300\textit{pe} cuts remove the small images which helps both angular resolution and Q$_{fact}$ but results in a loss of total event numbers. These larger image \textit{size} cuts would be more appropriate if applying hard cuts to observational data.
\end{itemize}
The optimum configuration for a 1.8 \rm{km} altitude site appears to be: a 500 \rm{m} telescope separation, a (6\textit{pe}, 3) triggering combination, an (8\textit{pe}, 4\textit{pe}) cleaning combination and a 60\textit{pe} image \textit{size} cut which is in agreement with results for the 0.22 \rm{km} altitude (Chapter 4).

\section{High and low altitude PeX comparison for Algorithm 1}
 \label{sec:comparison_alg1}

	With the optimised configurations for PeX placed at a 1.8 \rm{km} altitude site and a 0.22 \rm{km} altitude site, we can compare the results across the two altitudes.

	The angular resolution, energy resolution and Q$_{fact}$ parameters do not provide a significant difference between the two site altitudes (Figure~\ref{fig:multi_plot}). The RMS for the $\Delta{E}/E$ distribution shows that the 0.22 \rm{km} altitude site provides a tighter distribution. Therefore, the energy reconstruction for the 0.22 \rm{km} altitude site appears to provide improved energy reconstruction by roughly 25$\%$.
	The post-shape cut angular resolution for the 1.8 \rm{km} altitude site provides an improved angular resolution for E $<$ 10 TeV. For E $>$ 10 TeV, the post-shape cut angular resolution provides the same results for both sites. The results indicate that neither altitude provides a superior reconstructed shower direction. As long as the size of the image is large and enough pixels remain in the image for a proper ellipse to be fitted, then the reconstruction is not affected.
	The Q$_{fact}$ provides similar results for both altitude sites. Therefore, with the exception of one or two points the two Q$_{fact}$ results are consistent within errors. A slightly higher image \textit{size} cut could be applied since the number of photons per image should increase for a 1.8 \rm{km} altitude site. So the size cut could be adjusted. However, the image \textit{size} cut investigation at the 1.8 \rm{km} altitude shows that a 100\textit{pe} image \textit{size} cut does not provide much of an improvement compared to the 60\textit{pe} image \textit{size} cut although angular resolution might be improved. Therefore, the 60\textit{pe} image \textit{size} cut is not affecting the results. \\

\begin{figure}
\begin{centering}
\includegraphics[scale=.7]{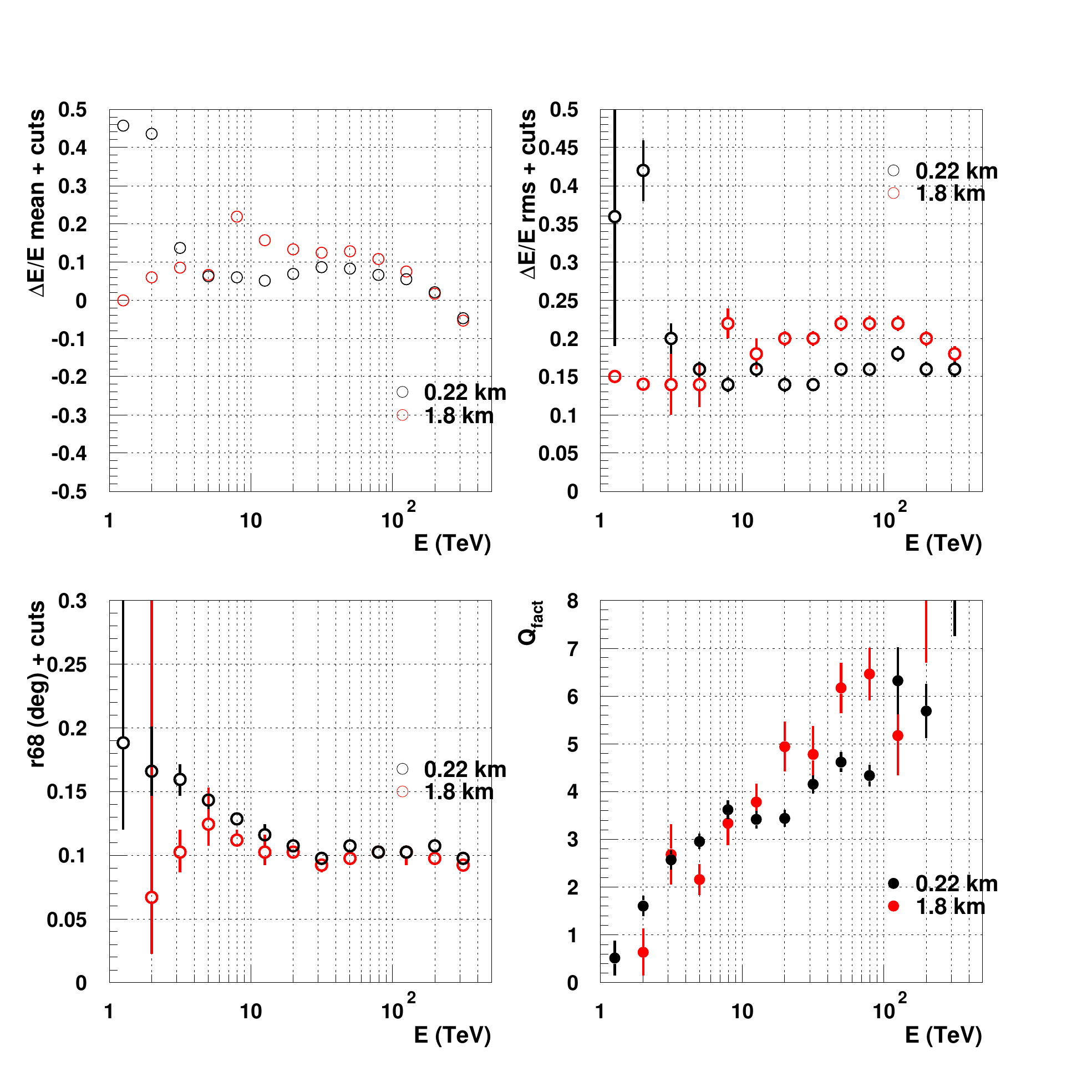}
 \caption{A panel of plots showing the comparison between parameters for both altitude sites. Top Left: The mean in the post-shape cut $\Delta{E}/E$ distribution. Top Right: The RMS for the post-shape cut $\Delta{E}/E$ distribution shows that the 0.22 \rm{km} altitude site produces a tighter distribution by roughly 25$\%$. Bottom Left: The post-shape cut angular resolution. A slight improvement is seen in angular resolution for E $<$ 10 TeV at the 1.8 \rm{km} altitude site. Bottom Right: The Q$_{fact}$ for both sites. The 1.8 \rm{km} altitude site appears to provide a slightly better rejection between $\gamma$-ray and proton events for mid-ranged energies.}
 \label{fig:multi_plot}
\end{centering}
\end{figure}

To explain the difference between the two altitude sites, the lateral distribution needs to be considered. For a 1.8 \rm{km} altitude site, the cell is closer to the shower maximum. The cell witnesses a brighter and narrower Cherenkov light cone with higher photon intensities. Just by comparing the Cherenkov light intensity as a function of core distance in plots Figure~\ref{fig:distance} and Figure~\ref{fig:distance_high} for 0.22 \rm{km} and 1.8 \rm{km} altitudes respectively, we can see that the light cones for the same energy events are different sizes. The photon intensities are higher for a 1.8 \rm{km} altitude site. This is due to the 1.8 \rm{km} observational site being closer to shower maximum. Less light is lost through atmospheric absorption or scattering and the light does not have enough altitude to spread out before reaching the observational level. Figure~\ref{fig:distance_high} shows that the lateral distribution for events detected at a 1.8 \rm{km} altitude site have an overall higher photon intensity, which helps the low energies events trigger the telescopes. This effect causes a larger effective area for low energies at the 1.8 \rm{km} altitude site (Figure~\ref{fig:relative_area}). The larger effective area for a 0.22 \rm{km} altitude site at high energies is due to the large core distance events. The 0.22 \rm{km} altitude site provides the shower with more atmosphere to travel through, which results in a larger light pool because the Cherenkov photons spread out more (Figure~\ref{fig:cherenkovshower}). The lateral distribution plots show the 0.22 \rm{km} altitude showers can trigger at large core distances which increases the pre- and post-selection cut effective area for the cell. This implies that the extra events gained by the 0.22 \rm{km} altitude cell provide reconstructions that pass cuts and provide useful information about a source.

	Figure~\ref{fig:relative_area} best represents the difference in post-selection cut effective areas. The 1.8 \rm{km} altitude effective area has been plotted relative to the 0.22 \rm{km} altitude effective area, A$_{eff_{1.8km}}$/A$_{eff_{0.22km}}$. The black line represents the 0.22 \rm{km} altitude post-selection cut effective area relative to the 0.22 \rm{km} altitude post-selection effective area and the red rings represent the 1.8 \rm{km} altitude post-selection cut effective area relative to the 0.22 \rm{km} altitude post-selection effective area. For E $<$ 10 TeV, the 1.8 \rm{km} altitude post-selection effective area provides a larger area than the 0.22 \rm{km} altitude site. For E $>$ 10 TeV, the 0.22 \rm{km} altitude site provides a larger post-selection cut effective area, by roughly 15$\%$. The effective area is the major difference between the two altitude sites.

\begin{figure}
\begin{centering}
\includegraphics[scale=0.65]{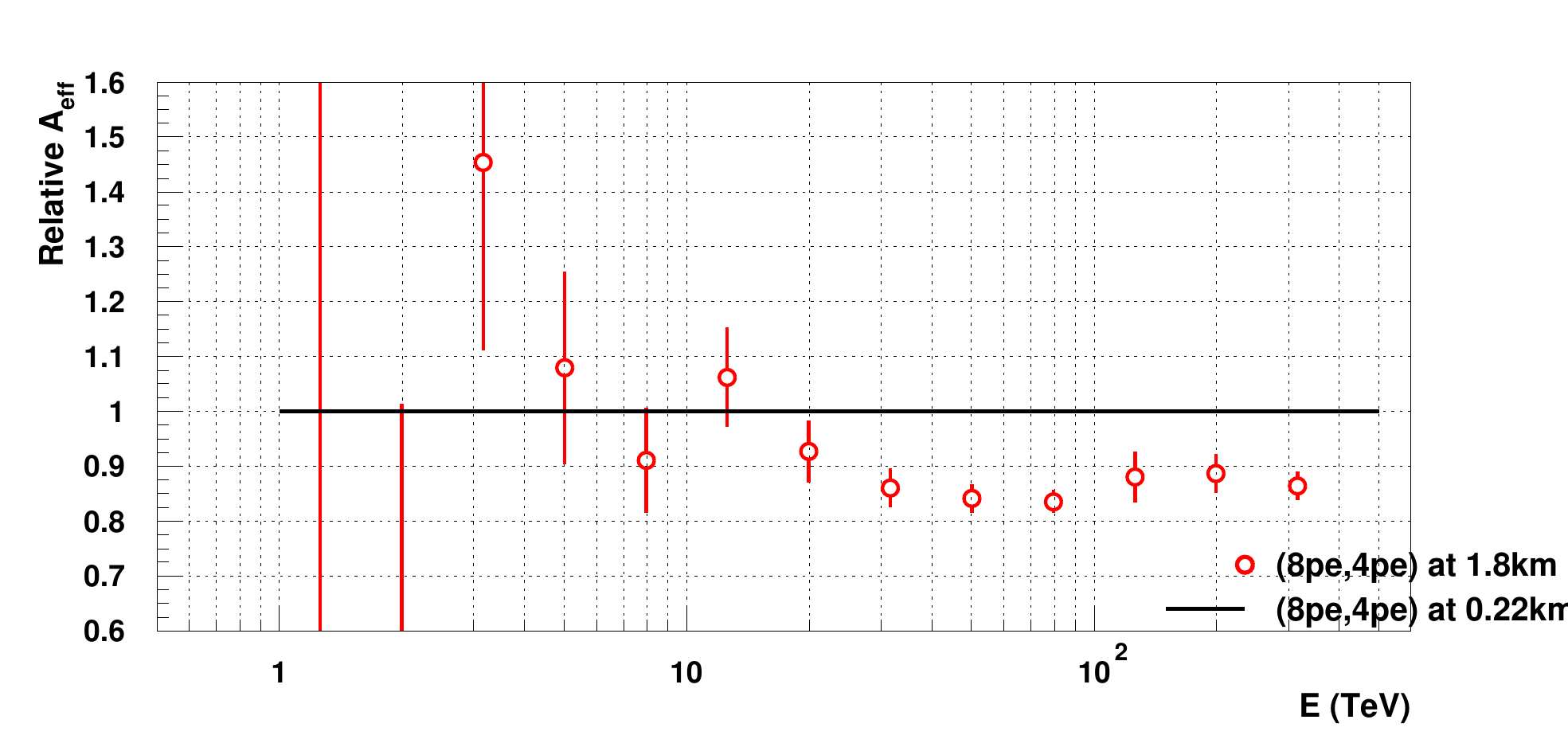}
 \caption{post-selection cut effective area for the best optimisations for a 0.22 \rm{km} and 1.8 \rm{km} altitude site. The plot shows the ratio of 1.8 \rm{km} altitude effective area over the 0.22 \rm{km} altitude effective area,  A$_{eff_{1.8km}}$/A$_{eff_{0.22km}}$. The effective area for the 0.22 \rm{km} altitude site provides an improvement if the ratio is $<$ 1. For E $>$ 10 TeV, the 0.22 \rm{km} altitude site provides a larger post-selection cut effective area by roughly 15$\%$.}
 \label{fig:relative_area}
\end{centering}
\end{figure}

The difference in effective areas is the major difference between the two altitude sites. The other parameters had little to no variation between a 1.8 \rm{km} and 0.22 \rm{km} altitude observational site. It appears that neither site improves the reconstruction of the shower direction. Both altitude sites detect events which have small image sizes or events which are truncated by the edge of camera. These factors limit the accuracy of the angular resolution and they occur for both sites. To improve the angular resolution, other factors need to be considered, see section~\ref{sec:pix_arrange}. \\



The 0.22 \rm{km} altitude cell provides a larger pre- and post-selection cut effective area, where the improvement in post-selection cut effective area is approximately 15$\%$ for E $>$ 10 TeV. The larger post-selection cut effective area for the 0.22 \rm{km} altitude site is the biggest improvement between the different altitude results. It is the main reason for choosing the 0.22 \rm{km} altitude site over a 1.8 \rm{km} altitude site. The 1.8 \rm{km} altitude site does provide a slightly larger post-selection cut effective area for E $<$ 10 TeV. However, the error bars are large, suggesting that few events are left. The other parameters provide similar results between altitudes. Therefore, the different altitude sites have no effect on event reconstruction or shape rejection.

Now that a 0.22 \rm{km} altitude site has been chosen to be an appropriate altitude for the PeX cell, we can investigate an improvement to the image cleaning algorithm. The idea was to further constrain the pixels used for the event reconstruction. This constraint involves applying a cut on the arrival times of the pixels to help remove the pixels that contain night sky background. This arrival time cut can be applied to the Algorithm 1 reconstruction code and the robustness of the time cut can be compared with varying levels of night sky background.

\chapter{Image Cleaning based on Pixel Timing}
 \label{sec:timing_cleaning}

In Chapter~\ref{sec:optimise_low} and ~\ref{sec:optimise_high}, several PeX parameters were individually optimised. These included the thresholds of the image cleaning algorithm, used to reduce the effect of the night sky background (NSB) on the Cherenkov images. The cleaning algorithm selects clumps of isolated pixels in the camera that have strong photoelectron counts and removes isolated pixels with weak photoelectron counts. 

The motivation to further alter the cleaning algorithm comes from the time development of Cherenkov showers and the time gradient studied in \cite{Stamatescu}. Figure~\ref{fig:victor_time_height} shows the time development of a 100 TeV $\gamma$-ray shower and a 130 TeV proton shower, which have similar image \textit{sizes}. The figure illustrates the small time difference between Cherenkov photons as the shower develops in the atmosphere. This time development shown includes the arrival time of the muons produced predominantly within proton showers. This figure has been taken from \cite{Stamatescu}. There is a strong time correlation between the time the photon was produced in the shower and the time it would arrive at ground level. This can be used to an advantage in the cleaning algorithm. A cut on the arrival time of photons in the camera can be added to the cleaning algorithm. If a time cut is applied to the boundary pixels (the weaker pixels at the edge of an image) then it may increase the probability that the boundary pixel will contain Cherenkov signal by helping to remove the randomly arriving NSB.
The time development shows the production of muons in a proton shower (Figure~\ref{fig:victor_time_height} right panel) and that the muons are observed at times generally earlier than the main part of the shower and lower in the atmosphere. Therefore, when finding an appropriate time cut the arrival time of the muons also need to be taken into account since they may help distinguish proton showers from $\gamma$-ray showers. As long as the time cut is not too narrow then the muons should be included. 


In this Chapter, we will investigate a cut on arrival time of pixel pulses. We will introduce this new timing cut to the cleaning algorithm and show the outcome for different situations including varying levels of NSB applied to a standard PeX configuration.


\begin{figure}[p]
\begin{centering}
\includegraphics[scale=0.68]{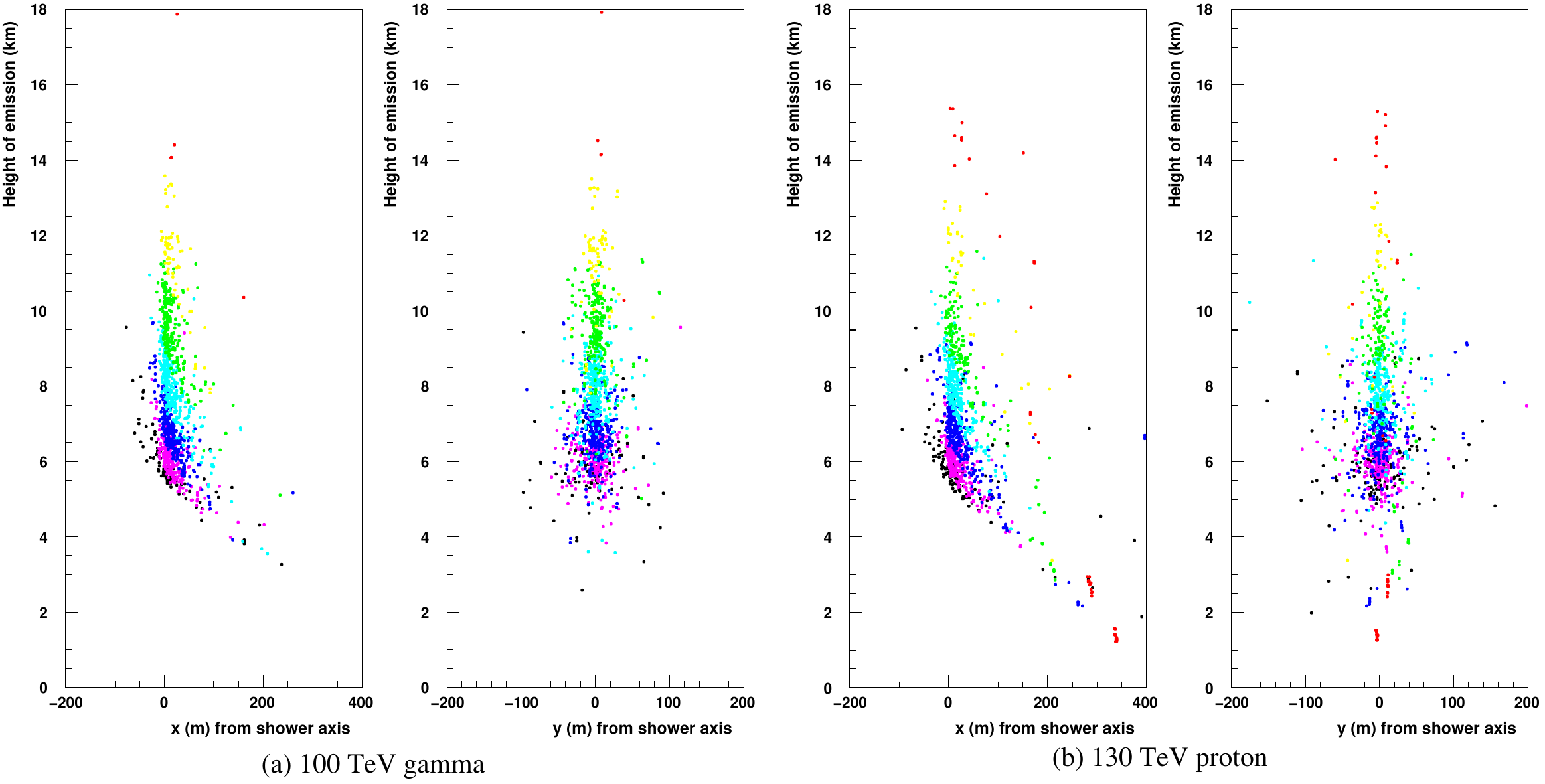}
\begin{minipage}[bottom]{6cm}
\includegraphics[scale=0.7]{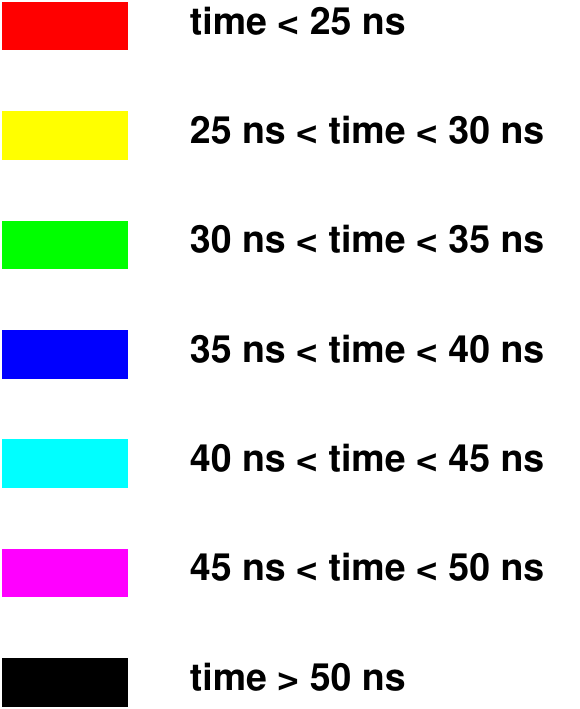}
\end{minipage}
\begin{minipage}[right]{8cm}

 \caption{The x and y projected emission positions from the shower axis for photons from a vertical 100 TeV $\gamma$-ray shower (a) and a 130 TeV proton shower (b). The telescope is located at 400 \rm{m} away from the shower core. The colours represent the arrival time of the photons with respect to the time at which the primary particle would reach ground level. Taken from \cite{Stamatescu}}
 \label{fig:victor_time_height}
\end{minipage}
\end{centering}
\end{figure}

\section{Night sky background}

The main contributions to the night sky background or NSB have been discussed in section~\ref{sec:NSB_cont}. In this Chapter we want to consider the variation in NSB across the Galatic plane. 

Preu$\beta$ et al \cite{NSB} produced multiple observations of the NSB at an 1.8 \rm{km} altitude for on and off the Galactic plane in s of 2$^{o}$ using a 0.8$^{o}$ diameter pixel. A sky map of the NSB was created and is shown in Figure~\ref{fig:preuss_plot}. The sky map shows multiple regions that have NSB values greater than the off-Galactic plane level that is currently being used. The robustness of the PeX $\gamma$-ray reconstruction needs to be tested against large variations in the NSB. To achieve this, two 8.2$^{\circ}$ by 8.2$^{\circ}$ regions have been taken from Figure~\ref{fig:preuss_plot} to represent regions on the Galactic plane and towards the Galactic centre. The first region, centred at (b, l) = (-4$^{\circ}$,-15$^{\circ}$), is within the Galactic plane. The level of NSB is a factor of two higher than the off-Galactic plane level. The second region, centred at (b, l) = (-4$^{\circ}$,-4$^{\circ}$), is towards the Galactic centre, which has a higher NSB flux by a factor of four compared to the off-Galactic plane region.

	The on-Galactic plane region represents the most common region for NSB since Galactic plane surveys are a key motivation for PeX. We have taken these regions of NSB at a 1.8 \rm{km} altitude and extrapolated these values to a 0.22 \rm{km} altitude, which gives appropriate NSB values for the 0.22 \rm{km} altitude. The NSB range in Figure~\ref{fig:preuss_plot} \cite{NSBthesis} is 2 $\times$ 10$^{12}$ \rm{photons (sr s m$^{2}$)$^{-1}$} to 7 $\times$ 10$^{12}$ \rm{photons (sr s m$^{2}$)$^{-1}$}, where the isolated red squares above 7 $\times$ 10$^{12}$ \rm{photons (sr s m$^{2}$)$^{-1}$} represent stars. These isolated pixels which contain stars are saturated. However, they would provide photon fluxes large enough to trigger the camera upper threshold. Practically, in hardware, this might switch off the pixel due to photon counts being larger than the upper threshold limit. For the purpose of use, the red squares will represent 7.5 $\times$ 10$^{12}$ \rm{photons (sr s m$^{2}$)$^{-1}$}. 

Converting the NSB to a 0.22 \rm{km} altitude site, the scaled value becomes 2 $\times$ 10$^{12}$ \rm{photons (sr s m$^{2}$)$^{-1}$} to 6 $\times$ 10$^{12}$ \rm{photons (sr s m$^{2}$)$^{-1}$}. These regions will be used to test the PeX $\gamma$-ray robustness of the reconstruction and to determine whether improvements are gained by including a time cleaning cut.\\
\\
\begin{figure}[h]
\begin{centering}
\includegraphics[scale=0.9]{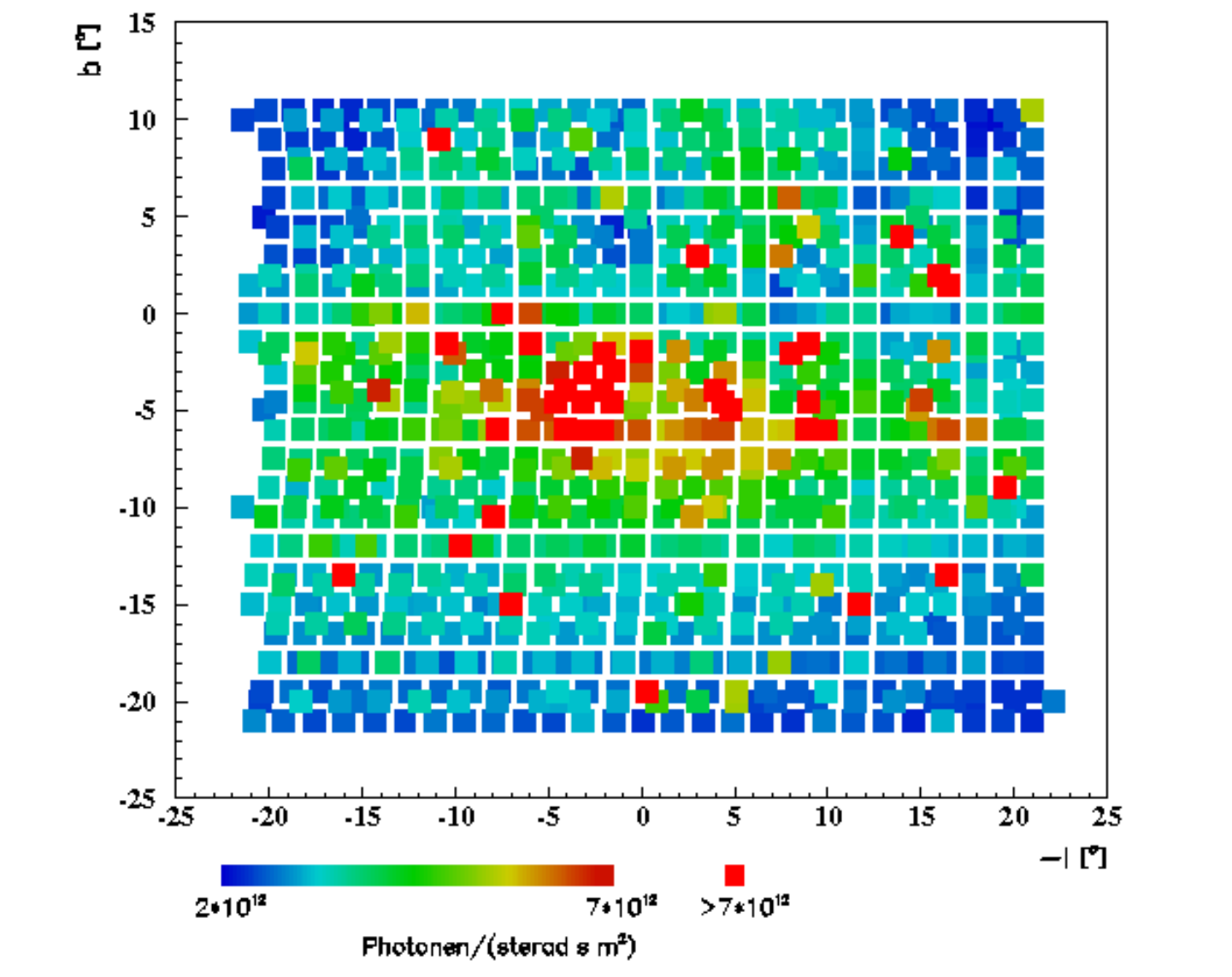}
\captionsetup{width=13cm}  
 \caption{Photon flux measurements for multiple regions in the Galactic plane and towards the Galactic centre (from \cite{NSBthesis}). Each square represents an 0.8$^{\circ}$ field of view with measurements take in 2$^{\circ}$ step sizes. The photon flux is largest towards the Galactic centre and is roughly a factor of four \cite{NSBthesis} higher than off-plane photon fluxes. The two regions used to represent my NSB regions are centred at (b, l) = (-4$^{\circ}$,-15$^{\circ}$) and (-4$^{\circ}$,-4$^{\circ}$) for on-Galactic plane and towards Galactic centre respectively.}
 \label{fig:preuss_plot}
\end{centering}
\end{figure}
\\

\section{Why add a time cut to the cleaning algorithm?}

The standard configuration for PeX has been chosen by using results from Chapter 4 and 5. The next step is to determine if any of the standard parameters can be altered with the addition of time cleaning to provide improvements in performance.

At present, a two level tail cut cleaning algorithm is used to clean images. Section~\ref{sec:cleaning_algorithm} introduced the cleaning algorithm but we summarise the cleaning conditions again here:
\begin{itemize}
\item Pixels that contain more photoelectrons than the \textit{picture} value are kept in the final image regardless of where they are located
\item Pixels that contain more photoelectrons than the \textit{boundary} value but less than the \textit{picture} value are kept if the pixel is adjacent to a picture pixel
\end{itemize}
The \textit{picture} value is 8\textit{pe} and is significantly higher than the current level of NSB noise. The \textit{boundary} value is 4\textit{pe} which is closer to the NSB level and therefore provides a higher chance that an NSB fluctuation can pass the \textit{boundary} value.


If only NSB photoelectrons are detected in the buffer then the peak in that pixel usually comes from the first NSB photoelectrons in the flash analogue to digital converter (FADC) buffer. Why does the peak come from the first NSB photoelectrons in the FADC? This is due to the peak finding algorithm accepting the first peak in the buffer. If there are more than one identical peaks in the buffer, then the first peak is taken as the peak in the FADC buffer. In a 100 \rm{ns} FADC buffer, the total contribution from NSB will be on average 4.5 \textit{pe} in the whole 100 \rm{ns}. These NSB photons are found anywhere within the 100 \rm{ns} FADC buffer. There is a low probability that all 4 NSB \textit{pe} will arrive in the same bin or at least within a 10 \rm{ns} section of the buffer. The 10 \rm{ns} section of the buffer is used since if the first NSB peak is chosen by the peak finding algorithm, it guarantees that all NSB \textit{pe} will be included in the integration of the pixel sum. It will also provide a \textit{pe} count that is equivalent to the \textit{boundary} value.

The probability that a 10 \rm{ns} time period contains \textit{x} number of \textit{pe} is shown in Table~\ref{table:prob}. The probability, P, comes from the Poisson distribution;
\begin{equation}
P(x:\mu) = \frac{\mu^{x}}{x!} \exp{-\mu}
\end{equation}
where $\mu$ is 0.045. This probability is shown for both the off-plane level of NSB and four times the off-plane NSB level, which represents the NSB towards the Galactic Centre \cite{NSB}. The off-plane region is a region pointed away from the Galactic plane.
\\

\begin{table}[h]
\centering
\begin{tabular}{lrr}
\hline
\textit{x} value & off-plane NSB prob. & four $\times$ off-plane NSB prob.\\
\hline
0 & 6.37 $\times$ 10$^{-1}$ & 1.65 $\times$ 10$^{-1}$\\
1 & 2.87 $\times$ 10$^{-1}$ & 2.98 $\times$ 10$^{-1}$\\
2 & 6.46 $\times$ 10$^{-2}$ & 2.68 $\times$ 10$^{-1}$\\
3 & 9.68 $\times$ 10$^{-3}$ & 1.61 $\times$ 10$^{-1}$\\
4 & 1.08 $\times$ 10$^{-3}$ & 7.23 $\times$ 10$^{-2}$\\
5 & 9.81 $\times$ 10$^{-5}$ & 2.60 $\times$ 10$^{-2}$\\
\hline
\end{tabular}
\captionsetup{width=13cm}
\caption{The probability that \textit{x} number of \textit{pe} arrive in a 10 \rm{ns} section of the 100 \rm{ns} FADC buffer. Two NSB levels are presented, an off-plane NSB level of 0.045\rm{\textit{pe} (ns pixel)$^{-1}$} and towards the Galactic Centre which has 4 $\times$ higher NSB.}
 \label{table:prob}
\end{table}

For the optimum cleaning combination, (8\textit{pe}, 4\textit{pe}), the probability that a pixel containing pure NSB will surpass the \textit{boundary} value is low, $\approx$ 10$^{-3}$. However, if the pixel contains half NSB and half Cherenkov signal from the tail of the shower, the pixel could pass the \textit{boundary} value. In this case the contribution from NSB is 2\textit{pe} which has a probability of 6.46 $\times$ 10$^{-2}$. Even a single \textit{pe} from the tail of the shower combined with NSB could surpass the \textit{boundary} value. If multiple pixels with more NSB than Cherenkov signal are included as boundary pixels in an image, it could worsen the ability to separate $\gamma$-ray and proton events.

The pixels from the tail of the shower could influence the lever arm for the major axis, so NSB influences on this tail may increase errors in the reconstruction as discussed in section~\ref{sec:cleaning_op}.


	Previous studies \cite{Heb} have shown that TeV $\gamma$-ray and proton events have small time evolution across the image. M. He$\beta$ et al \cite{Heb} stated that the key feature for proton showers is an almost linear relationship between time gradient along the major axis and the core distance, while the $\gamma$-rays only establish this feature for core distances $>$ 100 \rm{m}. The time gradient steepens with core distance for proton events to more than 5 \rm{ns/deg} \cite{Heb}. This indicates that, at a large core distance, the time difference between adjacent pixels would be small, approximately a few \rm{ns} and even less for $\gamma$-ray showers which are more compact. Thus the arrival time of Cherenkov photons in a shower would be similar for adjacent pixels, within a few \rm{ns}.

\begin{figure}[h]
\begin{centering}
\includegraphics[scale=0.7]{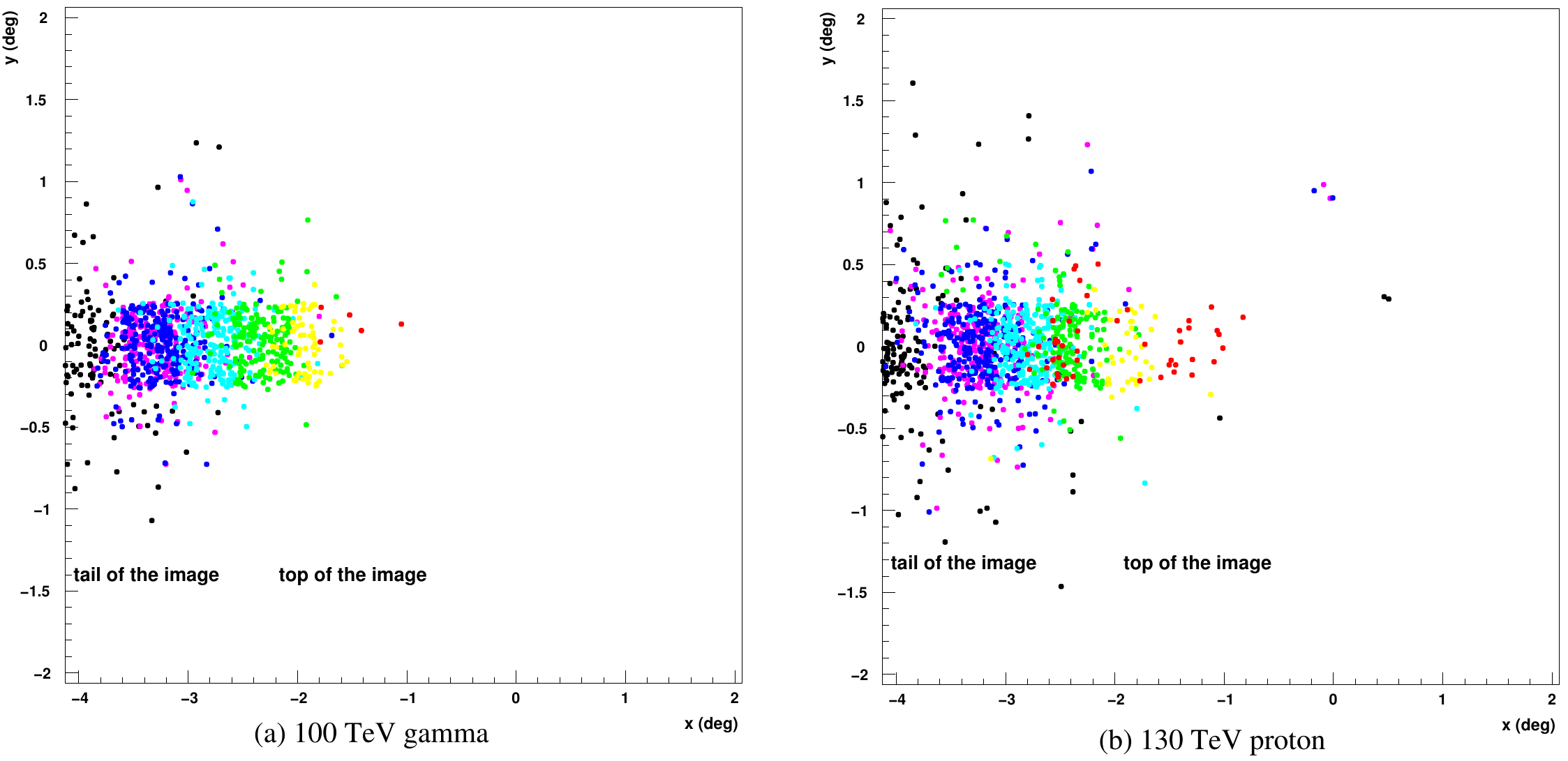}
\captionsetup{width=13cm}  
 \caption{Arrival times of detected Cherenkov photons in the x and y angular positions which represent the position in the camera. The colours represent the arrival times of the Cherenkov photons. The `top of the image' represents the photons that are close to the centre of the camera and arrive earliest. The `tail of the image' represents photons which are close to the edge of the camera and have large angles in the x axis. These photons arrive latest. Colour code is the same as in Figure~\ref{fig:victor_time_height}. Taken from \cite{Stamatescu}.}
 \label{fig:victor_time_development}
\end{centering}
\end{figure}

	Stamatescu \cite{Stamatescu}, showed an example of the shower time development for a 100 TeV $\gamma$-ray and a 130 TeV proton shower at a 400 \rm{m} core distance. Figure~\ref{fig:victor_time_development} shows the time development of a $\gamma$-ray and proton shower in the camera. The figure has been taken from \cite{Stamatescu}. The time increases along the major axis, along the x-axis in this case. Therefore, the time difference between adjacent pixels would be 2 to 4 \rm{ns} while the whole image would span roughly 25 \rm{ns}. The time difference between the adjacent pixels aligned along the y-axis of the camera, perpendicular to the major axis in this case, range from 0 to 2 \rm{ns}. From this information, a small time cut of the order of a few ns applied to adjacent pixels could be added to the cleaning algorithm in an effort to reduce the number of NSB dominated pixels.

	Figure~\ref{fig:timing_1} shows the pixel arrival times for the Cherenkov signal for a 78 TeV $\gamma$-ray shower and the NSB. A systematic time progression can be seen for the main Cherenkov signal, which agrees with the results from \cite{Heb} and \cite{Stamatescu}. The NSB pixels away from the Cherenkov image clearly have random arrival times. 

\begin{figure}
\begin{centering}
\includegraphics[width=\textwidth]{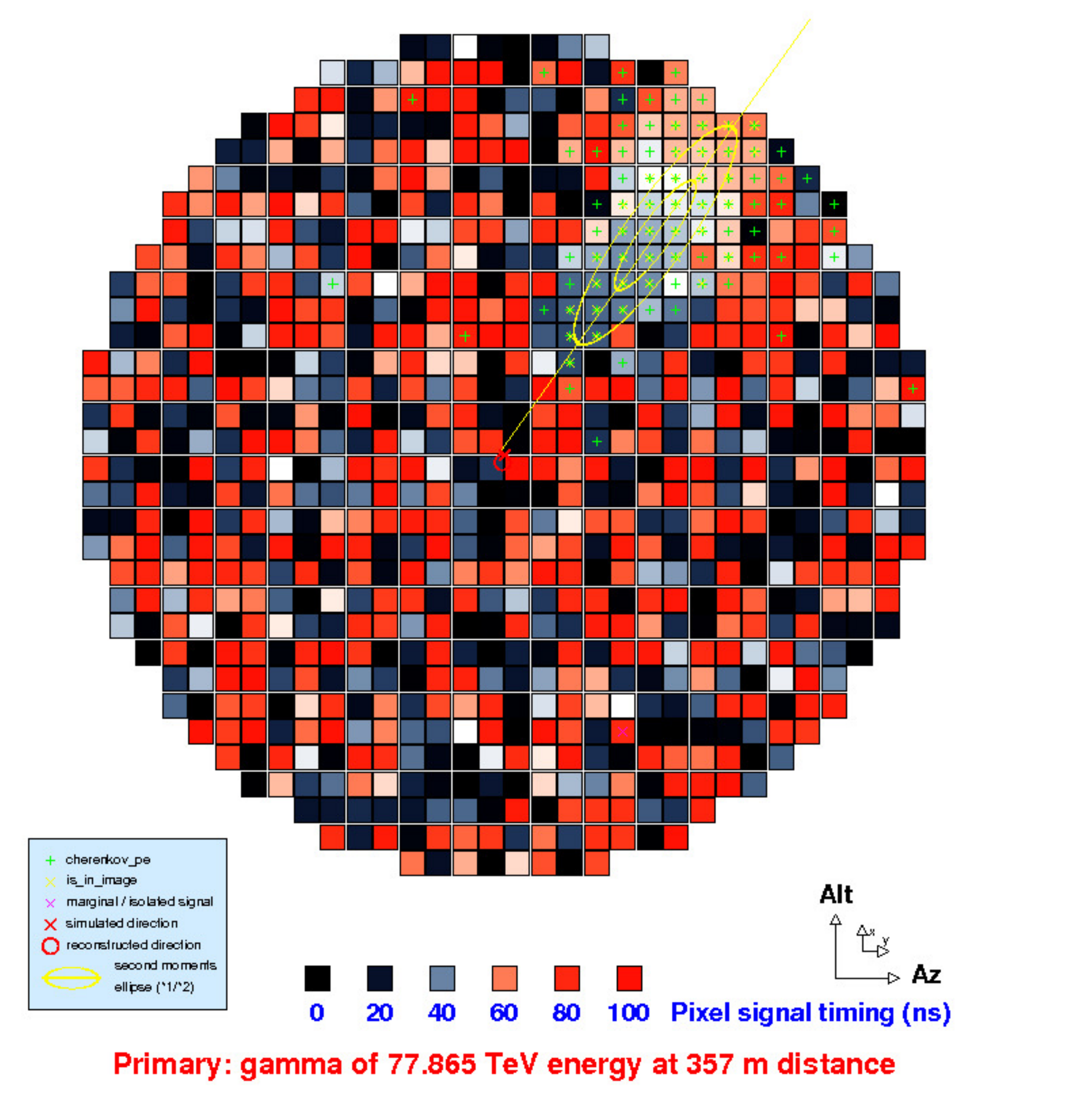}
\captionsetup{width=13cm}  
\caption{A single image from a 78 TeV $\gamma$-ray event recorded at a core distance of 357 \rm{m} with the standard level of NSB. The colour scale represents the arrival time for each pixel in the camera. The green pluses represent pixels which contain Cherenkov \textit{pes} from the shower. The major axis is represented by the yellow line. There are also pixels with green crosses that have arrival times that differ by more than 20 \rm{ns} compared to the main image. These pixels are undesired, since they might contain more NSB than Cherenkov signal. The other pixels which contain no Cherenkov signal have random arrival times and will be removed by a tight time cleaning cut.}
 \label{fig:timing_1}
\end{centering}
\end{figure}

To take advantage of the random NSB timing, compared to the compact timing profile of Cherenkov light (Figure~\ref{fig:victor_time_height} and~\ref{fig:victor_time_development}), an extra cleaning condition based on pixel timing was considered. The condition is that the time difference between adjacent picture and boundary pixels must be small ($\approx$ 10 \rm{ns} or less), but we will look at this in detail below.


\section{Time difference between core and boundary pixels in PeX}
 \label{sec:time_core_boundary}

	To find the time difference between picture and boundary pixels, the arrival time in the pixel, with respect to the start of the FADC buffer is used. The Cherenkov signal provides a time gradient along the major axis of the image and the slope of the time gradient depends on the core distance of the shower. For pixels which contain Cherenkov signal, the arrival times should be within a few ns of each other. To calculate the time difference, $\Delta t_{arrival}$, we take the time of the boundary pixel and compare it to the time of any adjacent picture pixel,
\\
\begin{eqnarray}
\Delta t_{arrival} = t_{boundary} - t_{picture}
\end{eqnarray}
where t$_{boundary}$ is the arrival time of a boundary pixel and t$_{picture}$ is the arrival time of the adjacent picture pixel. Both times are with respect to the start of the FADC buffer.  If the boundary pixel is adjacent to multiple picture pixels then the time difference is calculated for all combinations of boundary and picture pixels pairs. We consider all combinations of picture and boundary pixel pairs since we do not include the position or orientation of the pixels with respect to the major axis of the image. 

A negative $\Delta$t$_{arrival}$ indicates that t$_{picture}$ $>$ t$_{boundary}$, so the boundary pixel has an arrival time earlier than the picture pixel, although there are not as many instances where this occurs (Figure~\ref{fig:core-boundary_pe}). With respect to the time the extrapolated primary particle would reach ground level, the photons produced at the top of the shower arrive in the camera earlier than the photons produced at shower maximum. However, a majority of pixels have similar arrival times at the strongest part of the shower. 

A positive $\Delta$t$_{arrival}$ indicates that t$_{picture}$ $<$ t$_{boundary}$, so the picture pixel has an arrival time earlier than the boundary pixel. When plotting the time differences (Figure~\ref{fig:core-boundary_pe}), there are slightly more positive arrival time differences. 


	The best way to find the appropriate time cut is to consider the distribution of time differences between picture and boundary pixels for showers at different core distances. For this investigation, a (6\textit{pe}, 3) triggering combination and (8\textit{pe}, 4\textit{pe}) cleaning combination are used. To see the arrival time differences between picture and boundary pixels, simulations are performed with no NSB or electronic noise, so that the arrival time difference is associated only with the Cherenkov signal. The top panel in Figure~\ref{fig:core-boundary_pe} shows the arrival time difference between only picture and boundary pixels with varying core distance for no NSB. As the core distance increases, the arrival time difference between pixels increases. That is, the 100 \rm{m} core distance shower produces the tightest distribution, while the 500 \rm{m} showers produce the broadest distribution. The difference between the 100 \rm{m} and 500 \rm{m} distributions is small within a few ns, which indicates that the arrival times between pixels increases slightly with increasing core distance. A single picture-boundary arrival time cut can therefore be used for showers at all core distances. A time cut on the arrival time difference of roughly $\pm$5 \rm{ns} appears sufficient to contain most picture-boundary time differences.

\begin{figure}
\begin{centering}
\includegraphics[scale=0.8]{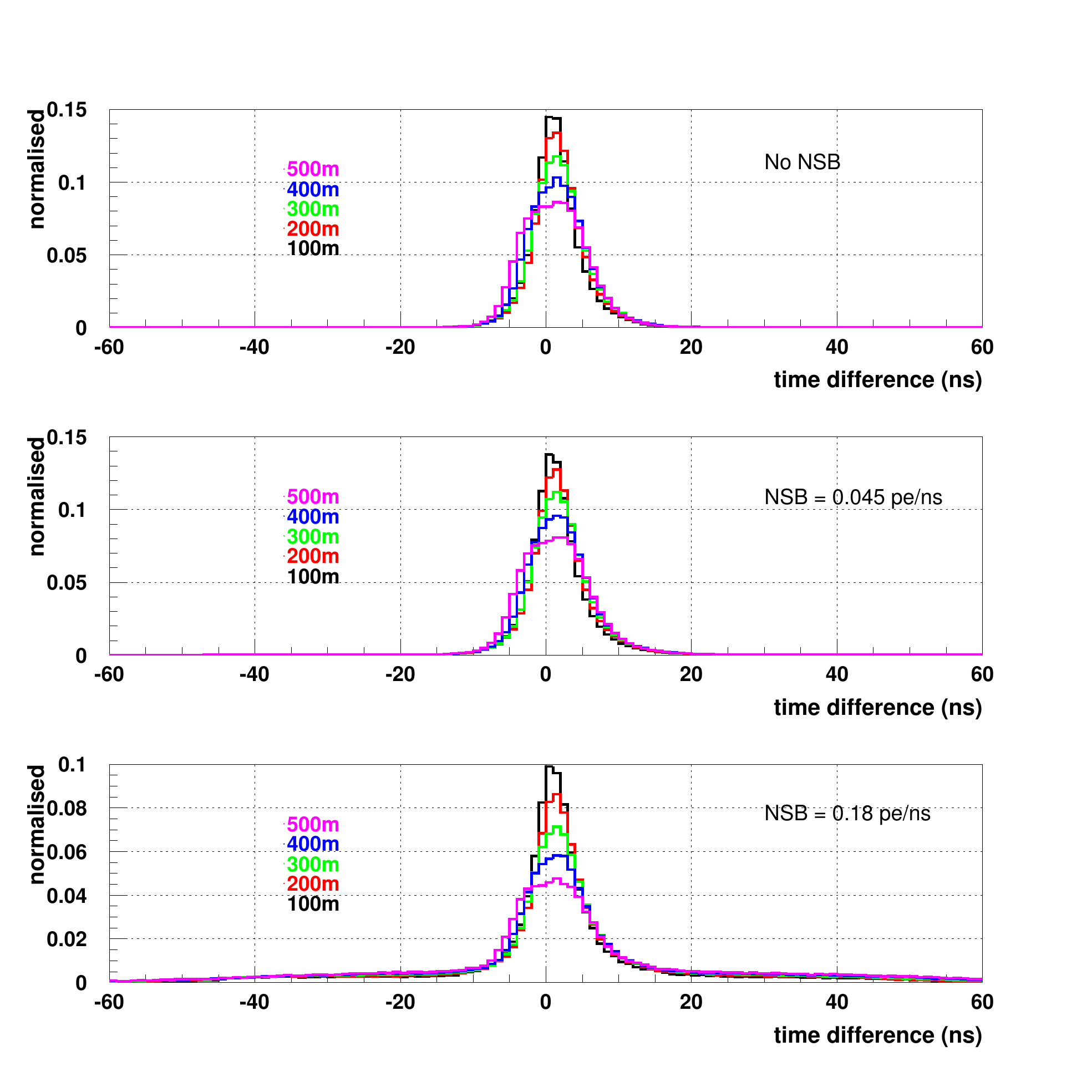}
\captionsetup{width=13cm}  
 \caption{The distribution of picture-boundary arrival time differences using a (6\textit{pe}, 3) triggering combination and (8\textit{pe}, 4\textit{pe}) cleaning combination. The distributions are shown for 5 different core distances ranging from 100 \rm{m} to 500 \rm{m}. The distributions are for all energies. Electronic noise is present in all.}
 \label{fig:core-boundary_pe}
\end{centering}
\end{figure}

As one adds NSB to the simulation, somewhat similar distributions are found. The middle panel in Figure~\ref{fig:core-boundary_pe} represents the arrival time differences between picture and boundary pixels with an off-Galactic Plane NSB. In the distributions for all core distance showers, the peak in the normalised distributions has decreased, due to the slight widening of the distributions. A combination of Cherenkov signal and NSB implies that fewer NSB \textit{pe} are required for the pixel to pass the \textit{boundary} value. If the pixel contains more NSB than Cherenkov signal, the NSB will dominate the arrival time. Since the NSB arrival time is random, the pixel time will not correspond to the time of the signal.

	Now we can consider what will happen if the NSB level was similar to that in a region towards the Galactic Centre. To test the arrival time difference for a Galactic Plane level of NSB, the NSB value was increased to 0.18 \rm{\textit{pe}/ns} over the entire field of view or 4$\times$ the off-plane value. From Figure~\ref{fig:preuss_plot}, there is no region of sky which has a constant 0.18 \rm{\textit{pe}/ns} NSB flux.

The bottom panel in Figure~\ref{fig:core-boundary_pe} represents the arrival time distributions with this Galactic Plane level of NSB (0.18 \rm{\textit{pe}/ns}). The distributions are wider than the previous distributions and the peak is smaller. The level of NSB is higher and the probability that the NSB can surpass the \textit{boundary} value increases significantly, see Table~\ref{table:prob}. The pixels around the true image could be pure NSB. If these pixels can be removed, then the reconstruction should improve. The widening of the distribution indicates that the arrival time of the boundary pixel has become more unstable with increasing NSB. A tight time cut around the no NSB distribution (Figure~\ref{fig:core-boundary_pe} top panel) should remove pixels heavily influenced NSB. The distribution shows long tails on both sides.



\section{Addition of a core-boundary arrival time cut}
 \label{sec:timing_cut}

	The optimised PeX configuration with and without timing cleaning will be compared for varying NSB fluxes. Two regions that are used to represent my high NSB levels come from Figure~\ref{fig:preuss_plot}, since they best represent varying NSB and provides a more realistic scenario. The three NSB levels are: 
\begin{itemize}
\item off-Galactic plane NSB $\approx$ 0.045 \rm{pe (ns pixel)$^{-1}$}
\item on-Galactic plane NSB that comes from (b, l) = (-4$^{\circ}$,-15$^{\circ}$) in Figure~\ref{fig:preuss_plot}
\item Galactic Centre NSB that comes from (b, l) = (-4$^{\circ}$,-4$^{\circ}$) in Figure~\ref{fig:preuss_plot}
\end{itemize}


	Figure~\ref{fig:plots_timing_nsb} top panels show the pre- and post-selection cut effective area for $\gamma$-ray events with (red) and without (black) time cleaning for off-Galactic plane NSB. No variation between the curves is noticed for the pre-cut effective areas. A similar trend is seen for the post-selection cut effective area which suggests that the same number of events are left. Therefore, the time cleaning cut does not affect the reconstruction of the events. This occurs since the off-Galactic plane NSB flux is low and the cleaning without a time cut still works effectively. 

\begin{figure}
\begin{centering}
\includegraphics[scale=0.8]{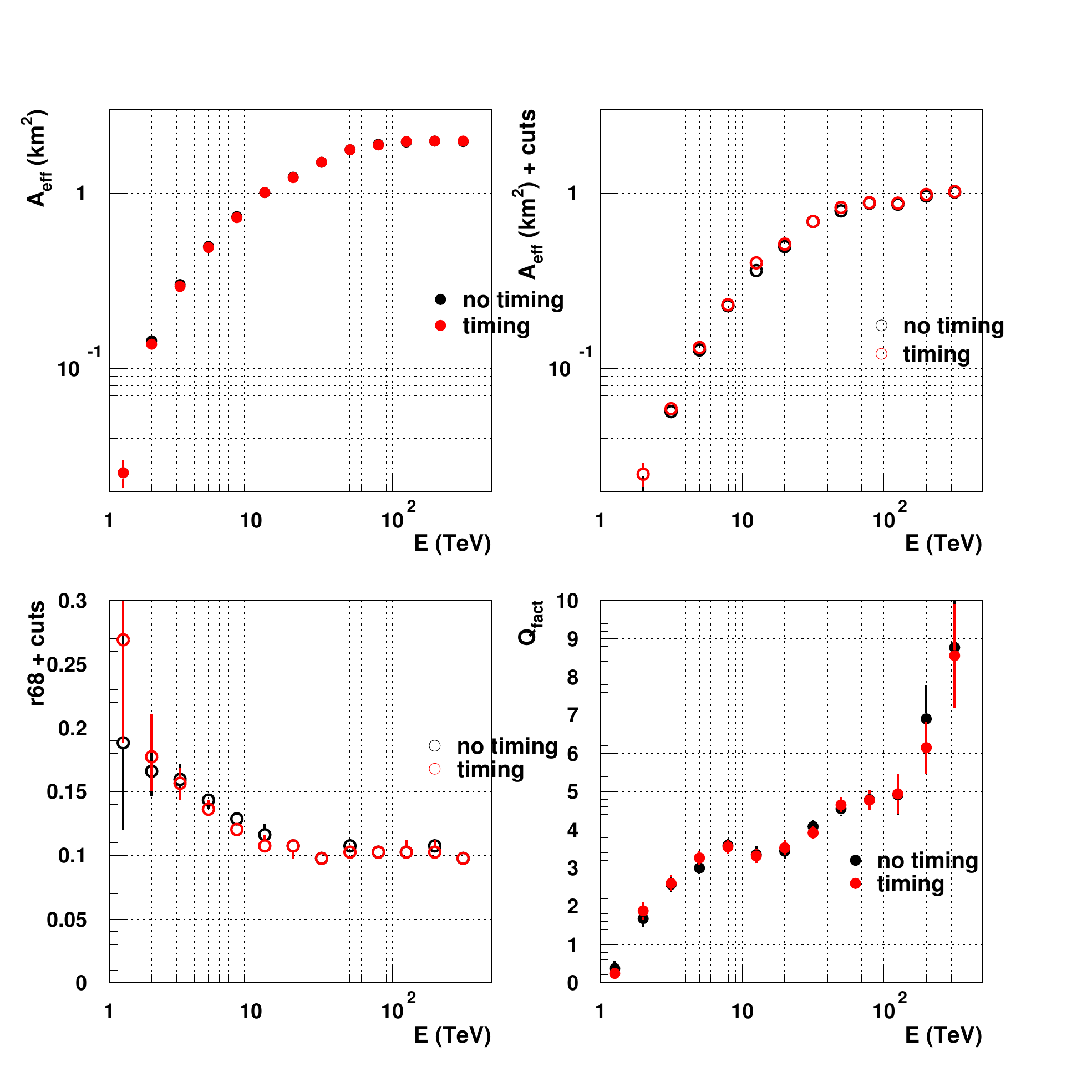}
\captionsetup{width=13cm}  
 \caption{Results with (red) and without (black) time cleaning cut for Algorithm 1 using off-Galactic plane NSB. The top left shows the effective area, the top right shows the post-selection cut effective area, the bottom left provides the post-shape cut angular resolution (r68) and the bottom right gives the Q$_{fact}$. There appears to be no difference between results. All plots use the standard (8\textit{pe}, 4\textit{pe}) cleaning combination.}

 \label{fig:plots_timing_nsb}
\end{centering}
\end{figure}


	Figure~\ref{fig:plots_timing_nsb} bottom panels show the post-shape cut angular resolution (r68) and Q-factor (Q$_{fact}$) for a standard PeX configuration with (red) and without (black) the time cleaning cut. The cleaning combination, (8\textit{pe}, 4\textit{pe}), is strong enough that the off-Galactic plane NSB does not affect parameterisation. The results indicate that the chosen time cut allows images to retain the vital information used to distinguish between $\gamma$-ray and proton events. The addition of the time cleaning cut does not appear to alter the performance.

The addition of a picture-boundary time cut into the cleaning algorithm has minimal change to the performance of a standard PeX configuration for an off plane NSB flux, where the standard configuration is 500 \rm{m} telescope separation, with a (6\textit{pe}, 3) triggering combination, an (8\textit{pe}, 4\textit{pe}) cleaning combination and a 60\textit{pe} image size cut for a 0.22 \rm{km} altitude site.\\
	

	
\begin{figure}
\begin{centering}
\includegraphics[scale=0.8]{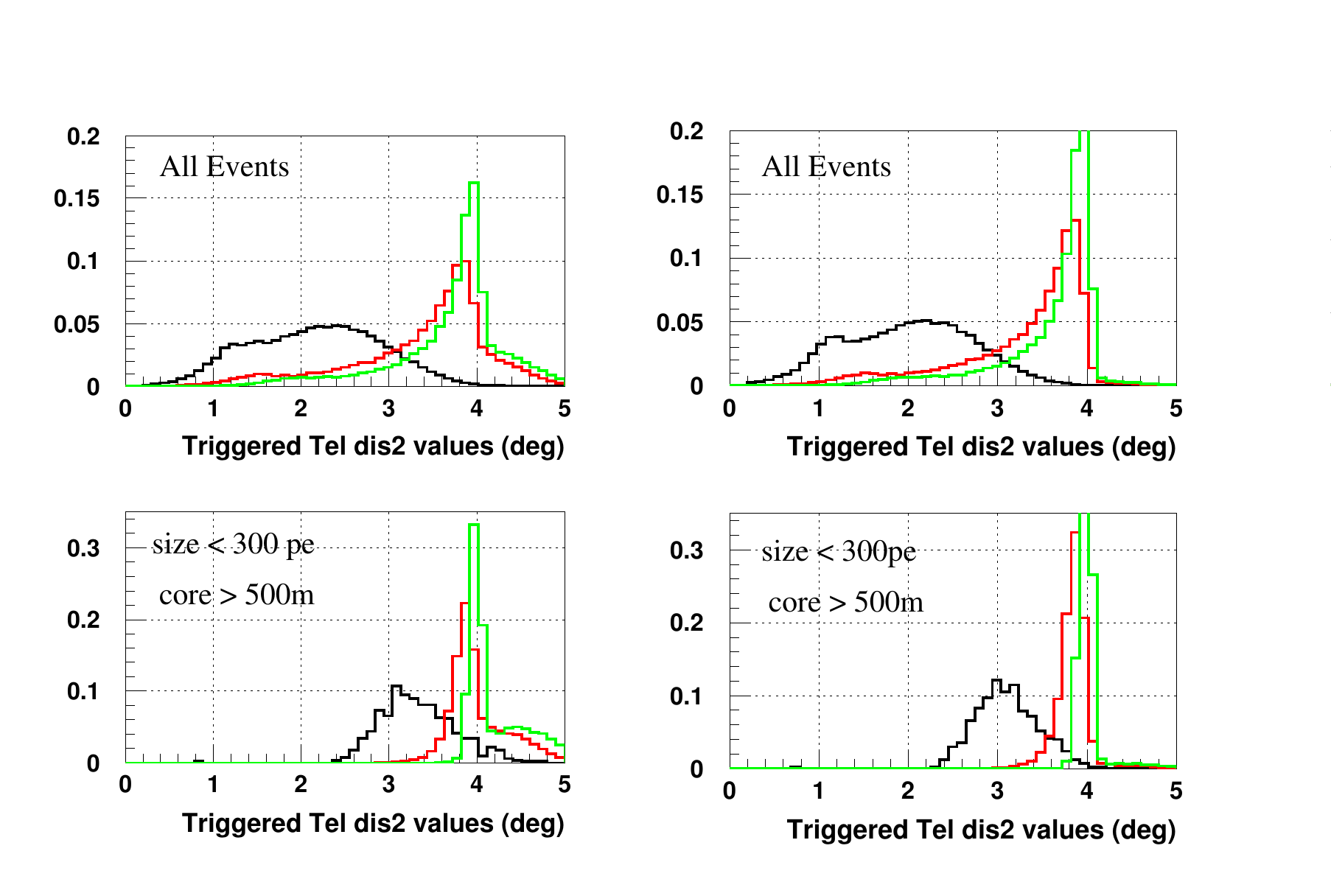}
\captionsetup{width=13cm} 
 \caption{The various \textit{dis2} values for events without time cleaning (left) and with time cleaning (right) for a Galactic Centre level of NSB using the optimised PeX cell. The events have been split into three different energy bands using Algorithm 1: 1 to 10 TeV (black), 10 to 100 TeV (red) and 100 to 500 TeV (green). The top panels show all events and the bottom panels show events with an image \textit{size} of 300\textit{pe} or less and a core distance greater than 500 \rm{m}. The edge of the physical camera is 4.1$^{\circ}$.}
 \label{fig:dis2_plots_nsbvar_comp}
\end{centering}
\end{figure}	
	

	The next step is to investigate the time cleaning cut with a higher level of NSB. To test this, a region toward the Galactic Centre was used. Instead of using a constant 0.18 \textit{pe}/ns over the entire field of view, a more realistic NSB region at (b, l) = (-4$^{\circ}$,-4$^{\circ}$) was taken from (Figure~\ref{fig:preuss_plot}).  \\

	Figure~\ref{fig:plots_timing_nsbvar} top panels show the pre- and post-selection cut effective area with (red) and without (black) the time cleaning cut for the standard PeX configuration with Galactic Centre NSB. Minimal difference is seen between the pre-cut effective areas at the increased NSB level. The black curve appears to contain more events for E $<$ 10 TeV and fewer events for E $>$ 100 TeV compared to the red curve. With an increased level of NSB, more NSB appears to pass the cleaning thresholds which helps more events pass the 60\textit{pe} stereo cut. 

As the level of NSB increases, the \textit{dis2} value appears to cut out more images and events for E $>$ 20 TeV (Figure~\ref{fig:plots_timing_nsbvar} top right). To best represent this, the \textit{dis2} values have been displayed in Figure~\ref{fig:dis2_plots_nsbvar_comp} for Galactic plane NSB level with no time cut (left panels). The results have been split into 3 energy bands: 1 to 10 TeV (black), 10 to 100 TeV (red) and 100 to 500 TeV (green). The results show that with an increased NSB level, more images have a \textit{dis2} value larger than 4.1 which is the physical size of the camera when no time cut is applied. This is due to the extra NSB passing the cleaning threshold which shifts the C.O.G. of the image. A time cleaning cut improves the image parameterisation. The effect is similar to the under-cleaning seen from a (5\textit{pe}, 2\textit{pe}) combination in Chapter 4.


\begin{figure}
\begin{centering}
\includegraphics[scale=0.8]{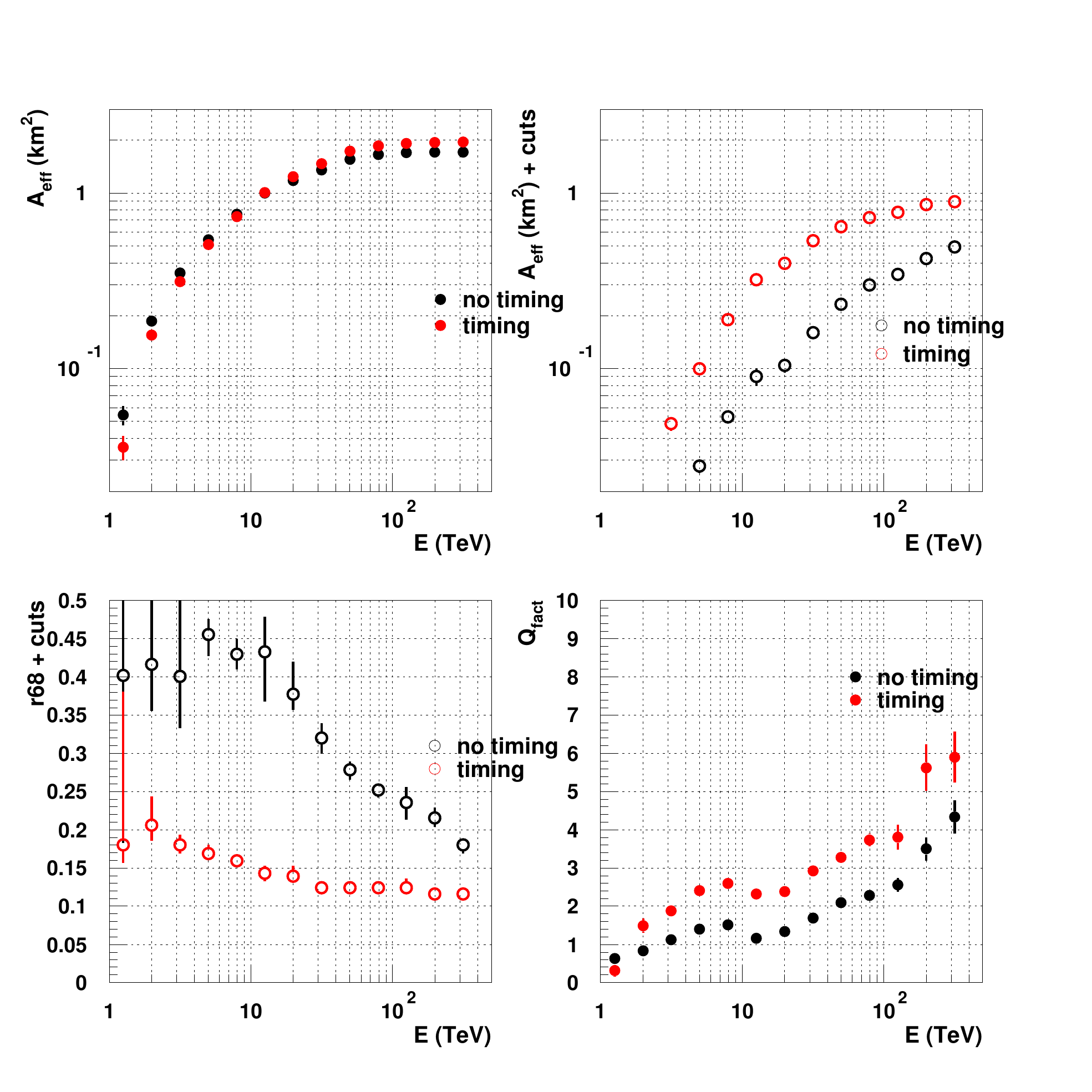}
\captionsetup{width=13cm}  
 \caption{Results with (red) and without (black) time cleaning cut for Algorithm 1 using a Galactic Centre NSB. The top left shows the effective area, the top right shows the post-selection cut effective area, the bottom left provides the post-shape cut angular resolution (r68) and the bottom right gives the Q$_{fact}$. There appears to be a big improvement to the results when a time cut is applied. All plots use the standard (8\textit{pe}, 4\textit{pe}) cleaning combination.}

 \label{fig:plots_timing_nsbvar}
\end{centering}
\end{figure}

	The pre-cut effective area for a Galactic Centre NSB level with no time cleaning shows a slightly lower area for E $>$ 100 TeV. The post-selection cut effective area with and without time cleaning shows a significant improvement for the time cleaning results. By including the extra time cut, the parameterisation and reconstruction for the events have considerably improved since a majority of the Galactic Centre NSB has been removed. The time cut therefore clearly helps remove the random NSB from the images.


	 Figure~\ref{fig:plots_timing_nsbvar} bottom left panel shows the post-shape cut angular resolution (r68) with (red) and without (black) time cleaning. The black curve shows poorly reconstructed events due to the Galactic Centre level of NSB affecting the true Cherenkov signal. The extra NSB can affect the parameterisation and major axis for images. An incorrect major axis will provide the wrong reconstructed direction, which for the standard cleaning algorithm gives an angular resolution value of approximately 0.4$^{\circ}$ for E $<$ 20 TeV. Above 20 TeV, the general size of the image increases, so including a few extra pixels around the edge of the image appears to have less of an effect. As the energy increases, the post-shape cut angular resolution without time cleaning does improve (Figure~\ref{fig:plots_timing_nsbvar} black curve).

For post-shape cut angular resolution with a time cleaning cut, the results have also strongly improved. With the time cleaning cut, the average angular resolution is $\sim$ 0.14$^{\circ}$. These results further indicate that a time cleaning cut appears to make the cleaning algorithm and event reconstruction more robust to different levels of NSB.

Figure~\ref{fig:plots_timing_nsbvar} bottom right panel shows the Q$_{fact}$. The time cleaning cut (red) produces a significant improvement over the no time cut cleaning (black). The results suggest that the separation between $\gamma$-ray and proton events is difficult when a high level of NSB is present. The Q$_{fact}$ with no time cleaning cut is roughly a factor of two worse than the time cut results. The larger error bars in the red curve indicate that the number of proton events remaining after cuts is smaller. This tends to indicate that a large fraction of proton events are still being rejected.

Figure~\ref{fig:nsbvar2_8_4} illustrates how the time cleaning cut works in the presence of Galactic plane level of NSB. The Galactic plane NSB level is appropriate for most observations while the Galactic Centre NSB level is an extreme case (Figure~\ref{fig:preuss_plot}). Even on-Galactic plane NSB affects the reconstruction algorithm without time cleaning, though the degree at which the reconstruction is affected is not as significant. The time cleaning results provide improvements mainly for the low energy or large core distance events, since doubling the off-plane level of NSB is enough to disrupt the small sized images. Therefore, a time cleaning cut provides considerable improvement in performance for a variety of NSB levels.

\begin{figure}
\begin{centering}
\includegraphics[width=15cm]{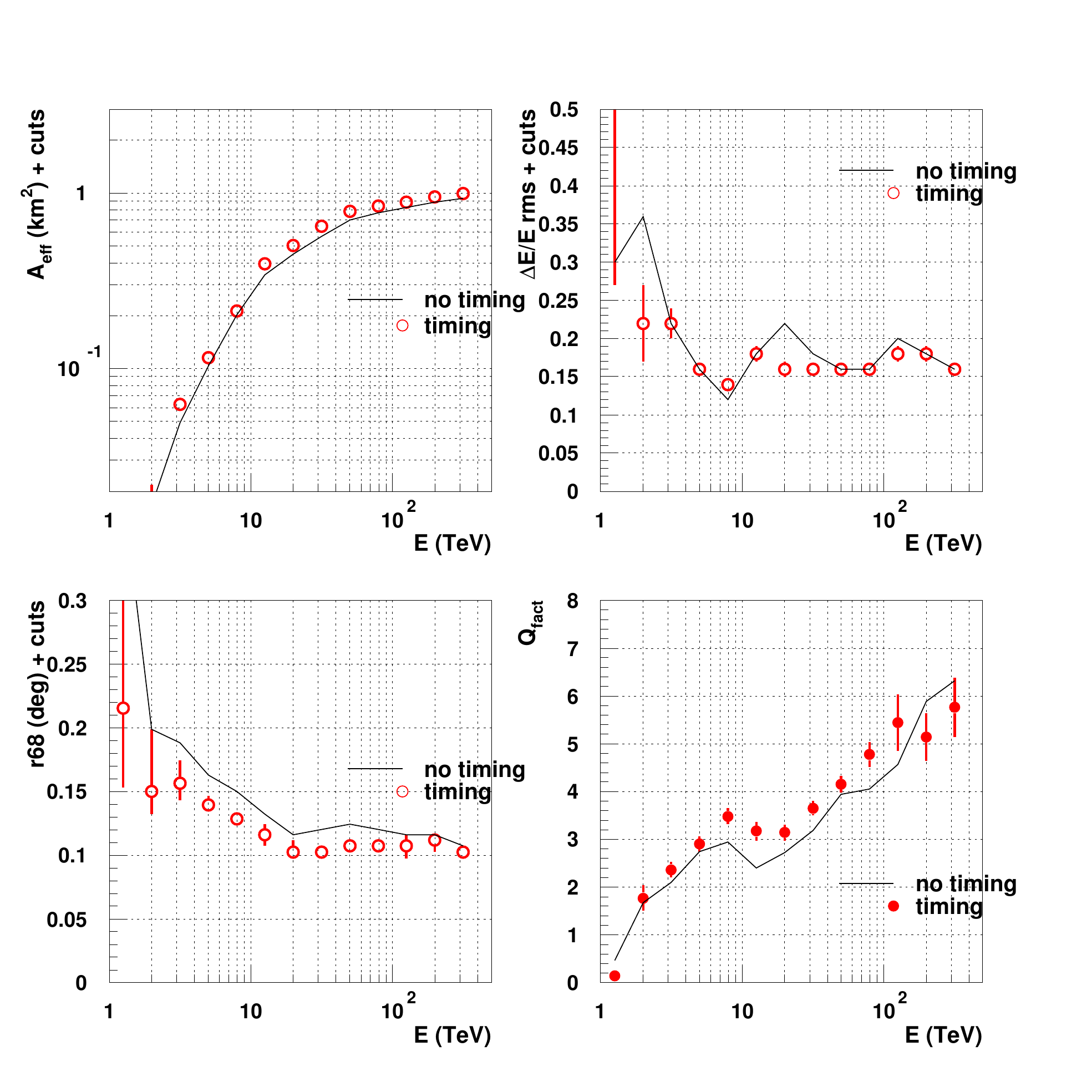}
\captionsetup{width=13cm} 
 \caption{Results with and without time cleaning cut for an (8\textit{pe}, 4\textit{pe}) cleaning combination and an on-Galactic plane level of NSB. The top left shows the post-selection cut effective area, the top right represents the energy resolution, the bottom left provides the angular resolution (r68) and the bottom right gives the Q$_{fact}$. A quick glance at the plots indicates that applying time cleaning provides an improvement in reconstruction for E $<$ 50 TeV in the presence an on-Galactic plane level of NSB. The error bars on the black line results are of a similar magnitude to the red line results.}
 \label{fig:nsbvar2_8_4}
\end{centering}
\end{figure}

\section{Concluding remarks on time cleaning}

	The inclusion of a time cleaning cut of $\pm$ 5 \rm{ns} appears to make the cleaning algorithm more robust towards variations in the NSB. This time-based cleaning can be added permanently to the analysis since results indicate that the time cut does not alter the event reconstruction of events (Figure~\ref{fig:plots_timing_nsb}). The improvement at a Galactic plane level of NSB is, on average, a factor of 1.5 times for effective area, angular resolution and Q$_{fact}$ compared to using a cleaning algorithm with no timing cut (Figure~\ref{fig:plots_timing_nsbvar}).

To make a comparison, we can consider a (10\textit{pe}, 5\textit{pe}) cleaning combination (Figure~\ref{fig:norm_10_5} in Appendix~\ref{sec:appendix_plot}). Applying this slightly higher cleaning combination, (10\textit{pe}, 5\textit{pe}), does help retain the same level of angular resolution, Q$_{fact}$ and effective area with and without time cleaning for an off-Galactic plane level of NSB. Altering the cleaning thresholds based on the NSB level within data may work but it can be tedious and time consuming. Our preferred analysis would use a constant cleaning algorithm instead of switching and being dependent on the level of NSB.

With the addition of the time cleaning cut, the Algorithm 1 reconstruction method has shown significant improvement when a high level of NSB is present. With this improvement to Algorithm 1, it is best to compare this reconstruction algorithm to the improved reconstruction algorithm, Algorithm 3, which has been investigated for the PeX cell. The next chapter will combine all aspects of the previous chapters but using Algorithm 3. The chapter will show optimisation for Algorithm 3, the difference between a 0.22 \rm{km} and 1.8 \rm{km} altitude observational site, with and without the time cleaning cut and then a final comparison between both algorithms to indicate the reconstruction method that provides the optimum reconstruction for PeX.

\chapter{Advanced Reconstruction (Algorithm 3): Application to PeX}

In Chapters 3, 4 and 5, the PeX 5 telescope performance has been investigated using Algorithm 1 (section~\ref{sec:event_reconstruction}) as the standard shower reconstruction method. Previous studies \cite{Stamatescu} conducted work on applying a new direction reconstruction algorithm for PeX. This approach is based on Algorithm 3 (section~\ref{sec:event_reconstruction}), \cite{hofmann}, which utilises the ratio of image \textit{width}/\textit{length} to estimate the distance from the C.O.G to the shower direction. Stamatescu et al \cite{Stamatescu} made improvements to Algorithm 3 based on the arrival time gradient of the image to help determine the distance from the C.O.G to the shower direction. This new algorithm has been shown to improve performance over the standard Algorithm 1 reconstruction especially at large core distances $>$ 300 \rm{m}. Algorithm 3 should therefore be used with PeX to further optimise performance. However, Algorithm 3 requires 3 $\times$ simulation runs to generate look-up tables, which was not required for the investigation with Algorithm 1 in Chapters 3, 4 and 5. For this reason, we chose to optimise the PeX configuration with Algorithm 1. Nevertheless, it is necessary to check that the optimal PeX configuration using Algorithm 1 is also optimal when using Algorithm 3. To shorten the simulation time, a more limited range of analysis parameters will be considered. This range of parameter values includes some of the extreme cases in the optimisations performed in Chapter 4 and 5. In this way we can check that the same trends that appeared in Algorithm 1 appear in Algorithm 3 results. 



In this chapter, the telescope separation, triggering combinations, cleaning combinations and image \textit{size} cut from Chapter 4 will be optimised individually for the new version of Algorithm 3. Then the results using Algorithm 3 will be compared with and without a time cleaning cut and NSB variations. This will show whether the improvement gained from time cleaning is present with Algorithm 3. Finally a comparison will be done between the best results for Algorithm 1 and the best results for Algorithm 3.

\section{Algorithm 3 application to PeX}
 \label{sec:alg3_op}

As a completion to the study of the PeX cell optimisation conducted in Chapter 4, the performance of Algorithm 3 needs to be investigated under varying parameters. Algorithm 3 utilises the Hillas parameterisation as a guide, while Algorithm 1 uses the Hillas parameterisation as the main reconstruction mechanism. Algorithm 3 reconstructs the shower in a different process. The time gradient and reconstructed light maximum (section~\ref{sec:event_reconstruction}) are used to provide a predicted angular distance to the shower arrival direction with respect to the image C.O.G. The predicted distance is calculated for all images and the final reconstructed direction comes from the combination of predicted distances weighted by uncertainties. The triggering combination, cleaning combination and image \textit{size} cut affect the Hillas parameterisation of each image. Since the Hillas parameterisation is the same for both algorithms, the Algorithm 3 results should in principle provide similar trends to Algorithm 1 with varying triggering, cleaning and image \textit{size} values. To confirm the trends with Algorithm 3, the same parameter variations have been investigated. \\

The standard PeX configuration for Algorithm 1 consists of a 500 \rm{m} telescope separation, a (6\textit{pe}, 3) triggering combination, a (8\textit{pe}, 4\textit{pe}) cleaning combination and a 60\textit{pe} image \textit{size} cut. This PeX configuration will also be considered as standard for Algorithm 3.

\subsubsection{Telescope Separation}

Figure~\ref{fig:telescope_sep_alg3} show the results from varying the telescope separation. The post-selection cut effective area indicates that as the separation increases so does the effective area at high energies (Figure~\ref{fig:telescope_sep_alg3} top panel). The smaller separation provides a larger effective area for E $<$ 50 TeV, while the larger separation provides a large effective area above 50 TeV. For low energies, the telescopes are required to be closer together so that the events can trigger the cell. For high energies, the improvement gained in effective area is due to the larger physical area of the cell. Since the telescopes are further apart, they detect events that are further from the cell and collect more of the image from events which could be truncated by a 300 \rm{m} separation. Therefore, the 500 \rm{m} separation provides a larger effective area. However, by using a larger separation the low energy events fail to trigger multiple telescopes and do not pass stereoscopic cuts.

\begin{figure}
\begin{centering}
\includegraphics[scale=0.65]{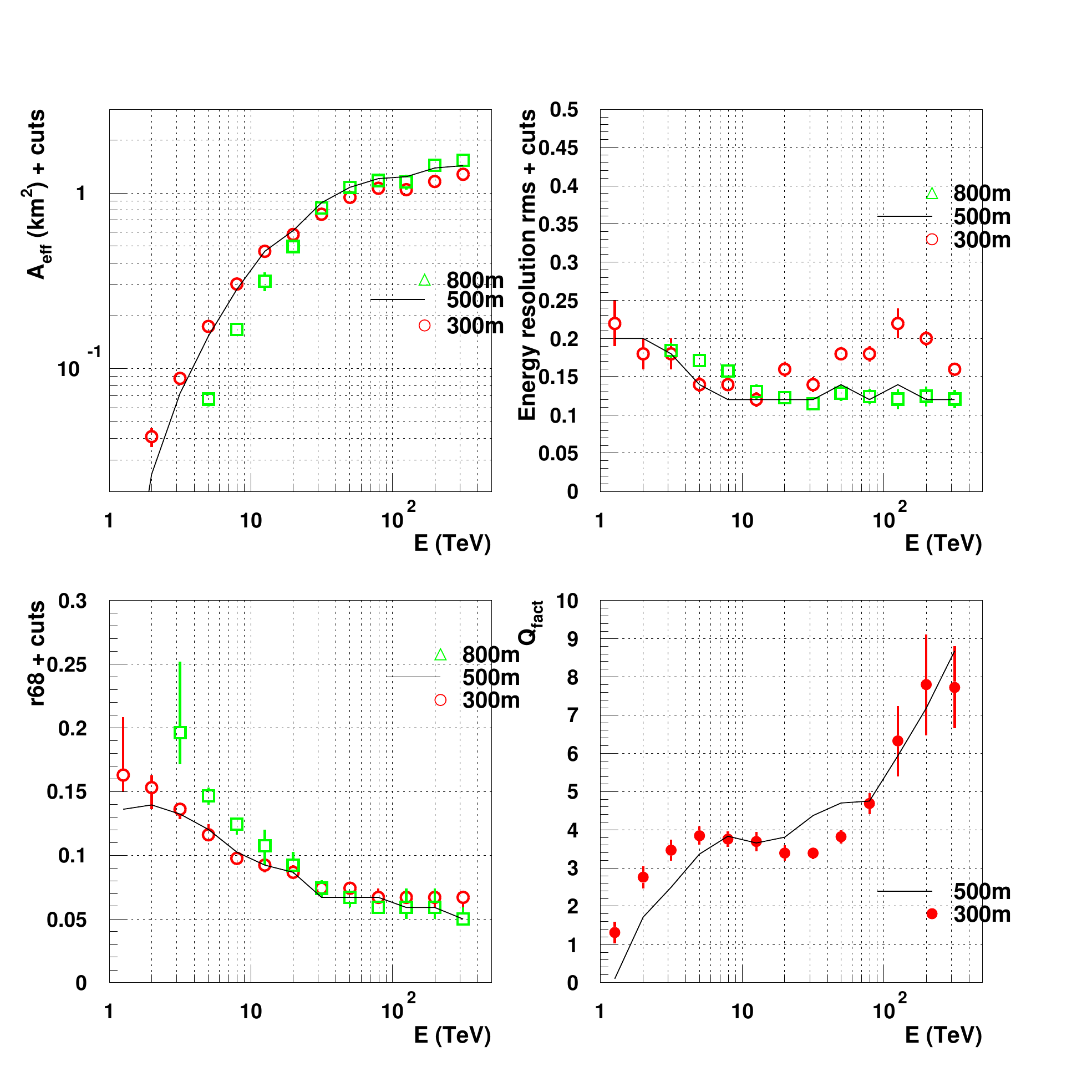}
\captionsetup{width=13cm}  
 \caption{Performance of PeX using Algorithm 3 for various telescope separations with standard triggering combination, cleaning combination and image \textit{size} cut. The top left shows the post-selection cut effective area, the top right represents the energy resolution, the bottom left provides the angular resolution (r68), and the bottom right gives the Q$_{fact}$. The error bars for the 500 \rm{m} line are of a similar magnitude to the error bars for the 300 \rm{m} points.}
 \label{fig:telescope_sep_alg3}
\end{centering}
\end{figure}

For the energy resolution in Figure~\ref{fig:telescope_sep_alg3}, the 500 \rm{m} separation provides an improved post-shape cut energy reconstruction over the smaller separation, especially for the high energy events (Figure~\ref{fig:telescope_sep_alg3} top right). The 300 \rm{m} separation provides fewer events at high energies and more events at low energies. 


The bottom left panel in Figure~\ref{fig:telescope_sep_alg3} shows that the post-shape cut angular resolution performance of the two telescope separations are similar. However, the improvement gained with Algorithm 1 over the same separations (Figure~\ref{fig:angres_sep_low_alt}) is not seen for Algorithm 3. Algorithm 1 results showed that a 500 \rm{m} separation provided a significant improvement to the angular resolution. Algorithm 3 does not require the images to have large angles between the major axes like Algorithm 1. The regions that provide poor event reconstruction (Figure~\ref{fig:core_recon}) for Algorithm 1 are not seen in Figure~\ref{fig:core_recon2} for Algorithm 3. These regions have less of an effect on Algorithm 3 when the telescope separation increases, which is due to the angle increasing between major axes. Algorithm 3 does not have the same regions which provide poor event reconstruction since it provides good reconstruction even for images that have small angles between major axes. 

The bottom right plot in Figure~\ref{fig:telescope_sep_alg3} shows the Q$_{fact}$ vs energy. No 800 \rm{m} separation results have been displayed due to low Monte Carlo statistics for Algorithm 3. The smaller telescope separation provides an improved Q$_{fact}$ at the low energies, E $<$ 10 TeV, since the events are more likely to trigger multiple telescopes. For energies between 10 and 100 TeV, the 500 \rm{m} separation provides improved Q$_{fact}$ since the separation provides improved images in the camera without them being truncated by the camera edge. As the telescope separation increases, the events at large core distance are now closer to the telescopes. 


For energies above 100 TeV, the Q$_{fact}$ for both separations are very similar. At the highest energies, both the 300 \rm{m} and 500 \rm{m} separations will detect events that provide truncated images due to the camera edge. The 500 \rm{m} telescope separation provides an improvement above 100 TeV for the energy resolution, angular resolution and effective area. 


Overall, the results indicate that the 500 \rm{m} telescope separation provides good results when used with Algorithm 3.

\subsubsection{Triggering Combination}

Figure~\ref{fig:triggering_alg3} shows the results of varying the triggering combinations with Algorithm 3, which indicates a similar trend to the Algorithm 1 results (Figure~\ref{fig:angres4_trigger},~\ref{fig:qfactor4_trigger},~\ref{fig:area_triggering_low_alt} and~\ref{fig:energy_res_triggering_low_alt}). The low triggering combination (4\textit{pe}, 2) provides the largest post-selection cut effective area along with the standard triggering of (6\textit{pe}, 3) (Figure~\ref{fig:triggering_alg3} top left). The other triggering combinations appear to reduce the number of events which can trigger the cell. The energy resolution indicates that the high triggering combination (12\textit{pe}, 5) provides higher quality energy reconstruction compared with the standard triggering (Figure~\ref{fig:triggering_alg3} top right). The angular resolution improves with the higher triggering combination (Figure~\ref{fig:triggering_alg3} bottom left). However, the high triggering combination removes a large majority of small or low energy events, as seen in the effective area. The bottom right panel shows the Q$_{fact}$ obtained by varying the triggering combination (Figure~\ref{fig:triggering_alg3}). The improvements seen with the high triggering combination comes from a selection effect. This improves angular resolution, Q$_{fact}$ and energy resolution.

The high triggering combination does provide an improved angular resolution and Q$_{fact}$. However, the large error bars suggest that fewer $\gamma$-ray events are present post-selection cuts and post-shape cuts. The effective area results (Figure~\ref{fig:triggering_alg3} top left) agree within the error bars. The effective area shows a large loss in $\gamma$-ray events for high \textit{threshold} triggering combination. Since a key aim of PeX is to maximise the effective area, the (6\textit{pe}, 3) combination provides the optimum balance between results.

	The final conclusion concerning the varying triggering combinations with Algorithm 3 is that the trends agree with those seen in Algorithm 1 (section~\ref{sec:tel_sep_op}). The standard triggering combination provides the optimum overall results with a large effective area, good Q$_{fact}$ and angular resolution.\\

\begin{figure}
\begin{centering}
\includegraphics[scale=0.65]{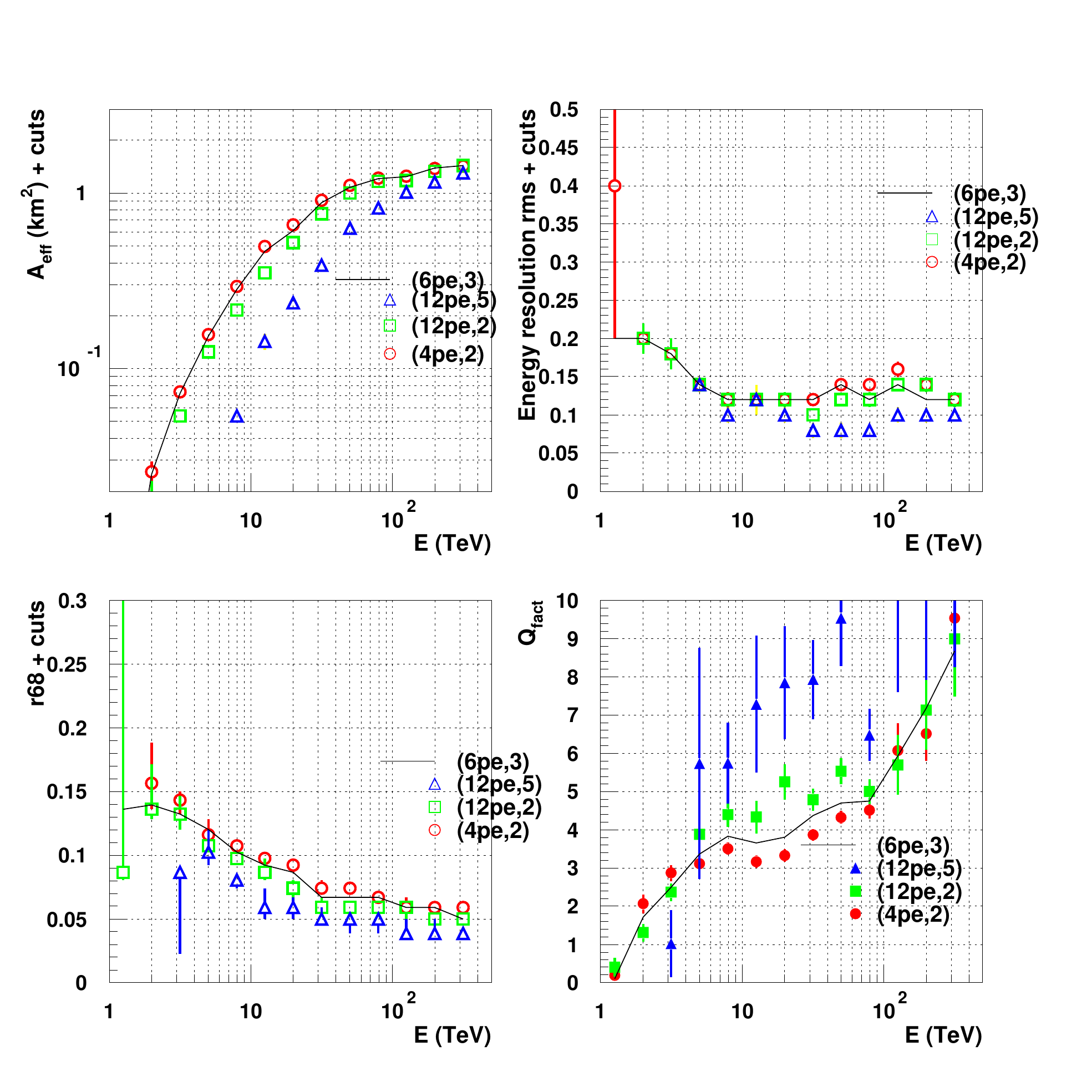}
\captionsetup{width=13cm}  
 \caption{Performance of PeX using Algorithm 3 for different triggering combinations using Algorithm 3 with standard telescope separation, cleaning combination and image \textit{size} cut. The top left shows the post-selection cut effective area, the top right represents the energy resolution, the bottom left provides the angular resolution (r68) and the bottom right gives the Q$_{fact}$. The black curve has error bars of the same order of magnitude as those seen for the (12\textit{pe}, 2) results.}
 \label{fig:triggering_alg3}
\end{centering}
\end{figure}

\subsubsection{Cleaning Combination}

Results for the four cleaning combinations (same as those for Algorithm 1 in Figure~\ref{fig:area_cleaning_low_alt}) are displayed in Figure~\ref{fig:cleaning_alg3_multi}. The standard cleaning combination is (8\textit{pe}, 4\textit{pe}). Results indicate that a \textit{picture} value $>$ 7\textit{pe} and a \textit{boundary} value $>$ 3\textit{pe} are best. The (5\textit{pe}, 2\textit{pe}) cleaning shows a significant reduction in performance for all parameters. The (10\textit{pe}, 0\textit{pe}) cleaning combination suggests that a \textit{boundary} value is required since the results fail to match the standard cleaning combination. The effect of having 0 or 1\textit{pe} for a \textit{boundary} value is more significant for E $<$ 20 TeV, especially for the angular resolution and energy resolution (Figure~\ref{fig:cleaning_alg3_multi} bottom left and top right). The (10\textit{pe}, 5\textit{pe}) cleaning combination provides similar results to the standard cleaning combination, although the Q$_{fact}$ shows that the rejection power decreases marginally (Figure~\ref{fig:cleaning_alg3_multi} bottom right). 

The (5\textit{pe}, 2\textit{pe}) cleaning combination under cleans the images leaving too much NSB within the image (Figure~\ref{fig:cleaning_alg3_multi} bottom left and right panels). This result is similar to that seen in the Algorithm 1 investigation (Figure~\ref{fig:angres4_clean},~\ref{fig:qfactor4_clean},~\ref{fig:area_cleaning_low_alt} and~\ref{fig:energy_res_cleaning_low_alt}). The effect is quite significant and it shows the importance of cleaning the images adequately. The (10\textit{pe}, 5\textit{pe}) cleaning combination illustrates over cleaning and appears to be too strong. Thus the standard cleaning combination (8\textit{pe}, 4\textit{pe}) appears to provide the optimum cleaning for Algorithm 3 as we found for Algorithm 1. \\

\begin{figure}
\begin{centering}
\includegraphics[scale=0.65]{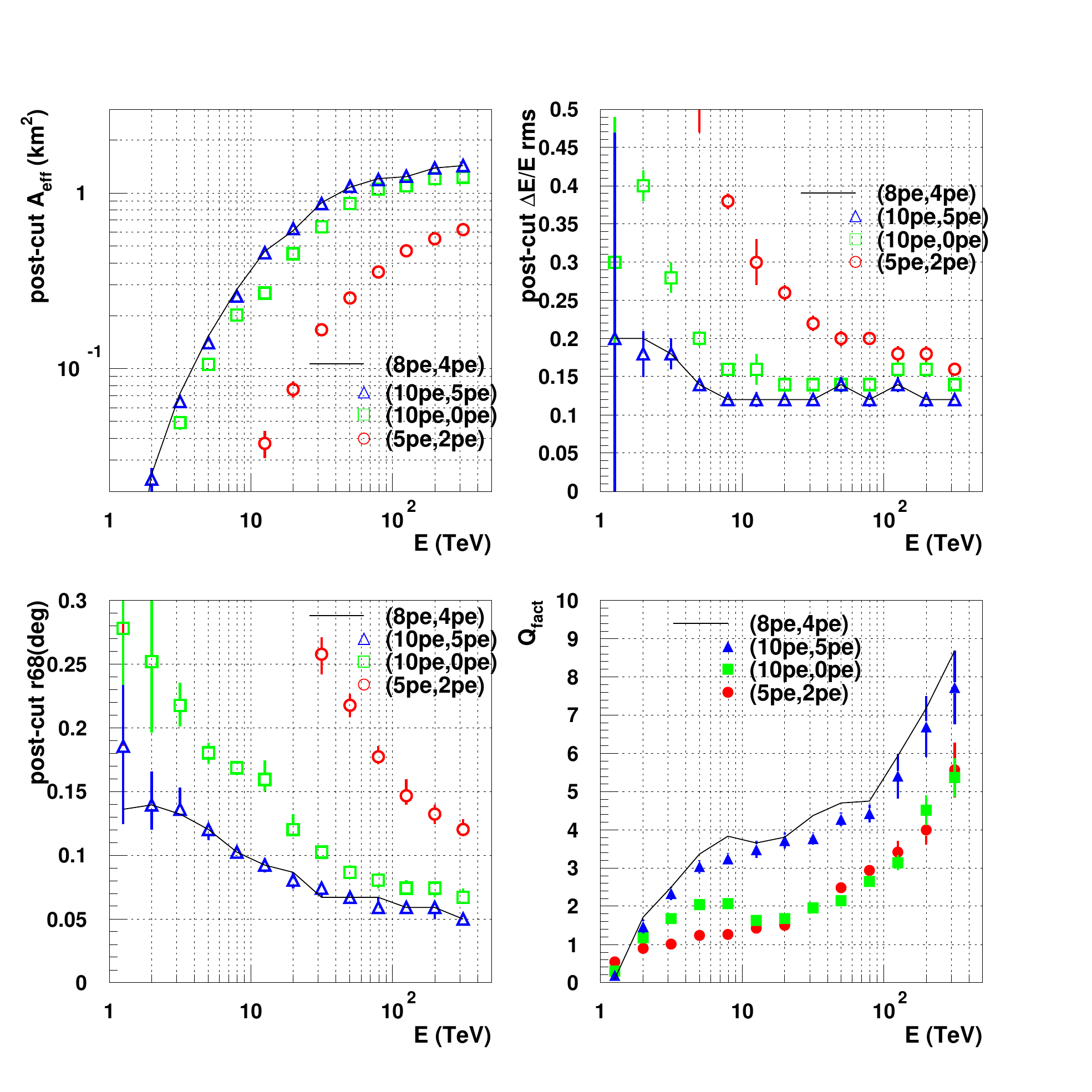}
\captionsetup{width=13cm}  
 \caption{Performance of PeX using Algorithm 3 for different cleaning combinations with telescope separation, triggering combination and image \textit{size} cut. The top left shows the post-selection cut effective area, the top right represents the energy resolution, the bottom left provides the angular resolution (r68) and the bottom right gives the Q$_{fact}$. The black curve has error bars of the same order of  magnitude as those seen for the (10\textit{pe}, 5\textit{pe}) results.}
 \label{fig:cleaning_alg3_multi}
\end{centering}
\end{figure}

\subsubsection{Image \textit{size} Cut}

Figure~\ref{fig:size_alg3} shows that a larger image \textit{size} cut of 200\textit{pe} provides improved energy resolution, angular resolution and Q$_{fact}$ compared with the standard cut of 60\textit{pe}. The improvement is seen for most of the energy range. The effective area sees a reduction in event numbers since the larger image \textit{size} cut removes a large number of small sized events (Figure~\ref{fig:size_alg3} top left). Given the loss of events, but improved reconstruction, the 200\textit{pe} cut is more appropriate for sources with high $\gamma$-ray fluxes where the analysis can be stricter with event cuts. These results provide similar trends when applied to Algorithm 3 as when applied to Algorithm 1 (section~\ref{sec:image_size_op}). The image \textit{size} cut depends on the strength of a source and the number of events that have been detected. The 200\textit{pe} image \textit{size} cut can be classed as a hard cut for strong sources.

\begin{figure}
\begin{centering}
\includegraphics[scale=0.65]{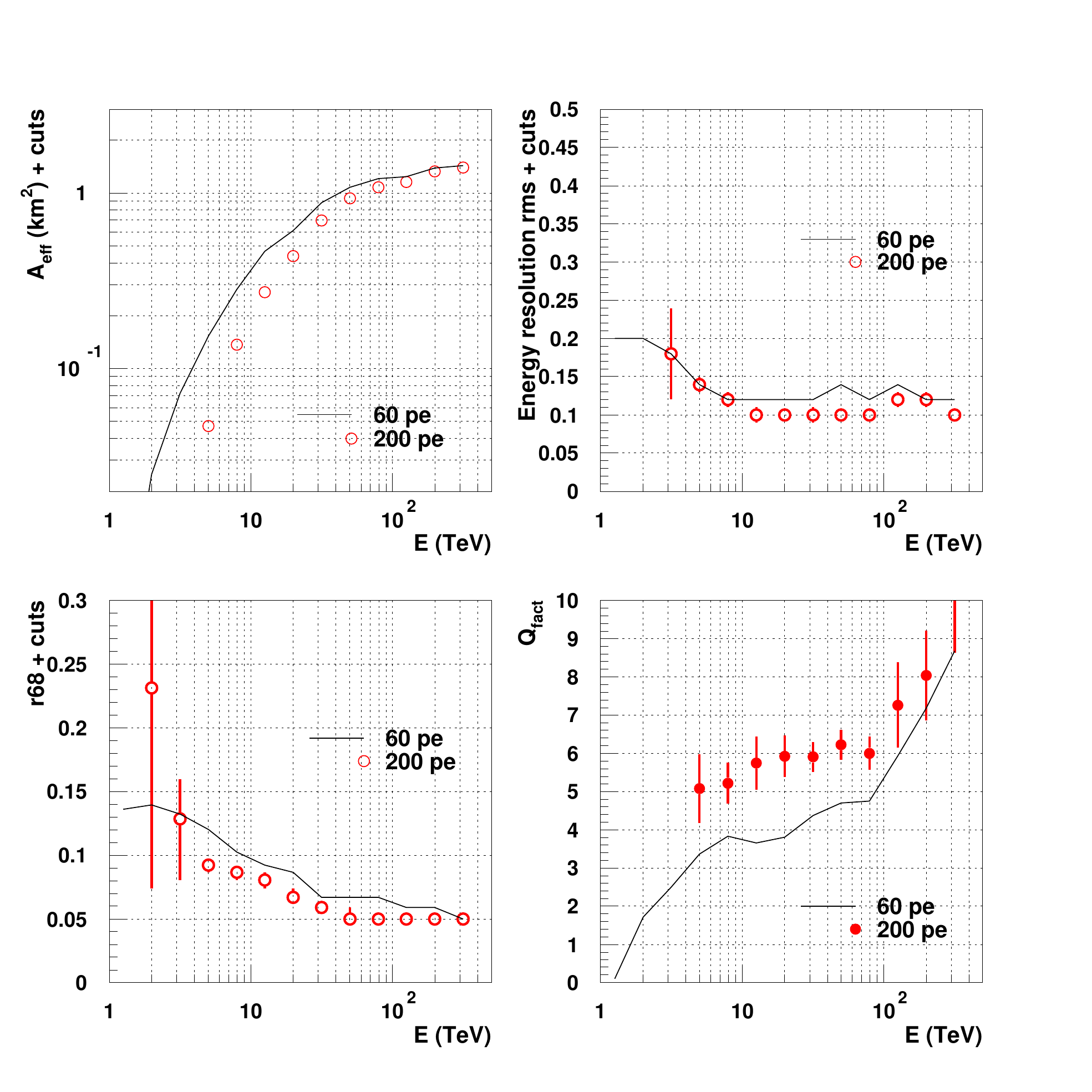}
\captionsetup{width=13cm}  
 \caption{Performance of PeX using Algorithm 3 for different image \textit{size} cuts with telescope separation, triggering combination and cleaning combination. The top left shows the post-selection cut effective area, the top right represents the energy resolution, the bottom left provides the angular resolution (r68) and the bottom right gives the Q$_{fact}$. The black curve has similar error bars than those seen for the 200\textit{pe} results.}
 \label{fig:size_alg3}
\end{centering}
\end{figure}

The chosen triggering, cleaning, image \textit{size} cut and telescope separation for Algorithm 1 appear to apply for Algorithm 3. In summary, the preferred PeX parameters for Algorithm 3 would use a 500 \rm{m} telescope separation, a (6\textit{pe}, 3) triggering combination, a (8\textit{pe}, 4\textit{pe}) cleaning combination and a 60\textit{pe} image \textit{size} cut, as found using Algorithm 1.

\section{Application of time cleaning with Algorithm 3}

The next step is to investigate the new time cleaning algorithm introduced in Chapter~\ref{sec:timing_cleaning} in combination with the event direction reconstruction from Algorithm 3. Algorithm 1 and Algorithm 3 reconstruct the shower in two different ways but the time cleaning cut affects the images before this reconstruction. We expect that the same trends in telescope separation, triggering combination, cleaning combination and image \textit{size} cuts should be seen for both algorithms with and without time cleaning.

The distribution of picture and boundary pixel arrival time differences is based on the arrival times of the photons in each pixel (Figure~\ref{fig:core-boundary_pe} in section~\ref{sec:time_core_boundary}). The picture and boundary pixels are defined by the normal cleaning algorithm. The distributions for picture and boundary arrival times will be the same for both algorithms. The time cut used for Algorithm 1 is the appropriate cut to use for Algorithm 3, so a time cut of $\pm$ 5\rm{ns} is applied to the cleaning algorithm. 

We investigated here, Algorithm 3 reconstruction with and without the time cleaning cut for both the off-Galactic plane level of NSB and the Galactic centre level of NSB. 

Figure~\ref{fig:alg3_time} shows the results for Algorithm 3 with and without time cleaning, where the level of NSB represents the off-plane value. The post-selection cut effective area shows that the time cleaning results provide no improvement for effective area (Figure~\ref{fig:alg3_time} top left). The RMS for the $\Delta{E}/E$ distribution shows no difference (Figure~\ref{fig:alg3_time} top right). The angular resolution (r68) indicates that the results with and without time cleaning are also similar (Figure~\ref{fig:alg3_time} bottom left). The Q$_{fact}$ shows no significant differences in results for the low energies, E $<$ 5 TeV, and the high energies, E $>$ 100 TeV (Figure~\ref{fig:alg3_time} bottom right) but are consistant within errors. The high energy differences are within statistical errors since the error bars on the normal cleaning, black line, are of the same magnitude as the time cleaning error bars, red points (Figure~\ref{fig:alg3_time}). 

Thus, when time cleaning is used in conjunction with Algorithm 3, it appears to provide no significant improvement in results. As stated in section~\ref{sec:timing_cut}, the time cleaning has no effect on the results for Algorithm 1, which suggests that the (8\textit{pe}, 4\textit{pe}) cleaning combination effectively mitigates the NSB in the image. Therefore, the time cleaning appears to affect Algorithm 3 the same way as it affected Algorithm 1.

\begin{figure}
\begin{centering}
\includegraphics[width=\textwidth]{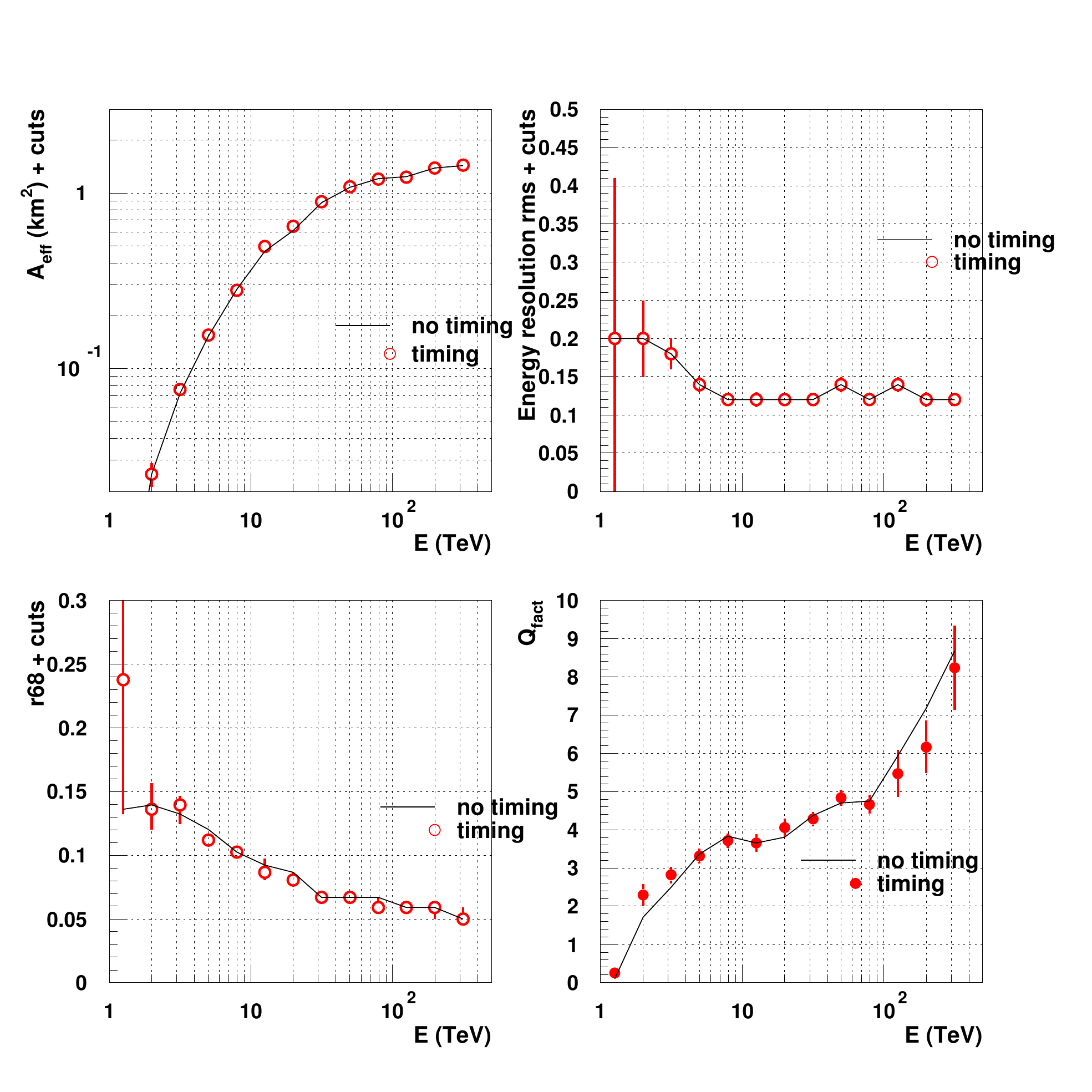}
\captionsetup{width=13cm}  
\caption{Performace of PeX for Algorithm 3 with and without time cleaning for a standard configuration using off-Galactic plane NSB. The top left shows the post-selection cut effective area, the top right represents the energy resolution, the bottom left provides the angular resolution (r68) and the bottom right gives the Q$_{fact}$. The black curve error bars as of a similar magnitude as the timing error bars on the red points.}
 \label{fig:alg3_time}
\end{centering}
\end{figure}

If the level of NSB is ramped up to a level appropriate for the Galactic centre as described in section~\ref{sec:timing_cut}, approximately four times the off-Galactic plane level of NSB, the robustness of Algorithm 3 and time cleaning can be investigated. The comparison between cleaning algorithms with a Galactic centre level of NSB is displayed in Figure~\ref{fig:alg3_time_nsbvar}. The effective area plot shows that the time cleaning method provides a significant improvement (Figure~\ref{fig:alg3_time_nsbvar} top left). The improvement gained by using a time cut suggests that the NSB pixels which disrupted the reconstruction have been effectively mitigated from the image. This trend confirms that the extra NSB in pixels with signal affects the parameterisation and event reconstruction. 

The energy resolution shows the same effect due to the Galactic centre NSB level (Figure~\ref{fig:alg3_time_nsbvar} top right). As the energy increases the Galactic centre NSB level has less effect on the energy reconstruction due to the larger images involved since the image \textit{size} generally increases compared to the amount of NSB in the image. Applying a time cut improves the energy resolution for all energies, especially the low energies where the small images suffer more from NSB. The time cut brings the average energy resolution down to 15$\%$ for E $>$ 10 TeV compared with 20$\%$ for no time cut for E $>$ 10 TeV.

The angular resolution is shown in Figure~\ref{fig:alg3_time_nsbvar} bottom left panel. The angular resolution for both cleaning algorithms is large for E $<$ 10 TeV. For E $>$ 10 TeV, the angular resolution improves with energy since the NSB contribution is smaller in comparison to some of the large image sizes at high energies. The angular resolution with a time cut has provided good event reconstruction.

The shape separation between $\gamma$-ray and proton events in a high NSB environment is shown by the black curve in the Q$_{fact}$ plot (Figure~\ref{fig:alg3_time_nsbvar}). The Galactic centre NSB reduces the quality of images. By applying the time cleaning cut, Q$_{fact}$ improves on average by a factor of roughly two for E $<$ 100 TeV. As energy increases, the difference between the Q$_{fact}$ decreases, as expected.

The general trends from Figure~\ref{fig:alg3_time_nsbvar} suggests that the low energy or small sized images suffer more for a high level of NSB. As the energy increases, the high NSB has a reduced effect on the results. This is due to larger sizes which makes the NSB contribution less significant or small when considering the ratio of NSB to \textit{pe} in an image.

The usefulness of the time cleaning cut was seen in section~\ref{sec:timing_cleaning} for Algorithm 1 and the results presented here show the same trends with Algorithm 3. The addition of a time cleaning cut in the cleaning algorithm provides an improved rejection of NSB pixels from the final image. The time cut provides a robust cleaning technique that copes with all levels of NSB. The time cleaning cut can therefore be a permanent addition to the analysis.

\begin{figure}
\begin{centering}
\includegraphics[width=\textwidth]{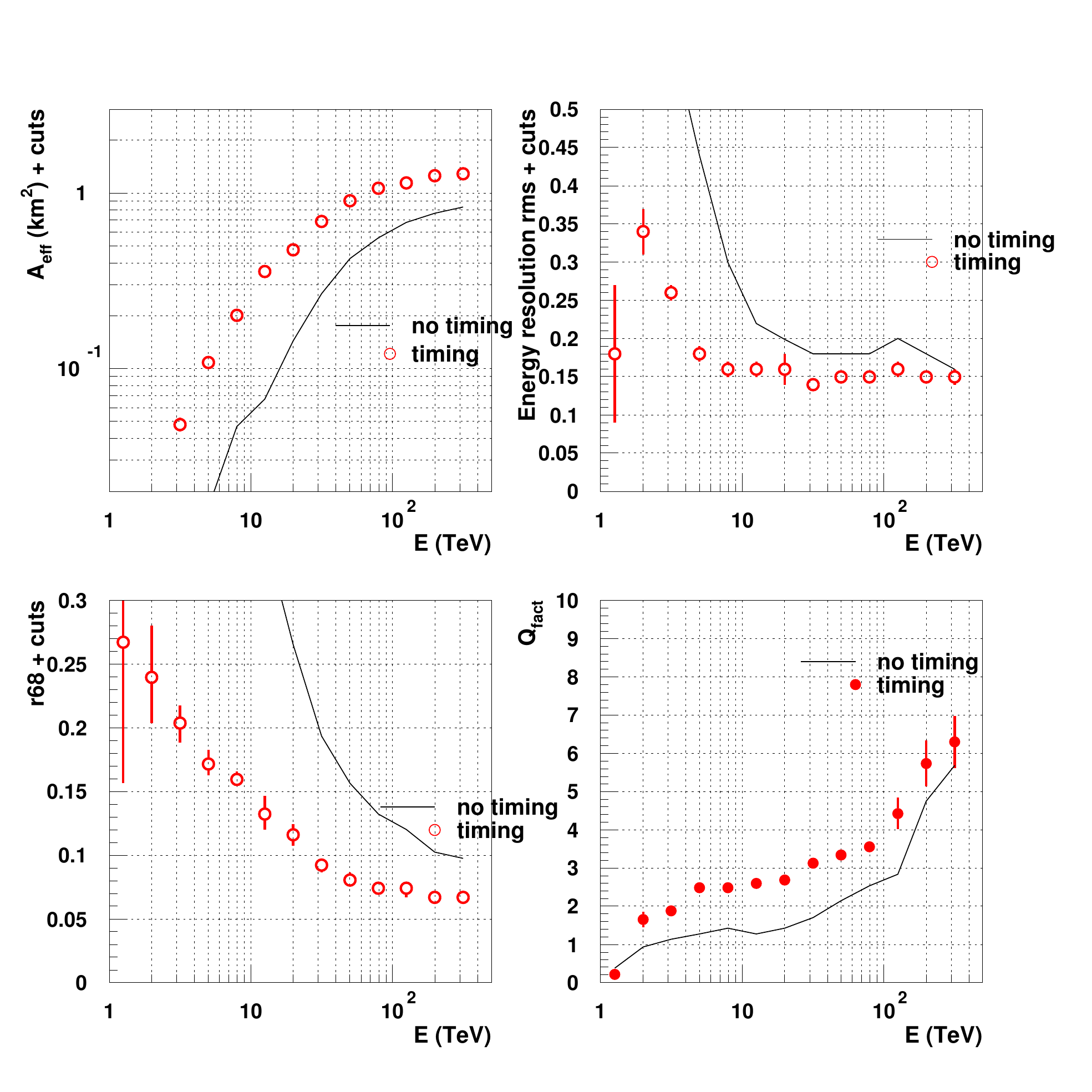}
\captionsetup{width=13cm}  
\caption{Performance of PeX using standard Algorithm 3 configuration with and without time cleaning for a Galactic centre level of NSB. The top left shows the post-selection cut effective area, the top right represents the energy resolution, the bottom left provides the angular resolution (r68) and the bottom right gives the Q$_{fact}$. Results indicate that a time cut is necessary to improve results at all energies and for all parameters. The black curve error bars are of a similar magnitude to the error bars on the red points.}
 \label{fig:alg3_time_nsbvar}
\end{centering}
\end{figure}

\section{The site altitude effect on Algorithm 3}

A site altitude study was previously conducted for Algorithm 1 in section~\ref{sec:comparison_alg1}. The results indicated that the effective area is somewhat affected by a varying the observational altitude and hence so is the flux sensitivity. Figure~\ref{fig:relative_area} showed that the 0.22 \rm{km} altitude site provides a 15$\%$ larger area for E $>$ 50 TeV. We will also investigate the effect of altitude on Algorithm 3 performance. 

Figure~\ref{fig:site_altitude_alg3_timing} shows the results for the two different site altitudes. The post-selection cut effective areas have been presented as a relative curve (Figure~\ref{fig:site_altitude_alg3_timing} top left). The 1.8 \rm{km} curve has been plotted relative to the 0.22 \rm{km} so A$_{eff_{1.8km}}$/A$_{eff_{0.22km}}$. It helps accentuate the difference in effective areas. The 0.22 \rm{km} altitude provides a larger area. For E $>$ 10 TeV, the 0.22 \rm{km} site has an area larger by roughly 20$\%$. The top right panel shows the energy resolution (Figure~\ref{fig:site_altitude_alg3_timing}). Only a marginal improvement is seen for the 0.22 \rm{km} site for E $>$ 10 TeV. The average energy resolution for all energies for the 1.8 \rm{km} site is 15$\%$ while the average energy resolution for all energies for the 0.22 \rm{km} site is 13$\%$. 

The angular resolution shows no significant improvement between the two observational sites (Figure~\ref{fig:site_altitude_alg3_timing} bottom left). The difference seen between 10 TeV to 80 TeV is mostly within errors. The Q$_{fact}$ shows multiple fluctuations over the entire energy range for both observational sites (Figure~\ref{fig:site_altitude_alg3_timing} bottom right). Both results are within the statistical errors of each other.

\begin{figure}
\begin{centering}
\includegraphics[width=\textwidth]{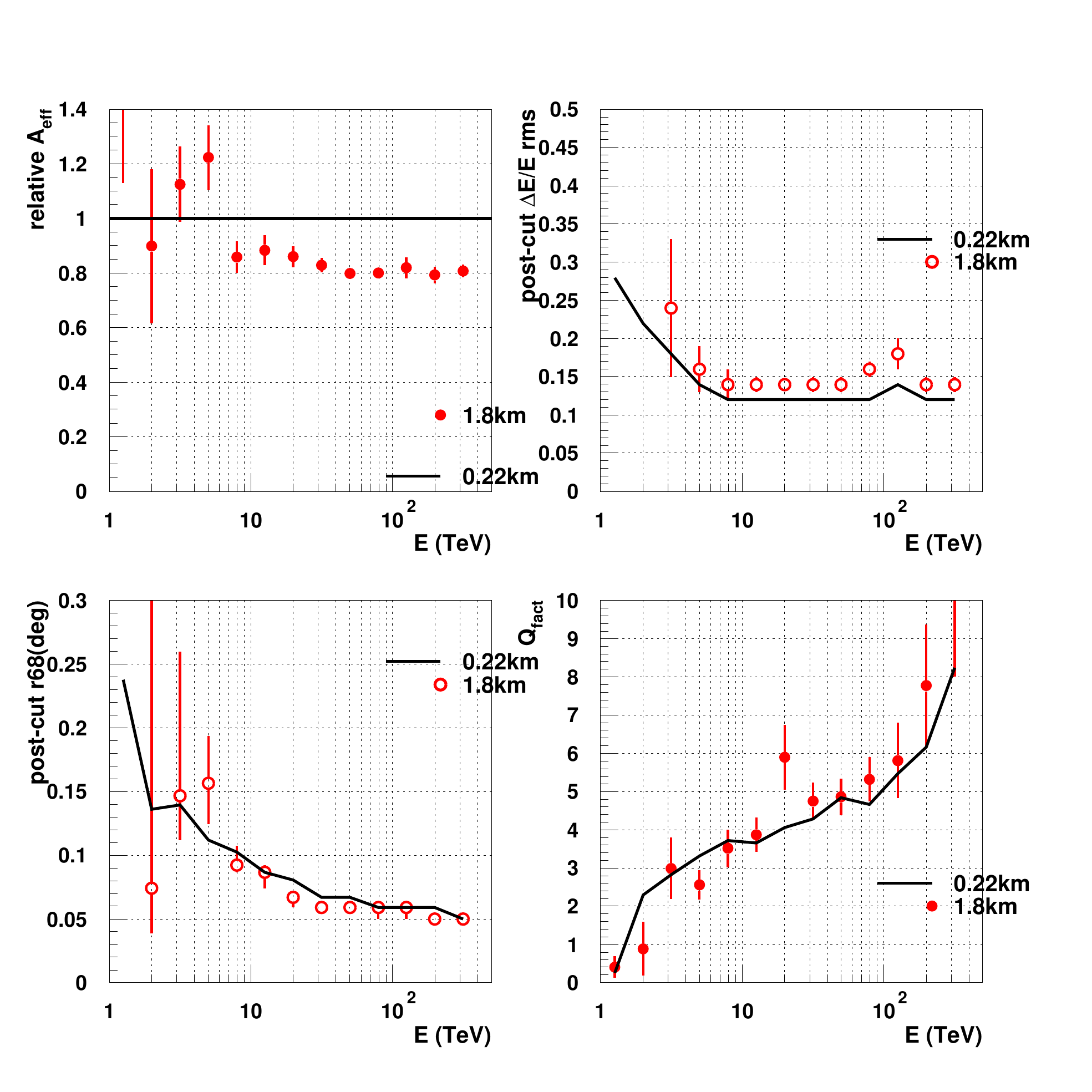}
\captionsetup{width=13cm}  
\caption{Performance of PeX for Algorithm 3 with two different observational altitudes, 1.8 {figure} and 0.22 {figure}. The top left shows the post-cut 1.8 \rm{km} effective area relative to the 0.22 \rm{km} result, the top right represents the energy resolution, the bottom left provides the angular resolution (r68) and the bottom right gives the Q$_{fact}$. The error bars on the black lines are of similar magnitude to the red circles.}
 \label{fig:site_altitude_alg3_timing}
\end{centering}
\end{figure}

The Algorithm 3 results for different observational altitudes show the same trends as for Algorithm 1 (Figure~\ref{fig:plots_timing_nsb}). These results are expected since the difference in altitude will affect the images in the same way. The 1.8 \rm{km} observational site produces a smaller effective area compared with the 0.22 \rm{km} observational site. The difference in altitude appears to affect the effective area more than the other parameters.

\begin{figure}
\begin{centering}
\includegraphics[width=\textwidth]{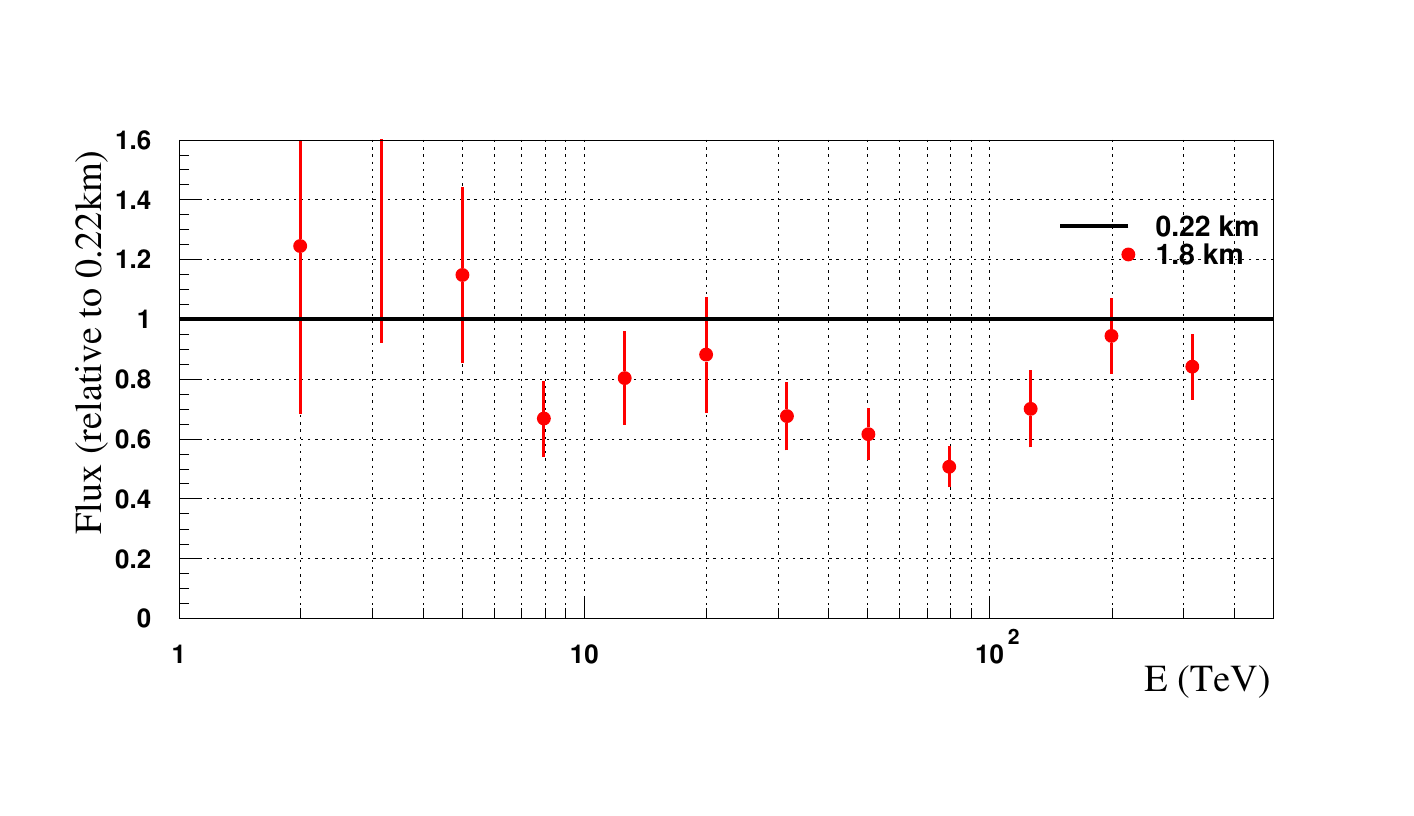}
\captionsetup{width=13cm}  
\caption{Flux sensitivity of PeX for Algorithm 3 with two different observational altitudes, 1.8 \rm{km} and 0.22 \rm{km}. The flux sensitivity has been plotted as the ratio of the 0.22 \rm{km} altitude flux over the 1.8 \rm{km} altitude flux, $F_{0.22km}$/$F_{1.8km}$. Therefore, if the 0.22 \rm{km} altitude flux provides an improvement in flux sensitivity over the 1.8 \rm{km} altitude flux then the plotted flux ratio will be $<$ 1.}
 \label{fig:flux_rel_altitude}
\end{centering}
\end{figure}

Figure~\ref{fig:flux_rel_altitude} illustrates the flux sensitivity ratio for the different altitude sites. The flux sensitivity has been plotted as the ratio of the 0.22 \rm{km} altitude flux over the 1.8 \rm{km} altitude flux, $F_{0.22km}$/$F_{1.8km}$. Therefore, if the 0.22 \rm{km} altitude flux provides an improvement in flux sensitivity over the 1.8 \rm{km} altitude flux then the plotted flux ratio will be $<$ 1. The 0.22 \rm{km} altitude site provides the better flux sensitivity. The improvement in effective area is apparent in the flux sensitivity. The other parameters show no variation between the results from both observational sites. Therefore, the improvement in flux sensitivity comes from the improvement in effective area.

All other parameters see little or no improvement with varying observational level. At both altitudes, the telescopes will trigger on small sized images, which limits the reconstruction quality from the algorithm. The only way to improve the reconstruction is to remove the small sized images. This was best displayed in the image \textit{size} variation results (Figure~\ref{fig:size_alg3}). 

The 0.22 \rm{km} altitude observational site provides an adequate result for Algorithm 3 for E $>$ 10 TeV compared with the 1.8 \rm{km} altitude observational site.

\section{Algorithm 1 vs Algorithm 3 and performance vs core distance}
 \label{sec:alg_comp}

After comparing Algorithm 3 performance with varying levels of NSB, time cleaning and different site altitudes, an adequate Algorithm 3 configuration is obtained. This configuration is: a 500 \rm{m} telescope separation, a (6\textit{pe}, 3) triggering combination, a (8\textit{pe}, 4\textit{pe}) cleaning combination with the time cleaning cut, a 60\textit{pe} image \textit{size} cut and situated at a 0.22 \rm{km} altitude site to improve collection area. The final comparison is done between the best PeX cell configuration for the Algorithm 1 analysis system and the Algorithm 3 analysis system. 

Algorithm 3 appears to provide the best results for the cell. Figure~\ref{fig:alg3_vs_alg1_timing} provides the comparison plots for Algorithm 1 vs Algorithm 3. The effective area curve indicates that the Algorithm 3 analysis code provides a larger collecting area over the Algorithm 1 analysis code (Figure~\ref{fig:alg3_vs_alg1_timing} top left). The improvement becomes larger as the energy increases. 
The energy resolution shows that the Algorithm 3 curve provides a tighter RMS for the $\Delta{E}/E$ distribution (Figure~\ref{fig:alg3_vs_alg1_timing} top right). The trend again indicates that the event reconstruction has improved with Algorithm 3 compared to Algorithm 1. The average RMS for Algorithm 3 is 12.5$\%$ and the average RMS for Algorithm 1 is 15$\%$. 

The angular resolution for Algorithm 3 provides significant improvement over Algorithm 1 (Figure~\ref{fig:alg3_vs_alg1_timing} bottom left). The effect is seen for all energies. The improvement is due to the way that Algorithm 3 reconstructs images being truncated by the edge of the camera at large core distances (Figure~\ref{fig:ang_res_alg1_v_alg3}). The figure illustrates the improvement gained from Algorithm 3 especially for the large core distance, $>$ 500 \rm{m}, showers. The reconstruction using Algorithm 1 replies heavily on the major axes from Hillas parameters. The accuracy of the major axis from images cut off by the edge of the camera is poor, which provides a poor reconstruction of the shower direction. Using the time gradient within the image provides improved reconstruction of the shower direction for Algorithm 3. The PeX cell needs to utilise the large core distance events to maximise the effective area at high energies. Algorithm 3 provides a significant improvement in angular resolution for core distances $>$ 300\rm{m} (Figure~\ref{fig:ang_res_alg1_v_alg3}). The improvement in large core distance events by Algorithm 3 was also seen in the post-selection cut effective area (Figure~\ref{fig:alg3_vs_alg1_timing} top left).

The improvements seen with Algorithm 3 come from the improved reconstruction of images truncated by the camera edge. These truncated images are usually from large core distance events. The improvement can be seen in Figure~\ref{fig:ang_res_alg1_v_alg3}. The reconstruction is improved since the algorithm does not rely on an accurate major axis from the image but utilises the major axis as a guide for where the true shower is located. Algorithm 3 uses the time gradient and reconstructed light maximum to find the predicted distance to the source position. The time gradient provides the biggest improvement to the results. This can be done accurately with only part of the image on the camera, which helps images cut off by the edge of the camera. The overall process of using time gradient and reconstructed light maximum to reconstruct the shower direction provides improved reconstruction compared with the intersection point between pairs of image major axes.

The Q-factor plot shows that there is minimal difference between algorithms 1 and 3 when it comes to Q$_{fact}$ vs core distance (Figure~\ref{fig:qfactor_alg1_v_alg3}). Both algorithms use the same image parameterisation, so it is expected that the algorithms should produce similar quality factors except for the most distant core showers. The performance of PeX shows that large core distance events can be utilised. From Figure~\ref{fig:qfactor_alg1_v_alg3} the high energy events can benefit out to $\sim$ 500 $-$ 700 \rm{m}, which is well outside the physical area of PeX. For core distances larger than 700 \rm{m} the Q$_{fact}$ decreases, the angular resolution improves and the effective area increases. The flux sensitivity, F $\propto$ $\sqrt{A_{eff}}$ Q$_{fact}$/r68, scales with these values. Therefore, events with core distance $>$ 700 \rm{m} will provide little improvement to the flux sensitivity for high energies.

\begin{figure}
\begin{centering}
\includegraphics[width=\textwidth]{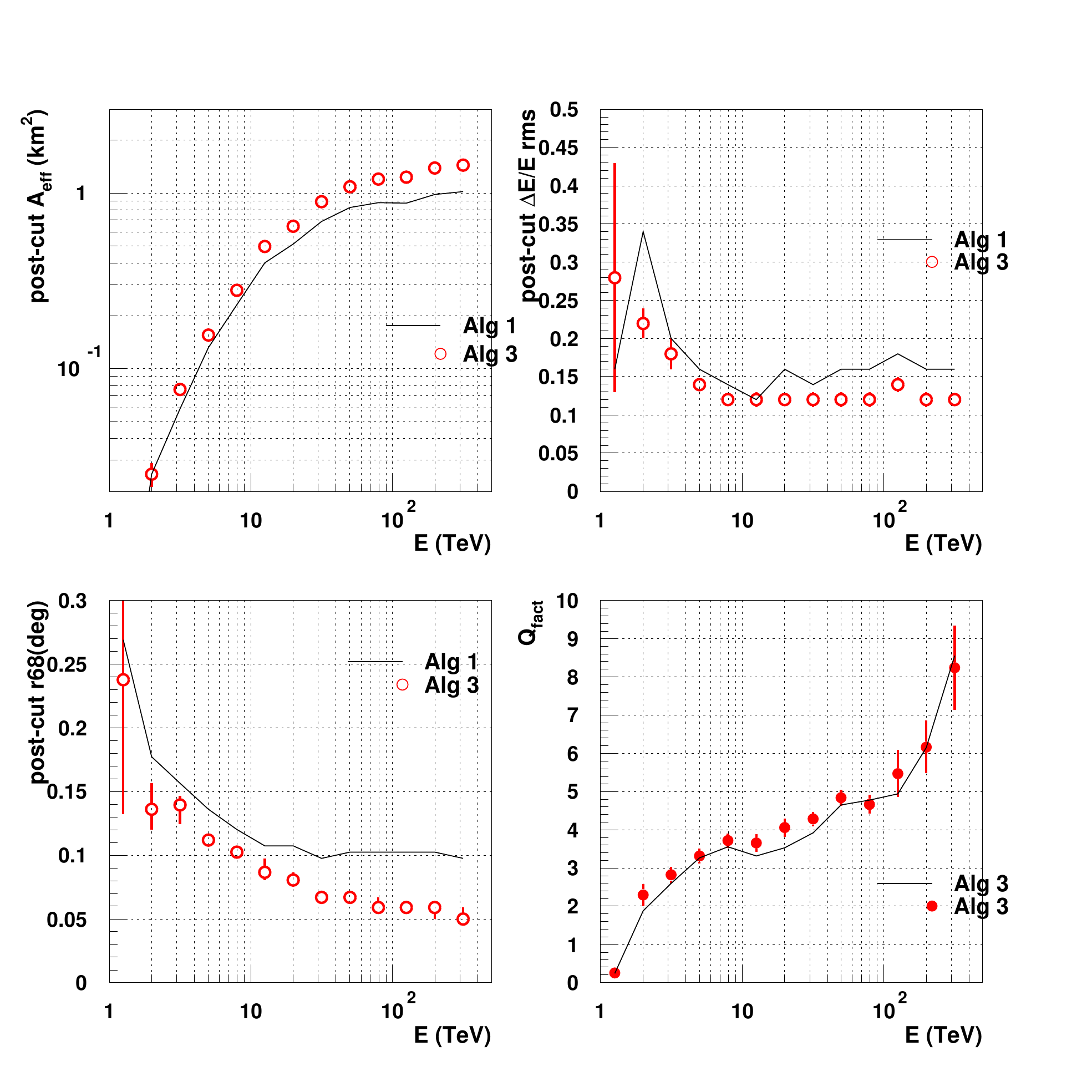}
\captionsetup{width=13cm}  
\caption{Comparing the results for Algorithm 1 and Algorithm 3 best configurations at 0.22 \rm{km} altitude. The top left shows the pre-cut effective area, the top right represents the energy resolution, the bottom left provides the angular resolution (r68) and the bottom right gives the Q$_{fact}$. Results indicate that Algorithm 3 improves results at all energies and for all parameters.}
 \label{fig:alg3_vs_alg1_timing}
\end{centering}
\end{figure}

\begin{figure}
\begin{centering}
\includegraphics[scale=0.8]{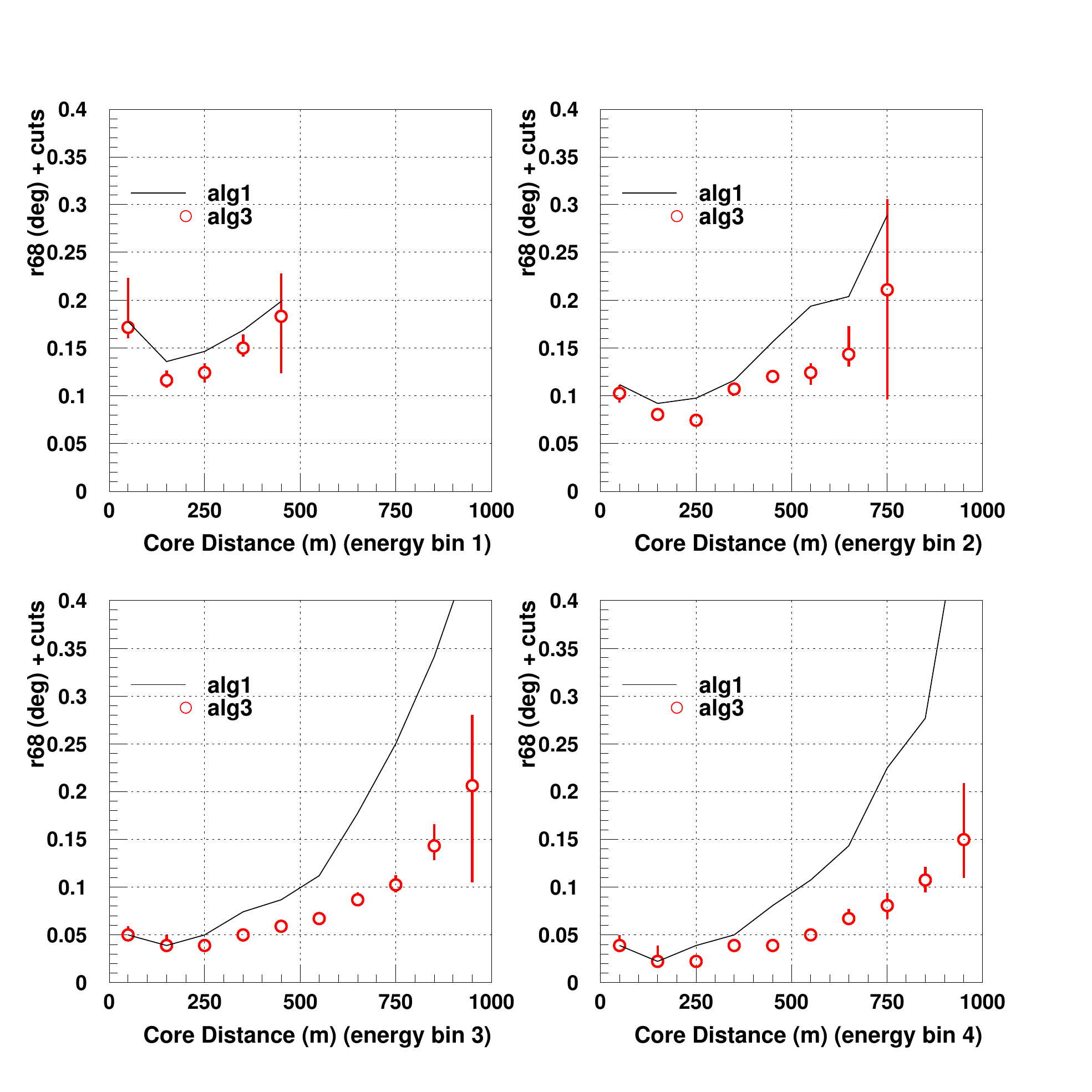}
\captionsetup{width=13cm}  
\caption{Comparing the post-shape cut angular resolution for Algorithm 1 and Algorithm 3 vs core distance. The results have been split into four energy bands: 1 - 4.2 TeV (bin 1), 4.2 - 22.3 TeV (bin 2), 22.3 - 105.7 TeV (bin 3) and 105.7 - 500 TeV (bin 4). The error bars on the Algorithm 1 black line are similar to the error bars for the Algorithm 3 red open circles. Algorithm 3 produces a superior reconstruction for showers with core distance $>$ 300 \rm{m}.}
 \label{fig:ang_res_alg1_v_alg3}
\end{centering}
\end{figure}

\begin{figure}
\begin{centering}
\includegraphics[scale=0.8]{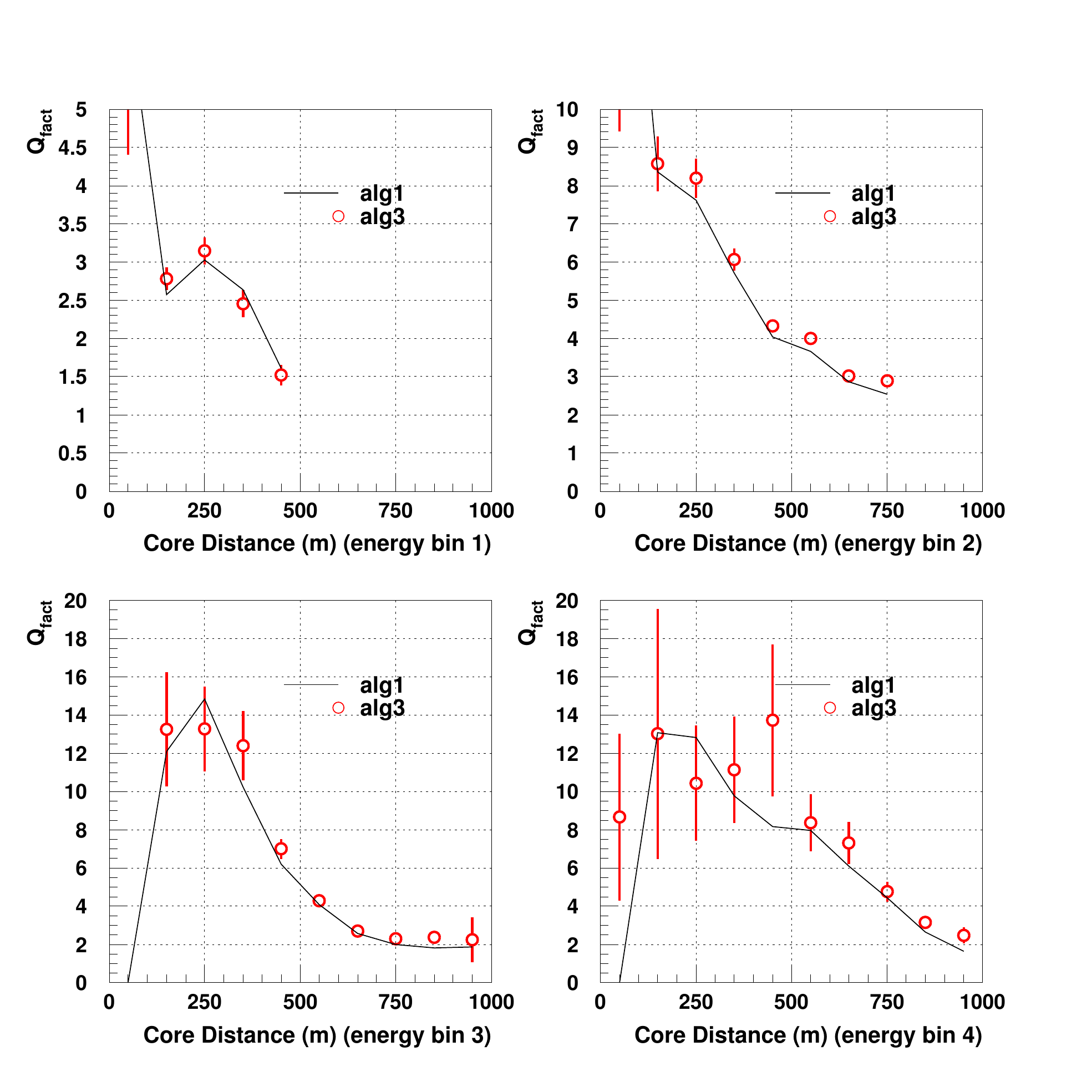}
\captionsetup{width=13cm}  
\caption{Comparing Q$_{fact}$ for Algorithm 1 and Algorithm 3 vs core distance. The results have been split into four energy bands: 1 - 4.2 TeV (bin 1), 4.2 - 22.3 TeV (bin 2), 22.3 - 105.7 TeV (bin 3) and 105.7 - 500 TeV (bin 4). The error bars on the Algorithm 1 black line are similar to the error bars for the Algorithm 3 red open circles. Algorithm 3 produces a superior reconstruction for showers with core distance $>$ 300 \rm{m}.}
 \label{fig:qfactor_alg1_v_alg3}
\end{centering}
\end{figure}

\section{Concluding remarks on Algorithm 1 vs Algorithm 3}

After optimising the telescope separation, triggering combination, cleaning combination, image \textit{size} cut, site altitude and time cleaning cut for Algorithm 3, we find that an adequate configuration for the Algorithm 3 cell is:

\begin{itemize}
\item a 500 \rm{m} telescope separation, a (6\textit{pe}, 3) triggering combination, an (8\textit{pe}, 4\textit{pe}) cleaning combination, a 60\textit{pe} image \textit{size} cut, a 0.22 \rm{km} observational site and using a $\pm$ 5 \rm{ns} time cleaning cut.
\end{itemize}

Comparing the optimised Algorithm 1 and Algorithm 3 analyses, the optimised Algorithm 3 provides the best results for the PeX. The effective area is a major component of the cell and Algorithm 3 provides the largest effective area without reducing the quality of the angular resolution, energy resolution or gamma/proton separation. Algorithm 3 will now be the preferred reconstruction algorithm for the cell.

Now that Algorithm 3 has been selected as the optimum reconstruction method for PeX, we can consider further enhancements beyond the standard configuration of PeX, and these will be considered in the next section. 


\chapter{Flux Sensitivity of PeX and Possible Enhancements}
 \label{sec:future_work}

A number of parameters for PeX such as those relating to image cleaning, event triggering, site altitude and reconstruction algorithm have been investigated in previous chapters. In this chapter, further enhancements to PeX will be investigated and concluding statements about flux sensitivity will be made. Some future enhancements to PeX could be changing the pixel arrangements in the camera, which allows more pixels to be added to the current sized camera and reducing the size of the pixel for PeX. The results will be displayed in terms of the final flux sensitivity. The flux sensitivity best represents the improvements in results since it incorporates all parameters which have been previously investigated. As a reminder, the current optimised PeX cell is shown in Table~\ref{table:stand_config}. 


\begin{table}[h]
\centering
\begin{tabular}{lrrrrrrrrrr}
\hline
Parameters & Standard configuration\\
\hline
Number of Telescopes & 5 \\
Pixels & 804 \\
Telescope Separation & 500\rm{m} \\
Triggering Combination & (6\textit{pe}, 3) \\
Cleaning Combination & (8\textit{pe}, 4\textit{pe}) \\
image \textit{size} Cut & 60 \textit{pe} \\
Altitude & 0.22\rm{km} \\
Time Cut & $\pm$5 time cleaning cut\\
Reconstruction & Algorithm 3 \\
\hline
\end{tabular}
\captionsetup{width=13cm}  
\caption{The configuration for the standard PeX cell.}
 \label{table:stand_config}
\end{table}

\section{Pixel size and pixel arrangement}
 \label{sec:pix_arrange}

The standard PeX camera consists of 804 square pixels arranged into a square grid arrangement with pixel size 0.24$^{\circ}$. The square grid provides only 4 axes of symmetry which could have an adverse effect on image parameterisation (Figure~\ref{fig:arrangement}). A circular pixel with a circular photocathode could be arranged easily into a hexagonal arrangement (Figure~\ref{fig:arrangement2}). The image parameterisation should improve given there are 6 axes of symmetry available. By keeping the total camera area the same, the new circular pixels arranged into a hexagonal pattern also provides a camera with 925 pixels compared to the 804 pixels in the standard camera.

\begin{figure}
\begin{centering}
\includegraphics[scale=0.6]{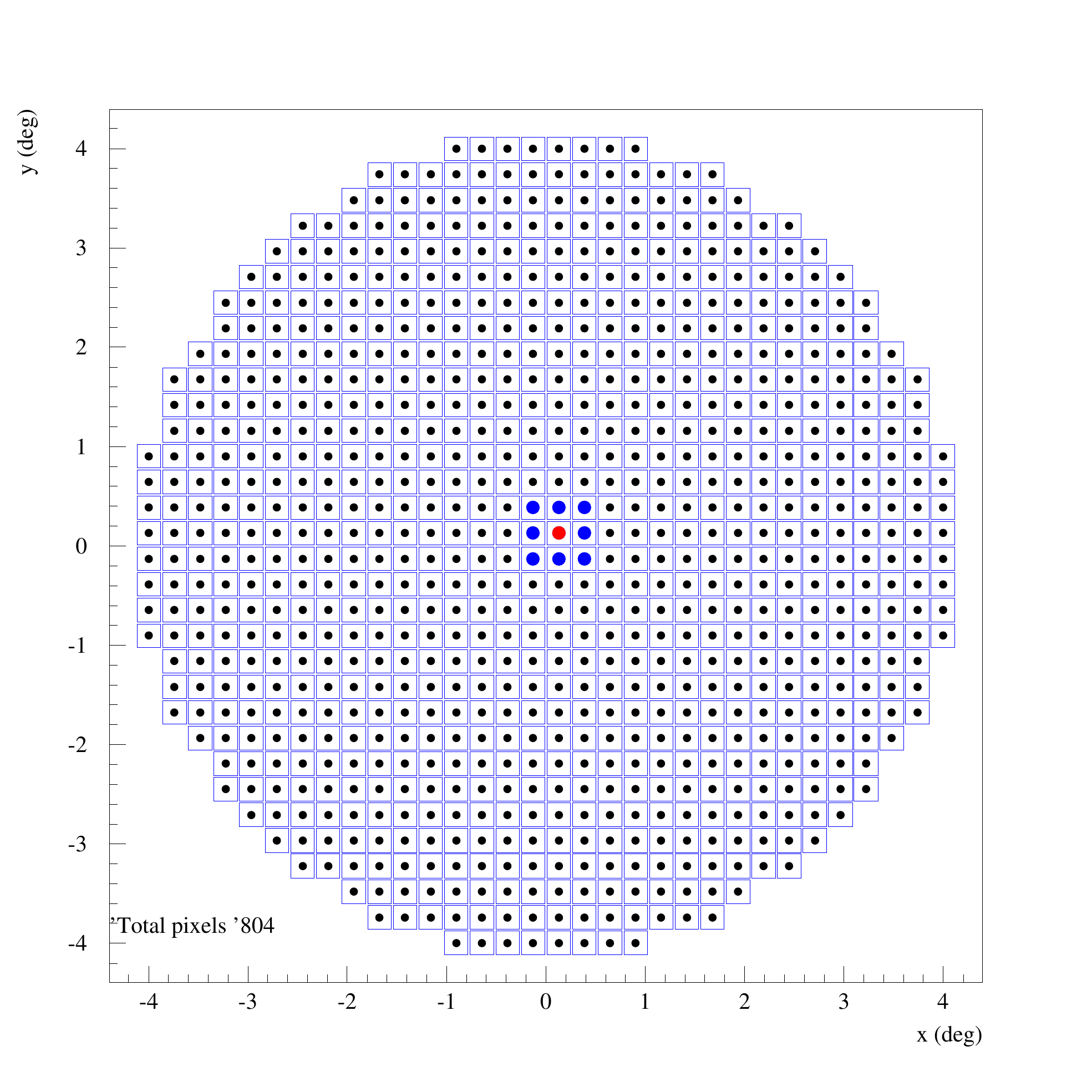}
\captionsetup{width=13cm} 
\caption{The pixel arrangement for the standard 5$^{\circ}$ field of view square grid camera with square pixels. The larger red circle represents a picture pixel and the larger blue circles represent the surrounding boundary pixels.}
 \label{fig:arrangement}
\end{centering}
\end{figure}

\begin{figure}
\begin{centering}
\includegraphics[scale=0.6]{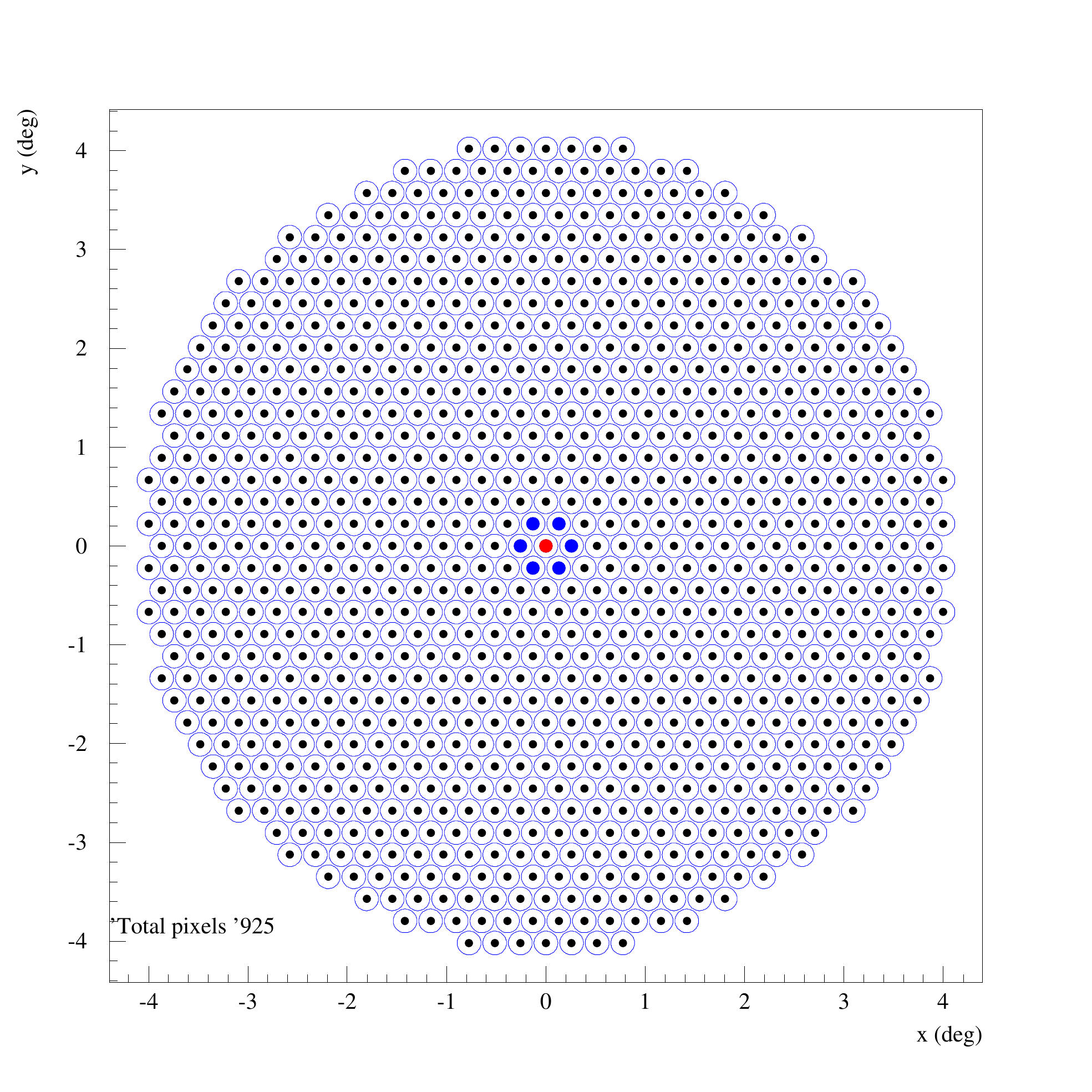}
\captionsetup{width=13cm} 
\caption{The pixel arrangement for the new 5$^{\circ}$ field of view hexagonal camera with the circular pixels. The thicker red circle represents a picture pixel and the thicker blue circles represent the surrounding boundary pixels.}
 \label{fig:arrangement2}
\end{centering}
\end{figure}

The area or solid angle of the circular pixel is slightly smaller than the square pixel. Therefore, the NSB per pixel, the cleaning combination and the triggering combination need to be adjusted for the circular pixel. The values are scaled in the same way as the H.E.S.S. values were scaled to provide the PeX values in section~\ref{sec:optimise_low}. The image \textit{size} cut will be left at 60\textit{pe} to make sure the quality of the images remains essentially the same. However, future studies could consider adjusting the \textit{size} cut.





\begin{figure}
\begin{centering}
\includegraphics[scale=0.7]{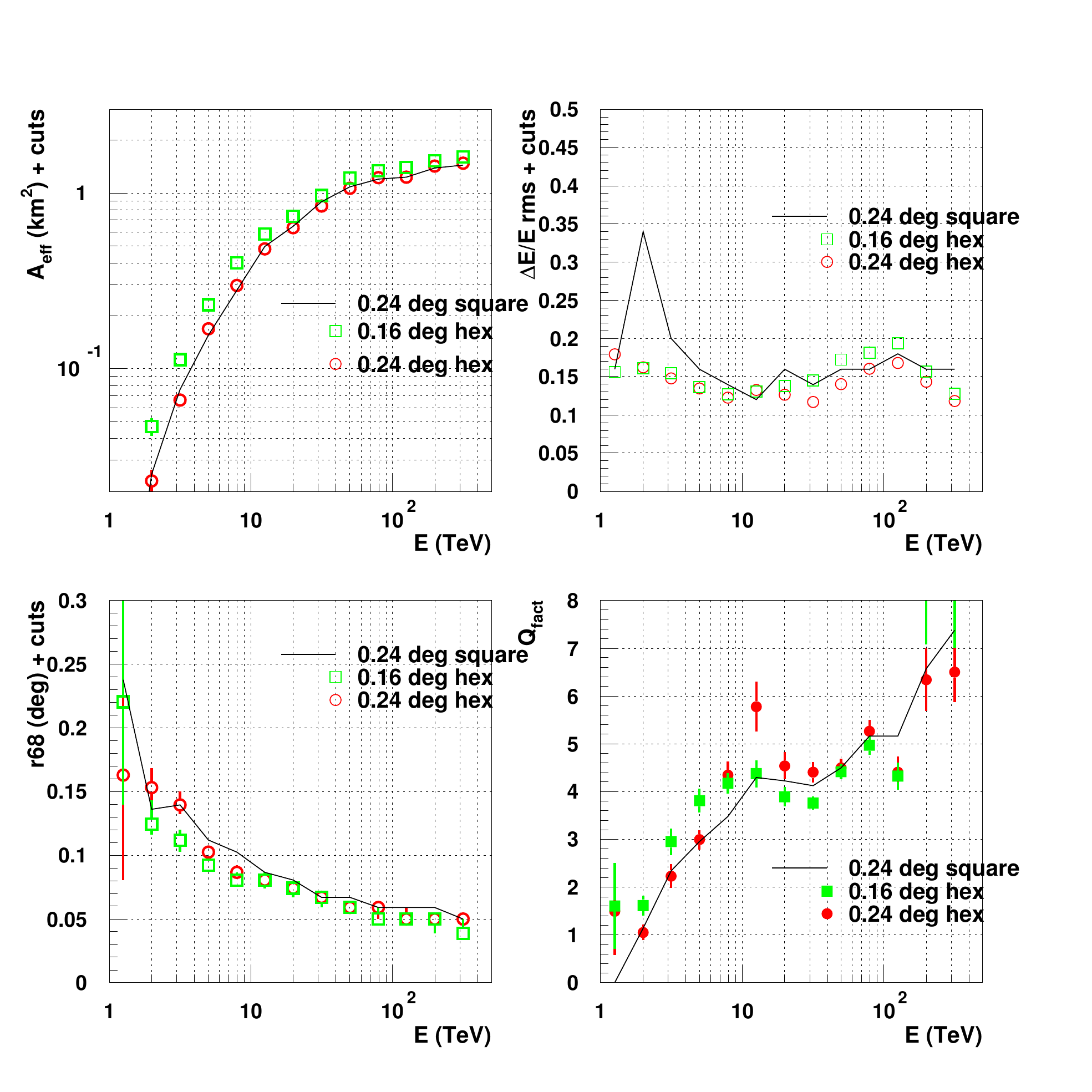}
\captionsetup{width=13cm}  
\caption{Post-selection cut effective area, post-shape cut energy resolution, post-shape cut angular resolution (Deg) and the post-shape cut Q$_{fact}$ for different camera arrangements for Algorithm 3. The line represents the standard camera using a 0.24$^{\circ}$ square pixels in a square grid arrangement, the red circles represent a new camera using 0.24$^{\circ}$ circular pixels in a hexagonal arrangement and the green squares represent a new camera using 0.16$^{\circ}$ circular pixels in a hexagonal arrangement. The error bars on the black line are of a similar magnitude to those of the red points.}

 \label{fig:pixel_arrangement_hex}
\end{centering}
\end{figure}


As well as changing the shape of the pixels, the size of the pixels can be altered. A smaller 0.16$^{\circ}$ circular pixel arranged into a hexagonal arrangement will be considered. The current PeX pixel is 0.24$^{\circ}$ and the current H.E.S.S. pixel is 0.16$^{\circ}$. Investigating pixel size has been considered for CTA since it will consist of 3 different sized arrays focusing on 3 different energy ranges \cite{CTAdesign}. If the pixel size decreases, then the reconstruction from small size images may improve especially for the large core distance events. To test the effect of decreasing the pixel size, simulations using a 0.16$^{\circ}$ pixel were conducted. Since the area or solid angle of the pixel has decreased, the triggering and cleaning combinations and NSB per pixel were scaled down to appropriate values (Table~\ref{table:pixel_value}), by the ratio of solid angles or pixel areas. Table~\ref{table:pixel_value} shows the triggering combination, cleaning combination, image \textit{size} cut and NSB level for each camera after the values have been scaled. The triggering and cleaning values have decreased according to the level of NSB in the smaller pixels. The image \textit{size} cut has been left at 60\textit{pe} since we still want the same minimum sized images to be present in the camera. Since this is a preliminary study the \textit{size} cut can be adjusted or reoptimised for the different sized pixels.

\begin{table}[h]
\centering
\begin{tabular}{lrrrrrrr}
\hline
parameter & 0.24$^{\circ}$ & 0.24$^{\circ}$ & 0.16$^{\circ}$ \\
\hline
Pixel shape & square & circular & circular \\
Arrangement & square & hexagonal & hexagonal \\
NSB (\rm{pe (ns pixel)$^{-1}$}) & 0.045 & 0.035 & 0.016 \\
Cleaning combination & (8\textit{pe}, 4\textit{pe}) & (6\textit{pe}, 3\textit{pe}) & (4\textit{pe}, 2\textit{pe}) \\
Triggering combination & (6\textit{pe}, 3) & (6\textit{pe}, 3) & (4\textit{pe}, 2) \\
image \textit{size} cut & 60\textit{pe} & 60\textit{pe} & 60\textit{pe} \\
Total pixels & 804 & 925 & 1921 \\
\hline
\end{tabular}
\captionsetup{width=13cm}  
\caption{A table showing all the parameters adjusted to suit three different camera configurations.}
 \label{table:pixel_value}
\end{table}
	
Considering the mirror PSF function in Figure~\ref{fig:abberation}, the PSF at the edge of the camera is equivalent to the size of the standard 0.24$^{\circ}$ pixel. If the pixel size is reduced to a value smaller than 0.16$^{\circ}$ then the PSF will be larger than the pixel size. The optics of the telescope should be improved before a smaller pixel size is seriously considered. 

Figure~\ref{fig:pixel_arrangement_hex} illustrates the results for varying pixel size over the four main parameters plus comparisons of a hexagonal vs square arrangement camera. The figure shows the configurations displayed in Table~\ref{table:pixel_value}. 

The triggering threshold values produce similar accidental single telescope trigger rates from NSB for all pixel sizes. The rough values for single pixel trigger rate and accidental telescope trigger rate are 10$^{3}$ Hz and 10$^{-2}$ Hz respectively for the standard camera \cite{Ricky} (section~\ref{sec:trigger_def}).




The top left panel in Figure~\ref{fig:pixel_arrangement_hex} represents the post-selection cut effective area. The results for the 0.16$^{\circ}$ pixel provides a slightly larger post-selection cut effective area. The top right panel represents the energy resolution RMS for the three camera configurations. The energy resolution does not appear to show a huge dependance on pixel shape and camera arrangements but the image \textit{size} cut could be a factor. The smaller sized pixels allow more events to pass the triggering condition and shape cuts, which is seen in Figure~\ref{fig:pixel_arrangement_hex} top left panel. A slightly larger image \textit{size} cut could provide an improved energy resolution.

The bottom left panel shows the angular resolution. The angular resolution for the camera using a 0.24$^{\circ}$ circular pixel arranged into a hexagonal arrangement provides an improvement in angular resolution over the standard camera. The smaller 0.16$^{\circ}$ circular pixel provides a similar angular resolution and slightly better at a few TeV energies. The Q$_{fact}$ in the bottom right panel shows that the best Q$_{fact}$ is obtained using a 0.24$^{\circ}$ circular pixel arranged in a hexagonal arrangement. 



The largest improvement for most parameters is gained in the 1 TeV $<$ E $<$ 50 TeV range for the 0.24$^{\circ}$ hexagonal arrangement. The hexagonal arrangement allows a more realistic shape of the image to be shown in the camera which improves the angular resolution and the separation between $\gamma$-ray and proton showers. The above energy range has more small sized images so they provide the largest gain with circular and smaller pixels. The smaller 0.16$^{\circ}$ pixels in a hexagonal arrangement provide a larger effective area, an improved angular resolution and a slightly better Q$_{fact}$ at low energies compared to the 0.24$^{\circ}$ pixel in a square grid arrangement. The hexagonal arrangement provides more axes of symmetry and the smaller pixel size provides improved Hillas parameterisation for small images. Therefore, we have shown that the hexagonal pixel arrangement provides improved performance over the square grid arrangement. 

The preliminary results indicate that the 0.16$^{\circ}$ pixel size appears to provide an improvement over the 0.24$^{\circ}$ pixel size. However, further simulations are required to re-optimise the image parameters for the new pixel size. 





\section{Applying an Algorithm 3 error cut and an n$_{tel}$ cut to PeX}

	To improve the performance as measured by effective area, energy resolution, angular resolution and Q$_{fact}$ further cuts on the events can be applied. Two ideas to improve these parameters is to apply a cut on the error associated with each reconstructed source position dervied from Algorithm 3 (\cite{Stamatescu2}, \cite{hofmann}) or to apply a cut on the number of telescopes, n$_{tel}$, required for stereoscopic reconstructions.

When reconstructing the shower with Algorithm 3, each predicted distance has an associated error ellipse which represents the uncertainty in the calculation (as discussed in more detail in section~\ref{sec:event_reconstruction}). The predicted distances are used to provide a weighted mean for the reconstructed source position. The reconstructed source position, X, is given by
\begin{equation}
\textbf{X} = \sigma \sum_a \sigma_{a}^{-1} \textbf{X}_{a}
\end{equation}
where \textit{a} represents the images which pass cuts or number of telescopes participating, \textit{$\sigma$} is the covariance matrix associated with the reconstructed source position, \textit{$\sigma_{a}$} is the matrix of weighting values or errors associated with each predictor and \textbf{\textit{X$_{a}$}} is the mean predicted position of a source predictor. The weighting values in $\sigma_{a}$ are the error values from the predicted source positions
\begin{equation}
\sigma_{a} = 
\left(
\begin{array}{cc}
   \sigma_{a_{x}}^{2} & \sigma_{a_{xy}}\\
   \sigma_{a_{xy}} & \sigma_{a_{y}}^{2}\\
\end{array}
\right)
\end{equation}

The errors for all source predictors are propagated into a single covariance matrix, $\sigma$. The covariance matrix can be used to provide the error on the reconstructed source position. The correct way to show the error is by taking the determinant of the covariance matrix. 

\begin{equation}
\sigma = \frac{1}{D}
\left(
\begin{array}{cc}
   \sum_a {\sigma_{a_{x}}^{2}}/{det(\sigma_{a})} & \sum_a {\sigma_{a_{xy}}}/{det(\sigma_{a})}\\
   \sum_a {\sigma_{a_{xy}}}/{det(\sigma_{a})} & \sum_a {\sigma_{a_{y}}^{2}}/{det(\sigma_{a})}\\
\end{array}
\right)
\end{equation}
where
\begin{equation}
D = \sum_a \frac{\sigma_{a_{x}}^{2}}{det(\sigma_{a})} \sum_a \frac{\sigma_{a_{y}}^{2}}{det(\sigma_{a})} - (\sum_a \frac{\sigma_{a_{xy}}}{det(\sigma_{a})})^{2}
\end{equation}

Since the matrix contains variances, the error will come from $\sigma_{x}$ = $\sqrt{det{\;\sigma}}$ which is the standard deviation for the reconstruction shower direction from Algorithm 3. Importantly, this error gives an indication of the quality of the shower reconstruction. Therefore, applying a cut on this error could improve the parameters such as angular resolution or Q$_{fact}$ and this has been demonstrated by HEGRA \cite{hofmann}. Figure~\ref{fig:alg3_det_g_v_p} represents the Algorithm 3 error distributions for both $\gamma$-rays and protons in four different energy bands after shape cuts (MSW, MWL and MSNpix). The cut on Algorithm 3 error has been chosen as 0.06$^{\circ}$ since a marginal separation can be seen between $\gamma$-ray and proton events (Figure~\ref{fig:alg3_det_g_v_p} and Figure~\ref{fig:alg3_det_g_v_p_all}). The separation in Algorithm 3 error is due to the lower telescope multiplicity for proton events which worsens the event reconstruction and produces a large error on the predicted distance. For $\gamma$-rays, the telescope multiplicity is larger, which improves the event reconstruction and lowers the error on the predicted distance (Figure~\ref{fig:multi_tel}).

\begin{figure}
\begin{centering}
\includegraphics[scale=0.7]{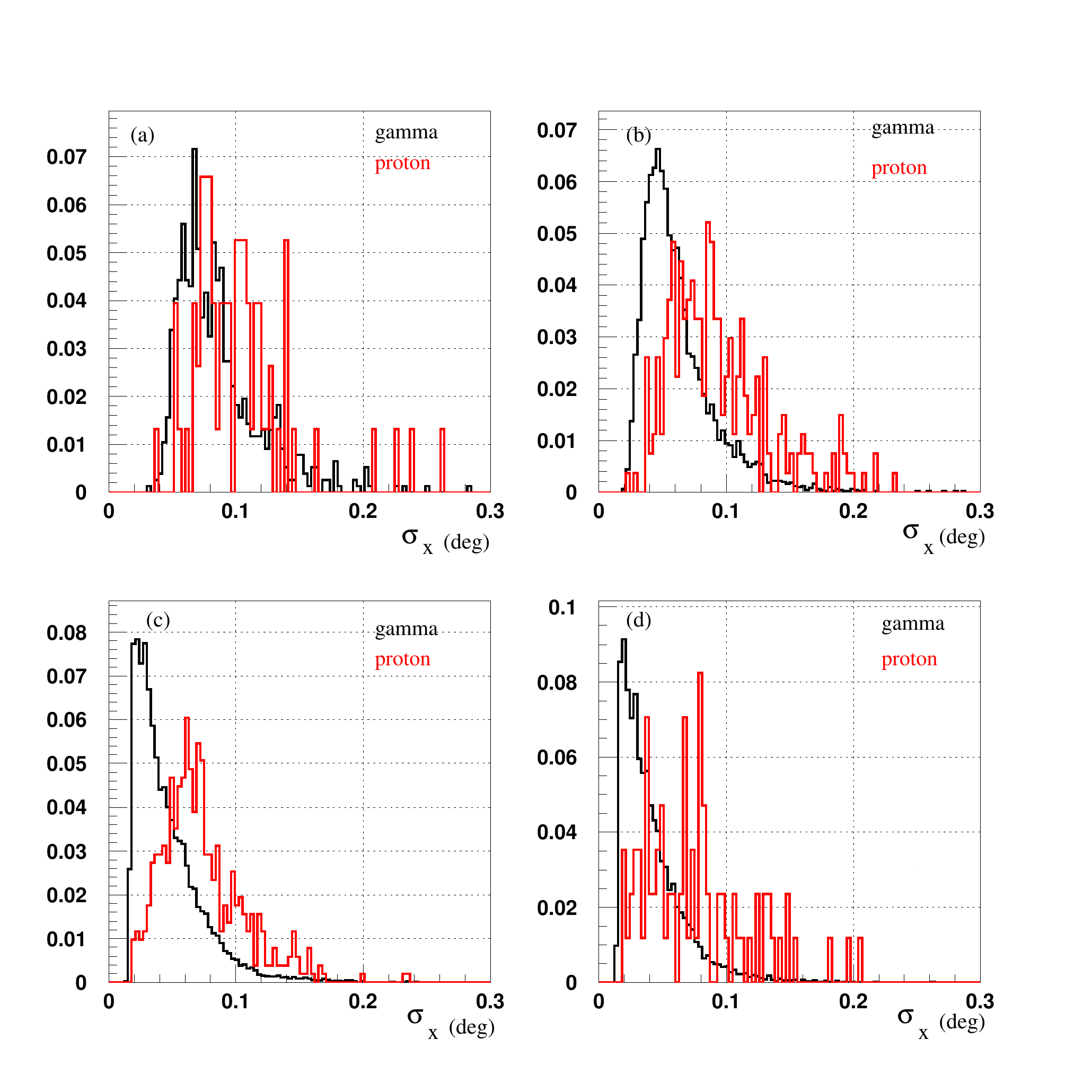}
\captionsetup{width=13cm}  
\caption{Post-shape cut distribution of Algorithm 3 error values for $\gamma$-ray and proton events. The results have been split into four energy bands: 1 - 4.7 TeV (a), 4.7 - 22.3 TeV (b), 22.3 - 105.7 TeV (c) and 105.7 - 500 TeV (d).
}
 \label{fig:alg3_det_g_v_p}
\end{centering}
\end{figure}

A cut on the number of telescopes required for an event to be reconstructed can also be applied. The standard number of telescopes required for an event to be reconstructed is usually 2 telescopes or n$_{tel}$ $\geq$ 2. A higher n$_{tel}$ cut can be placed over all energies or the cut can differ in different energy bands. The higher energy events have larger sized images and hence might be able to handle a larger n$_{tel}$ cut. Increasing the number of images required to reconstruct an event should provide improvements to angular resolution, energy resolution and Q$_{fact}$. Figure~\ref{fig:multi_tel} bottom right panel in Chapter 3 shows the telescope multiplicity as a function of energy. The results show that below 10 TeV, the average telescope multiplicity is less than 3. For these events, no telescope multiplicity cut can be applied. A cut on n$_{tel}$ only for E $>$ 10 TeV should improve results. 

\begin{figure}
\begin{centering}
\includegraphics[scale=0.7]{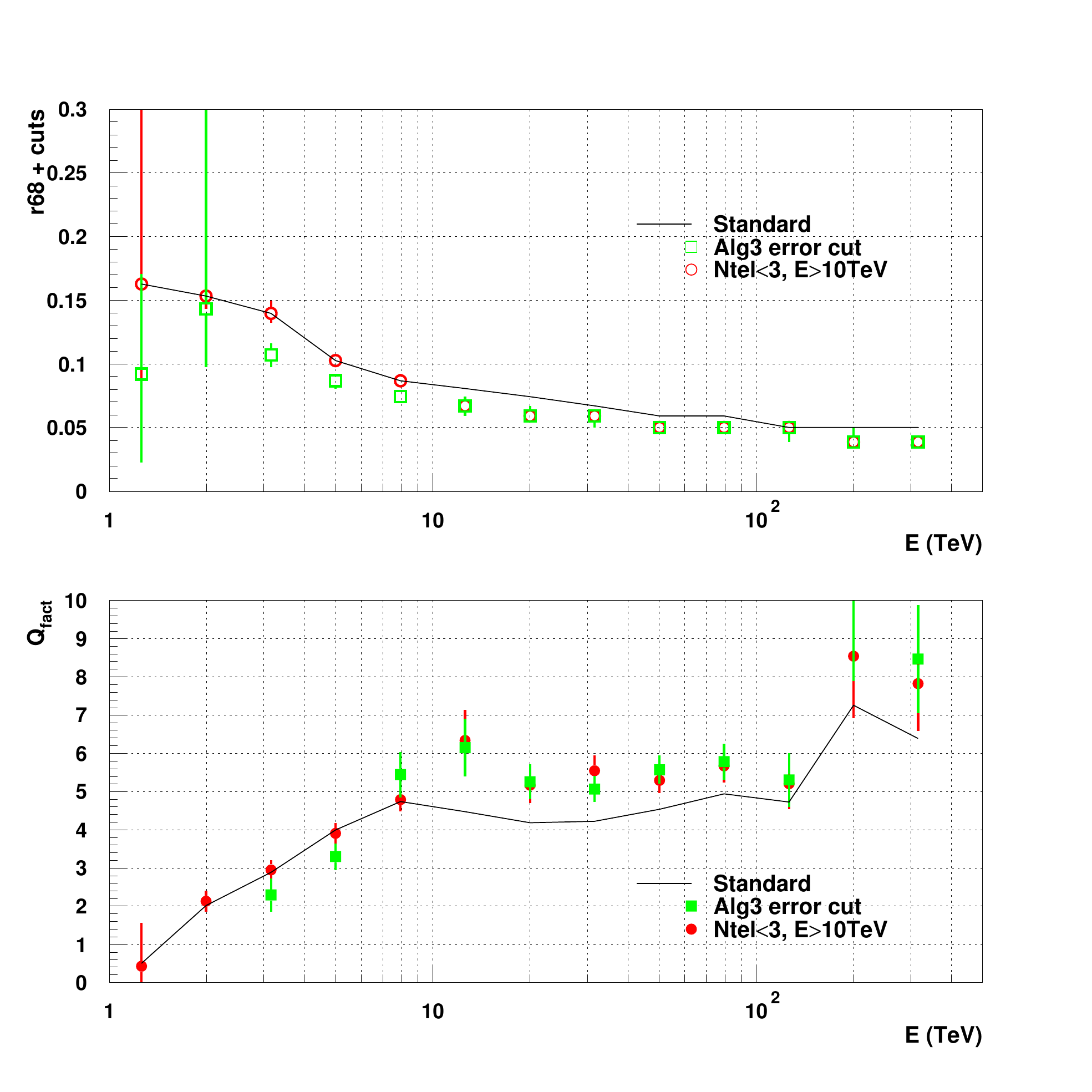}
\captionsetup{width=13cm}  
\caption{The post-cut angular resolution in degrees and post-shape cut Q$_{fact}$ for Algorithm 3 using hexagonally arranged 0.24$^{\circ}$ pixels. The figures show three results: a standard configuration (black), a standard configuration with an n$_{tel}$ $>$ 2 for E $>$ 10 TeV cut (red) and a standard configuration plus a 0.06$^{\circ}$ cut on the error in Algorithm 3 reconstruction (green).}

 \label{fig:q-factor_ang_cut}
\end{centering}
\end{figure}

Figure~\ref{fig:q-factor_ang_cut} shows the results after applying a cut on the number of telescopes and applying a cut on the error in Algorithm 3 reconstruction. The angular resolution and Q$_{fact}$ show some an improvement with an n$_{tel}$ or Algorithm 3 error cut. The angular resolution shows a small improvement in the $\sim$ 5 to 100 TeV range when using an n$_{tel}$ cut and the cut on the Algorithm 3 error. 

The angular resolution improves since both an n$_{tel}$ cut and a cut of the Algorithm 3 error select the better quality images. The event reconstruction improves with number of telescopes and a smaller Algorithm 3 error implies an improved event reconstruction.


The Q$_{fact}$ results are displayed in Figure~\ref{fig:q-factor_ang_cut} bottom panel. The cut on the Algorithm 3 error could be too strong for events with E $<$ 10 TeV (green squares). Since it was shown that telescope multiplicity is less than 3 for E $<$ 10 TeV, the event reconstruction is usually poorer. Therefore, a larger Algorithm 3 error is predicted. No n$_{tel}$ cut has been applied for E $<$ 10 TeV, so there is no change in results (red circles). Applying a cut on n$_{tel}$ or on the error in Algorithm 3 reconstruction appears to provide a significant improvement between 10 and 100 TeV. For E $>$ 100 TeV, there is no difference between the results. The cut on Algorithm 3 error could be optimised for each energy bin, since applying one cut for all energies appears to be too strong at low energies.



Overall, the results indicate that a cut on Algorithm 3 error or an n$_{tel}$ cut appear to provide improvements to the reconstruction and rejection power. The biggest improvement is seen between 10 and 100 TeV. The Algorithm 3 error cut could be optimised for smaller energy bands and the current cut of 0.06$^{\circ}$ is chosen for demonstration purposes only. 

\subsection{Cut on energy resolution}

As a side note, CTA \cite{CTAdesign} utilises a cut on the energy resolution per energy bin. Due to limited time, this study was not part of this thesis but Figure~\ref{fig:energy_resolution_comp} shows the RMS of the energy resolution per telescope for $\gamma$-rays and protons after shape cuts. The RMS of energies for the sample of triggered energies is given by
\begin{eqnarray}
RMS = \frac{1}{n_{tel}} \sum_i^{n_{tel}} \sqrt{\left(\frac{E_{i} - E_{recon}}{E_{recon}}\right)^2}
\end{eqnarray}
where E$_{i}$ is the expected energy for each telescope from look-up tables and E$_{recon}$ is the reconstructed energy for each event. The distributions for $\gamma$-ray and proton events shows that no further separation can be gained by applying a cut on the RMS of energies from each telescope, but further investigations are warranted.



\begin{figure}
\begin{centering}
\includegraphics[scale=1.0]{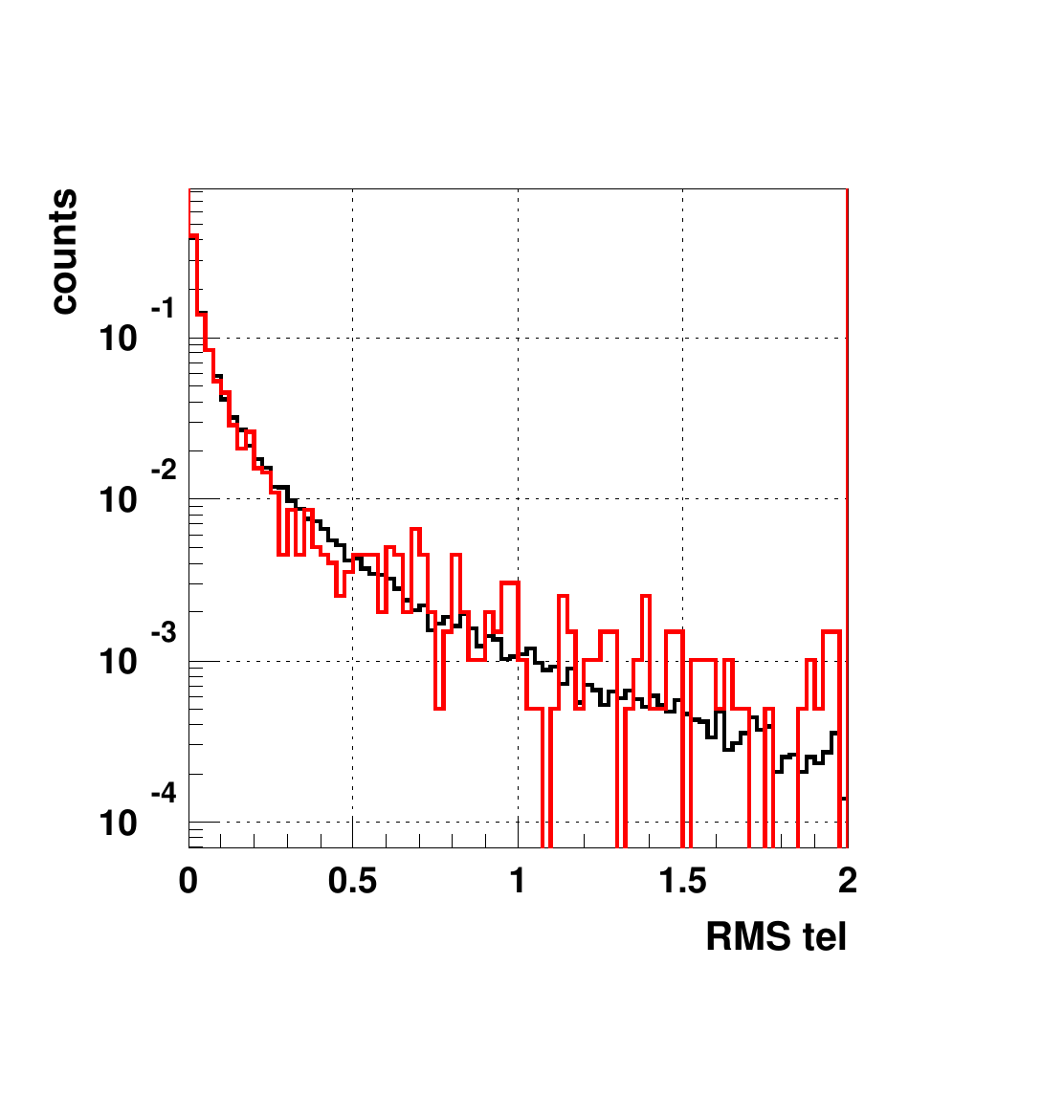}
\captionsetup{width=13cm}  
\caption{Post-shape cut RMS values for $\Delta{E}/E$ from each telescope for all event energies. The black line represents the $\gamma$-ray events and the red line represents the protons events. No further separation could be achieved by applying a cut on this value.}

 \label{fig:energy_resolution_comp}
\end{centering}
\end{figure}

\section{FADC trace signal separation between $\gamma$-rays and protons}
\label{sec:FADC}
	
There are differences between the time development along the shower axis for $\gamma$-ray and proton EAS. Recall Figure~\ref{fig:cherenkovshower} illustrates that proton showers have larger lateral distributions than $\gamma$-ray showers since the Cherenkov emission from a proton showers comes from a larger area which produces large Cherenkov light cones. 

Stamatescu \cite{Stamatescu} showed that $\gamma$-ray and protons showers provide different FADC traces. The wider FADC signal for protons is due to the muons present in proton showers. The proton events have longer shower developments along the major axis, which creates a wider or longer time profile in the FADC. It occurs more at large core distances since the muons may be visible more for smaller sized images. The time development shows the production of muons in a proton shower (Figure~\ref{fig:victor_time_height} right panel) and that their light precedes light from the main part of the shower. If the muon signal in the FADCs is predominant \cite{Stamatescu}, then considering FADC traces could be important. Therefore, further separation of $\gamma$-ray and proton showers could be obtained by looking at the pulse shapes of EAS in the FADCs.

To compare $\gamma$-ray and proton events, we consider the arrival times of photons in the FADCs. The events are split into narrow core distance and image \textit{size} bands. For core distance, the bands are: 100 to 120 \rm{m}, 200 to 220 \rm{m}, 300 to 320 \rm{m}, 400 to 420 \rm{m}, 500 to 520 \rm{m} and 600 to 620 \rm{m}. For the image \textit{size}, the bands are: 450 to 500 \textit{pe}, 1000 to 1120 \textit{pe}, 2840 to 3160 \textit{pe} and 6350 to 7050 \textit{pe}. These particular size and core bands are chosen to correspond to the bin sizes used in the error look-up tables mentioned in section~\ref{sec:event_reconstruction}. There are too many core and size bands to consider each bin in the look-up tables so a few different combinations have been chosen. Therefore, any image with a size between 6350 and 7050 \textit{pe} will use the same look-up values. Only the 1000 to 1120 \textit{pe} size band will be displayed since it highlights the reason for considering the FADC traces. The other size band results have been placed in Appendix~\ref{sec:appendix_plot} (Figures~\ref{fig:size_200_comparison_pe_norm_cut} to~\ref{fig:size_6000_comparison_pe_norm_cut}).

	To align the arrival times of each event, the shower trigger time is used as a time standard. The pulse has been shifted by 15\rm{ns} so that it can be seen clearly. The zero position on the x-axis represents the trigger time - 15\rm{ns}. Multiple events are summed up over all pixels in the image to obtain an average time distribution of the signal in the FADC from $\gamma$-ray and proton events (Figure~\ref{fig:size_1000_comparison_pe_norm} and Figure~\ref{fig:size_1000_comparison_pe_norm_cut}). By comparing these FADC signals, the events can be separated at the earliest stage in the detection process. Figure~\ref{fig:size_1000_comparison_pe_norm} illustrates the FADC signals for images sizes between 950 to 1050\textit{pe} at varying core distances. The red curves represents the $\gamma$-ray events while the blue dotted lines represent the proton events. The FADC signals in Figure~\ref{fig:size_1000_comparison_pe_norm} illustrates the event signal with the off-plane level of NSB and after the signal has undergone the cleaning algorithm. There appears to be a small separation between the $\gamma$-ray and proton events. The proton events appear to have slightly wider pulses so applying a fit to the FADC signal could help further separation of the events. Also, any differences between proton and $\gamma$-ray showers are more visible at large core distances since the showers are more elongated. As the core distance of the shower increases, the images get narrower in width and longer in length which will accentuate the small difference in time profiles seen at smaller core distances. This effect is shown in Figure~\ref{fig:core_distance3}, the angles $\phi_{B}$ and $\phi_{A}$ become larger so the image becomes elongated. However, it is best to see if any difference still exists after the standard shape cuts have been applied. The standard shape cuts remove a large majority of proton images and any residual difference on FADC traces can be investigated as evidence of independent information not exploited by the shape cuts.

\begin{figure}
\begin{centering}
\includegraphics[scale=0.5]{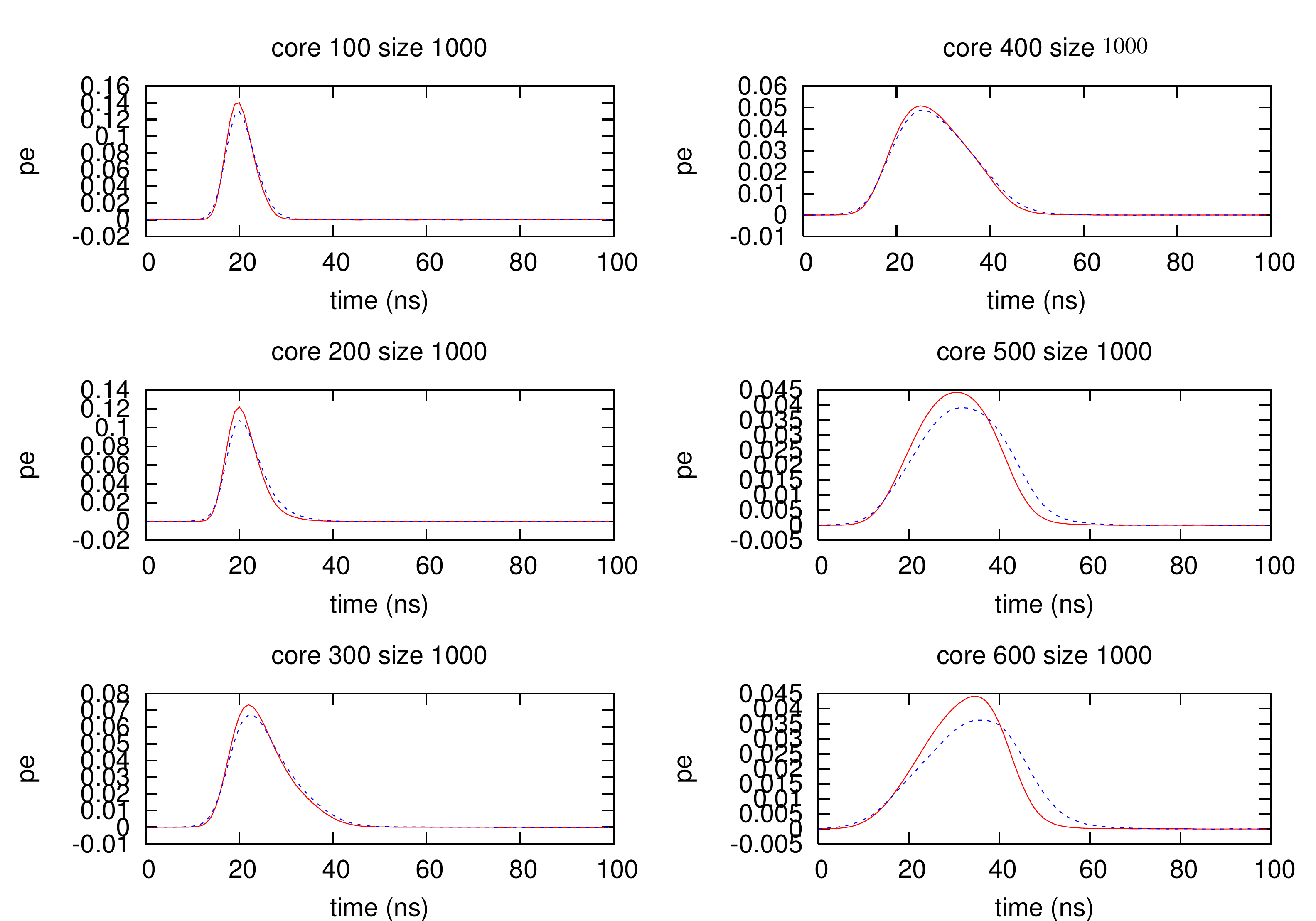}
\captionsetup{width=13cm}  
\caption{Average FADC signal for images split into core distance bands (see text for bands). The red line represents the $\gamma$-ray images and the blue dashed line represents proton images. The results here are for images with a size of 1000 \textit{pe}. The images have a standard level of NSB and are post cleaning. The pulse has been shifted by 15\rm{ns} so that it can be seen clearly. The zero position on the x-axis represents the trigger time - 15\rm{ns}. A small difference is seen between the normalised $\gamma$-ray and proton pulse shapes for most core distances.}

 \label{fig:size_1000_comparison_pe_norm}
\end{centering}
\end{figure}

Figure~\ref{fig:size_1000_comparison_pe_norm_cut} illustrates the same as Figure~\ref{fig:size_1000_comparison_pe_norm} but after applying post-shape cuts. The red lines represent the post-shape cut $\gamma$-ray events and the blue dotted lines represent the post-shape cuts proton events. Comparing the curves in Figure~\ref{fig:size_1000_comparison_pe_norm_cut}, it can be noticed that the small difference in the pre-shape cut results from Figure~\ref{fig:size_1000_comparison_pe_norm} no longer exist. Applying shape cuts has removed most non $\gamma$-ray like proton events, leaving the $\gamma$-ray like proton events which are hard to separate from true $\gamma$-ray events. A cut on the pulse shape after shape cuts appears to provide no improvement to the rejection power of PeX. We can conclude that there is no need to apply cut on the shape of the FADC signal. 

\begin{figure}
\begin{centering}
\includegraphics[scale=0.5]{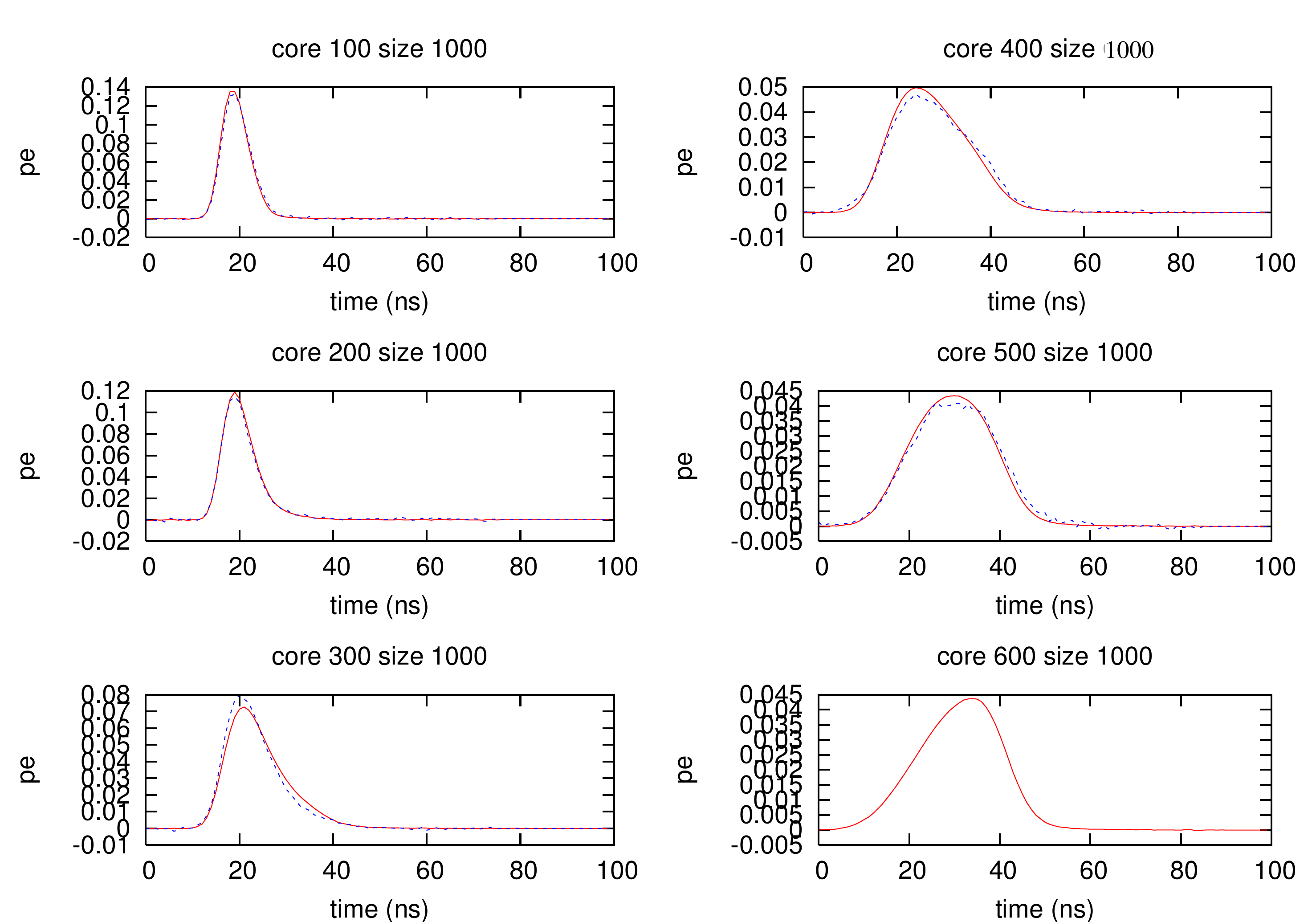}
\captionsetup{width=13cm}  
\caption{Same as for Figure~\ref{fig:size_1000_comparison_pe_norm} but post-shape cuts.}

 \label{fig:size_1000_comparison_pe_norm_cut}
\end{centering}
\end{figure}


\section{Flux Sensitivity}

In this last section we look at the flux sensitivity for PeX for a range of setup parameters; 50 hours, greater than 10 events per energy bin, a 5 $\sigma$ significance and 5 bins per energy decade. To test that our flux sensitivity calculation is reliable, we replicated the H.E.S.S. flux sensitivity. The H.E.S.S. flux sensitivity for 25 hour observations has been obtained from the H.E.S.S. website \cite{Hessflux} and is displayed in Figure~\ref{fig:flux_paper}. We must note that this flux is calculated using events at zenith, while our simulation flux is calculated using events at 30$^{\circ}$. This difference in zenith angle will definitely cause discrepancies between flux sensitivities.

\begin{figure}
\begin{centering}
\includegraphics[scale=0.8]{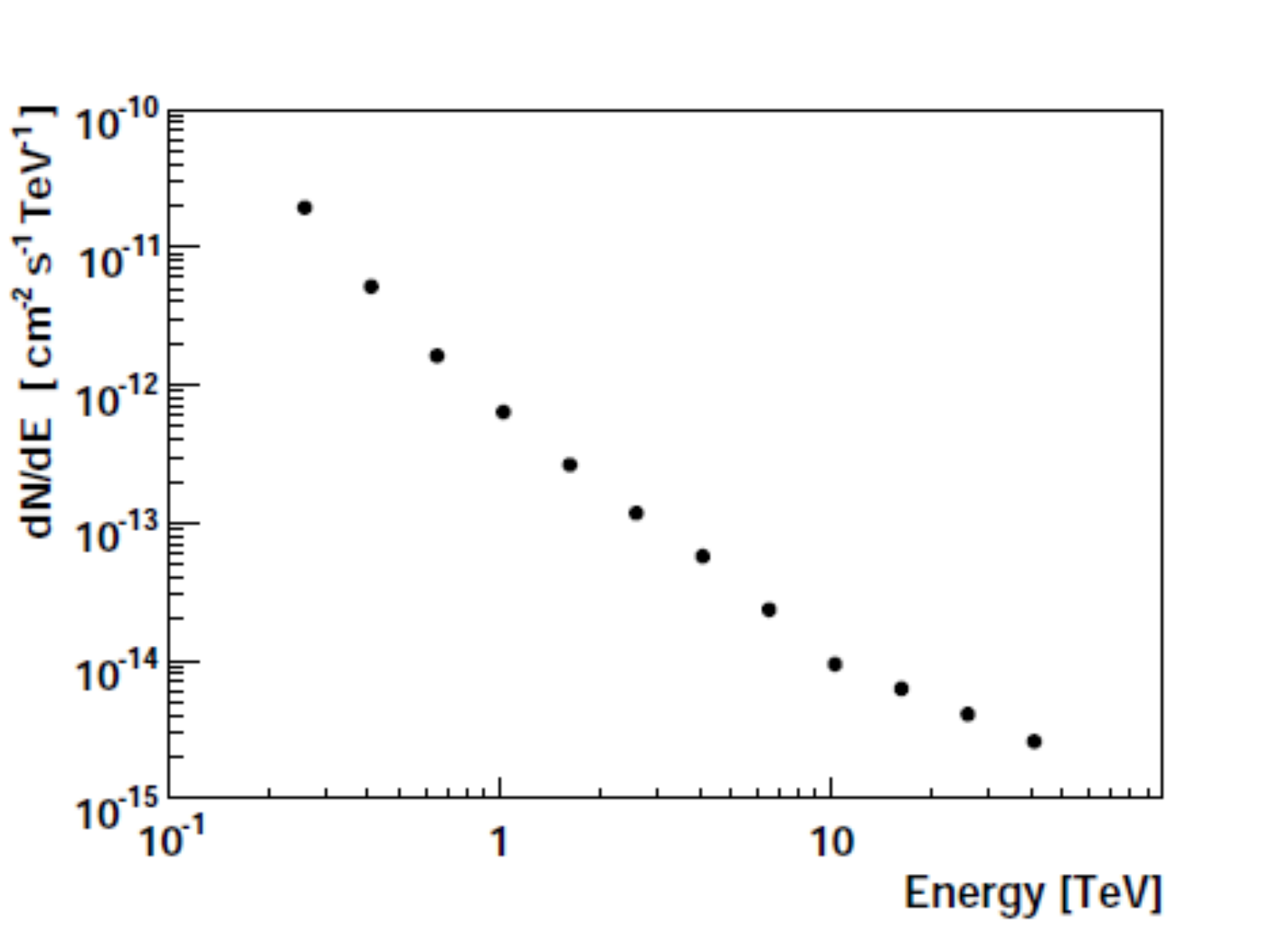}
\captionsetup{width=13cm}  
\caption{The H.E.S.S. flux sensitivity after hard cuts with 5 bins per energy decade, 25 hour exposure time, 5 $\sigma$ significance and for more than 10 events (from H.E.S.S. website \cite{Hessflux}) This flux has been produced from events at zenith.}

 \label{fig:flux_paper}
\end{centering}
\end{figure}

This flux sensitivity can be converted to units more relevant to our calculations. To convert the flux sensitivity from a differential flux in (cm$^{2}$ s TeV)$^{-1}$ binned at 5 bins per decade to a differential flux in ergs (cm$^{2}$ s)$^{-1}$, we multiply the flux by 1.602 E$^{2}$. The flux sensitivity for H.E.S.S., HAWC and CTA are calculated using 50 hours of observation time, so to convert the 25 hour exposure to a 50 hour exposure we need to multiply the flux sensitivity by $\sqrt{0.5}$ since it is related to the exposure time by $\sqrt{t}$. So the new flux sensitivity becomes:

\begin{eqnarray}
1.602\;E^{2}\;\sqrt{0.5}\; dN/dE\; \; [\rm{ergs\;cm^{-2} s^{-1}}]
\end{eqnarray}

As well as using this simulated flux sensitivity, the flux sensitivity can be calculated from observational data taken from a measured point source, in this case, the high mass X-ray binary LS 5039 \cite{Aharonian6}. Hard cuts, including a 200\textit{pe} \textit{size} cut, were applied to the LS 5039 data set. The LS\,5039 flux and errors were converted to a 50 hour exposure and 5 sigma significance. 
Both of these flux sensitivity calculations for H.E.S.S. will be compared to the flux sensitivity calculated from our simulations. To calculate our flux sensitivity, we use the Li and Ma equation (Eq~\ref{eqn:sig_li_ma}) for significance. 

If the background flux is known then we can calculate the minimum detectable $\gamma$-ray flux above the background flux. To do this, we iterate over a simple power law which represents the $\gamma$-ray flux, $F_{\gamma}$. Using the background flux and the first iteration of the $\gamma$-ray flux, we calculate the number of on-source and off-source counts (Eq~\ref{eqn:ngam} and~\ref{eqn:nprot}). The $\alpha$ value is typically 0.2, which is equivalent to 5 background regions per on-source region. The Li and Ma equation (Eq~\ref{eqn:sig_li_ma}) can be used to provide the signal significance. We continue to iterate over a simple power law until the signal significance reaches 5$\sigma$. Once the 5$\sigma$ significance is reached, it indicates that the minimum detectable $\gamma$-rays flux for PeX has been obtained. So the final F$_{\gamma}$ would be the flux sensitivity. 

For our simulations, the standard H.E.S.S. configuration with hard cuts has been used (Table~\ref{table:hess_config}).

\begin{table}[h]
\centering
\begin{tabular}{lrrrrrrrrrr}
\hline
Parameters & H.E.S.S. configuration\\
\hline
Mirror Area & 107 m$^{2}$ \\
Pixels & 960 \\
Field of View & 5$^{\circ}$ \\
Number of Telescopes & 4 \\
Telescope Separation & 120 \rm{m} \\
Cleaning Combination & (10\textit{pe}, 5\textit{pe}) \\
Trigger Combination & (5.3\textit{pe}, 3) \\
image \textit{size} Cut & 200 \textit{pe} \\
Reduced Scaled Width & $<$ 0.9 \\
Reduced Scaled Length & $<$ 2.0 \\
\hline
\end{tabular}
\captionsetup{width=13cm}  
\caption{The H.E.S.S. configuration used to calculate the flux sensitivity.}
 \label{table:hess_config}
\end{table}

Hard cuts use a \textit{size} cut of 200 \textit{pe} and reduced mean scaled width and length values values \cite{RMSW}. Plotting the H.E.S.S. flux sensitivity in Figure~\ref{fig:hess_sensitivity_comparison}, our simulated sensitivity appears to match that from \cite{Hessflux} and from LS 5039. This gives us confidence in our calculation. 

\begin{figure}
\begin{centering}
\includegraphics[scale=0.7]{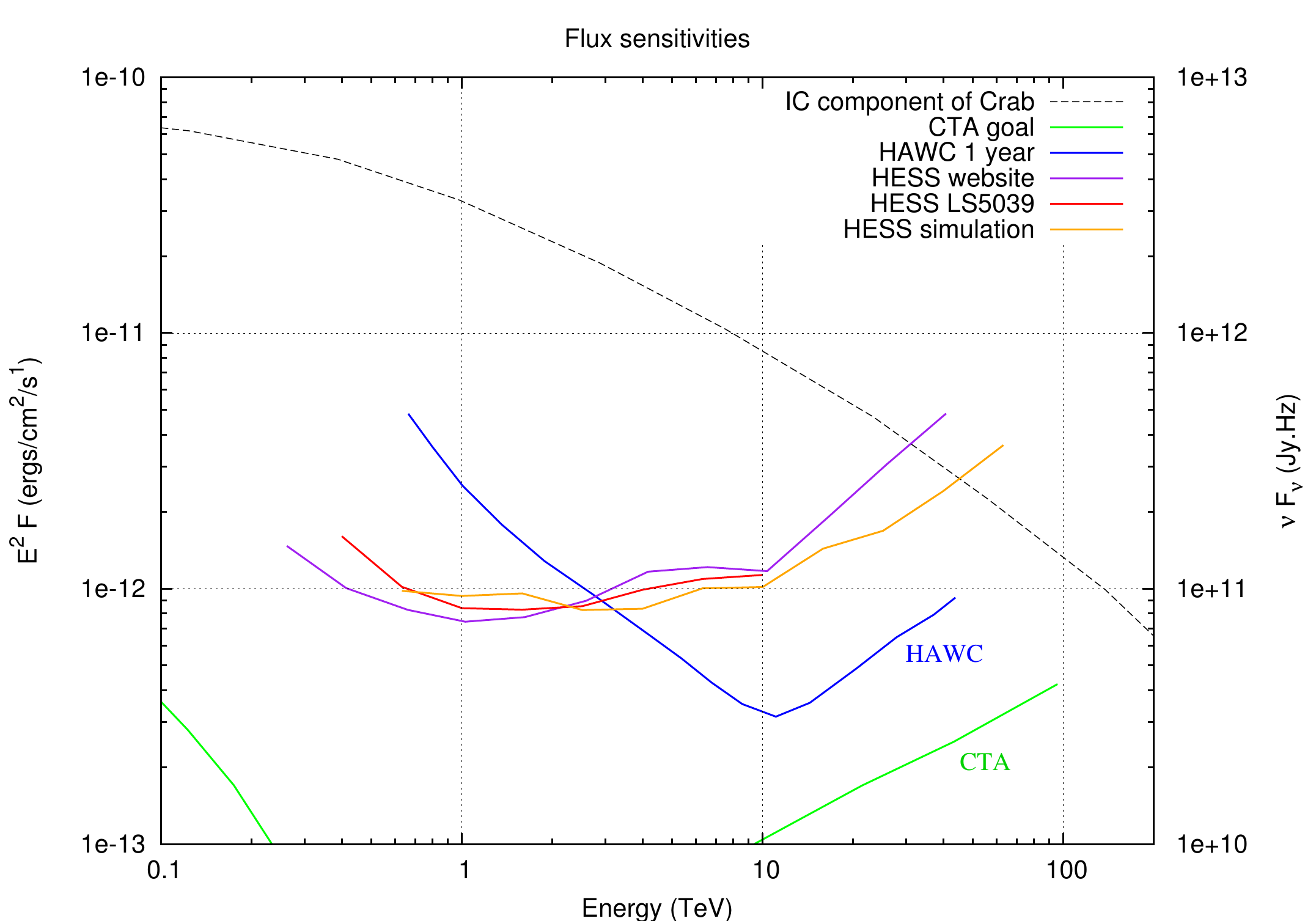}
\captionsetup{width=13cm}  
\caption{A comparison of the HESS flux sensitivity. The black dashed line represents the IC component of the Crab Nebula, the green line represents the CTA goal flux \cite{CTAcurve}, the blue line represents the HAWC 1 year flux sensitivity \cite{HAWC}, the purple line represent the HESS flux sensitivity from \cite{Hessflux}, the red line represent the HESS flux sensitivity from LS 5039 \cite{Aharonian6} and the orange line represent the HESS flux sensitivity curve from our simulations.}

 \label{fig:hess_sensitivity_comparison}
\end{centering}
\end{figure}

The flux sensitivity for the PeX cell has been broken down into four stages to best represent the improvements from the investigations presented in this thesis (Table~\ref{table:flux_config}). 

\begin{table}[h]
\centering
\begin{tabular}{lrrrrrrrrrr}
\hline
Parameters & Config A & Config B & Config C & Config D \\
\hline
Number of Telescopes & 5 & 5 & 5 & 5\\
Telescope Separation & 500\rm{m} & 500\rm{m} & 500\rm{m} & 500\rm{m}\\
Triggering Combination & (6\textit{pe}, 3) & (6\textit{pe}, 3) & (6\textit{pe}, 3) & (6\textit{pe}, 3) \\
Cleaning Combination & (8\textit{pe}, 4\textit{pe}) & (8\textit{pe}, 4\textit{pe}) & (6\textit{pe}, 3\textit{pe}) & (6\textit{pe}, 3\textit{pe})\\
image \textit{size} Cut & 60 \textit{pe} & 60 \textit{pe} & 60 \textit{pe} & 60 \textit{pe}\\
Altitude & 0.22\rm{km} & 0.22\rm{km} & 0.22\rm{km} & 0.22\rm{km}\\
Time Cleaning Cut & no & $\pm$ 5 \rm{ns} & $\pm$ 5 \rm{ns} & $\pm$ 5 \rm{ns}\\
Number of Pixels & 804 & 804 & 925 & 925\\
Pixel Arrangement & square & square & hexagonal & hexagonal\\
Reconstruction & Algorithm 1 & Algorithm 3 & Algorithm 3 & Algorithm 3\\
Algorithm 3 error cut & no & no & no & 0.06$^{\circ}$ \\
\hline
\end{tabular}
\captionsetup{width=13cm}  
\caption{The PeX configurations used in the flux sensitivity calculations.}
 \label{table:flux_config}
\end{table}

The results are presented in Figure~\ref{fig:final_flux_sensitivity}. Config B, config C and config D show improvement over config A but there is very little difference in the flux sensitivities of config B, config C and config D. Perhaps config D provides a slightly better flux sensitivity at low energies. These differences come from the improved parameterisation of images using the hexagonal pixel arrangement in config B. With a hexagonal arrangement the images can provide better representation of the image shape which allows for a superior parameterisation. The superior parameterisation leads to an improved Q$_{fact}$ and an improved reconstructed direction seen in Figure~\ref{fig:pixel_arrangement_hex}. The other improvement is due to applying a cut of the Algorithm 3 error value. Applying this extra cut removes events with poor reconstruction due to large error values. As well as displaying the flux for each configuration, the config D set-up has been used for very deep observations. For 200 hours of observations, the config D flux sensitivity approaches the HAWC 1 year flux sensitivity. This is a positive outcome for the PeX design study.

\begin{figure}
\begin{centering}
\includegraphics[scale=0.6]{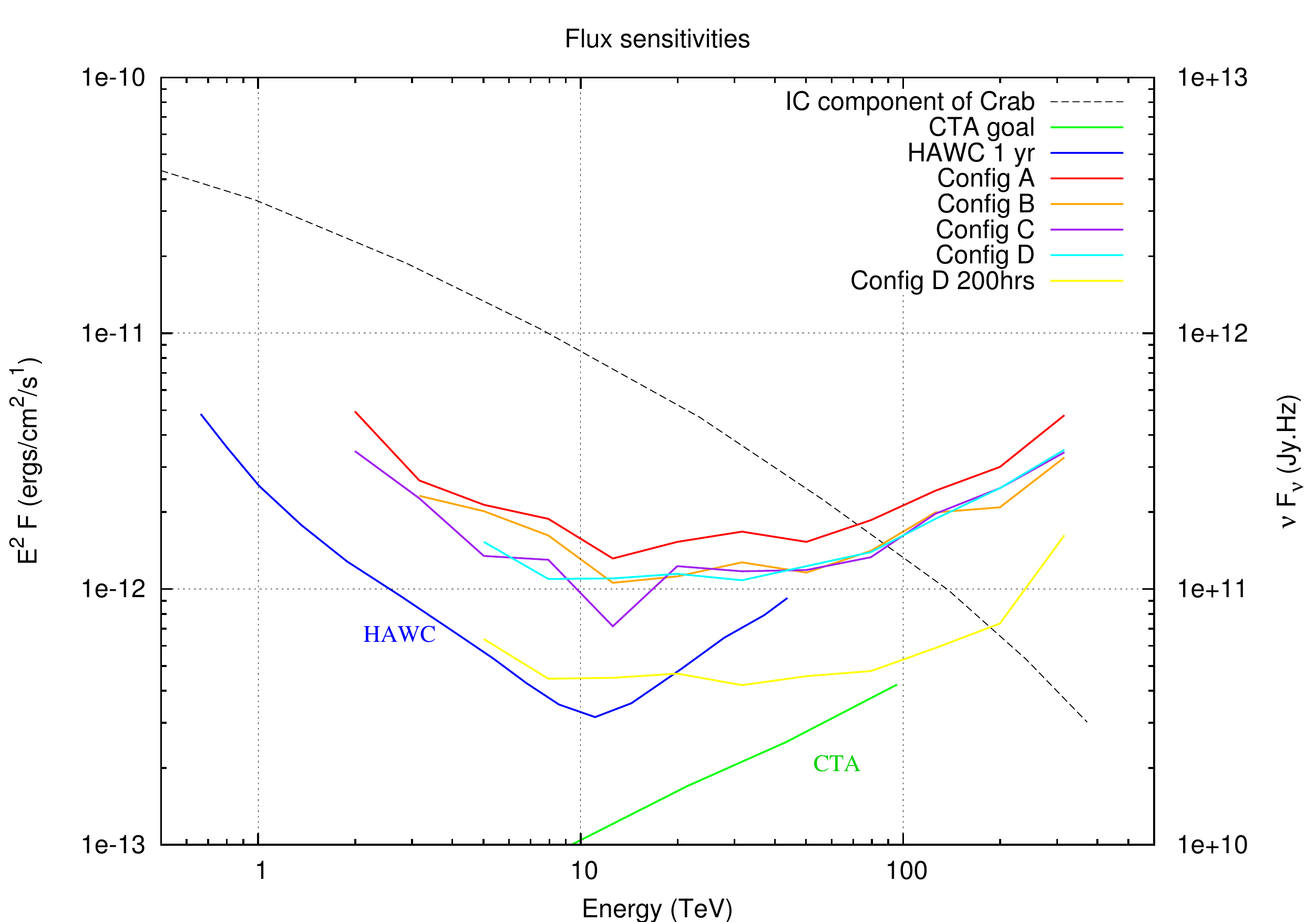}
\captionsetup{width=13cm}  
\caption{A comparison of the PeX flux sensitivity. The black line represents the IC component of the Crab Nebula, the green line represents the CTA goal flux \cite{CTAcurve}, the blue line represents the HAWC 1 year flux sensitivity \cite{HAWC}. Config A through D show the changes as improvements are made to the system.}

 \label{fig:final_flux_sensitivity}
\end{centering}
\end{figure}

We also have to consider the flux sensitivity with varying levels of NSB. Figure~\ref{fig:sensitivity_nsb} in appendix~\ref{sec:appendix_plot} shows how the flux sensitivity changes with a Galactic Centre level of NSB. The time cut from Chapter 6 has definitely improved the results since it helps bring the results for the Galactic Centre level of NSB closer to the results with an off-Galactic plane level of NSB (Figure~\ref{fig:plots_timing_nsb}). The low energy events suffer the most from a Galactic Centre level of NSB even with a time cut added to the cleaning algorithm. 

One issue that needs to be considered is the constant angular cut applied to the data set. Since the angular cut value was defined by \cite{Buckley} and based on Algorithm 1 results we can readjust its value or we could apply a angular cut that varies with energy. Using results from \cite{Buckley}, the optimum cut, for a Gaussian point-like source, is a radius of 1.26$\sigma$ where $\sigma$ is the angular r68 resolution for $\gamma$-rays. Figure~\ref{fig:effective_area_var_ang} in Appendix~\ref{sec:appendix_plot} shows the effective area for a variable angular cut across all energies. The post-selection cut effective area shows that more events are retained at lower energies up until 20 TeV since the variable angular cut is larger than the constant angular cut (Figure~\ref{fig:effective_area_var_ang} open red squares). Above 20 TeV, the post-selection cut effective area starts to decrease since the variable angular cut is smaller than the constant angular cut. Therefore, the variable angular cut allows more events to past for low energies and fewer events at high energies. After applying a variable angular cut to the proton events, we can not calculate a reliable flux sensitivity due to limited Monte Carlo statistics.

Another option is to leave the angular cut as 0.1$^{\circ}$ until the angular resolution is better than 0.1$^{\circ}$, then start applying a variable angular cut. The only parameter that will be affected by the variable angular cut is the effective area of PeX, and subsequently the flux sensitivity. No results have been displayed since it is not a feasible option for the limited Monte Carlo statistics. \\

Future work will consider extended sources and off-axis response. Some work has been done on off-axis performance for PeX \cite{Stamatescu} but further investigations can be conducted. The results from \cite{Stamatescu} indicate that with Algorithm 3, the off-axis performance of PeX is quite good for off-axis angles up to 3$^{\circ}$. A high zenith angle has not been considered, zenith $>$ 60$^{\circ}$. To complete the investigation for PeX, all zenith angles must be looked at. In this thesis only 30$^{\circ}$ zenith angles have been considered.

The PeX cell has shown improvements when the main parameters have been varied. When other cuts or alterations are applied such as a cut on Algorithm 3 error and smaller hexagonal pixels, there is further improvement to the flux sensitivity for PeX. We have seen that the individual parameters that determine the flux sensitivity show variation and improvement throughout the last few chapters, so to confirm the improvements by showing the overall flux sensitivity of the PeX cell is satisfying. Therefore, all improvements made in Chapters 4, 5 $\&$ 6 provide a better flux sensitivity for PeX, while the time cleaning cut will improve the performance in the presence NSB fluctuations. After considering many different parameters, cuts, altitudes and algorithms throughout this thesis, a final configuration for PeX can be produced which incorporates all changes made to the standard configuration presented in Chapter 3. The final configuration for the PeX cell is config D or config C if the cut on Algorithm 3 error is not desired. This configuration indicates that a new multi-TeV detector can provide adequate results that benefit this new energy regime.

\chapter{Final Remarks}

This thesis has outlined the design of a new Imaging Atmospheric Cherenkov Telescope (IACT) to be known as `Pevatron Explorer' or PeX. The initial design incorporated 5 small sized telescopes arranged each with a 6 \rm{m} mirror providing a 23.8 \rm{m$^{2}$} mirror area. Each camera consisted of 804 pixels arranged into a square grid with a pixel gap of 0.3 \rm{cm} between each 0.24$^{\circ}$ pixel. PeX will provide an 8.2$^{\circ}$ by 8.2$^{\circ}$ field of view. The desired operational energy range is a few TeV to 500 TeV, which will allow energy overlap with current IACTs. PeX may be a pathfinder for a larger array known as TenTen, with an effective area of 10\rm{km$^{2}$} at 10 TeV. The TenTen detector could consist of 30 - 50 telescopes or 6 - 10 PeX cells. With multiple PeX cells combined into one array, the sensitivity and operational capabilities will improve. Another future IACT in design is CTA, which will have multiple telescopes ranging in size to cover the largest possible energy range. We could consider the PeX cell as a sub-array of the CTA SST. The results from PeX can provide an indication of how well the full CTA SST could perform. With staged funding, the SST could be built in PeX sized sub-arrays which motivates the development and investigation of the PeX sized cell. 

Many scientific investigations can be conducted with a PeX sized cell such as; providing a new look at the Galactic plane, identifying the origin of $\gamma$-ray emission in current TeV sources, working towards uncovering the origin of Galactic cosmic ray acceleration and providing a new look at unidentified sources in a new energy range. These astrophysical motivations have provided a strong case for developing a moderately sized multi-TeV $\gamma$-ray detector such as PeX.

The work presented in this thesis has optimised parameters such as telescope separation, triggering combination, cleaning combination, image \textit{size} cut and site altitude. The optimum parameters appear to be close to the initial values that were scaled down from the H.E.S.S. parameters. The 0.22 \rm{km} altitude site provides the best results for the PeX cell compared to the 1.8 \rm{km} altitude site. 

A new time cut was introduced, which utilises the time difference between core and boundary pixels to further reduce interference from NSB. The time cut showed that it does not affect the off-Galactic plane NSB observations since the level of NSB is low compared to the signal. The time cut provides an improvement in all parameters when a Galactic plane or Galactic centre level of NSB is present. This new time cut will be permanently added to the cleaning algorithm. 

Optimisation was conducted with Algorithm 1 but the same parameters were investigated with Algorithm 3. The final conclusion suggested that Algorithm 3 provides the optimum results. The algorithm improved the reconstruction of events at large core distances, which is a big benefit to PeX. Additional studies have considered smaller pixel sizes and different pixel arrangement in the camera. The results showed that a hexagonal arrangement with a 0.24$^{\circ}$ pixel provides an improved event reconstruction. Further cuts can be applied to the PeX in the form of an Algorithm 3 cut on the error in reconstructed direction. This cut is preliminary but shows that further separation between $\gamma$-ray and proton events can occur. 

The final conclusion from this thesis is the optimised PeX configuration, which consists of: 5 telescopes, a 500 \rm{m} telescope separation, a (6\textit{pe}, 3) triggering combination, a (6\textit{pe}, 3) cleaning combination, a 60\textit{pe} size cut, a $\pm$ 5 \rm{ns} time cut, 925 pixels in a hexagonal arrangement, Algorithm 3 reconstruction, a 0.06$^{\circ}$ Algorithm 3 error cut, and a 0.22 \rm{km} altitude observation site. This final conclusion shows that a design like PeX would provide a flux sensitivity equivalent to current detectors but for energies up to 500 TeV.
\appendix

\chapter{Appendices}

\section{Hillas parameterisation}
 \label{sec:hillas_a}

The reconstruction of the shower can be broken down into different moments as follows (Figure~\ref{fig:hillasplot}):

The zeroth order moment of the light distribution, size, is the sum over all pixel intensities:
\begin{eqnarray}
S = \sum_i I_{i}
 \label{eqn:size}
\end{eqnarray}
where $I$ is the pixel amplitude and $i$ is $i$th pixel in the camera.

The first order moment provides the centre of gravity, C.O.G, of the light distribution with coordinates $<$x$>$ and $<$y$>$ in degrees:
\begin{eqnarray}
<x> = \frac{\sum_i I_{i} x_{i}}{\sum_i I_{i}} , <y> = \frac{\sum_i I_{i} x_{i}}{\sum_i I_{i}}
 \label{eqn:first_order}
\end{eqnarray}
where $x$ and $y$ are the positions of the pixels in the camera.

The second order moments $<$x$^{2}>$, $<$y$^{2}>$ and $<$xy$>$ provide the width, W, Eq~\ref{eqn:width}, and length, L, Eq~\ref{eqn:length}, of the image. The width corresponds to the RMS of the light distribution along the minor axis while the length corresponds to the RMS of the light distribution along the major axis:
\begin{eqnarray}
<x^{2}> = \frac{\sum_i I_{i} x_{i}^{2}}{\sum_i I_{i}} , <y^{2}> = \frac{\sum_i I_{i} y_{i}^{2}}{\sum_i I_{i}}
 \label{eqn:second_order}
\end{eqnarray}
\begin{eqnarray}
W = \sqrt{\frac{1}{2}tr(C) - \sqrt{\frac{1}{4}(tr(C))^{2} - Det(C)}}
 \label{eqn:width}
\end{eqnarray}
\begin{eqnarray}
L = \sqrt{\frac{1}{2}tr(C) + \sqrt{\frac{1}{4}(tr(C))^{2} - Det(C)}}
 \label{eqn:length}
\end{eqnarray}
where tr(C) is the trace of the matrix C given by the sum of the main diagonal tr(C) = $\sigma_{x}^{2} + \sigma_{y}^{2}$. 
\begin{eqnarray}
\sigma^{2}_{x} = <x^{2}> - <x>^{2}\;\;\sigma^{2}_{y} = <y^{2}> - <y>^{2}\;\;\sigma_{xy} = <xy> - <x><y>
 \label{eqn:sigmas}
\end{eqnarray}
These moments produce the covariance matrix C:


\begin{equation}
C =
\left(
\begin{array}{cc}
   \sigma_{x}^{2} & \sigma_{xy}\\
   \sigma_{xy} & \sigma_{y}^{2}\\
\end{array}
\right)
\end{equation}

The first and second order moments are combined to provide other shower information. The nominal distance, d, is the distance between the pointing direction and the C.O.G of the image:
\begin{eqnarray}
d = \sqrt{(x - <x> )^{2} + (y - <y>)^{2}}
 \label{eqn:nomial_dis}
\end{eqnarray}

The major axis represents the shower axis for the EAS. The major axis is given by the vector $\vec{u}$ 
\begin{eqnarray}
\vec{u}= (\hat{u}, \hat{v}) = (\sqrt{\frac{z - d}{2z}}, sign(\sigma_{xy})(\sqrt{\frac{z + d}{2z}})
 \label{eqn:major_axis}
\end{eqnarray}
where $d = \sigma_{y^{2}} - \sigma_{x^{2}}$ and $z =\sqrt{d^{2} + 4\sigma_{xy}}$.

The direction, $\phi$, and orientation, $\gamma$, are given by the second order moments:
\begin{eqnarray}
\phi = \arctan(\frac{W^{2} - \sigma_{x}^{2}}{\sigma_{xy}})
 \label{eqn:direction}
\end{eqnarray}
and
\begin{eqnarray}
\gamma = \arctan(\frac{<y>}{<x>})
 \label{eqn:orientation}
\end{eqnarray}
The angle $\phi$ is the angle between the major axis and the x-axis of the camera (Figure~\ref{fig:hillasplot}) while the angle $\gamma$ is the angle between the nominal distance and the x-axis of the camera (Figure~\ref{fig:hillasplot}). 

The \textit{miss} parameter provides the perpendicular distance between the major axis and the centre of the camera:
\begin{eqnarray}
miss = \sqrt{\frac{[(1+\frac{d}{z})<x>^{2} + (1-\frac{d}{z})<y>^{2}]}{2} - \frac{2<x><y>\sigma_{xy}^{2}}{z}}
 \label{eqn:miss}
\end{eqnarray}

The significance equation comes from the li and ma paper and is used as the standard significance:

\begin{eqnarray}
S=\sqrt{-2\ln{\lambda}}=\sqrt{2}\Big{[}\rm{N_{on}}\ln{\Big{[}\frac{1+\alpha}{\alpha}\Big{(}\frac{N_{on}}{N_{on}+N_{off}}\Big{)}\Big{]}}} + \rm{N_{off}}\ln{\Big{[}(1+\alpha)\Big{(}\frac{N_{off}}{N_{on}+N_{off}}\Big{)}\Big{]}\Big{]}^{1/2}
 \label{eqn:sig_li_ma}
\end{eqnarray}
where $\alpha$ is the ratio of on source time or solid angle, $t_{on}$, and off source time or solid angle, $t_{off}$, $N_{on}$ is the number of on source counts and $N_{off}$ is the number of off source counts. The $N_{on}$ counts consist of both counts from signal and background while the $N_{off}$ counts only consist of counts from background.





\section{Additional plots}
 \label{sec:appendix_plot}

\begin{figure}
\begin{centering}
\includegraphics[width=13cm]{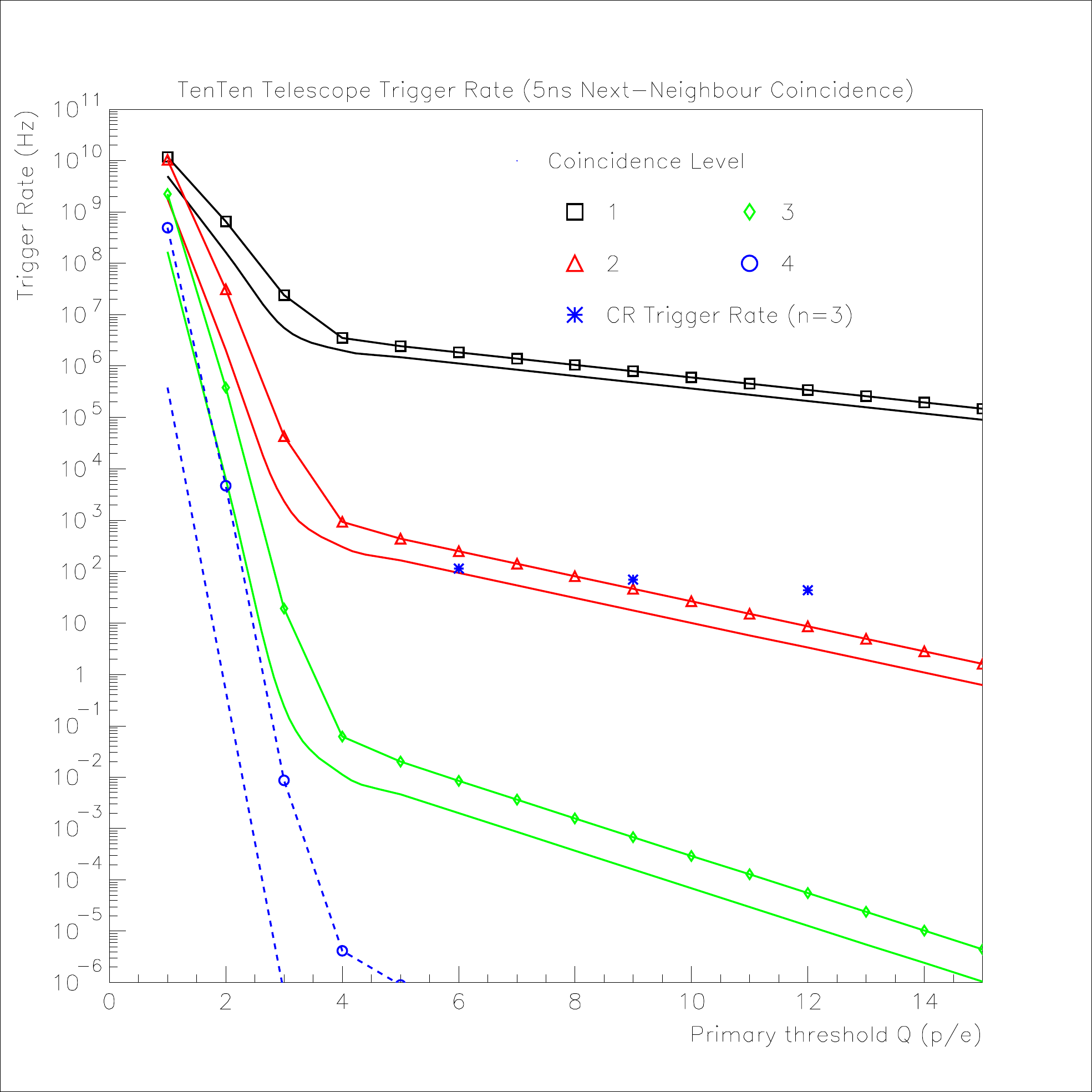}
\captionsetup{width=13cm}  
 \caption{The telescope trigger rate due to NSB at various \textit{threshold} values (Q) and \textit{n} pixel or coincidence values. The spacing between the solid lines represents the uncertainty in the camera trigger rates due to systematic errors in the NSB rate. The spacing between the dashed lines represents the same uncertainties. Taken from \cite{Ricky}.}
 \label{fig:telescope_trigger}
\end{centering}
\end{figure}

\begin{figure}
\begin{centering}
\includegraphics[width=13cm]{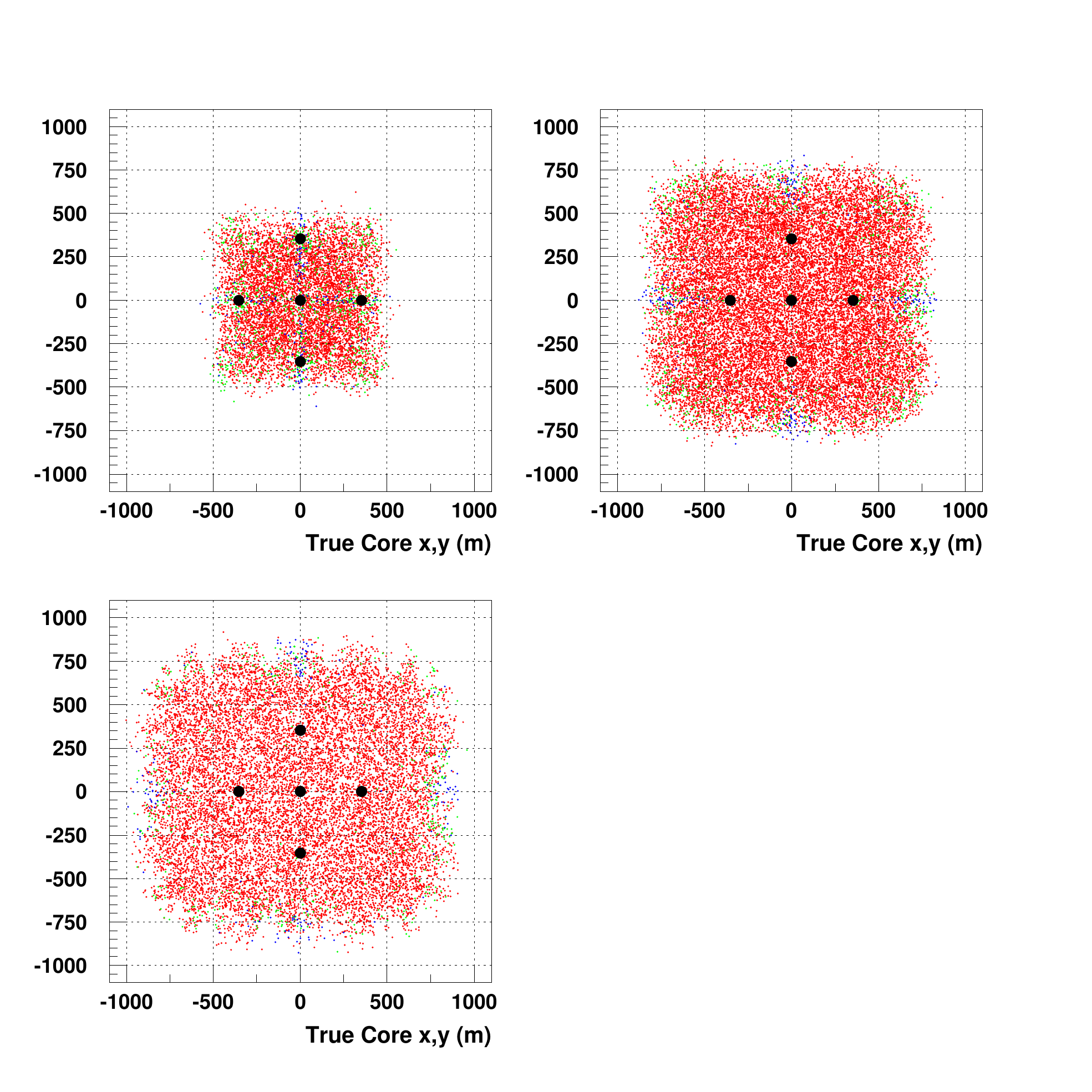}
\captionsetup{width=13cm}  
 \caption{Scatter plot for simulated true core locations for $\gamma$-ray events split into different energy bands for algorithm 3. The colours represent different angular distance bands: $\theta$ $<$ 0.2$^{\circ}$ (red), 0.2$^{\circ}$ $\leq$ $\theta$ $<$ 0.6$^{\circ}$ (green) and 0.6$^{\circ}$ $\leq$ $\theta$ (blue). The blue sections indicate the controversial regions where the event reconstruction is poor. These regions correspond to positons between adjacent telescopes or behind each telescope outside of the cell. These regions are smaller for alogirthm 3 compared to the algorithm 1 results.}
 \label{fig:core_recon2}
\end{centering}
\end{figure}




\begin{figure}
\begin{centering}
\includegraphics[scale=0.75]{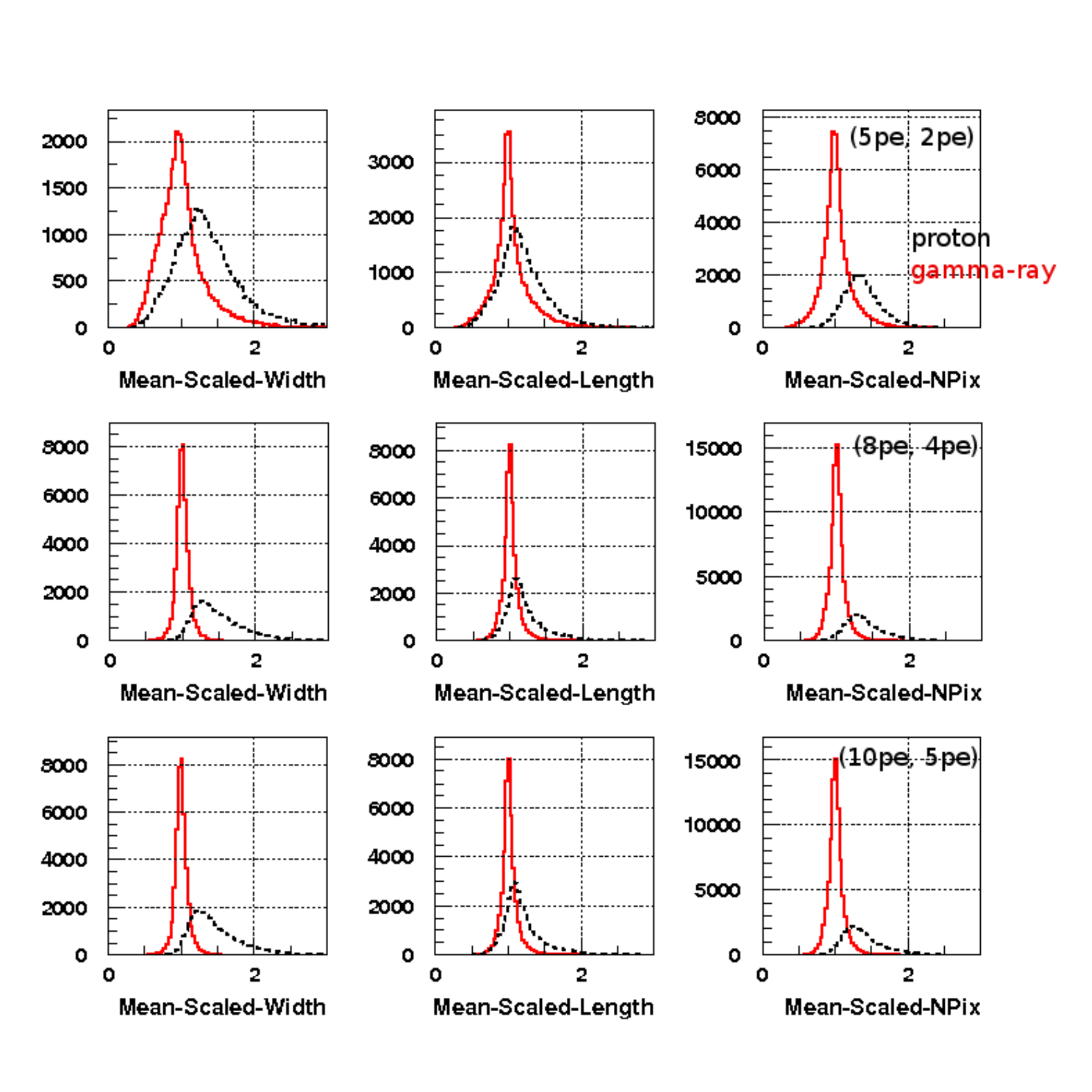}
\captionsetup{width=13cm}  
 \caption{The MSW, MSL and MSNpix distributions for $\gamma$-ray and proton events. The top panel illustrates the (5\textit{pe}, 2\textit{pe}) cleaning results, the middle panel illustrates the (8\textit{pe}, 4\textit{pe}) and the bottom panel illustrates the (10\textit{pe}, 5\textit{pe}) cleaning results. The separations between $\gamma$-ray and proton events is less distinguished for the low cleaning combination compared with the other two cleaning combinations.}
 \label{fig:rejection_cleaning}
\end{centering}
\end{figure}

\begin{figure}
\begin{centering}
\includegraphics[scale=0.6]{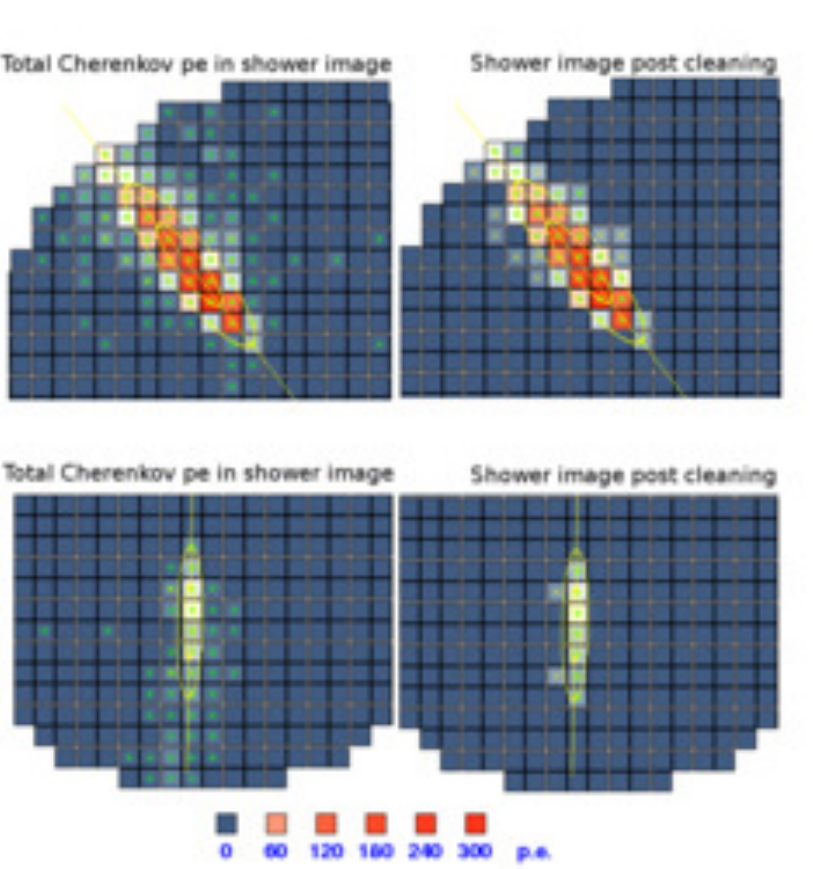}
\captionsetup{width=13cm}  
 \caption{Two $\gamma$-ray shower images from different events. The top panel illustrates a 70 TeV shower that produces a large image in the camera and the bottom panel illustrates a 20 TeV shower that produces a small image in the camera. The left panels show all the Cherenkov photons produced by the shower and the right panels show the images after they have undergone the standard cleaning combinations. The green triangles represent all the Cherenkov photons from a shower and a large fraction of the Cherenkov photons that does not pass the cleaning combination is situated after the C.O.G..}
 \label{fig:tail_of_shower}
\end{centering}
\end{figure}



















\begin{figure}
\begin{centering}
\includegraphics[width=15cm]{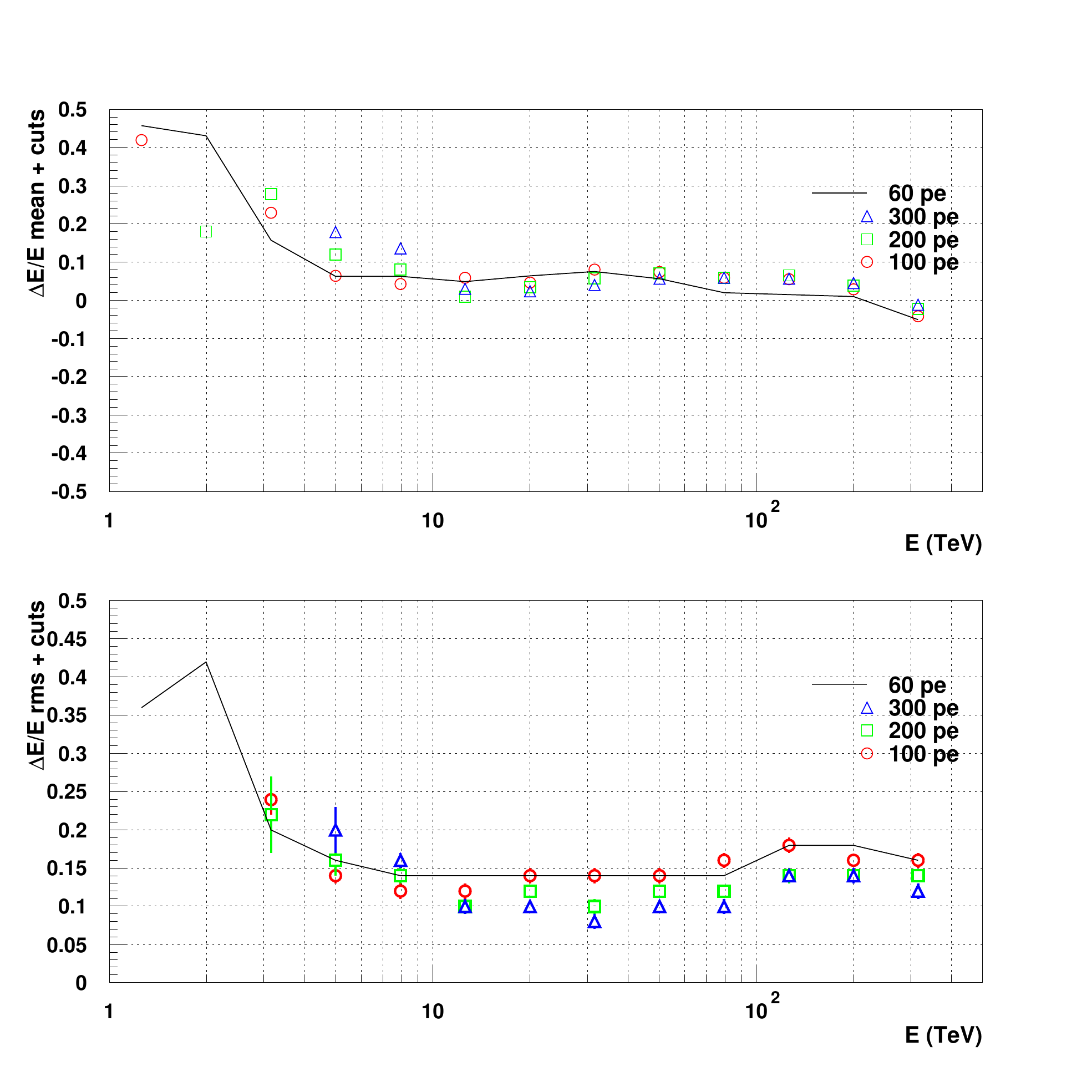}
\captionsetup{width=13cm}  
 \caption{The mean and rms of the $\Delta{E}/E$ distribution for varying image \textit{size} cuts for algorithm 1 with standard telescope separation, triggering combination and cleaning combination. Top: The mean in post-shape cut energy resolution is not affected by the varying image \textit{size} cut.
Bottom: The rms in post-shape cut energy resolution improves slightly with increasing image \textit{size} cut. The 300\textit{pe} image size cuts provides a tighter distribution for 10 TeV $<$ E $<$ 80TeV. }
 \label{fig:energy_res_size_low_alt}
\end{centering}
\end{figure}

\begin{figure}
\begin{centering}
\includegraphics[width=15cm]{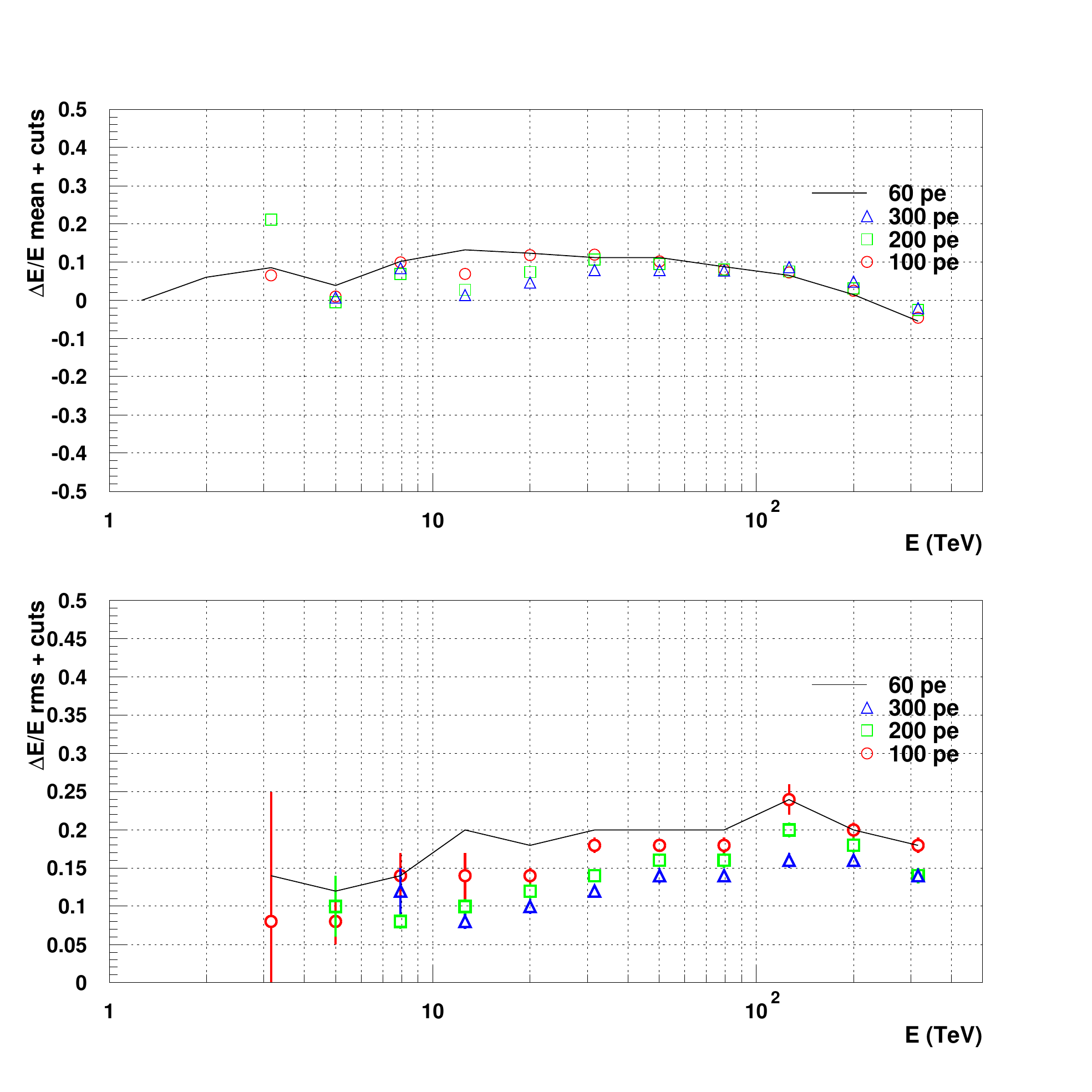}
\captionsetup{width=13cm}  
 \caption{The mean and rms of the $\Delta{E}/E$ distribution for varying image \textit{size} cuts at a 1.8 \rm{km} altitude site for algorithm 1 with standard triggering combination, cleaning combination and telescope separation. Top: The mean in the $\Delta{E}/E$ distribution.
Bottom: The rms of the $\Delta{E}/E$ distribution. The 300\textit{pe} image \textit{size} cuts provides a tighter distribution for 10 TeV $<$ E $<$ 80TeV. }
 \label{fig:energy_res_size_high_alt}
\end{centering}
\end{figure}

\begin{figure}
\begin{centering}
\includegraphics[width=15cm]{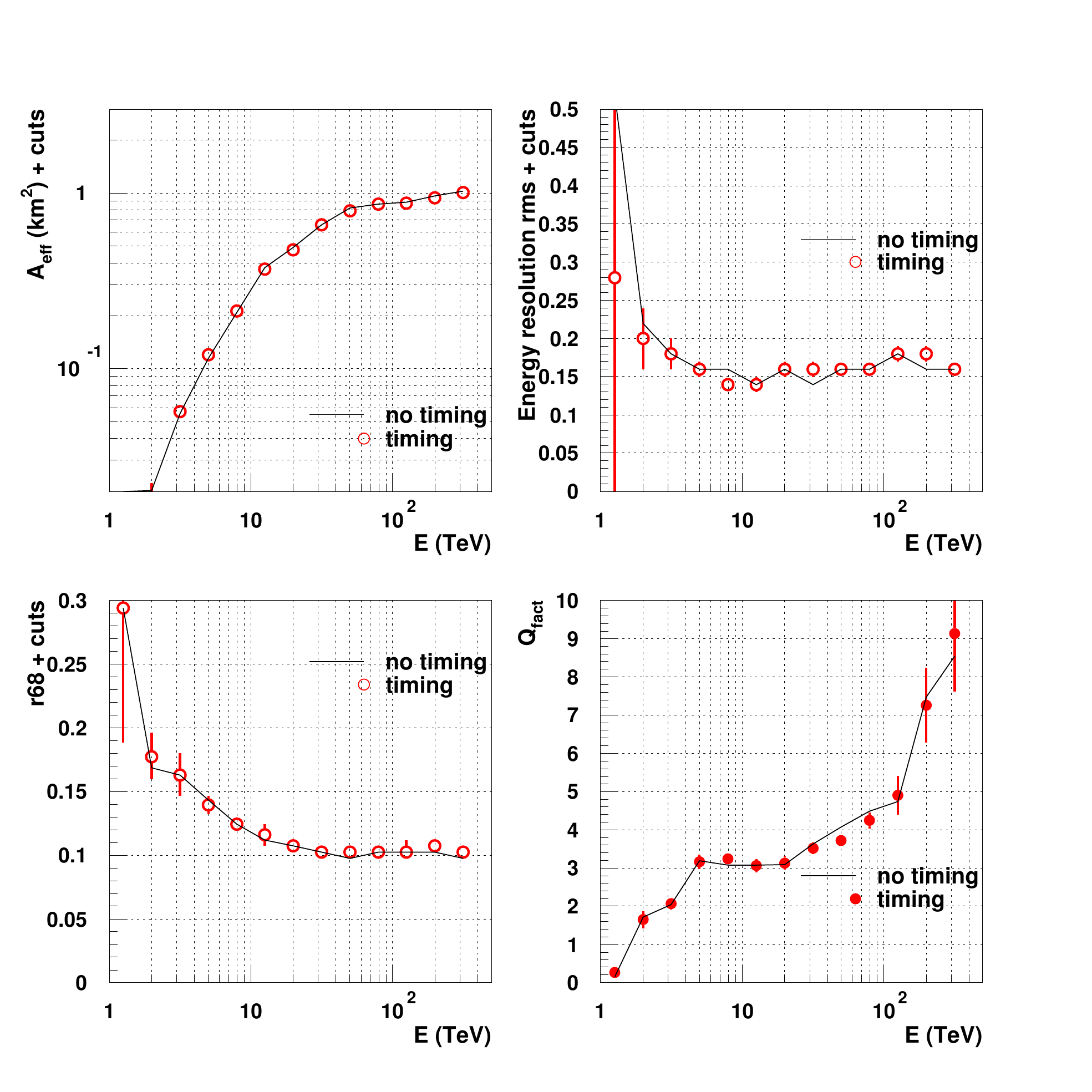}
\captionsetup{width=13cm}  
 \caption{The results for algorithm 1 with and without time cleaning for a (10\textit{pe}, 5\textit{pe}) cleaning combination and a off-Galactic plane level of NSB. The top left shows the post-selection cut effective area, the top right represents the energy resolution ($\Delta{E}/E$ rms), the bottom left provides the angular resolution (r68) and the bottom right gives the Q$_{fact}$. A quick glance at the plots indicates that applying time cleaning to the analysis code provides no improvement for an off-Galactic plane level of NSB.}
 \label{fig:norm_10_5}
\end{centering}
\end{figure}




\begin{figure}
\begin{centering}
\includegraphics[scale=0.75]{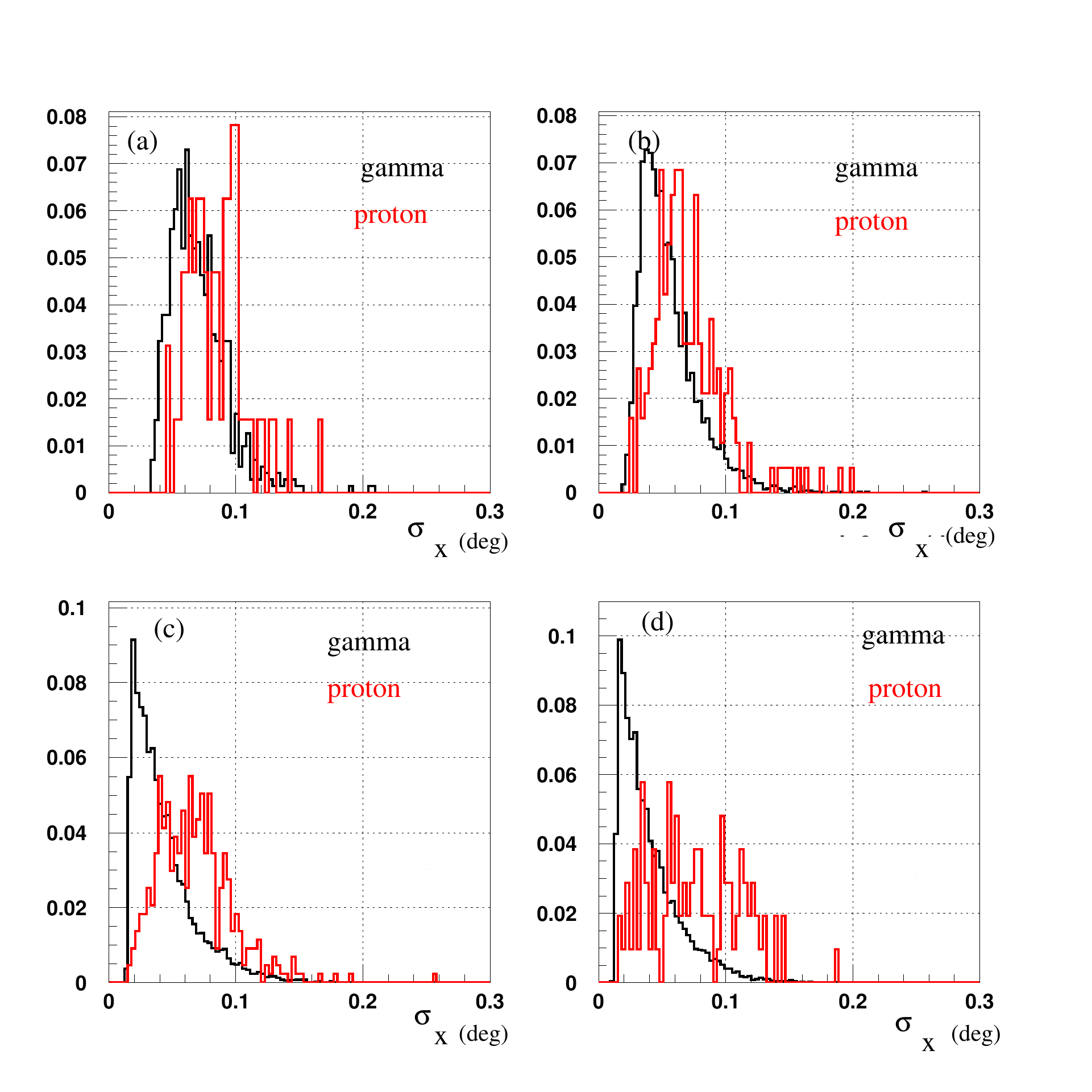}
\captionsetup{width=13cm}  
\caption{Distribution of algorithm 3 error values for all $\gamma$-ray and proton events. The results have been split into four energy bands: 1 - 4.7 TeV (a), 4.7 - 22.3 TeV (b), 22.3 - 105.7 TeV (c) and 105.7 - 500 TeV (d).}

 \label{fig:alg3_det_g_v_p_all}
\end{centering}
\end{figure}

\begin{figure}
\begin{centering}
\includegraphics[scale=0.6, angle=270]{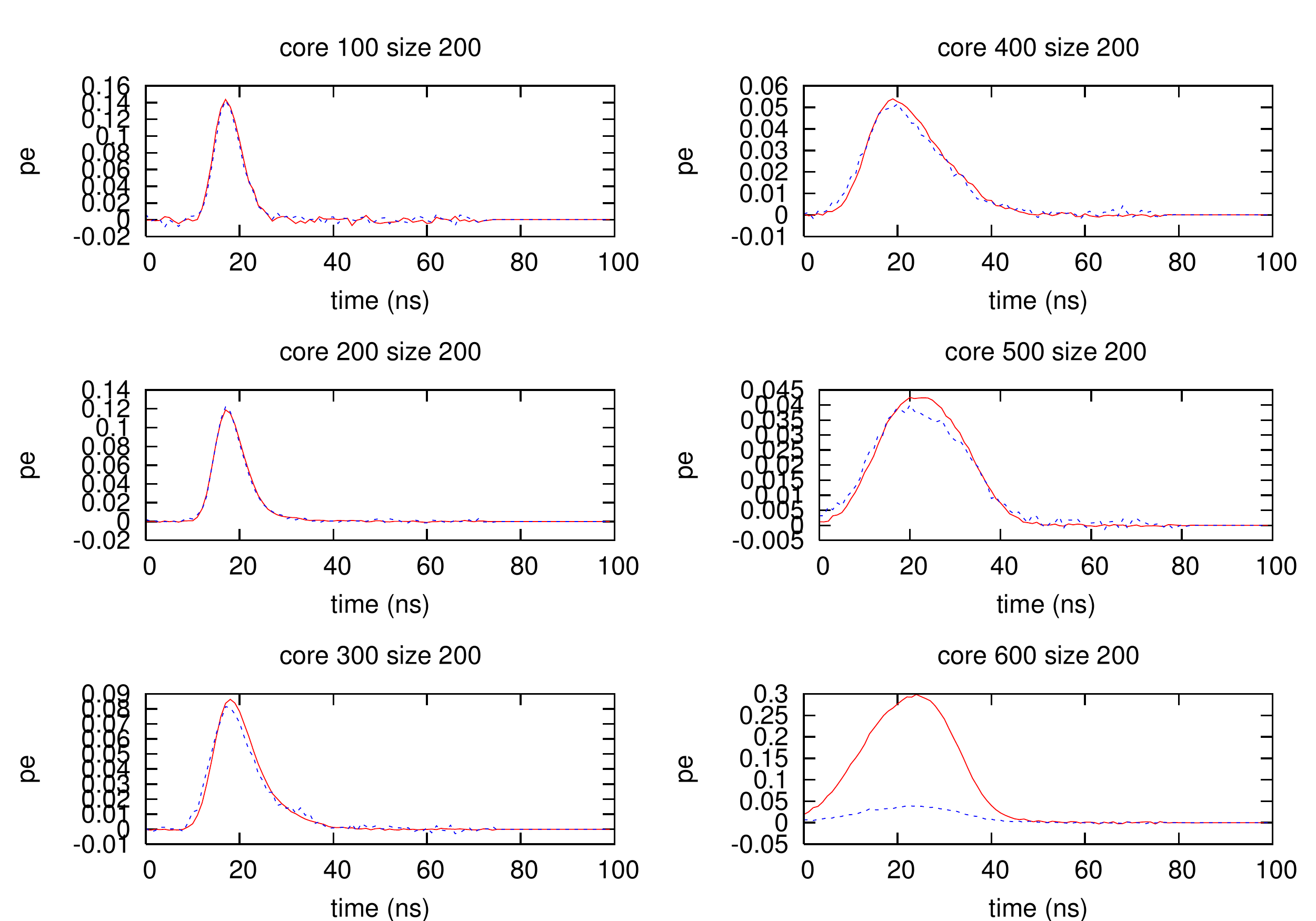}
\captionsetup{width=13cm}  
\caption{Average FADC signal for images split into core distance bands (see Section~\ref{sec:FADC}). The red line represents the $\gamma$-ray images and the blue dashed line represents the proton images. The results here are for images with a size of 200 \textit{pe}. The images have a standard level of NSB, are post cleaning and post-shape cuts. This way we can see whether any further separation can be gained. }

 \label{fig:size_200_comparison_pe_norm_cut}
\end{centering}
\end{figure}



\begin{figure}
\begin{centering}
\includegraphics[scale=0.6, angle=270]{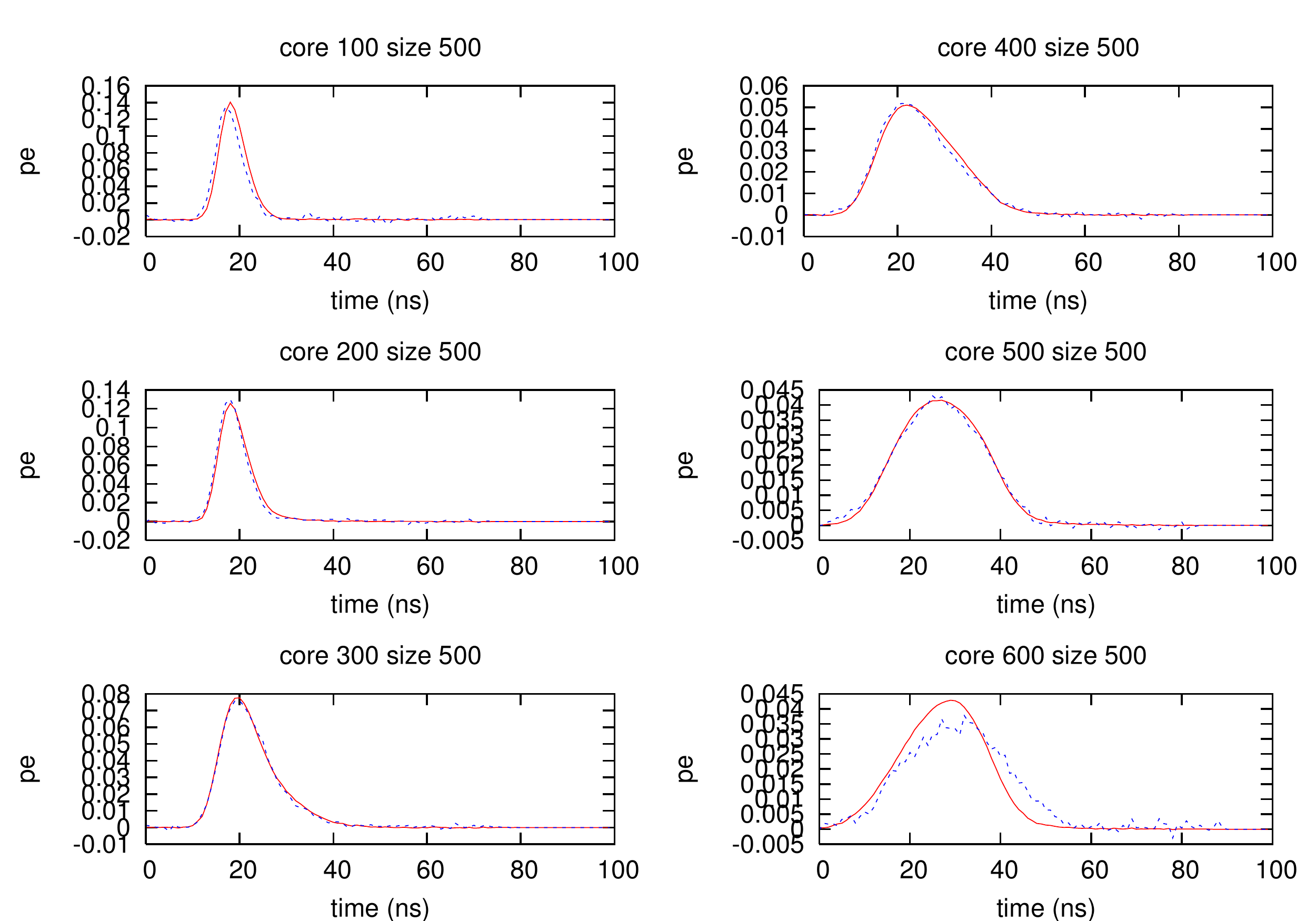}
\captionsetup{width=13cm}  
\caption{Average FADC signal for images split into core distance bands (see Section~\ref{sec:FADC}). The red line represents the $\gamma$-ray images and the blue dashed line represents the proton images. The results here are for images with a size of 500\textit{pe}. The images have a standard level of NSB, are post cleaning and post-shape cuts. This way we can see whether any further separation can be gained. }

 \label{fig:size_500_comparison_pe_norm_cut}
\end{centering}
\end{figure}



\begin{figure}
\begin{centering}
\includegraphics[scale=0.6, angle=270]{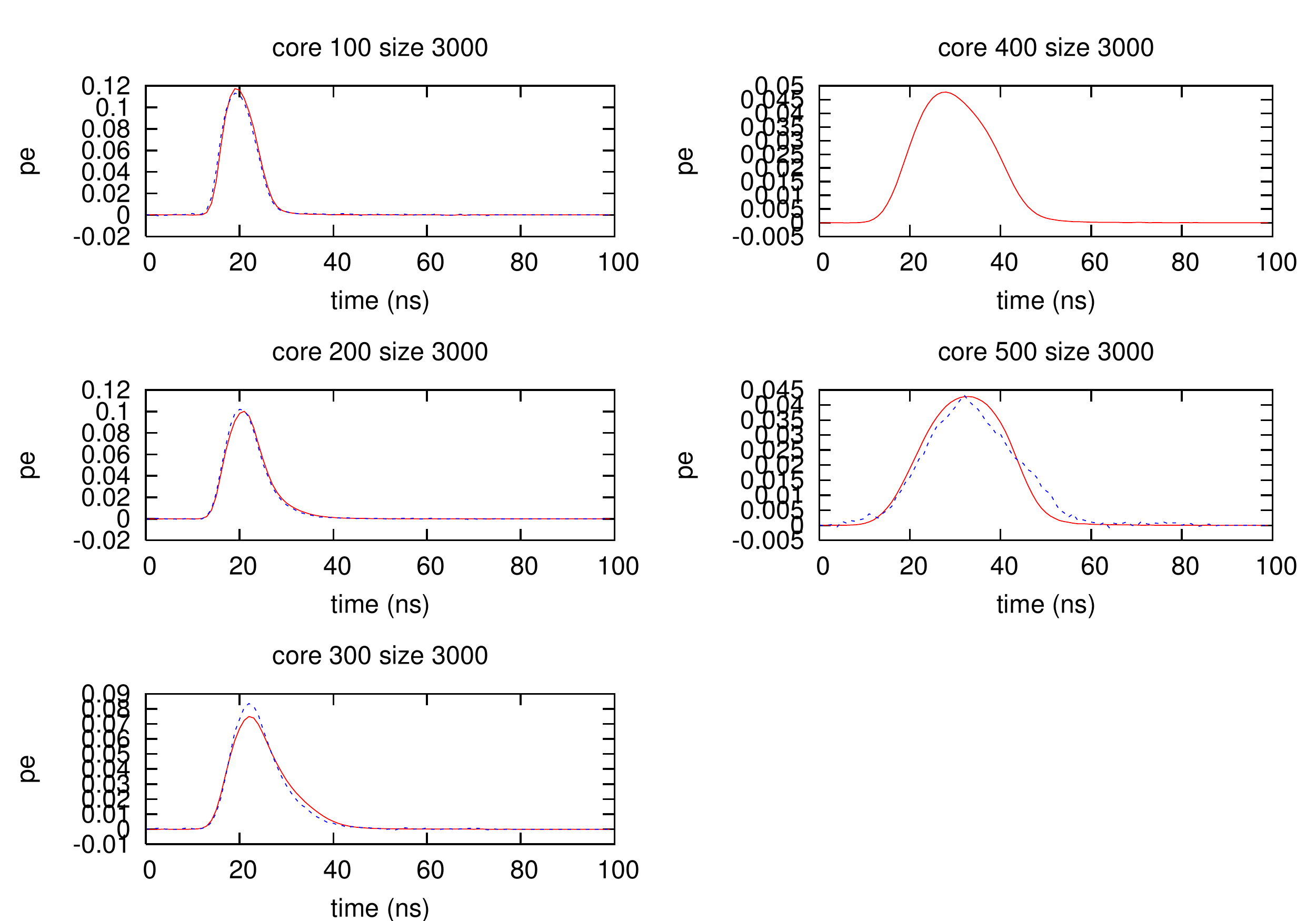}
\captionsetup{width=13cm}  
\caption{Average FADC signal for images split into core distance bands (see Section~\ref{sec:FADC}). The red line represents the $\gamma$-ray images and the blue dashed line represents the proton images. The results here are for images with a size of 3000\textit{pe}. The images have a standard level of NSB, are post cleaning and post-shape cuts. This way we can see whether any further separation can be gained. }

 \label{fig:size_3000_comparison_pe_norm_cut}
\end{centering}
\end{figure}



\begin{figure}
\begin{centering}
\includegraphics[scale=0.6, angle=270]{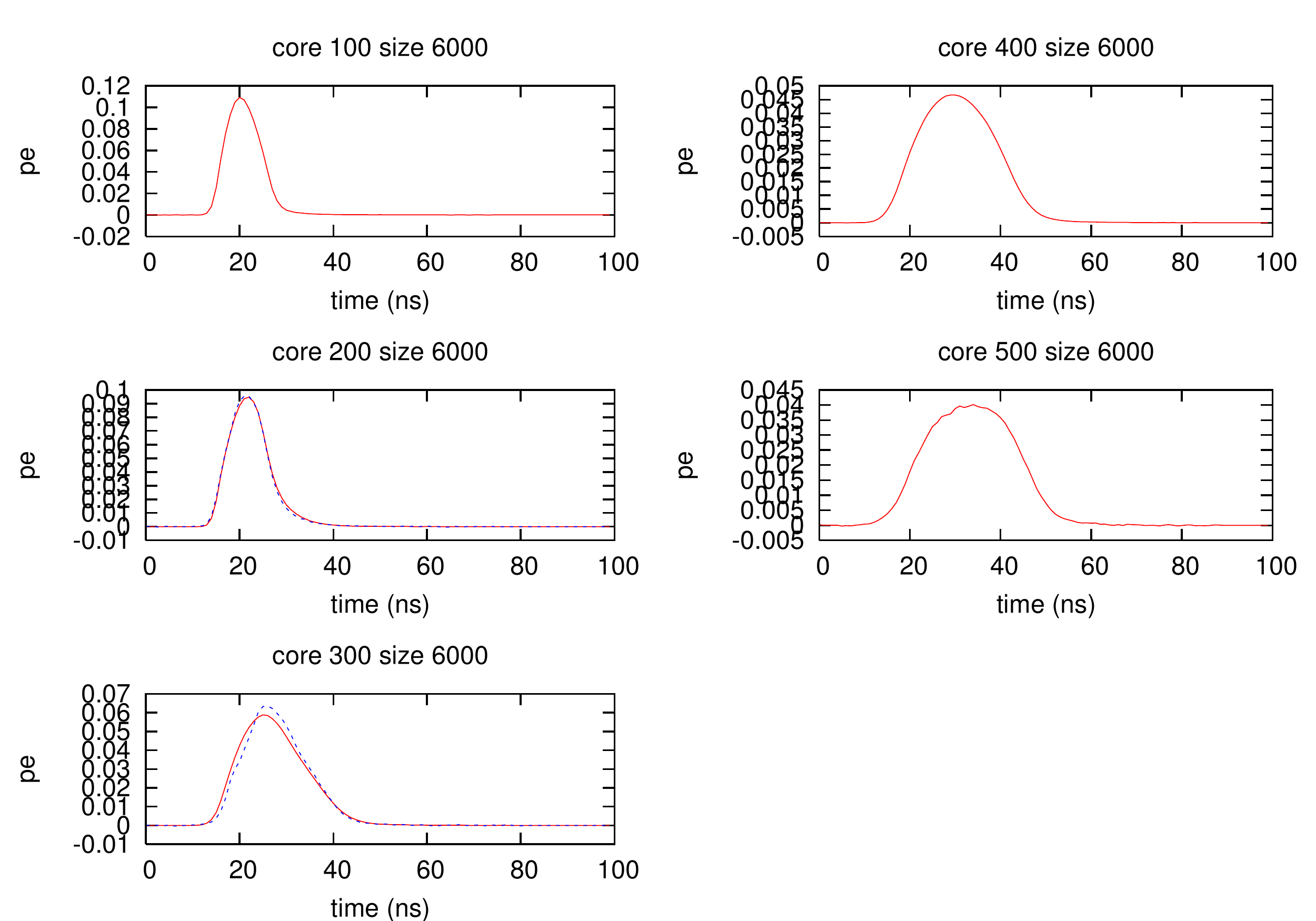}
\captionsetup{width=13cm}  
\caption{Average FADC signal for images split into core distance bands (see Section~\ref{sec:FADC}). The red line represents the $\gamma$-ray images and the blue dashed line represents the proton images. The results here are for images with a size of 6000\textit{pe}. The images have a standard level of NSB, are post cleaning and post-shape cuts. This way we can see whether any further separation can be gained. }

 \label{fig:size_6000_comparison_pe_norm_cut}
\end{centering}
\end{figure}

\begin{figure}
\begin{centering}
\includegraphics[scale=0.67]{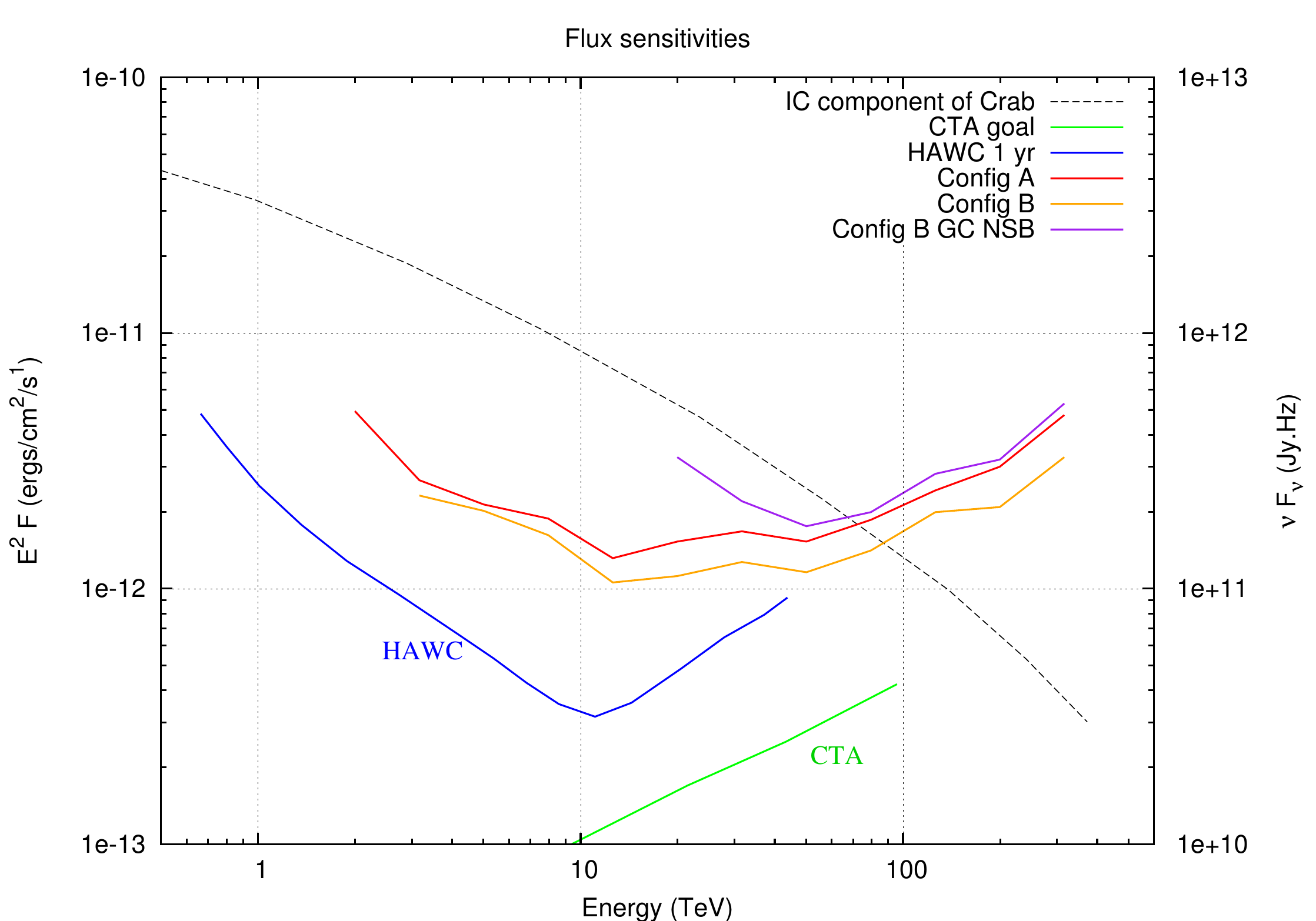}
\captionsetup{width=13cm}  
 \caption{The best PeX flux sensitivity before and after applying an extra variable angular cut over the entire energy range for algorithm 3. The red line represents the IC component of the Crab Nebula, the green line represents the CTA goal flux \cite{CTAcurve}, the blue line represents the HAWC 1 year flux sensitivity \cite{HAWC}. Config B has been used to show how a Galactic Centre level of NSB will affect the flux sensitivity.}
 \label{fig:sensitivity_nsb}
\end{centering}
\end{figure}

\begin{figure}
\begin{centering}
\includegraphics[scale=0.8]{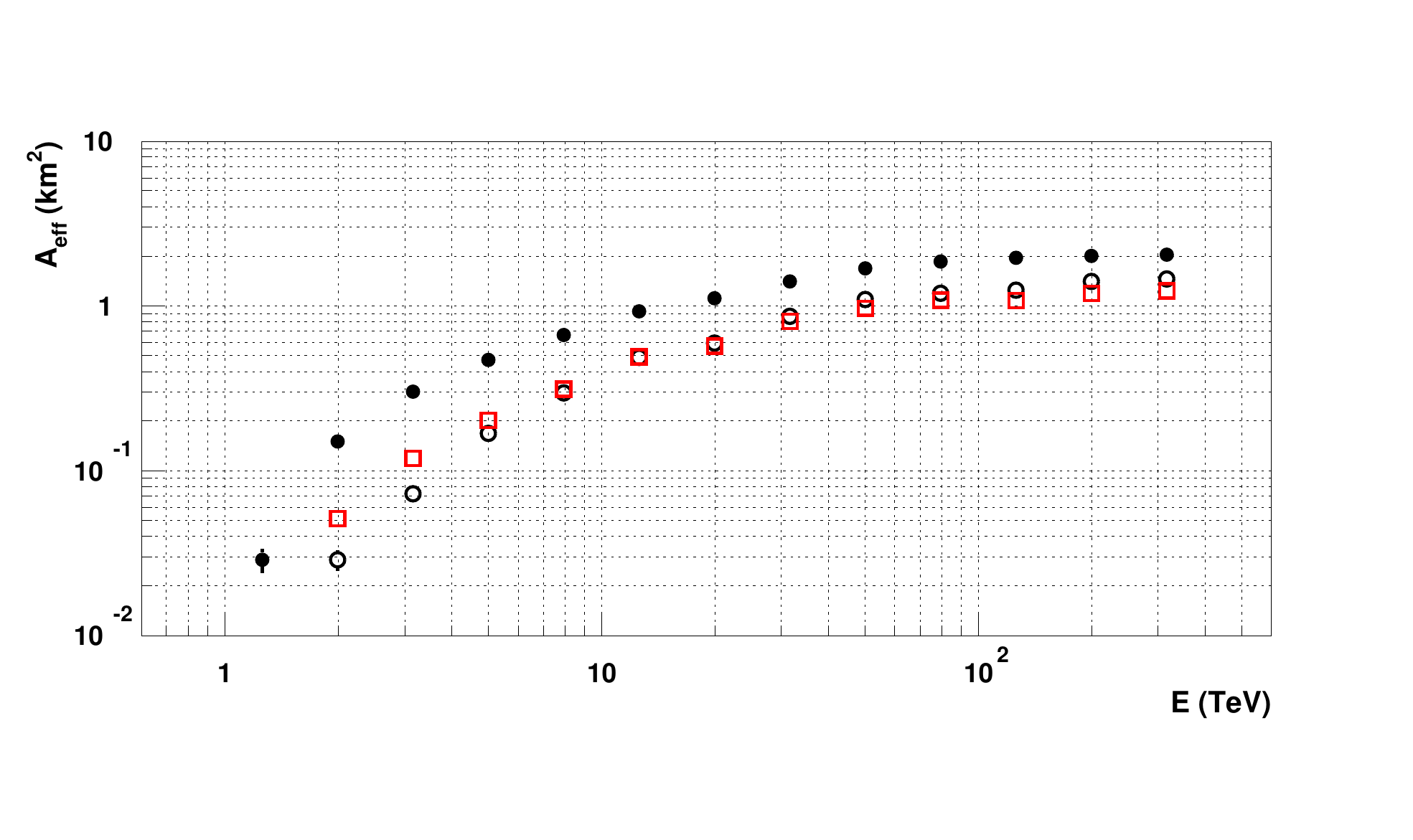}
\captionsetup{width=13cm}  
 \caption{The effective area with and without a variable angular cut. The filled black circles represent the effective area using the best PeX configuration, the open black circles represent the best PeX configuration post-selection cuts and the open red squares represent the best PeX configuration post-selection cut with a variable angular cut.}
 \label{fig:effective_area_var_ang}
\end{centering}
\end{figure}


\bibliography{references}

\begin{thebibliography}{100}

\bibitem{Pacini}
Pacini, D.,
\newblock Nuovo Cimento 1 (1912) 93.

\bibitem{HESS}
Hess, V.,
\newblock Physics 13 (1912) 1084.

\bibitem{CRtalk}
Birkl, R.,
\newblock Cosmic ray and sources lectures, 2006.

\bibitem{DSAenergy}
Shibata, M., Katayose, Y., Huang, J., and Chen, D.,
\newblock Astro. Journal 716 (2010) 1076.

\bibitem{spectrum}
Dova, M., Epele, L., and Swain, J.,
\newblock Proceedings of ICRC, 2001.

\bibitem{Auger}
The Pierre Auger Collaboration,
\newblock Physics Letters B 685 (2010) 239.

\bibitem{spectrum2}
Yoshida, S., and Dai, H.,
\newblock Astro. Journal 506 (1998) 495.

\bibitem{GZK}
Greisen, K.,
\newblock Phys. Rev. Lett. 16 (1966) 748.

\bibitem{GZK2}
{Zatsepin}, G.~T., and {Kuz'min}, V.~A.,
\newblock Soviet Journal of Experimental and Theoretical Physics Letters 4
  (1966) 78.

\bibitem{CRspectrum}
Gaisser, T.~K.,
\newblock Workshop on Energy Budget in the High Energy Universe, 2006.

\bibitem{Stamatescu}
Stamatescu, V.,
\newblock \textit{New Timing Analysis Techniques in Multi-teV Gamma-Ray
  Astronomy},
\newblock PhD thesis, The University of Adelaide, School of Chemistry and
  Physics, 2010.

\bibitem{PAO}
Pierre Auger Observatory website,
\newblock \textit{http://www.auger.org/index.html}.

\bibitem{Augeragn}
The Pierre Auger Collaboration,
\newblock Science 318 (2007) 939.

\bibitem{Hayakawa}
Hayakawa, S.,
\newblock Prog. Theo. Phys. 8 (1952) 571.

\bibitem{Morrison}
Morrison, P.,
\newblock Nuovo Cimento 7 (1958) 858.

\bibitem{longair}
Longair, M.~S.,
\newblock \textit{High Energy Astrophysics} (Cambrigde University Press, 1997).

\bibitem{pion}
Amsler, C.,
\newblock Physics Letters B667 (2008).

\bibitem{whipple}
{Weekes}, T.~C., {Cawley}, M.~F., {Fegan}, D.~J., {Gibbs}, K.~G., {Hillas},
  A.~M., {Kowk}, P.~W., {Lamb}, R.~C., {Lewis}, D.~A., {Macomb}, D., {Porter},
  N.~A., {Reynolds}, P.~T., and {Vacanti}, G.,
\newblock Astro. Journal 342 (1989) 379.

\bibitem{Sourcelist}
Wakely, S., and Horan, D., TeVCat Online Gamma-ray Catalog,
\newblock \textit{$http://tevcat.uchicago.edu$}.

\bibitem{HESSplane}
Funk, S., et al. (H.E.S.S. Collaboration),
\newblock 29th ICRC 00 (2005) 101.

\bibitem{veritasplane}
Weinstein, A., et al. (VERITAS Collaboration),
\newblock arXiv09124492  (2009) 6.

\bibitem{CANGAROOplane}
Ohishi, M., et al. (CANGAROO Collaboration),
\newblock Astro. Phys. 30 (2008).

\bibitem{Milagroplane}
Casanova, S., et al. (Milagro Collaboration),
\newblock AIP Conference Proceedings 966 (2007) 55.

\bibitem{teraelectron}
Hinton, J., and Hofmann, W.,
\newblock Astron. and Astrophys. 47 (2009) 523.

\bibitem{Lagage}
Lagage, P., and Cesarsky, C.,
\newblock Astron. and Astrophys. 125 (1983) 249.

\bibitem{Aharonian1}
Aharonian, F., et al. (H.E.S.S. Collaboration),
\newblock Astron. and Astrophys. 449 (2006) 223.

\bibitem{CANG1713}
Muraishi, H., Tanimori, T., Yanagita, S., Yoshida, T., Moriya, M., and Kifune,
  T., et al. (CANGAROO Collaboration),
\newblock Astron. and Astrophys. 354 (2000) 57.

\bibitem{Koyama}
Koyama, K.,
\newblock Advances in Space Research 25 (2000).

\bibitem{FUKUI}
Fukui, Y., et al.,
\newblock Astro. Journal 746 (2012) 18.

\bibitem{fermi-latSNR}
Abdo, A.~A., et al. (Fermi-LAT Collaboration),
\newblock ApJ 734 (2011) 28.

\bibitem{Aharonian2}
Aharonian, F., et al. (H.E.S.S. Collaboration),
\newblock Astron. and Astrophys. 460 (2006) 365.

\bibitem{Aharonian3}
Aharonian, F., et al. (H.E.S.S. Collaboration),
\newblock Astron. and Astrophys. 441 (2005) 465.

\bibitem{Aharonian4}
Aharonian, F., et al. (H.E.S.S. Collaboration),
\newblock Astron. and Astrophys. 455 (2006) 461.

\bibitem{CANGAROOmkn421}
Okumura, K., et al. (CANGAROO Collaboration),
\newblock Astro. Journal 579 (2002).

\bibitem{BlazarsCANG}
Mizumura, Y., Kushida, J., and Nishijima, K.,
\newblock Astro. Phys. 35 (2012) 563.

\bibitem{veritasIBL}
Hanna, D., et al. (VERITAS Collaboration),
\newblock J. of Phys.: Conf. series 203 (2010).

\bibitem{wcomae}
Acciari, V.~A., et al. (VERITAS Collaboration),
\newblock Astro. Journal 684 (2008).

\bibitem{3C66A}
Acciari, V.~A., et al. (VERITAS Collaboration),
\newblock Astro. Journal 693 (2009).

\bibitem{microquasar}
{de Naurois}, M., Dubus, G., Funk, S., Rowell, G., and Chadwick, P., et al.
  (H.E.S.S. Collaboration),
\newblock 29th International Cosmic Ray Conference Vol.~4, p. 101.

\bibitem{Aharonian6}
Aharonian, F., et al. (H.E.S.S. Collaboration),
\newblock Astron. and Astrophys. 460 (2006) 743.

\bibitem{LSIbinary}
Albert, J., et al. (MAGIC. Collaboration),
\newblock ApJ 684 (2008) 1351.

\bibitem{1259pulsar}
Aharonian, F., et al. (H.E.S.S. Collaboration),
\newblock Astron. and Astrophys. 442 (2005d) 1.

\bibitem{Aharonian7}
Aharonian, F., et al. (H.E.S.S. Collaboration),
\newblock Astron. and Astrophys. 467 (2007) 1075.

\bibitem{Aharonianwester1}
Aharonian, F., et al. (H.E.S.S. Collaboration),
\newblock Astron. and Astrophys. 537 (2012) 114.

\bibitem{cygnusob2}
Anchordoqui, L.~A., Goldberg, H., Moore, R.~D., Palomares-Ruiz, S., Torres,
  D.~F., and Weiler, T.~J.,
\newblock Phys. Rev. D 80 (2009) 103004.

\bibitem{Galactic_Centre}
Goldwurm, A.,
\newblock arXiv:1007.4174v1  (2010).

\bibitem{Gabici_2008}
Gabici, S.,
\newblock arXiv:0811.0836v1  (2008).

\bibitem{molecularcloud}
Aharonian, F., et al. (H.E.S.S. Collaboration),
\newblock Astron. and Astrophys. 481 (2008).

\bibitem{M82}
Acciari, V.~A., et al. (VERITAS collaboration),
\newblock Nature 462 (2009) 770.

\bibitem{Karlsson_Collaboration_2009}
Karlsson, N., et al. (VERITAS Collaboration),
\newblock 2009 Fermi Symposium Washington DC  (2009) 4.

\bibitem{NGC253}
Acera, F., et al. (H.E.S.S. Collaboration),
\newblock Science 326 (2009) 1080.

\bibitem{unidentified}
Aharonian, F., et al. (H.E.S.S. Collaboration),
\newblock Astron. and Astrophys. 499 (2009) 723.

\bibitem{MilagroUn}
{He}, H.,
\newblock 38th COSPAR Scientific Assembly Vol.~38, p. 2708, 2010.

\bibitem{attenuation}
Moskalenko, I.~V., Porter, T.~A., and Strong, A.~W.,
\newblock ApJ 640 (2006) 155.

\bibitem{Aharonian5}
Aharonian, F., et al. (H.E.S.S. Collaboration),
\newblock Astron. and Astrophys. 464 (2007) 235.

\bibitem{veritas1908}
Ward, J.~E., et al. (VERITAS Collaboration),
\newblock AIP Conference Proceedings  (2008) 301.

\bibitem{veritasstatus}
Maier, G., et al. (VERITAS Collaboration),
\newblock AIP Conference Proceedings 81 (2008) 187.

\bibitem{nuclear}
Williams, W. S.~S.,
\newblock \textit{Nuclear and Particle Physics} (Oxford Science Publications,
  1991).

\bibitem{KNenergy_spectrum}
{Schlickeiser}, R., and {Ruppel}, J.,
\newblock New Journal of Physics 12 (2010).

\bibitem{2012arXiv1202.6439K}
{Kohri}, K., {Ohira}, Y., and {Ioka}, K.,
\newblock ArXiv e-prints  (2012), 1202.6439.

\bibitem{Casanova}
Casanova, S., and Jones, D.~I.,
\newblock PASJ 62 (2010) 1127.

\bibitem{GABICI}
Gabici, S., Aharonian, F., and Casanova, S.,
\newblock Mon. Not. R. Astron. Soc. 396(3) (2009) 1.

\bibitem{Plyasheshnikov}
Plyasheshnikov, A., Aharonian, F., and V$\ddot{o}$kl, H.,
\newblock J. of Phys. 26 (2000) 183.

\bibitem{backmodel}
Berge, D., Funk, S., and Hinton, J.,
\newblock Astron. and Astrophys 466 (2007) 1219.

\bibitem{Tentenrowell}
Rowell, G., Stamatescu, V., Denman, J., Thornton, G., Smith, A., Clay, R.,
  Dawson, B., Protheroe, R., and Wild, N.,
\newblock Proceedings of the 4th International Meeting on High Energy Gamma Ray
  Astronomy Vol. 1085, pp. 813--817, AIP conference proceedings, 2008.

\bibitem{tenten2007}
Rowell, G., et al.,
\newblock 30th International Cosmic Ray Conference Vol.~3, p. 1293, 2008.

\bibitem{CTAdesign}
The CTA Consortium,
\newblock Experimental astron. 32 (2011) 193.

\bibitem{OSO}
Kraushaar, W.~L., Clark, G.~W., Garmire, G.~P., Borken, R., Higbie, P., Leong,
  C., and Thorsos, T.,
\newblock Astro. Journal 177 (1972) 341.

\bibitem{SAS}
Fichtel, C.~E., Hartman, R.~C., Kniffen, D.~A., Thomson, D.~J., and Bignami,
  G.~F.,
\newblock Astro. Journal 198 (1975) 163.

\bibitem{COS}
Bignami, G., Boella, G., Burger, J.~J., Taylor, B.~G., Keirle, P., Paul, J.~A.,
  Mayer-Hasselwander, H.~A., Pfeffermann, E., Scarsi, L., and Swanenburg,
  B.~N.,
\newblock Space Science Instrucmentation 15 (1975) 245.

\bibitem{EGRET}
Kanbach, G., Bertsch, D.~L., Favale, A., Fichtel, C.~E., Hartman, R.~C.,
  Hofstadter, R., Hughes, E.~B., Hunter, S.~D., Hughlock, B.~W., Kniffen,
  Y.~C., Lin, Y.~C., Mayer-Hasselwander, H.~A., Nolan, P.~L., Pinkau, K.,
  Rothermel, H., Schneid, E., Sommer, M., and Thompson, D.~J.,
\newblock Space Science Review 49 (1988) 69.

\bibitem{FERMIwebsite}
Fermi-LAT website,
\newblock \textit{http://www-glast.stanford.edu/}.

\bibitem{Blackett}
Blackett, P.~M.,
\newblock Phys. Abstr. 52 (1949) 4347.

\bibitem{Lidvansky}
Lidvansky, A.,
\newblock Radiation Phys. and Chem. 75 (2006) 891.

\bibitem{Jelley}
Galbraith, W., and Jelley, J.,
\newblock Nature 171 (1953) 349.

\bibitem{HEGRAsite}
HEGRA website,
\newblock \textit{$http://www.mpi-hd.mpg.de/hfm/HEGRA/HEGRA.html$}, 2004.

\bibitem{ground-based}
Aharonian, F., Buckley, J., Kifune, T., and Sinnis, G.,
\newblock Reports on Progress in Physics 71 (2008) 56.

\bibitem{Bezak}
Bezak, E.,
\newblock Honours nuclear and radiation lectures, 2006.

\bibitem{Heitler}
Heitler, W.,
\newblock \textit{Quantum Theory of Radiation} (Dover Press, 1954).

\bibitem{Berge}
Berge, D.,
\newblock \textit{A detailed study of the gamma-ray supernova remnant RX
  J1713.7-3946},
\newblock PhD thesis, Ruperto-Carola University of Heidelberg, Germany, 2006.

\bibitem{Hiller}
Hiller, R.,
\newblock \textit{Gamma-ray Astronomy} (Oxford Science Publications, 1984).

\bibitem{Stanev}
Stanev, T.,
\newblock \textit{High Energy Cosmic Rays} (Praxis, 2004).

\bibitem{Gaisser}
Gaisser, T.~K.,
\newblock \textit{Cosmic Rays and Particle Physics} (Cambridge University
  Press, 1990).

\bibitem{rowellhon}
Rowell, G.,
\newblock Private communications and honours lectures, 2010.

\bibitem{Reviewpt}
Engel, R., Heck, D., and Pierog, T.,
\newblock Annu. Rev. Nucl. Part. Sci. 61 (2011) 467.

\bibitem{Giancoli}
Giancoli, D.,
\newblock \textit{Physics for Scientists and Engineers} (Prentice Hall, 2000) .

\bibitem{FrankTamm}
Mead, C.~A.,
\newblock Phys. Rev. 110 (1958) 359.

\bibitem{hillas1996}
Hillas, A.~M.,
\newblock Space Science Reviews 75 (1996) 17.

\bibitem{MAGICwebsite}
MAGIC website,
\newblock \textit{$http://magic.mppmu.mpg.de$}.

\bibitem{CATdetector}
Barrau, A., et al. (CAT Collaboration),
\newblock Nucl. Instrum. Meth. A416 (1998) 278.

\bibitem{Hesswebsite}
H.E.S.S. website,
\newblock \textit{$http://www.mpi-hd.mpg.de/hfm/HESS$}.

\bibitem{Cangaroowebsite}
CANGAROO website,
\newblock \textit{$http://icrhp9.icrru.tokyo.ac.jp$}.

\bibitem{Veritaswebsite}
VERITAS website,
\newblock \textit{$http://veritas.sao.arizona.edu$}.

\bibitem{1994AAS...185.1604S}
Shoup, A., Barwick, S., Chumney, P., and Yodh, G.~B.,
\newblock American Astronomical Society Meeting Abstracts Vol.~26, p. 1335,
  1994.

\bibitem{HAWCconcept}
Smith, A.~J., et al. (HAWC Collaboration),
\newblock Journal of Physics: Conference Series 60 (2007) 131.

\bibitem{2010cosp382333T}
{Takita}, M.,
\newblock 38th COSPAR Scientific Assembly Vol.~38, p. 2333, 2010.

\bibitem{2007ApSS309435A}
{Amenomori}, M., et al. (Tibet AS Collaboration),
\newblock Astro. and Space Science 309 (2007) 435.

\bibitem{tibetnew}
{Amenomori}, M., et al. (Tibet AS Collaboration),
\newblock Proceedings of the 32nd ICRC Vol.~6, p. 336.

\bibitem{CORSIKA}
{Heck}, D.,
\newblock Corsika v6.204, \textit{$http://www-ik3.fzk.de/corsika$}, 2008.

\bibitem{Fletcher}
Fletcher, R.~S., Gaisser, T.~K., Lipari, P., and Stanev, T.,
\newblock Phys. Rev. D 50 (1994) 5710.

\bibitem{simtelarray}
Bernl$\dot{o}$hr, K.,
\newblock sim-telarray version 24th oct, \textit{http://www.mpi-hd.mpg.de/hfm},
  2005.

\bibitem{Modtran}
Berk, A., et al,
\newblock Modtran4 version3 user's manual, 2003.

\bibitem{Stamatescu2}
Stamatescu, V., Rowell, G., Denman, J., Clay, R., Dawson, B., Smith, A.,
  Sudholz, T., Thornton, G., and Wild, N.,
\newblock Astro. Particle 34 (2011).

\bibitem{Denman}
Denman, J., Rowell, G., Stamatescu, V., Thornton, G., Dunbar, R., Clay, R.,
  Dawson, B., Smith, A., Wild, N., and Protheroe, R.,
\newblock Proceedings of the Fourth International Meeting on High Energy
  Gamma-Ray Astronomy, p. 838, 2008.

\bibitem{andrewsmith}
Smith, A.,
\newblock Internal pex communications, 2010.

\bibitem{NSBbook}
Roach, F.~E., and Gordon, J.~L.,
\newblock \textit{The light of the night sky} (D.Reidel Publishing Company,
  1973) pp. 5--6.

\bibitem{NSBthesis}
Preu$\beta$, S.,
\newblock \textit{Photometric measurement of the night sky background light on
  La Palma and in Namibia},
\newblock PhD thesis, Max-Planck-Institut f$\ddot{u}$r Kernphysik, internal
  report, 2000.

\bibitem{NSB}
Preu$\beta$, S., Hermann, G., Hofmann, W., and Kohnle, A.,
\newblock Nuclear Instruments and Methods in Physics Research 481 (2002) 229.

\bibitem{1997APh}
{Daum}, A., et al. (HEGRA Collaboration),
\newblock Astro. Phys. 8 (1997) 1.

\bibitem{Schliesser}
Schliesser, A., and Mirzoyan, R.,
\newblock Astro. Phys. 24 (2005) 382.

\bibitem{PMTnightskybackground}
Hermann, G., K$\ddot{o}$hler, C., Kutter, T., and Hofmann, W.,
\newblock arXiv:astro-ph/9508028  (1995).

\bibitem{Ricky}
Dunbar, R.,
\newblock \textit{Honours Thesis}, The University of Adelaide .

\bibitem{HESSclean}
{Benbow}, W.,
\newblock High Energy Gamma-Ray Astronomy, edited by {F.~A.~Aharonian,
  H.~J.~V{\"o}lk, \& D.~Horns}, , AIP Conference Proceedings Vol. 745, pp.
  611--616, 2005.

\bibitem{Hillas1985}
Hillas, A.~M.,
\newblock Proceedings of 19th ICRC Vol.~3, pp. 445--448, 1985.

\bibitem{hofmann}
Hofmann, W., Jung, I., Konopelko, A., Krawczynski, H., Lampeitl, H., and
  P$\ddot{u}$hlhofer, G.,
\newblock Astro. Phys. 122 (1999) 135.

\bibitem{modelanalysis}
{de Naurois}, M.,
\newblock Towards a Network of Atmospheric Cherenkov Detectors VII, pp.
  149--151, 2005.

\bibitem{Heb}
He$\beta$, M., et al. (HEGRA Collaboration),
\newblock Astro. Phys. 11 (1999) 362.

\bibitem{MSWref}
Konopelko, A., Hemberger, M., Aharonian, F., Daum, A., Hofmann, W., Kohler, C.,
  Krawczynski, H., and Volk, H.~J., et al. (HEGRA collaboration),
\newblock Astro. Phys. 10 275.

\bibitem{Puhlhofer2003267}
P$\ddot{u}$hlhofer, G., et al. (HEGRA collaboration),
\newblock Astro. Phys. 20 (2003) 267 .

\bibitem{Buckley}
Buckley, J., et al.,
\newblock Astron. and Astrophys. 329 (1998) 639.

\bibitem{liandma}
{Li}, T.~P., and {Ma}, Y.~Q.,
\newblock {ApJ} 272 (1983) 317.

\bibitem{HESShardcut}
Rowell, G.,
\newblock Journal of Physics: Conference Series 47 (2006) 21.

\bibitem{Hampf}
Hampf, D., et al.,
\newblock Adv. in Space Reseach 48 (2011) 1017.

\bibitem{Hessflux}
H.E.S.S. Flux Sensitivity,
\newblock \textit{http://www.mpi-hd.mpg.de/hfm/HESS/pages/home/proposals}.

\bibitem{RMSW}
Benbow, W., et al. (H.E.S.S. Collaboration),
\newblock AIP Conference Proceedings 745 (2005) 611.

\bibitem{CTAcurve}
{Martinez}, M., et al.,
\newblock Flux curve
  \textit{http://www-conf.slac.stanford.edu/vhegra/CTA-SLAC$\%$202007-Manel$\%20$Martinez.pdf},
  2007.

\bibitem{HAWC}
{Dingus}, B., et al.,
\newblock Flux curve
  \textit{http://www-conf.slac.stanford.edu/vhegra/Dingus$\_$HAWC$\_$VHEGRA.pdf},
  2007.

\end{thebibliography}
\bibliographystyle{h-elsevier2}

\end{normalsize}
\end{document}